\documentclass[a4paper,12pt]{article}
\usepackage[francais]{babel}
\usepackage{amssymb}
\usepackage{latexsym}
\usepackage{amsmath}

\begin{document}
\title {\bf  The Universe Expansion and Energy Problems}
\author{{\bf\large Moukaddem Nazih}
\hspace{2mm}\vspace{3mm}\\
{\it Departement of Mathematics},\\
{\it Lebanese University}, {\it Tripoli-Lebanon}}
\date{January $2007$-March $2009$}
\maketitle \vskip -15cm \vskip 14cm \vspace{1cm}

\textbf{Abstract} In this paper we first construct a mathematical
model for the universe expansion that started up with the original
Big Bang. Next, we discuss the problematics of the mechanical and
physical laws invariance regarding the spatial frame exchanges. We
then prove the (theoretical) existence of a variable metric $g_t$,
depending on time and satisfying a simplified Einstein equation,
so that all free ordinary trajectories are geodesics. This is done
by considering the classical Galileo$-$Newtonian space and time
notions, by using generalized Newtonian principles and adding the
approved physical new ones (as covariance principle, Mach
principle, the Einstein equivalence principle \ldots) in order to
establish a new cosmological model of the dynamical universe as
being $(U(t))_{t > 0}=(B_e(O,R(t)),g_t)_{t > 0}$, where
$B_e(O,R(t))$ is the Euclidean ball of radius $R(t)$ in $\mathbb
R^3$ and $R(t) \sim t$ when \emph{t} $\gg$ 0 and $c=1$. The
cosmological metric $g_t$ is totally determined, at time $t,$ by
the mass$-$energy distribution $E_t(X)$ on $B_e(O,R(t))$. We also
study the black holes phenomenon and we prove that the total and
global cosmological energy distribution $E_t(X)$ satisfies a wave
equation whose solutions are characterized by pseudo-frequencies
depending on time and related to the spectrum of the Dirichlet
problem on the unit ball $B_e(O,1)$ for the Laplace$-$Beltrami
operator $-\Delta$. Our model is consistent in the sense that all
Newtonian and classical physical laws are valid as particular
cases in classical situations. We end this construction by
introducing, possibly, the most important feature of the
expansion$-$time$-$energy triangle that is the
temperature$-$pressure duality factor and so achieving the
construction of our real physical model of the expanding universe.
Then, we show that all basic results of modern Physics are still
valid without using neither the erroneous interpretation of the
special relativity second postulate nor the uncertainty principle.
Moreover, we give a mathematical model that explains
the matter$-$antimatter duality and classifies the fundamental particles and we conclude that there exist only two privileged fundamental forces.\\
We then show that our model results in a well posed initial value
formulation for the most general Einstein's equation and leads to
a well determined solution to this equation by using a constraint
free Hamiltonian system that reduces, according to our model, to
twelve equations relating twelve independent
unknown functions.\\
We also adapt the Einstein's general relativity theory to our
setting thus freeing it from several obstacles and constraints and
leading to the unification of general relativity with quantum
Physics and Newton - Lagrange - Hamilton's Mechanics.\\
We end this paper by determining (within the framework of our
model) the age, the size and the total energy of our universe and
proving that only the energy \emph{E}, the electromagnetic
constant $ke^2$ , the Boltzmann characteristic $K_B \hskip 0.05cm
T$ (where $T$ is the cosmic temperature) and the speed of light \emph{c} (to which we add a quantum
Statistics' constant \emph{A}) are time - independent universal
constants. The other fundamental constants (such as \emph{G},
$\overline{h}$, $K$, $\alpha$...) are indeed time - dependent and
naturally related to the previous ones proving, in that way, the
unity of the fundamental forces and that of all Physics' notions.
This essentially is done by adapting the Einstein - de Sitter
model (for the Hubble homogeneous
and isotropic Cosmology) and the Einstein - Friedmann equations to our setting.\\\\\\\\

{\bf{\large{0$\hskip 0.20cm$ Introduction, Summary and Contents}}}\\\\

\normalsize{In the beginning of the 21$^{st}$ century a crisis,
which seems to be structural, reappears inside modern Physics.
Physics seemed to have resolved, in the first half of the
preceding century, all the problems that appeared at the end of
the 19$^{th}$ century with the discovery of many phenomena and
laws that were considered as being contradictory to the classical
Galileo-Newtonian Mechanics
and Physics.\\
In our days, the modern cosmology is based on the Big Bang theory
(universe expansion) supported with multiple evidences. More
recently, our comprehension of the matter is based on the
atom-nucleus-electrons and nucleons-quarks model on one side, and
on the hadrons-leptons classification and the matter-antimatter
duality on the other side. Everything obeys a rigorous
quantification of charges, energy levels and precise laws, in
which the most important are the energy and momentum conservation
and the Pauli exclusion laws. With the Einstein's famous formula
$E =mc^2$, we better understood the equivalence of all energy
forms. The quantization and unification of energy forms were also
better understood after the discovery of the photon and the
photo-electric effect, equally by Einstein. Also, the discovery of
the quantum Mechanics and Schr\"{o}dinger's equations led to a
great progress in the comprehension of the waving nature of matter
and in the understanding of a large number of natural phenomena
without giving any precise (theoretical and
experimental) explanation.\\
On the light of these great discoveries, a part of which is due to
the quantum theory and the other part to the relativity theory,
two important questions above many others have
raised:\\

1- Is there a compatibility or a complementarity between these two
theories, which seem have contributed together in solving
classical Physics impasses?\\

2- After the definitive comprehension of the electromagnetic
phenomenon and the progress in the unification theory of the
electromagnetic force with the weak and strong two interaction
forces, can all the forces (including the gravitational one) be unified inside a global theory?\\

The answer to the first question seems to be negative. Many
unification theories run into a reality that seems to be
inexplicable. The mentioned reasons of failure in such attempts
are various, such as, our technical incapability to execute
infinitely small or infinitely large measures,  the existence of
imperceptible dimensions or the nonexistence of objective
realities that would be governed by precise laws or, in case of
such an existence, our incapability of understanding their true
nature and real functioning.\\\\

With the help of our global model, we propose resolving a large
number of open problems and removing the apparent contradictions
related to the interpretation of the new results and facts,
without contradicting the fundamental principles of classical and
modern Mechanics or Physics, as long as they are scientifically
(theoretically and experimentally) valid. Among these principles
we can mention, as examples, the energy and momentum conservation
laws, Maxwell laws, Mach-Einstein law on the equivalence between
matter, energy and space curvature, the constancy of
electromagnetic waves' speed in the absolute vacuum, the wavy
nature of the matter, the quantized nature of waves and finally,
the indissociability of the expansion process, temperature,
pressure, interaction forces and energetic equilibriums' notions.
A certain number of these principles are reconfirmed, a
posteriori, inside the framework of our
model.\\
However, our model excludes certain principles which have been
introduced and used with the only justification (half-intellectual
and half-experimental): getting rid of some apparent weaknesses in
classical Physics. In fact, one can find at the base of our model
the refutation (perfectly justified) of the erroneous
interpretation of the special relativity second postulate which
consists of supposing that the light speed does not depend on the
inertial referential that is being used in order to measure it,
without putting into question the light speed independence of the
source movement. Another fundamental aspect in our model is to
situate Schr\"{o}dinger's equations and the quantum Statistics
into their proper context and within their fair limits. They
actually consist of a sort of important approximate and predictive
approach towards the studied phenomena and the explanation of the
obtained experimental results; to this we associate the
reinspection of the
uncertainty principle.\\
Evidently the reinspection of these postulates is based on logical
reasoning and rigorous mathematics, offering at the same time a
coherent alternative, in order to explain the phenomena whose
apparent contradiction with better established physical principles
was at the base
of their adoption.\\\\

Concerning special relativity, we show that none of the
experiments, (real or imaginary) which led to the spacetime
relativistic notion, justifies the alteration of the natural
(Galileo-Newtonian) space and time relationship. All these
experiments admit coherent and simple explanations. This is the
case, for example, of the train, the two observers, the emitter
and the mirror experiment, the experiment of Michelson-Morley or
the one of the emitter in the middle of a truck with two
mirrors on both sides....\\
In addition, we show that the covariance law is totally respected
by Maxwell's equations by demonstrating that the wave equation is
transformed, in a canonical way, for all inertial referential
exchanges. This simply needs the use of a perfectly natural
derivation notion that integrates the relative referential frame
movement. In fact, a more general derivation notion serves to
demonstrate the covariance properties (or tensoriality) for
different types of moving frames and it coincides in habitual
cases with the usual derivation notion. We also demonstrate that
the arbitrary cleavage between the relativistic and non
relativistic particles leads to evident
contradictions.\\\\

Similarly, based on a hardly contestable mathematical logic, we
show that some experiments, like those of energetic particles
arriving on a screen, after having passed into two thin slightly
spaced slits, do not let us conclude that the fact of knowing
which one of the two slits the particles passed in, is by itself
sufficient for altering the real and objective physical result.
These results can be altered uniquely by technical and
circumstantial means used in order to arrive to this recognition.
We demonstrate equally that the really noticed uncertainties
anywhere in Nature are actually caused by the dynamic, complex,
and evolutive nature of natural phenomena (movement trajectories,
interactions, energetic equilibriums...) and by the imperfection
and limits of our technical means, which are essentially
circumstantial. These technical means are luckily more and more
efficient and precise; which explains the permanent progress on
the level of our understanding of the universe and the matter
structure. Schr\"{o}dinger's equations give evidently a strong
method to determine, for example, the probability of finding
particles in a given space region and to explain phenomena that
seemed to be classically unexplainable, but this does not allow us to give
quantum Statistics a theoretical or exact
status.\\
Thus, we demonstrate by taking the simple pendulum example in a
stable vertical equilibrium and the example of a ball at rest in a
box (supposed to be at rest too) that the wavy nature of Matter
does not allow us to talk neither about frequency nor about the
wave length corresponding to oscillations in the space and thus we
can not talk about their minimal energy that would not be null, in
a flagrant contradiction with Newton's principles. Consequently
the fact of invoking the uncertainty principle has no place to be.
In a similar way, the electron ground state energy in a hydrogen
atom and its associated Bohr's approximate radius are the results
of an energy equilibrium between many forms of energy and many
internal and external interaction forces and have nothing to do
with the uncertainty principle. The total and the minimal
potential energies have to be finite. The energy equilibrium is
naturally traduced by the orbital clouds about the Bohr radius. We
prove also that wavelength and frequency notions, when attributed
to pointlike material particles into movement such as electrons
inside
atoms, lead to some contradictions.\\

Likewise, we consider that the use of quantum Mechanics and
quantum Statistics methods is only justified for the analysis of
infinitesimal subatomic cases where our capacity to carry out
accurate (or even approximate) measurements with our presently
technical means are so limited to make the analysis of these
situations, within the framework of classical Mechanics, inefficient.\\
The successful use of the quantum approach in order to obtain, in
the macroscopic cases, nearly the same results as those obtained
by Lagrange and Hamilton's laws by the intermediate of
Hamilton-Jacobi equations, must only lead to a sort of a
justification (or a legitimacy) for the use of these methods in
the infinitesimal cases and does not allow us to conclude that
Nature's laws obey only the Quantum Mechanics rules. For our part,
we think that the uncertainties that are inherent into these
methods (and are in fact a legitimate consequences of them)
reflect the approximate aspect of this approach and it is not
excluded that, using other experimental or theoretical analysis
means, we can better optimize these
approximations and uncertainties.\\

We will give in the following some of the strong points of our
model. Note that, to start, our model is based upon all the
mathematical Physics laws and principles whose (theoretical and
experimental) validity is unquestionable, together with submitting
those that were partially and circumstantially admitted (in order
to resolve some unexplainable problems) to an attentive
examination. Those that have not resisted the mathematical logic
have been abandoned, with all their consequences, after
establishing the necessary justifications and the clearly more
natural alternatives. After that, the model has been constructed
on the base of some simple ideas which are far from being
simplistic. We can resume them by the expansion theory (which is
gaining ground since Hubble) and the use of a Riemannian metric,
that is variable with time and position, reflecting Mach's
principle which was retaken by Einstein: matter = curvature; to
which we add a scientific and philosophical principle that
consists in the unity and coherence of many of Nature's laws
including: the conservation laws, the covariance laws, the
equivalence laws, and the original
conflicting unity of forces.\\\\

Thus, the universe at time \emph{t} ($t > $0) consists, according
to our model, of a ball $B_e(O,R(t))$ of $\mathbb{R}^3$ (with
$R(t)\sim t$ when $t \gg $0) equipped with a Riemannian metric
$g_t(X).$ This metric reflects at every instant, by the
intermediate of its variable curvature, the energy distribution
and all its effects. This metric contracts the distances and
volumes around material agglomerations of high density level, and
especially around the black holes, which are characterized by an
extremely high energy density level. On the other hand, this
metric measures the distances with respect to our conventionally
(Euclidean) scale in a place that is almost far from all matter
influences (especially gravitational influence). All trajectories
which describe free movements (i.e. under only the action of
natural forces) in the universe would be (with respect to this
metric) geodesic as the trajectories associated to free Newtonian
movements (i.e. not submitted to exterior forces) which are
straight lines covered in a constant velocity or, in other words,
geodesics relative to our flat Euclidean metric.\\
The free fall from a reasonable distance from Earth describes a
geodesic \emph{X}(\emph{t}) for a metric $g_t$ (i.e.
$\nabla_{X^{'}(t)}^{g_t}X^{'}(t)\;=\;0$) which can be determined
easily in both cases; either we assume that the gravity is uniform
or central. This notion helps in solving numerically the $n$
bodies'
problem for example.\\

However, instead of trying to determine the metric in question by
resolving the Einstein tensorial equation, we decided to follow
another way. In fact, the dependence of this equation on a large
number of factors, in addition to time, makes the resolution
inextricable, even after all possible simplifications and
reductions. Our way is progressive, beginning by a purely
theoretical mathematical modeling of the virtual space expansion,
followed by the progressive introduction of physical realities
passing from the idealization to the regularization to the
quasi-linearization, ending up by integrating all the factors
which make our real universe
in an essentially simultaneous and non dissociable way.\\

In a first step, we can prove, using the (generalized) Newtonian
fundamental principles of Mechanics, that the creation and the
expansion of space in which lives the universe should (starting
from a certain time) be produced at a quasi-constant speed which
tends to $c$ (supposed to be $1$). We introduce then for every $t$
the distribution of matter mass $m_t(X)$ and that of generalized
potential energy $E_t(X)$ on the ball $B_e(O,R(t))$ to which we
successively associate the measure $\rho_t\;:=\;m_t(X)dX$ and
$\nu_t\;:=\;E_t(X)dX,$ and we consider the measure $\mu_t$
associated to the physical metric $g_t$ determined by the
distribution $E_t(X)$ by the following relation:
$$\mu_t = dv_{g_t} = v_t(X)dX = dX-\nu_t(X) = dX - E_t(X)dX $$
The measure $\nu_t$ measures the failure caused by the energy in
order for the volume to be Euclidian and $\mu_t$ measures the real
physical volume on the universe at time $t,$ taking into
consideration all energy manifestations. We consider then the
semi-cone of time and space:

$$C^{'} = \{(x,y,z,t) \in \mathbb{R}^4; x^2 + y^2 + z^2 \leq R^2(t); t
\geq 0\} = \bigcup_{t\geq0} B_e(O,R(t))\times \{t\}$$ which we
consider during our construction as being
$$C = \{(x,y,z,t) \in \mathbb{R}^4; x^2 + y^2 + z^2 \leq t^2; t
\geq 0\} = \bigcup_{t\geq0} B_e(O,t)\times \{t\}$$ in order to
make it easier and simpler. This, in fact, means that we assume
that the electromagnetic waves' speed were always equal to the
speed of light in the absolute vacuum (i.e. $c=1$) and that the
expansion speed were always the same as the empty geometric space
one, which is determined ,according to our model, by the
electromagnetic propagation. The general case will be discussed at
the end of this
paper.\\

The universe at time $t_0$ will be the intersection of this
semi-cone with the hyperplane of equation $t\;=\;t_0$ of $\mathbb{R}^4$
equipped with the Riemannian metric $g_{t_0}.$ Then we apply the
Stokes theorem on the semi-cone provided with the flat metric of
Minkowsky (considering the empty virtual space in which the
physical geometric space evolves with the time progress) on one
side and on the same semi-cone equipped with the metric $h_t\;=\;
dt^2\;-\;g_t$ on the other side, in order to demonstrate that the
generalized energy (covering the matter) $E$ verifies the
canonical wave equation:
$$\square E(t,X) = \frac{\partial^2}{\partial{t^2}}E(t,X) -
\Delta E(t,X) = 0 \hskip 1cm \mbox{ for } X \in
B_e(O,t)$$with$$E(t,X)|_{S_e(O,t)} = 0 \mbox { for every } \hskip
0.2cm t,$$whose solutions are pseudo-periodic functions admitting
pseudo-frequencies decreasing with time. According to
Planck-Einstein principle, we can write (along the propagation line):
$$E_\mu(t,X) = g_\mu(t)\psi(\frac{X}{t}) =
h_{\mu}(t)f_\mu(t)$$where $\psi$ and $\mu$ are respectively the
eigenfunctions and the eigenvalues associated with the Dirichlet
problem on the unit ball $B_e(O,1),$ $f_\mu(t)$ is the frequency
of the solution and $h_\mu(t)$ is a sort of
a Planck's constant.\\
Introducing the temperature factor, which is (with the pressure)
non dissociable from the universe expansion, we can prove that for
every free movement (geodesic for $g_t$) $X(t),$ the energy
$E_\mu(t,X(t))$ is a decreasing function with respect to time (via
the decreasing of cosmic temperature) and
depends on $\mu$ in a purely conventional manner.\\
Finally, we recover, in the framework of our model, the famous
equation $E = mc^2$($=m$) and we prove that, for every material
particle of initial mass (at rest) $m_0=m(0)$ circulating at a
speed $v<1,$ the total energy $E(t)$ is equal to $\gamma(t) m_0
c^2 + \frac{1}{2} \gamma_1(t) m_0 v^2$ where $\gamma(t)$ is the
Lorentz factor and $\gamma_1(t)$ is a factor that comes from the
loss in mass energy by the intermediate of radiations depending on
velocity fluctuation and temperature. This factor can be
calculated theoretically or experimentally in many ways. We show
that it decreases from 1 to 0 when the velocity goes from $0$ to
$1.$ Then we prove that our formulation concerning the energy and
the momentum of particles coincide approximately with the
relativistic formulations.\\

We can continue in this way and reexamine all the modern Physics
formulas and results (such as their validity is approved
experimentally) for which one has used either the relativistic
notion of spacetime or the quantum statistics methods or also the
uncertainty principle, in order to give them an interpretation
that is more solid and (why-not) more precise, from the moment
they do not give other than approximate results established from
experiments. This naturally requires a collective hard and
assiduous work. However, this reexamination needs the readjustment
of some notions and the reestablishment of the time dependence for
some notions and constants. We show for example that, the redshift
phenomenon is explained by the increasing, with time and distance,
of the wavelength and not by the velocity of
the wave source.\\

Moreover, it is clear that within the framework of our model we
can recover, even more precisely, all confirmed results in modern
Cosmology that are based on Hubble and Friedmann works and on the
Einstein-de Sitter model. Our model conforms with the second
statement of the cosmological principle concerning the relative
speed of galaxies but not with the first postulate; the universe
can not actually look rigorously the same for any observer on any galaxy.\\\\

We continue our study by establishing a mathematical model that
leads to a global classification of all (material and
antimaterial) fundamental particles. This classification is
achieved by using a wave equation where the Laplacian is replaced
by the Dirac operator $D$ defined by the spinorial structure
associated with the Riemannian space $(B(O,1),g_e)$ and we
conclude that there exist only two privileged fundamental forces
that are both essentially related to the original unity of the
matter-energy, the original expansion movement and the natural
unity of the universe, on one hand, and to the fundamental
antagonistic aspects of the natural forces which essentially are
related to the matter, i.e. attraction and repulsion, on the other
hand. They are the gravitational and the
electromagnetic forces.\\

In other respects we demonstrate that, within the framework of our
modeling, the solution to the most general Einstein's equation can
be obtained by means of a well posed constraint free initial value
formulation leading to a well defined maximal Cauchy development.
In the same manner, we prove that the resolution of Einstein's
equation is equivalent to the resolution of a constraint free
Hamiltonian system that reduces to twelve equations of twelve
independent unknown functions $g_{ij}$ and $\pi^{ij}$
corresponding to an initial metric $g_{t_0}$ defined on any Cauchy
surface $\Sigma_{t_0}$ (or equivalently, on the universe
\emph{B}(\emph{O},$t_0$)) at an initial time $t_0$ such that its
derivative components with respect to time ${\dot{g}}_{ij}$($t_0$)
identify to twice the extrinsec curvature components
$K_{ij}$($t_0$) of ($\Sigma_{t_0}$, $g_{t_0}$) within the space -
time manifold \emph{M} = \emph{C}(\emph{t}) provided with its
space - time lorentzian metric, $h_t = dt^2 - g_t$.\\

Finally, we notice that our model is totally consistent as it is
compatible with classical Physics in the Newtonian and
quasi-Newtonian situations where the metric $g_t$ becomes so close
to $g_e$ and the measure $\mu_t$ becomes so close to the Lebesgue
measure as soon as the $E(t,X)$ distribution becomes approximately
null in a given region of space. The metric $g_t(X)$ integrates
and explains all the real and approximate situations as well as
the singular situations (black holes) and shows that physical
reality is almost
continuous without being differentiable (except for the original singularity).\\

We end this study by readjusting the Einstein's general relativity
theory to our model. For doing that, we use both the macroscopic
model of the homogeneous isotropic cosmology (considering the
universe as a dust of galaxies), which leads to the Friedmann -
Einstein equations, together with the presently reliable
experimental values of some fundamental constants in order to
correctly determine the age, the size and the total energy of our
universe. We then use some results originated in the quantum
Statistics to show that only the energy \emph{E}, the
electromagnetic constant $k e^2$, the Boltzmann characteristic
$K_B \hskip 0.05cm T$ and the speed of light \emph{c} are
universal time - independent constants; the other fundamental
constants (the gravitational constant \emph{G}, the Planck
constant $\overline{h}$, the electromagnetic force factor $\alpha$
and the curvature parameter \emph{K}(\emph{t})) are indeed time -
dependent. This fact gives, by the way, a new viewpoint on the
quantization process showing its limits and its relative
character. Finally, we establish some relations involving all
these "constants" showing in this way the unity of all Physics
theories: Electromagnetism, general relativity, quantum Physics,
Thermodynamics and the Newton - Lagrange - Hamilton's Mechanics.
These relationships lead also to the unification of the
fundamental forces. All of our results conform with the well
confirmed classical and modern Physics' results. Nevertheless, we
note some deviations with respect to other approximate results
that have been expected in a general (and sometime hypothetical)
way without being rigourously established and which are far from
making the unanimity of the scientific community. Otherwise, our
model clearly confirms that the universe laws and the expansion
process are well governed by the (slightly reviewed) Einstein's
general relativity
theory.\\\\

At the end, we think that, for understanding furthermore our
universe, we should combine the theory with practices, Mathematics
with Physics and adding some imagination, philosophy and
confidence.\\\\
\begin{center}
{\bf\Large{Contents}}
\end{center}
\vspace{0.5cm}
$$
\begin{array}{lll}
{1.}&{\mbox{Moving frames and Isometries}......................................................}&{11}\\
{2.}&{\mbox{Mathematical modeling of the expanding universe}.......................}&{18}\\
{3.}&{\mbox{General tensoriality with respect to the frame exchanges}.............}&{30}\\
{4.}&{\mbox{Physical modeling of the expanding universe}................................}&{35}\\
{5.}&{\mbox{Matter-Energy, black holes and inertial mass}...............................}&{60}\\
{6.}&{\mbox{Energy, Pseudo-waves and Frequencies}........................................}&{76}\\
{7.}&{\mbox{Some repercussions on modern Physics}........................................}&{86}\\
{8.}&{\mbox{The limits of Quantum theory}.....................................................}&{102}\\
{9.}&{\mbox{Matter, antimatter and fundamental forces}.................................}&{130}\\
{10.}&{\mbox{The reviewed Einstein's General Relativity Theory}.....................}&{150}\\
{11.}&{\mbox{Introduction to a reviewed Cosmology}........................................}&{166}\\
{12.}&{\mbox{Fundamental constants of modern Physics}..................................}&{181}\\
{13.}&{\mbox{Commentaries and open issues}.....................................................}&{190}\\
\end{array}
$$
\vspace{1cm}
\section{Moving frames and Isometries}
We start this paper by noticing that the three first sections are
only devoted to establish some tensoriality properties concerning
the moving frame exchanges and to construct a (purely theoretical)
mathematical model corresponding to the space creation.This
geometrical space is filled up simultaneously by the physical real
universe whose modeling will be achieved progressively through the
remaining sections.The definitive model that is characterized by
the real physical metric $g_{t}(X)$ will be achieved in the
seven$^{th}$ section when we introduce the factor that consists on
the temperature-pressure duality.The metric introduced before
constitutes a reasonable approximation for the real metric
on finite time intervals $[t_0,t]$ for $t_0\gg1.$\\\\
We assume that there exists on $\mathbb R^3$ a family of
Riemannian metrics $g_t$ continuously differentiable with respect
to $t\in ]0,+\infty[$ (this will be the case of all mathematical
objects indexed by $t$ in what follows) and that, for fixed
$t_0\geq 0$, there exists a continuous family of isometries
$\varphi _{(t_0,t)}=:\varphi_t$ from $(\mathbb R^3,g_{t_0})$ onto
$(\mathbb R^3,g_{t})$. We suppose that $\mathbb R^3$ is provided
with a referential frame ${\cal R}_{0}(t_0)$ and we consider a
moving frame ${\cal R}(t)$ which coincides at $t=t_0$ with a frame
${\cal R}(t_0)$ having the same origin as ${\cal R}_{0}(t_0)$ and
makes, in the same time, a family of linear transformations $A_t$
with respect to ${\cal R}_{0}(t_0)$. Let $a_0(t)$ denote the curve
described by the origin of ${\cal R}(t)$ relatively to ${\cal
R}_{0}(t_0)$. A model of this situation will be given together
with many consequences in the second section of this paper.
Finally, we consider a moving punctual particle that coincides at
$t=t_0$ with the origin of ${\cal R}_{0}(t_0)$ and we suppose that
its trajectory
is determined in ${\cal R}_{0}(t_0)$ by $x_0(t)$ for $t\geq t_0$.\\
For $t_1>t_0$ and $t_0\leq t \leq t_1$, let$$y_0(t)=\varphi
_{t_1}(x_0(t)),$$ $$b_0(t)= \varphi _{t_1}(a_0(t)),$$
$$\hskip 4cm u_0(t)=\varphi _t(x_0(t)) \hskip 3.6cm ({\cal R}_0)$$$$\alpha _0(t)=\varphi _t(a_0(t)).$$
Next we denote by $x_1(t)$, $y_1(t)$, $\alpha _1(t)$ and $u_1(t)$
the new coordinates of the curves $x_0(t)$, $y_0(t)$, $\alpha
_0(t)$ and $u_0(t)$ with respect to the frame ${\cal R}(t_1)$
whose origin is $b_0(t_1)= \alpha _0(t_1)=\varphi _{t_1}(a_0(t_1))$ (fig.1). \\
Then, for $t_0 \leq t\leq t_1$, we have:
$$x_0(t)-b_0(t_1)=A_{t_1}.x_1(t),$$
$$\hskip 3.9cm y_0(t)-b_0(t_1)=A_{t_1}.y_1(t), \hskip 3cm ({\cal R}_1)$$
which yields$$y_1(t)-x_1(t)=A_{t_1}^{-1}(y_0(t)-x_0(t))$$ and
$$y_1(t_1)-x_1(t_1)=A_{t_1}^{-1}(y_0(t_1)-x_0(t_1))=
A_{t_1}^{-1}.u_0(t_1)-A_{t_1}^{-1}.x_0(t_1)$$and then\\
$$
u_1(t_1)=x_1(t_1)+ A_{t_1}^{-1}.u_0(t_1)-A_{t_1}^{-1}.x_0(t_1)$$
$$=x_1(t_1)+ A_{t_1}^{-1}.u_0(t_1)-A_{t_1}^{-1}.b_0(t_1)-x_1(t_1)$$
$$=A_{t_1}^{-1}(u_0(t_1)-b_0(t_1)),$$
\\
since (using (${\cal R}_0$) and (${\cal R}_1$)) we have:
$u_0$($t_1$) = $y_0$($t_1$),$\;\;\;$ $u_{1}$($t_{1}$) =
$y_{1}$($t_{1}$) $\;\;\;\;\;\;$ and $\;\;\;$ $x_{0}$($t_{1}$) =
$b_{0}$($t_{1}$) + $A_{t_{1}}.x_{1}$($t_{1}$).\\\\
This equality can be written
as$$u_1(t_1)-\alpha _1(t_1)=A_{t_1}^{-1}(u_0(t_1)-\alpha _0(t_1))$$ since $\alpha _1(t_1)= 0$.\\
Therefore we obtain, for $t\geq t_0$, the following formula
relating the coordinates in ${\cal R}_{0}(t_0)$ to those in ${\cal
R}(t)$:
\begin{equation}\label {r1}
u(t)-\alpha (t)=A_{t}^{-1}(u_0(t)-\alpha _0(t)).
\end {equation}
Here, $u_0(t)$ and $\alpha_0(t)$ respectively specify the
trajectories (in ${\cal R}_{0}(t_0)$) of the particle and the
origin of the moving frame ${\cal R}(t)$ into the space $\mathbb
R^3$ provided (at any time $t\geq t_0$) with the variable metric
$g_t$, i.e. into $(\mathbb R_t^3, g_t)_{t\geq t_0}$, whereas
$u(t)$ and $\alpha (t)=0$ are the coordinate vectors of the
trajectories $u_0(t)$ and $\alpha _0(t)$ in the moving frame (along $\alpha_0(t)$) ${\cal R}(t)$. Thus, $u_0(t)$ and $\alpha_0(t)$ modelize
trajectories, with respect to a fixed frame $\mathcal{R}_0(t_0)$, into an
evolving universe that is permanently provided with an evolving
curved metric $g_t$, whereas $u(t)$ modelizes the punctual
particle trajectory with respect to the moving frame $\mathcal{R}(t)$ along $\alpha_0(t)$.\\
From equation (1), we deduce:
\begin{equation} \label {r2}
u'(t)-\alpha '(t)= A_{t}^{-1}(u'_0(t)-\alpha
'_0(t))+(A_{t}^{-1})'(u_0(t)-\alpha _0(t))
\end{equation}
or (using (1) again )
$$
u'(t)=A_{t}^{-1}(u'_0(t)-\alpha '_0(t))+(A_{t}^{-1})'\circ
A_t(u(t))\hskip 3cm (2').$$In particular, if the motion of the
frame ${\cal R}(t)$ is uniform with respect to ${\cal R}_{0}(t_0)$
(i.e. $\alpha '_0(t)=\overrightarrow {V_0}$ and $A_t =A_{t_0}=:A$
for $t\geq t_0$), we get
$$u'(t)=A^{-1}(u'_0(t)-\overrightarrow {V_0})$$and if, in
addition, we have $\varphi_t=Id _{\mathbb R^3}$, we obtain
$$x'(t)=
A^{-1}(x'_0(t)-\overrightarrow {V_0})$$
and$$x''(t)=A^{-1}.x''_0(t)$$i.e.$$\Gamma (t)=A^{-1}.\Gamma
_0(t),$$where we have denoted by $x(t)$ the coordinates of
$x_0(t)$ in the moving frame ${\cal R}(t)$.\\On the other hand, if
$a_0(t)=0=\alpha _{0}(t)$ for $t\geq t_0$, then (using ($2^{'}$)
and (1))
$$u'(t)=A_{t}^{-1}.u'_0(t)+(A_{t}^{-1})'.u_0(t), \hskip 2cm (2'')$$ and if,
in addition, we take $A_t\equiv A$, then we have ${\cal
R}(t)=A{\cal R}_{0}(t_0)$ for $t\geq t_0$ and
$$u'(t)=A^{-1}.u'_0(t),$$
$$u''(t)=A^{-1}.u''_0(t),$$
and finally, for $\varphi _t=Id _{\mathbb R^3}$, we get
$$x'(t)=A^{-1}.x'_0(t)$$as well as
$$\Gamma(t) = A^{-1}.\Gamma_0(t).$$

\subsection*{A new time derivation operator}
In the general
case, let $$v_0(t)=u_0(t)-\alpha _0(t),$$
$$v(t)=u(t)-\alpha (t)=u(t).$$So we have
$$\frac {d}{dt} v_0(t)=\frac {d}{dt}u_0(t)-\frac {d}{dt}\alpha _0(t).$$
Now we put
$$\hskip 2.4cm \frac{d_1}{dt}v(t):=
\frac {d}{dt} v(t)-(A_{t}^{-1})'(u_0(t)-\alpha _0(t))
\;\;\;\;\;\;\;\;\;\;\;\;\;\;\;\;\;\;\;\;\;\;\;\;\;\;\;\;
(d_1)$$(or equivalently, using (1)),$$\frac{d_1}{dt}v(t)=\frac
{d}{dt} v(t)-(A_{t}^{-1})' \circ A_t.v(t)) .
\;\;\;\;\;\;\;\;\;\;\;\;\;\;\;\;\;\;\;\;\;\;\;(d^{'}_{1})$$Then,
 equation (2) shows that:
\begin{equation}\label {r3}
\frac{d_1}{dt}v(t)= A^{-1}_{t}.\frac
{d}{dt}v_0(t)=A^{-1}_{t}.\frac {d_1}{dt}v_0(t),
\end{equation}
since $\frac{d_1}{dt}$ coincides with $\frac{d}{dt}$ when using
the coordinate vectors with respect to $\mathcal{R}_{0}$($t_{0}$).\\
This formula points out the tensoriality of the coordinate
exchange of the speed vector defined by this derivation that takes
into account the speed of the moving frame origin and its
rotation. For the acceleration vector defined when using the
variable metric $g_t$ and this derivation, we have
$$\nabla^{g_t}_{\frac{d_1}{dt}v(t)}\frac{d_1}{dt}v(t)= \nabla ^{g_t}_{A_{t}^{-1}.\frac {d_1}{dt}v_0(t)}A_{t}^{-1}.\frac {d_1}{dt}v_0(t)
$$$$=\nabla ^{g_t}_{A_{t}^{-1}.\frac {d}{dt}v_0(t)}A_{t}^{-1}.\frac
{d}{dt}v_0(t)=\nabla_{A_{t}^{-1}.v'_0(t)}^{g_t}A_{t}^{-1}.v'_0(t).$$
If $A_t=A$ for $t\geq t_0$, we get (using ($d_1$))
$$ \hskip 4.5cm \frac
{d_1}{dt}v(t)=\frac {d}{dt}v(t)  \hskip 5.1cm (3')$$and
\begin{equation}\label{r4}
\nabla^{g_t}_{\frac{d_1}{dt}v(t)}\frac{d_1}{dt}v(t)=
\nabla^{g_t}_{\frac{d}{dt}v(t)}\frac{d}{dt}v(t)= \nabla
^{g_t}_{A^{-1}.\frac {d}{dt}v_0(t)}A^{-1}.\frac
{d}{dt}v_0(t)=\nabla ^{g_t}_{A^{-1}.v'_0(t)}A^{-1}.v'_0(t).
\end{equation}
If, in addition, we have $\alpha'_{0}(t)=0$, we obtain (using $(3)$
and $(3')$)
$$ \frac{d_1}{dt} u(t) = A^{-1}.\frac{d_1}{dt} u_0(t) =
A^{-1}.\frac{d}{dt} u_0(t)$$and
$$\nabla^{g_t}_{\frac{d_1}{dt}u(t)}\frac{d_1}{dt}u(t)=
\nabla ^{g_t}_{A^{-1}.\frac {d}{dt}u_0(t)}A^{-1}.\frac
{d}{dt}u_0(t)$$or
\begin{equation}\label {r5}
\nabla^{g_t}_{u'(t)}u'(t)= \nabla
^{g_t}_{A^{-1}.u'_0(t)}A^{-1}.u'_0(t).
\end{equation}
If now we suppose that the metrics $g_t$ are flat, then the
identities (4) and (5) yield successively:
$$v''(t)=\nabla^{g_t}_{v'(t)}v'(t)=\nabla
^{g_t}_{(A^{-1}.v_0(t))'}(A^{-1}.v_0(t))'=A^{-1}.v''_0(t)$$and

\begin{equation}\label{r6}
u''(t)=A^{-1}.u''_0(t).
\end{equation}
Finally,
for $\varphi _t=Id _{\mathbb R^3}$ and $a^{'}_0(t) \equiv
\overrightarrow{V}$ (a constant, non necessarily vanishing, vector), we get
\begin{equation}\label{r7}
x''(t)=\nabla^{g_{t_0}}_{x'(t)}x'(t) = A^{-1}.x''_0(t),
\end{equation}
i.e.$$ \Gamma (t)= A^{-1}.\Gamma _0(t).$$
We notice that the
equality (6) (resp.(7)) holds when we assume only that $A_t\equiv
A$, $g_t$ is flat for any $t$ and $\alpha _0(t)$ (resp. $a_0(t)$)
is a geodesic. Therefore, if we assume in
addition that $u_0(t)$ (resp. $x_0(t)$) is a geodesic, then $u(t)$ (resp. $x(t)$) itself is a geodesic.\\\\
More generally, let us consider two moving frames ${\cal R}_1(t)$
and ${\cal R}_2(t)$ coming from ${\cal R}_{0}(t_0)$ in the same
manner as ${\cal R}(t)$. Using the following obvious notations
$$v_0(t)=u_0(t)-\alpha _0(t),$$  $$w_0(t)=u_0(t)-\beta _0(t),$$
$$u_1(t) = v_1(t) \mbox{ (coord.of } v_0\mbox{(}t\mbox{) in } \mathcal{R}_1
\mbox{(}t\mbox{))},$$
$$u_2(t) = w_2(t) \mbox{ (coord.of } w_0\mbox{(}t\mbox{) in } \mathcal{R}_2
\mbox{(}t\mbox{))},$$

$$\frac {d_1}{dt}u_1(t)=\frac {d}{dt}u_1(t)-(A_{t}^{-1})'\circ A_t.u_1(t),$$

$$\frac {d_2}{dt}u_2(t)=\frac {d}{dt}u_2(t)-(B_{t}^{-1})'\circ B_t.u_2(t),$$
we obtain
$$\frac {d_1}{dt}u_1(t)=A_t^{-1}.\frac {d}{dt}v_0(t)$$or equivalently
$$\frac{d}{dt} v_0(t) = A_t . \frac{d_1}{dt}u_1(t))$$and
$$\frac {d_2}{dt}u_2(t)=B_t^{-1}.\frac {d}{dt}w_0(t)= B_t^{-1}.\frac {d}{dt}(u_0(t)-\beta _0(t))
 = B_t^{-1}.\frac {d}{dt}(v_0(t)+\alpha _0(t)-\beta _0(t)) $$
 $$\hskip 3 cm=B_t^{-1}.\frac {d}{dt}v_0(t) +B_t^{-1}.\frac {d}{dt}(w_0(t)-v_0(t)).$$
$$\hskip 3 cm
=B_t^{-1}\circ A_t.\frac {d_1}{dt}u_1(t)+B_t^{-1} .\frac
{d_1}{dt}(w_0(t)-v_0(t)).$$ 
So, we have

\begin{equation}\label{r8}
\frac {d_2}{dt}u_2(t)=B_t^{-1}\circ A_t.\frac
{d_1}{dt}u_1(t)+B_t^{-1}. \frac {d}{dt}(\alpha_{0}(t) -
\beta_{0}(t))
\end{equation}
For $\alpha'_0(t)=\beta'_0(t),$ we get
$$\frac {d_2}{dt}u_2(t)=B_t^{-1}\circ A_t.\frac
{d_1}{dt}u_1(t)$$and
$$\nabla ^{g_t}_{\frac{d_2}{dt}u_2(t)}\frac{d_2}{dt}u_2(t)=
 \nabla ^{g_t}_{B_t^{-1}\circ A_t.\frac {d_1}{dt}u_1(t)}B_t^{-1}\circ A_t.\frac {d_1}{dt}u_1(t).$$
If furthermore we assume that $A_t\equiv A$ and $B_t\equiv B$, we
obtain
 $$\frac {d}{dt} u_2(t)=B^{-1}\circ A.\frac {d}{dt}u_1(t)$$i.e.$$u'_2(t)=B^{-1}\circ
 A.u'_1(t)$$and
 $$\nabla ^{g_t}_{u'_2(t)}u'_2(t)=\nabla ^{g_t}_{(B^{-1}\circ A.u_1(t))'}(B^{-1}\circ A.u_1(t))'.$$
If, in addition, $g_t$ is flat, we get$$u''_2(t)=B^{-1}\circ
A.u''_1(t).$$Finally, if we assume that $A _t \equiv A$,
$B_t\equiv B$, $\varphi _t\equiv Id _{\mathbb R^3}$ and $a' _0(t)=
b'_0(t)$, then we clearly obtain
$$x'_2(t)=B^{-1}\circ A.x'_1(t)$$and
$$x''_2(t)=B^{-1}\circ A.x''_1(t)$$i.e.$$\Gamma _2(t)=B^{-1}\circ
A.\Gamma _1(t),$$where here, we have denoted by $x_1(t)$ and
$x_2(t)$ respectively the coordinates of $x_0(t)$ in the two
moving frames ${\cal R}_1(t)$ and ${\cal R}_2(t)$.\\ Notice that,
in order to obtain the equality $u''_2(t)=B^{-1}\circ A.u''_1(t)$
(resp. $ x''_2(t)=B^{-1}\circ A.x''_1(t)$), it is sufficient that
$\alpha _0(t)$ and $\beta _0(t)$ (resp. $a_0(t)$ and $b_0(t)$) be
geodesics for the flat metric. If furthermore $u_1(t)$ (resp.
$x_1(t)$) is a geodesic, then $u_2(t)$ (resp. $x_2(t)$) itself is
a
geodesic.\\\\
The above relations show the tensoriality of the acceleration
vector when exchanging two frames having the same speed vector or
having both constant speed vectors that can be obtained from
each other by means of a fixed linear transformation.\\\\
If now we assume that the transformations $A_t$ and $B_t$ are
isometries of $(\mathbb R^3,\;g_{t})$ and if $\alpha '_0(t)=\beta
'_0(t)$ for any $t$, we obtain (using (8))
$${\parallel \frac {d_2}{dt} u_2(t)\parallel}_{g_t}=
{\parallel \frac{d_1}{dt} u_1(t)\parallel}_{g_t},$$ and if
furthermore we have $A_t\equiv A$ and $B_t\equiv B$, then we have
necessarily
$$g_t=g_{t_0}$$
and
$${\parallel u'_2(t)\parallel}_{g_{t_0}}={\parallel u'_1(t)\parallel}_{g_{t_0}}$$
$$||\widetilde {\Gamma }_2(t)||_{g_{t_0}}=||\widetilde {\Gamma
}_1(t)||_{g_{t_0}}$$where, for flat metric, 
$$\widetilde {\Gamma }_2(t):=\nabla^{g_{t_0}}_{u'_2(t)}u'_2(t)=u''_2(t) \hskip 1cm \mbox{ and }
\hskip 1cm  \widetilde {\Gamma }_1(t):=\nabla^{g_{t_0}}_{u'_1(t)}u'_1(t)=u''_1(t).$$ 

Finally, if in addition we take $ \varphi _t=Id _{\mathbb R^3}$, then we get naturally
$${\parallel x'_2(t)\parallel}_{g_{t_0}}={\parallel x'_1(t)\parallel}_{g_{t_0}}={\parallel
x'_0(t)\parallel}_{g_{t_0}}$$and, for flat metric,
$${\parallel x''_2(t)\parallel}_{g_{t_0}}={\parallel x''_1(t)\parallel}_{g_{t_0}}=
{\parallel x''_0(t)\parallel}_{g_{t_0}}$$i.e.
$${\parallel{\Gamma}_{2}(t)\parallel}_{g_{t_0}}={\parallel{\Gamma}_{1}(t)\parallel}_{g_{t_0}}={\parallel{\Gamma}_{0}(t)\parallel}_{g_{t_0}}.$$\\

\section {Mathematical modeling of the expanding universe}

We consider a function $\lambda \in C^0([0,\;\infty[)$ that
satisfies:
\begin{enumerate}
\item $\lambda \in C^2(]0,\;\infty[)$, \item $\lambda(t)\not= 0$,
for $t$ $\in [0,\;\infty[$.
\end{enumerate}
Next, we consider, for $t\geq 0$, the metric $g_t$ defined on the (theoretical) virtual space $\mathbb R^3$ by:$$g_t:=\frac {1}{\lambda ^2(t)}g_e,$$and the
function from $\mathbb R^3$ onto $\mathbb R^3$  defined by:
$$\varphi _t=\lambda (t)Id_{\mathbb R^3}.$$So we have
$$^{t}\varphi _t\circ g_t\circ \varphi _t= g_e$$and then
$\varphi _t:\;(\mathbb R^3,\;g_e)\longrightarrow (\mathbb R^3,\;g_t)$ is an isometry for any $t\geq 0$.\\
Finally, we consider, in $(\mathbb R^3,\;g_e)$ provided with an
orthonormal Euclidean frame $(O,\vec{i},\vec {j},\vec{k})$, the
geodesic ${exp}_O(t \overrightarrow {V_0})$, for $\overrightarrow
{V_0} \in \mathbb R^3$, which can be considered as the $Ox$
coordinate axis parametrized by $t\longrightarrow x(t)=v_0 t$,
where $v_0= {\parallel\overrightarrow {V_0}\parallel}_{g_e}$. Let,
for $t\geq 0$, $U(t)= \varphi _t (t\overrightarrow {V_0})$,
incorrectly written as
$$ u(t)=\varphi _t (tv_0).$$Assuming that $v_0=1$, we can consider
$$u(t)=\varphi _t (t)=t \lambda(t)$$as a trajectory in (${\mathbb R}_t^3, g_t)_{t \geq 0}$ of a particle lying,
at $t=0$, on the origin $O$.\\Under these conditions, we have, for
$t > 0$:
$$u'(t)=t
\lambda'(t)+\lambda(t)$$and
$$\widetilde {\Gamma}(t):=\nabla ^{g_t}_{u'(t)}u'(t)=\frac {d}{dt}u'(t)=u''(t)=
t\lambda''(t)+2\lambda'(t).$$Let $m$ be the mass of a fundamental
material particle supposed to be, for a while, independent of time
(although the
volume depends obviously on $g_t$ and then on time).\\\\
\textbf{Remark}: We must however notice that a particle having a
non vanishing mass $m$ (even for $m\ll1$) can only move at a
Euclidean speed $v<1$ even though $v$ can be as
close to $1$ as we wish. $1$ is here the speed of light in the absolute vacuum.\\\\
Let now $F(t)=m\widetilde {\Gamma}(t)$ and $E(t)$ be a primitive
of the function $ g_t(F(t),u'(t))$ i.e.
$$\frac {dE}{dt}=E'(t)=g_t(m\widetilde {\Gamma}(t),u'(t)).$$We
then consider the differential equation
$$g_t(m\widetilde
{\Gamma}(t),u'(t))= g_e(m\Gamma(t),x'(t))\equiv 0,$$where
$\Gamma(t)=x''(t)=0.$ This equation yields
$$ \frac{1}{\lambda
^2(t)}m u''(t)u'(t)\equiv 0$$or
$$u''(t)u'(t)\equiv 0$$i.e.
$$
(t\lambda ''+2\lambda ')(t\lambda'+\lambda)=0 \hskip 3cm (E)
$$
Any solution $\lambda$ of $(E)$ satisfying the pre-requested
conditions gives, in a way, a generalization to the space $\mathbb
R^3$, provided with the variable metric $g_t$, of the fundamental
laws of classical Mechanics.\\
Let us prove that solutions of $(E)$ actually are the set of all
non zero constants. Indeed, the identity $u''(t)u'(t)=0$ gives
$\frac{d}{dt}u'^2(t)=0$ which implies $u'^2(t)=C^2$ and
$|u'(t)|=C$. Now $C$ can not be zero as
$$u'=0\Rightarrow
t\lambda'+\lambda=0\Rightarrow\frac{\lambda'}{\lambda}=\frac{-1}{t},\mbox{
for $t >0$ }$$
$$\Rightarrow ln \frac
{\lambda}{C_1}=-lnt=ln\frac{1}{t}\Rightarrow\lambda=\frac
{C_1}{t},$$which contradicts our hypothesis on the regularity at
the origin for $\lambda$. Equation (\emph{E}) is therefore
equivalent, on $[0,\;\infty[$, to
$$t\lambda''+2\lambda'=0 \hskip 3 cm (E').$$
We take a local solution $\lambda$ of $(E')$ that is not
identically equal to a non zero constant and we assume, for
instance, that $\lambda(1)=a>0$ and $\lambda'(1)=b>0$. This
solution is, in fact, analytical and defined on $]0,\;\infty[$
since, for $t_0>0$, ($E^{'}$) shows that $$\lambda
'(t_0)=0\Leftrightarrow \lambda ''(t_0)=0$$and then, if such a
$t_0$ exists, the only solution of $(E')$ on $]0,\;\infty[$ is
$\lambda\equiv\lambda(t_0)=\lambda(1)$ that obviously satisfies
$\lambda'(1)=0$, which is contradictory. This also proves that
$\lambda'$ and $\lambda''$ can not vanish at any $t \in
]0,\;\infty[$. Therefore $(E')$ is equivalent on $]0,\;\infty[$ to
$$t\lambda ''=-2\lambda'\Leftrightarrow\frac{\lambda''}{\lambda'}=
\frac {-2}{t}\Leftrightarrow \int _{1}^{t}
\frac{\lambda''}{\lambda'}ds=-\int _{1}^{t}\frac {2}{s}ds$$ $$\Rightarrow ln\lambda'(t)-lnb=-2lnt=ln\frac{1}{t^2}\\
$$$$\Rightarrow \lambda'(t) =\frac{b}{t^2}\Rightarrow
\lambda(t)=\frac{-b}{t}+c$$ with $c=a+b$.\\ Since $b$ is $\not
=0$, this solution $\lambda$(\emph{t}) can not extend, not only to
a continuous function on $[0,\;\infty[$, but even to a
distribution on $[0,\;\infty[$ . We obtain the same contradiction
when we assume that $\lambda'(1)=b<0$.\\
Therefore, the only families of $g_t$ and $\varphi_t$ satisfying
the above conditions are defined by$$g_t=\frac{1}{\lambda ^2}g_e
\;\mbox{ and }\;\varphi _t=\lambda Id_{\mathbb R^3}$$ for
$t\geq 0$ and a constant $\lambda \not = 0.$\\
Choosing $\lambda=1$, we determine a specific Riemannian metric and a
privileged scale so that $g_t=g_e$ and $\varphi _t= Id_{\mathbb
R^3}$ for $t\geq 0.$ Using this natural choice, we obtain
$u'(t)=1$ for $t\geq 0$ and then $u'=H$ (Heaviside function).\\
Thus, it is shown that any Riemannian metric $g$ on $\mathbb{R}^3$
in the conformal class of the Euclidean metric $g_e$ that
satisfies the fundamental laws of Mechanics ($F =
m\widetilde{\Gamma}$ and $\frac{dE}{dt} = g(F,u^{'})$)is, up to a
positive constant, the Euclidean metric itself. Choosing this
constant equal to 1 amounts to fix a given scale for all physical
objects. This metric will characterize all regions in the physical
real universe that are assumed of being devoid of matter as well
as of all its effects. Matter modifies the Euclidean distances and
volumes and creates a physical (real) metric that contracts these
latter. It will be constructed along the following sections.\\
We notice that the choice of the conformal factor $\lambda$ as
being a function of $t$ only reflects the global
homogeneity and isotropy properties of the universe and the choice
of $\lambda = 1$ corresponds to a nearly null energy density
$\rho$ into a universe of a nearly infinite Euclidean volume
within which we usually measure distances with the Euclidean
metric.\\\\

\subsection*{Consequences}
According to the preceding results, we can assimilate the spatial
universe, at any time $t_0 > 0$ (when being theoretically
considered as an empty geometrical space),to the Euclidean ball of
radius $\displaystyle R(t_0)=\int _{0}^{t_0}{\parallel
u'(r)\parallel}_{g_e}dr=t_0$, provided with the Euclidean metric
$g_e$:$$ U(t_0):=(B(O,t_0),g_e).$$Then, we can take the Euclidean
ball $B(O,1)$ equipped with the variable metric $
g_t:=t^*g_e:=t^2g_e$ as a mathematical model of the expanding
universe (reduced to a virtual geometrical space) at the time $t
> 0$:
$$U_1(t):=(B(O,1),t^2 g_e).$$Here, the function
$X\longrightarrow t_0X$ is an isometry from $U_1(t_0)$ onto $U(t_0)$, for any $t_0 > 0.$\\
We notice that, in order to study a motion taking place between
time $t_1$ and time $t_2$, we can also use any of the following
spatial models:
$$\left(B(O,1),t^2g_e\right)\hskip 2 cm t_1\leq t \leq t_2,$$
$$\left(B(O,t_1),\frac {t^2}{t_1^2}g_e\right)\hskip 2 cm t_1\leq t \leq t_2,$$
$$\left(B(O,t_2),\frac {t^2_2}{t^2}g_e\right)\hskip 2 cm t_1\leq t \leq t_2.$$
However, we notice that the real physical universe at the time $t$
provided with its real curved metric is studied afterwards.\\
The properties $u=tH$, $u'=H$ and $\Gamma=u^{''}=\delta$ (Dirac
measure on ${\mathbb R^+}$) are generalized to $\mathbb R^3,$ for
$X(t)=t{\overrightarrow {V}}$ where $t\geq 0$ and
${\overrightarrow {V}}\in \mathbb R^3$ with ${\parallel
{\overrightarrow {V}}\parallel}_{g_e}=1,$ as
\begin{eqnarray*}
  &{}& t=|X(t)|:={\parallel X(t)\parallel}_{g_e}=d(O,X(t)),  \\
  &{}& u^{'}=1 \;\mbox { on the half-line }\;t{\overrightarrow {V}},\;t\geq 0,\\
   &{}& u=Id_{t{\overrightarrow {V}}}\;\mbox{and}\;u(t)=|X(t)| = t, \\
   &{}& \Gamma _{t{\overrightarrow {V}}} =\delta
_{t{\overrightarrow {V}}} (\mbox{Dirac measure on the half-line}\;
t{\overrightarrow {V}}).
\end{eqnarray*}
These properties can be stated as:\\
\begin{description}
    \item[$\;\;$]The time is the Euclidean distance.
    \item[$\;\;$]The speed is the unit of time and distance.
    \item[$\;\;$] The acceleration $\Gamma _{t{\overrightarrow {V}}}$ is the potential
of motion in the direction $t{\overrightarrow {V}}$, concentrated
at the origin of space and time.
    \item[$\;\;$]$E_{0}=m_{0}$ is the original potential mass energy of a pointlike particle (or the
original inertial mass) that is perpetual and eternal.
    \item[$\;\;$]$F=m_0\Gamma_{t\overrightarrow {V}} = m_0 \delta_{t \overrightarrow{V}}$ is the original energy (or
mass) $m_0$ provided with the potential of motion in the
${\overrightarrow{V}}$ direction.
\end{description}

Moreover, for an original hypothetical motion of a pointlike material
particle of mass $m$ described by $X(t)=t{\overrightarrow {V}}$
with ${\parallel {\overrightarrow {V}}\parallel}_{g_e}=V<1$, we
have $X'(t)=\overrightarrow {V}$ and $X''(t)=0$ and
so$$E(t)-E(0)=\int _{0}^{t}\frac {d}{ds}E(s)ds=\int
_{0}^{t}g_e(mX''(s),X'(s))ds=0.$$ Therefore, we have
$$E(t)=E(0) \hskip 0.3cm \mbox  {  for any }\hskip 0.1cm t$$and
the particle energy for such a motion is unchanged.\\
Recall that,
the above statements are a purely theoretical approach to the
physical space expansion of the universe and can
be used only as a macro-approximation of the Newtonian physical universe for large $t.$\\
We also notice that all mechanical and physical quantities are
essentially perceptible and measurable when using moving frames
with respect to a virtual original one and with respect to each
other. But, we can eliminate the first kind of mobility in the
following way:\\
If the motion of a particle, relatively to an original frame
$(O,\vec {i},\vec{j},\vec{k})$, is described by the vector
$\overrightarrow {OM}=X(t)$ and the motions of two other frames by
$a(t)$ and $b(t)$, then the motion of this particle is described
in the first frame ${\cal R}_1(t)$ by the vector $\overrightarrow
{O_1M}=Y(t)=X(t)-a(t)$ and in the second one ${\cal R}_2(t)$ by
the vector $\overrightarrow {O_2M}=Z(t)=X(t)-b(t).$ So
$$Y(t)=Z(t)+b(t)-a(t)=Z(t)+\overrightarrow {O_1O_2}.$$This can be
used when we have to show that a mechanical or a physical law does
not depend on the choice of any one of these two frames; it is
sufficient to assume that one of them is at rest. So, we can use
any Euclidean referential frame $(O_1,\vec
{i_1},\vec{j_1},\vec{k_1})$ where $O_1$ is a fictive point that
coincides at a given time $t_0$ (the present instant, for example)
with a given point (on the earth, for example).\\\\
Finally, let us consider a light emitter whose trajectory is
described in $\mathbb{R}^3$, provided with a fixed frame
$R_0(t_0)$, by $Y(t)$ whereas the receiver trajectory is described
by $X(t)$ (fig.2). Then the light ray that reaches the receiver at
time $t_1$ is, in fact, emitted by the emitter at a time $t_0 <
t_1$ from a point $Y(t_0)$ such as
$$ || X(t_1) - Y(t_0)||_{g_e} = d = t_1 - t_0.$$

\subsection*{Space and time half-cone}
Let $C$ be
the half-cone defined by
$$C=\{ (x,y,z,t)\in \mathbb R^4;\;x^2+y^2+z^2\leq t^2,\;t\geq 0\}=\bigcup _{t\geq 0} B(O,t) \times \{ t \}.$$
Let us say that, if $B(O,t)$ is the geometrical space within which
lives the real physical universe at time $t$, then $C$ constitutes
the geometrical virtual space within which takes place the process
of the expansion and the creation of the physical real space
intrinsically related to the progress of time.\\\\
For an event $P_0=(x_0,y_0,z_0,t_0) \in C$, $U(t_0)$ is considered
as being the set of all events $P$ that are taking place
simultaneously with $P_0$, that is$$P=(x,y,z,t_0)\in U(t_0)\times
\{t_0\}\subset C.$$ The future of $P_0$, denoted by ${\cal
F}(P_0)$, is the set of all events $Q\in U(t)\times \{t\}$, for
$t\geq t_0$, that can be situated, at a moment $t\geq t_0$, on a
trajectory of origin $P_0$. The set ${\cal F}(P_0)$ is (into our
theoretical context) the upper half-cone with vertex $P_0$ having
a span of $\frac{\pi}{2}$ (the light half-cone). The points $P'$
on the half-cone surface can be reached only by trajectories of
speed 1 joining the point $P_0 = (x_0,y_0,z_0)_{t_0}$ of
$B(O,t_0)$ to the point $P^{'} = (x_1,y_1,z_1)_{t_1}$ of
$B(O,t_1)$. A point $Q\in U(t_1)\times \{t_1\} \simeq B(O,t_1)$
within this half-cone can be joined to $P_0 \in B(O,t_0)$ only by
trajectories $\gamma$ of Euclidean length (in $\mathbb{R}^3$),
$l(\gamma)$, less than $t_1-t_0$ (fig.3).\\We denote by ${\cal
P}(P_0)$ (the past of $P_0$) the set of events $Q\in U(t)\times
\{t\}$, when $t\leq t_0$, that can be joined to $P_0$ by
trajectories of origin $Q$. This set is the intersection of the
lower half-cone of vertex $P_0$ with the half-cone of space and
time $C$ (fig.3). Any trajectory $\gamma$ joining an event $Q\in
U(t_2)\times \{t_2\} \simeq B(O,t_2)$ to $P_0 \in B(O,t_0)$, for
$t_2\leq t_0$, have to satisfy the inequality $l(\gamma)\leq
t_0-t_2$. Only the trajectories with speed 1 that pass through
$P_0$ come from the half-cone surface of this
intersection.\\

\subsection*{The train, mirror and two observers experiment}

 Let us now show that the experiment of the train, the
mirror and the two observers can be naturally interpreted and does
not justify the twisting of the natural pre-relativistic
conception of the space and time. Indeed assume that we have a
fixed virtual frame $(O_1,\vec {i_1},\vec{j_1},\vec{k_1})$ and
another frame $(O_2,\vec {i_2},\vec{j_2},\vec{k_2})$ (the rest
frame of the train) which is inertial with respect to the first
one (i.e. moving uniformly in the $O_1x$ direction, for example)
and coinciding at $t=0$ with it, just where the passenger was
originally lying, as well as the emitter and the mirror above it
(on the $O_1z$ axis) were lying. Then the ray, or more exactly the
photon emitted at $t=0$ that headed vertically (i.e. in the $O_1z$
direction) continued its way and finds itself at the instant $2t$
= $2h$ on a distance $2h$ on $O_1z$ ($h$ is the distance from the
emitter to the mirror), because the mirror can be supposed as
small and as distant as we wish. On the other hand, the photon
emitted at the same instant ($t=0$) that headed towards another
direction in order to cross the mirror which, meanwhile, has moved
horizontally (in the $O_1x$ direction) making the distance $d$,
and then to reflect in order to recover the passenger, who has,
meanwhile, moved horizontally to make a distance $2d$; this second
photon has made the distance $2\sqrt{d^2+h^2}$. The passenger then
recovers the second photon at the time $2t_1= 2\sqrt{d^2+h^2}$,
when the first ray finds itself at the same distance $2t_1$ at the
same time $2t_1$. If we have put initially a fictive fixed mirror
(with respect to $(O_1,\vec {i_1},\vec{j_1},\vec{k_1})$) a little
higher, the first photon would have recovered its original
position at the same time as the time at which the second one has
recovered the passenger. Besides, if we suppose that the emission
of light is strictly instantaneous (i.e. happening during a time
less than any fraction of second) and unidirectional (i.e.
producing only one vertical beam) and if the mirror is
sufficiently high and sufficiently small, then our passenger would
never receive the reflected ray. The same situation occurs when we
consider the fixed original frame and two other frames (on the
earth for example) moving relatively to each other with a constant
relative speed (the computation of distances
becomes a little more complicated (fig.4)).\\
If now we suppose that both emitter and passenger are continually
located at the point $O_2$, origin of the rest frame with respect
to the passenger, and suppose that the mirror above them is
macroscopic, then the photon that is vertically emitted at time
$t=0$ cross the mirror at a point that is not the same as the
point which were originally located vertically above $O_2$.
Moreover the downward vertically reflected photon does not exactly
reach the point $O_2$ which meanwhile has moved horizontally.
Thus, this photon has not been at any time located vertically
above $O_2$.\\
Moreover, concerning the passenger and its proper frame of origin
$O_2$, he can claim that the photon (or the signal) he has
received (after reflexion) at time $2t$ has actually travelled the
distance $2h$ only if the photon has been located, along its
travelling, vertically above him. Now, for a given speed $v$, the
hight $h$ is determined by the time $t$ of the crossing between
the photon and the mirror and determines the direction of the
emitted photon at $t=0$ in order to satisfy the above property.
Another photon that is emitted just after the first one in the
same direction and does cross the mirror at time $t$ does satisfy
this property only if we modify adequately the hight $h$.
Similarly, another photon that is emitted at time $t=0$ toward
another direction needs also a modification of $h$ even though the
cross point will not occur above the passenger. So, for $v$, $h$
and $t$ already given, the photon that would cross the mirror
above the passenger has probably been emitted after the time $t=0$
and would probably not be the same as that it would reach the
passenger at time $2t$. Can we then state that the light ray has
travelled a distance $2h$
with respect to the passenger proper frame?\\
The introduction of the proper frame notion, which is moving with
respect to another inertial frame, is canonically related to the
flow of time notion and leads, in case of our experiment, to the
following situation:\\
The relative passenger speed $v$ being arbitrary chosen, then the
choice of the crossing time $t$ implies that a unique photon,
having a well determined direction, can be all time vertically
located above the passenger for only one specific hight $h$ of the
mirror. All other photons of the light ray, having this same
direction, as well as all photons of other rays of the beam, can
not satisfy this property. Consequently all of these photons
travel, with respect to the passenger proper frame, different
distances. Therefore, this frame can not be canonically used for
measuring the distances that are travelled by photons and by light
rays which are made up by an "infinite" succession of photons.
Therefore, it can not be used for measuring neither the distance
that is travelled by "light" nor the light propagation speed which
is an intrinsic feature of light. The canonicity of the constancy
of the light speed i.e. the speed of all photons of an arbitrary
light ray will be established later on within a proper context
i.e. with respect to a virtually fixed frame (using the derivative
operator $\frac{d_*}{dt}$) and to any other inertial frame (using
the galilean transformation and
the derivative operator $\frac{d_1}{dt}$).\\\\

\subsection*{Remark on the relativistic spacetime notion}
\bigskip
It is well known that prerelativistic notions (adopted by Euclid,
Descartes, Galileo and Newton between many others) confer to the
three dimensional space and to the time, which is progressing
continuously, an absolute character. The Euclidean distance
$\Delta x$ between two punctual bodies at a given time $t_0$ and
the time interval $\Delta t$ between two non simultaneous events
have an intrinsic reality that is time and observer independent.
We will show below that the introduction of the relativistic
spacetime
notion has no reason to take place.\\
For that, we consider two inertial obsevers $O_1$ and $O_2$
respectively located at the origin of two Euclidean referential
frames $R_1 = (O_1, e_1, e_2, e_3)$ and $ R_2 = (O_2, e_1, e_2,
e_3)$ that co\"{\i}ncide at a time $t_0$ and such as $O_2$ is
moving on the $O_1 x$ axis of $R_1$ with a constant relative speed
$v$. Both observers would easily agree, at every time $t \geq 0$,
on the measure of the Euclidean distance separating two arbitrary
points $A_1$ and $A_2$ of the space $\mathbb{R}^3$. Indeed, we can
assume (without loss of generality) that the frame $R_1$ is fixed,
both points $A_1$ and $A_2$ are fixed on the $O_1 x$ axis and that
the frame $R_2$ is moving along the $O_1 x$ axis with a constant
speed $v$ with respect to $R_1$. Thus, if $x_1$ and $x_2$
designate the abscissas of $A_1$ and $A_2$ in $R_1$, then the
Euclidean length of the segment $A_1A_2$ is given by $x_2-x_1$
when being measured by both observers using the two frames $R_1$
and $R_2$ at every time $t \geq 0$. Actually, at every time $t
\geq 0$, the distance $A_1A_2$ measured by $R_2$ is always given
by $(x_2-tv)-(x_1-tv) = x_2-x_1$. So, a metallic bar, for
instance, having $A_1$ and $A_2$ as extremities has a length $l =
A_1A_2 = x_2-x_1$ when being measured by the frame $R_1$ and by
all inertial frames $R$ moving with any constant speed $v$ with
respect to $R_1$. Likewise, if the same bar is slipping on the
$O_1 x$ axis with a constant speed $v_0$ with respect to $R_1$,
then, at any time $t$, its length measured by $R_1$ is
$x_2+v_0t-(x_1+v_0t) = x_2-x_1$ and it is equal to
$x_2+v_0t-vt-(x_1+v_0t-vt)= x_2-x_1 = l$ when it is measured by
any frame $R_2$ moving along the $O_1 x$ axis with an arbitrary
constant speed $v$ with respect to $R_1$.\\
Moreover, when we consider the Galileo - Newtonian space-time
$\mathbb{R}^4$ provided with the frame $R^{'}_1 = (O_1, e_1, e_2,
e_3, e_4)$ where $e_4$ corresponds to the time axis $O_1 t$ that
is orthogonal to the hyperplane that is determined by $R_1 = (O_1,
e_1, e_2, e_3)$, then the metal bar has always the same length,
namely $l = x_2-x_1$, when it is measured by both Euclidean frames
$R_1$ and $R^{'}_1$ at any arbitrary fixed time $t$. But, when
we introduce the fourth dimension, i.e. the time, represented by
the $O_1 t$ axis, the points $A_1$ and $A_2$ should be labelled in
$\mathbb{R}^4$, at every time $t \geq 0$, by the frame $R^{'}_1$
with the coordinates $(x_1,0,0,t)$ and $(x_2,0,0,t)$ which would
be designated as $(x_1,t)$ and $(x_2,t)$ neglecting the two other
dimensions. The length of the bar is then given in this frame by
$\sqrt{(x_2-x_1)^2+(t-t)^2}=x_2-x_1=\Delta x$. When the observer
$O_2$ measures this length in the space-time at time $t$, one
should take into account the fact that $O_2$ would be then
labelled by $R^{'}_1$ with the coordinates $(vt,t)$ and that it
would be itself located in the hyperplane of hight $t$ in
$\mathbb{R}^4$, which is the same in which are located the points
$A_1$ and $A_2$ at time $t$, that is the points $(x_1,t)$ and
$(x_2,t)$ in the frame $R^{'}_1$. So, the observer $O_2$ obtains
the same length $\Delta x = l$ when using as well as its frame
$R_2$ or its four-dimensional frame $R^{'}_2 = (O_2, e_1, e_2,
e_3, e_4)$. The wrong issue that justified the introduction of the
relativistic spacetime notion is the fact that the spatial
interval $\Delta x$ separating two non simultaneous events $E_1$
(having $(x_1,t_1)$ as coordinates in $R^{'}_1$) and $E_2$ (having
$(x_2,t_2)$ as coordinates in $R^{'}_1$) at distinct times $t_1 <
t_2$ depends on the inertial observers. Indeed, the event $E_1$ is
labelled by $O_2$, using the frame $R^{'}_2$ at time $t_1$, with
$(x_1-vt_1,t_1)$ and $E_2$ is labelled by $O_2$, using $R^{'}_2$
at time $t_2$, with $(x_2-vt_2,t_2)$ when we assume that $O_2$ is
still lying on the $O_1 x$ axis. Let us denote, respectively, by
$A^{'}_1$, $A^{'}_2$, $A^{''}_1$ and $A^{''}_2$ the points of
coordinates $(x_1,t_1)$, $(x_2,t_1)$, $(x_1,t_2)$ and $(x_2,t_2)$
in the frame $R^{'}_1$. The spatial interval from $A^{'}_1$ to
$A^{'}_2$ measured by $O_2$, using the frame $R^{'}_2$ at time
$t_1$ is $x_2-vt_1-(x_1-vt_1)=x_2-x_1$ and the spatial interval
from $A^{''}_1$ to $A^{''}_2$ measured by $O_2$, using the frame
$R^{'}_2$ at time $t_2$, is $x_2-vt_2-(x_1-vt_2)=x_2-x_1.$ These
results are the same as those obtained by this observer when using
its frame $R_2$ when it is located in the space-time
$\mathbb{R}^4$ respectively at the points $O^{'}_2(t_1) =
(O_2,t_1)$ and $O^{''}_2(t_2) = (O_2,t_2)$ at times $t_1$ and
$t_2$
(see fig.4$^{'}$).\\
Unlike the preceding results, the spatial interval $\Delta_1 x$
between $A^{'}_1$ and $A^{''}_2$ measured by $R^{'}_1$ is
$x_2-x_1$ and the spatial interval $\Delta_2 x$ between $A^{'}_1$
and $A^{''}_2$ measured by $R^{'}_2$ is $x_2-vt_2-(x_1-vt_1) \neq
x_2-x_1$. Likewise, the Euclidean interval between $A^{'}_1$ and
$A^{''}_2$ measured by $R^{'}_1$ is
$$ I_1 = \sqrt{(x_2-x_1)^2+(t_2-t_1)^2}$$and this same interval
measured by $R^{'}_2$ is
$$ I_2 = \sqrt{((x_2-vt_2)-(x_1-vt_1))^2+(t_2-t_1)^2}$$
which implies $I_2 \neq I_1$.\\
Now, neither the $\Delta_i x$ intervals nor the $I_i$ intervals
(\emph{i} = 1,2) do correspond to a physical reality. Indeed, if
we consider the bar $A_1A_2$ that lies at time $t_1$ at
$A^{'}_1A^{'}_2$ in the space-time $\mathbb{R}^4$ and at
$A^{''}_1A^{''}_2$ in the same space at time $t_2$, then its real
spatial length measured by $O_2$ using the frame $R^{'}_2$ is
equal to $x_2-vt_1-(x_1-vt_1) = x_2-x_1$ at time $t_1$ and to
$x_2-vt_2-(x_1-vt_2) = x_2-x_1$ at time $t_2$. Likewise, the two
Euclidean intervals in the space-time $\mathbb{R}^4$ measured by
$O_2$ using the frame $R^{'}_2$ at times $t_1$ and $t_2$ also are
equal :
$$
\sqrt{((x_2-vt_1)-(x_1-vt_1))^2+(t_1-t_1)^2}=$$
$$\sqrt{((x_2-vt_2)-(x_1-vt_2))^2+(t_2-t_2)^2}=x_2-x_1$$
The two extremities that are labelled, in the physical universe,
by ($x_1, y_1, z_1$) (resp. ($x_2,y_1,z_1$)) in the first frame
have to be physically labelled, in the "moving" second frame, by
($x_1-vt, y_1, z_1$) (resp. ($x_2-vt,y_1,z_1$)) since this frame
has physically made the displacement $vt$ in the positive $x$
direction during the time $t$.\\
The two intervals $A^{'}_1A^{''}_2$ and $A^{'}_2A^{''}_1$
measurements in the frames $R^{'}_1$ and $R^{'}_2$ do not have any
physical meaning. They have nothing to do with the bar neither at
time $t_1$ nor at time $t_2$. We can say the same thing concerning
the area and the volume of any two dimensional or three
dimensional body in the real space $\mathbb{R}^3$. Such a body is
located entirely, at every time $t$, in the hyperplane of hight
$t$ in the space-time $\mathbb{R}^4$. Lengths, areas and volumes
are the same when measured by $R_1$ and $R_2$ in $\mathbb{R}^3$ or
by $R^{'}_1$ and $R^{'}_2$ in $\mathbb{R}^4$ at any time $t$.\\
In other respects, if the permanently expanding universe is
represented by $U(t_1)$ at time $t_1$ and by $U(t_2)$ at time $t_2
> t_1$ and if we assume that it is provided at every time $t$ with
a metric $g_t$ that evolves with time, we can not measure the
distance of a point $A_1$ in $U(t_1)$ to a point $A_2$ in $U(t_2)$
because we can not adequately use neither $g_{t_1}$ nor $g_{t_2}$
in order to carry out this measurement. At any time $t$, the
universe is not static and the metric $g_t$ is never the same at
any distinct times $t_1$ and $t_2$. Consequently the problem of
measuring the spatial distance $\Delta x$ between two events $E_1$
and $E_2$ at different times, $t_1$ for $E_1$ and $t_2$ for $E_2$,
or the Euclidean interval $I$ in the space-time $\mathbb{R}^4$
between $E_1$ and $E_2$ should not be posed because it is
physically meaningless.\\\\
We now suppose that the motion of a particle is described using
two frames so that the relative speed of each of them with respect
to the other is constant and that they can be obtained from each
other by means of an orthogonal transformation $A$. So, if
$\overrightarrow {OM}=X(t)$ and $\overrightarrow {OO'}=a(t)$ in
the first frame and $\overrightarrow {O'M}=Y(t)$ in the second
one, then we have $\overrightarrow {O'M}=\overrightarrow {O^{'}O}+
\overrightarrow {OM}=\overrightarrow {OM}-\overrightarrow
{OO^{'}}$, and so$$Y(t)=A^{-1}(X(t)-a(t))$$
$$Y'(t)=A^{-1}(X'(t)-a'(t))$$and
$$Y''(t)=A^{-1}(X''(t)-a''(t)).$$If, in addition we have
$a'(t)=0$, then
$$Y'(t)=A^{-1}.X'(t)$$and therefore$${\parallel Y'(t)\parallel}_{g_e}={\parallel X'(t)\parallel}_{g_e}.$$
If we have $a''(t)=0,$ then$$Y''(t)=A^{-1}.X''(t)$$and therefore
$${\parallel Y''(t)\parallel}_{g_e}={\parallel
X''(t)\parallel}_{g_e}.$$This yields, for instance,
$$\triangle E_1(t):=\int _{t_0}^{t}X''(r).X'(r)dr=\frac {1}{2}\int _{t_0}^{t}\frac{d}{dr}
{\parallel X'(r)\parallel}^{2}_{g_e}dr$$
$$= \frac{1}{2}({\parallel X'(t)\parallel}^{2}_{g_e}-{\parallel X'(t_0)\parallel}^{2}_{g_e})$$
$$= \triangle E_2(t):=\frac {1}{2}\int _{t_0}^{t}Y''(r).Y'(r)dr$$
$$= \frac{1}{2}({\parallel Y'(t)\parallel}^{2}_{g_e}-{\parallel Y'(t_0)\parallel}^{2}_{g_e}).$$

\section{General tensoriality with respect to the frame exchanges}

$\hskip 0.5cm$We begin by noticing that, the hypothesis of the
$g_t$-family independence of the virtual position in $\mathbb R^3$
is perfectly justified when considering the procedure of modeling
the theoretical universe expansion. Contrary to this situation,
the study of a motion or a physical phenomenon taking place
between two times $t_1$ and $t_2$ in the real dynamic space must
take into account the existence of gravitational fields created by
the matter distribution, the black holes, the energetic,
electro-magnetic and quantum phenomena scattered in the universe
at every time $t \in [t_1,t_2]$. This explains, for instance, the
deviation undergone by the light propagation with respect to
classical geodesics
(straight lines).\\\\
So, we now assume that the real physical universe at time $t$ (for $t\gg1$) is
assimilated to$$U(t)=(B(O,t),g_t(X))$$where $g_t(X)$ is a variable
metric (depending on both position and time) which is determined
entirely by all manifestations of the matter-energy that is
filling the
space at time $t$.\\\\
In the following, we then consider the general case and some
particular cases (real, approximate or virtual) concerning the
families $g_t$, $A_t$ and the curve $a_0(t)$ (defined through the
preceding sections) that will be specified later on. Indeed $g_t$
can be depending on time or on position (locally or globally),
$a_0(t)$ can be a geodesic or not and $A_t$ can be an
isometry that depends or not on time....\\
So, let us consider within $B(O,T)$, a trajectory $X_0(t),$ for $0
< t < T,$ with respect to a virtual fixed referential frame ${\cal
{R}}_0=(O_0,\vec{i_0},\vec {j_0},\vec{k_0}).$ In order to study
the tensoriality of the speed vector and the acceleration vector
relatively to frame exchanges, we consider two moving frames
${\cal {R}}_1(t)$ and ${\cal{R}}_2(t)$ so that their origins
describe the trajectories $a_0(t)$ and $b_0(t)$ (with respect to
${\cal {R}}_0$) and so that their passage matrices are given by
$A_t$ and $B_t$. We notice that the choice of the frame $\mathcal{R}_0(t_0)$
does not have any influence on the nature of the results that will
be established below since we aim to study the passage from one to
another frame that are both moving with respect to
$\mathcal{R}_0(t_0)$.\\
Let
$$X_1(t)=A_t(X_0(t)-a_0(t))$$and
$$X_2(t)=B_t(X_0(t)-b_0(t)),$$ be, respectively, the expressions of the
trajectory $X_0(t)$ in ${\cal {R}}_1(t)$ and ${\cal
{R}}_2(t)$.\\

\subsection*{Another time derivation operator}
We now put
$$\frac{d_*}{dt}X_1(t):=X_1'(t)+A_t.a'_0(t)-A'_t(X_0(t)-a_0(t))=A_t.X'_0(t) \;\;\;\;\;\;\;(d_\ast)$$
and
$$\Gamma _{1*}(t):=\nabla _{\frac{d_*}{dt}X_1(t)}^{g_t}\frac{d_*}{dt}X_1(t)
=\nabla _{A_t.X'_0(t)}^{g_t}A_t.X'_0(t)$$as well as

$$\frac{d_*}{dt}X_2(t)=B_t.X'_0(t)$$and

$$\Gamma _{2*}(t):=\nabla _{B_t.X'_0(t)}^{g_t}B_t.X'_0(t).$$This
derivation as well as the one we have denoted previously by
$\frac{d_1}{dt}$ takes into account the speed and the rotation of
the moving frames. In these conditions, we have
$$\frac{d_*}{dt}X_2(t)=B_t\circ A_t^{-1}.\frac{d_*}{dt}X_1(t)$$
and
$$\Gamma _{2*}(t):=\nabla _{\frac{d_*}{dt}X_2(t)}^{g_t}\frac{d_*}{dt}X_2(t)
=\nabla _{B_t\circ A_t^{-1}\frac{d_*}{dt}X_1(t)}^{g_t}B_t\circ
A_t^{-1}\frac{d_*}{dt}X_1(t).$$So, if $B_t$ and $A_t$ are
isometries, then these notions depend only on the geometry (with
respect to $g_t$) of the space and we have
$${\parallel\frac{d_*}{dt}X_2(t)\parallel}_{g_{t}}
={\parallel\frac{d_*}{dt}X_1(t)\parallel}_{g_{t}}$$and
$${\parallel\Gamma _{2*}(t)\parallel}_{g_{t}}=
{\parallel\Gamma _{1*}(t)\parallel}_{g_{t}}.$$In the general case,
if $B_t=A_t$, we have
$$\Gamma _{2*}(t)=\Gamma _{1*}(t).$$On the other hand, if $A_t\equiv A$
and $B_t\equiv B$, we obtain
$$\frac{d_*}{dt}X_1(t)= \frac{d}{dt} X_1(t) + A. a'_0(t) = A.X'_0(t),$$
$$\frac{d_*}{dt}X_2(t)= \frac{d}{dt} X_2(t) + B. b'_0(t) = B.X'_0(t) = B\circ A^{-1}.\frac{d_\ast}{dt} X_1(t)$$and
$$\Gamma _{2*}(t)= \nabla^{g_t}_{B\circ A^{-1}.\frac{d_\ast}{dt} X_1(t)} B\circ A^{-1}\frac{d_\ast}{dt}X_1(t).$$If, in addition,
$g_t$ is flat, we have
$$\Gamma_{1 \ast}(t) = \nabla _{(AX_0(t))^{'}} (AX_0(t))^{'} =
AX_0^{''}(t) = A \Gamma_0(t)$$
$$\Gamma_{2 \ast}(t) = \nabla _{(BX_0(t))^{'}} (BX_0(t))^{'} =
BX_0^{''}(t) = B \Gamma_0(t)$$which yields
$$\Gamma_{2 \ast }(t) = B \circ A^{-1} . \Gamma_{1 \ast} (t).$$
Furthermore, in the same conditions, we have
$$\Gamma _1(t):=\nabla _{X'_1(t)}^{g_{t}}X'_1(t)=\nabla
_{A.(X'_0(t)-a'_0(t))}^{g_t}A.(X'_0(t)-a'_0(t))$$$$
=X''_1(t)=A.(X_0(t)-a_0(t))''$$and
$$\Gamma _2(t):=\nabla _{X'_2(t)}^{g_{t}}X'_2(t)=X''_2(t)=B.(X_0(t)-b_0(t))''.$$
Therefore, if $a_0(t)$ and $b_0(t)$ are geodesics (or if $a_0(t)
=b_0(t)=0$), we obtain
$$\Gamma
_{1*}(t)=\nabla_{A.X'_0(t)}^{g_{t}}A.X'_0(t)=A.X''_0(t)=A.\Gamma_0(t)=\Gamma_1(t),$$
$$\Gamma
_{2*}(t)=\nabla_{B.X'_0(t)}^{g_{t}}B.X'_0(t)=B.X''_0(t)=B.\Gamma_0(t)=\Gamma_2(t),$$
$$\Gamma_2(t)=B\circ A^{-1}.\Gamma_1(t),$$
and 
$$\Gamma_2(t)=\Gamma_1(t)\quad\mbox{for}\;\; A=B\quad\mbox{or, in particular, for}\quad A=B=\mbox{Id}_{\mathbb{R}^3}.$$
All acceleration vectors are then null if $X_0(t)$ is a
geodesic.\\
If, in addition, $A$ and $B$ are isometries, then we have
$${\parallel\Gamma _{2}(t)\parallel}_{g_{t_0}}=
{\parallel\Gamma _{1}(t)\parallel}_{g_{t_0}}.$$We conclude that,
in the general case the speed vector transforms in a tensorial
manner in the following sense: If the vector $X'_0(t)$ is
multiplied by a function $h(X_0(t))=:h(t)$ where $h$ is a
differentiable function on a neighborhood of the trajectory
$X_0(t)$, then the vectors $\frac{d_*}{dt}X_1(t)$ and
$\frac{d_*}{dt}X_2(t)$ are multiplied by the same function of $t$.
For the acceleration vector, we have
\begin{eqnarray*}
  \nabla _{hX^{'}_0(t)}^{g_t}hX^{'}_0(t)&=& h(h
  \nabla^{g_t}_{X^{'}_0(t)} X^{'}_0(t) +
  dh(X_0(t)).X^{'}_0(t)X^{'}_0(t)\\
  &=& h^2\Gamma_{0}(t)+\frac{1}{2}dh^2(X_0(t)).X^{'}_0(t)X^{'}_0(t) \\
  &=& h^2\Gamma_{0}(t)+\frac{1}{2}\frac{d}{dt}h^2(X_0(t))\frac{d}{dt}X_0(t)
\end{eqnarray*}
and
\begin{eqnarray*}
  \nabla
_{h\frac{d_*}{dt}X_1(t)}^{g_t}h\frac{d_*}{dt}X_1(t) &=& \nabla_{h A_t.X'_0(t)}^{g_t}h A_t.X'_0(t) = h \nabla_{A_t.X'_0(t)}^{g_t} h A_t.X^{'}_0(t) \\
   &=& h^2 \nabla_{A_t.X'_0(t)}^{g_t} A_t.X'_0(t) + h
   ((A_t.X'_0(t)).h)A_t.X'_0(t)\\
   &=& h^2(t) \Gamma_{1*}(t) + h
   (dh(X_0(t)).(A_t.X'_0(t))A_t.X'_0(t)\\
   &=& h^2(t) \Gamma_{1*}(t) + \frac{1}{2}
   dh^2(X_0(t)).(A_t.X'_0(t))A_t.X'_0(t)\\
   &=& h^2(t)\Gamma_{1*}(t)+\frac{1}{2}dh^2(X_0(t)).(\frac{d_*}{dt}X_1(t))\frac{d_*}{dt}X_1(t). \\
\end{eqnarray*}
and
\begin{eqnarray*}
  \nabla _{h\frac{d_*}{dt}X_2(t)}^{g_t}h\frac{d_*}{dt}X_2(t) &=& h^2(t)\Gamma_{2*}(t)+\frac{1}{2}dh^2(X_0(t)).(\frac{d_*}{dt}X_2(t))\frac{d_*}{dt}X_2(t) .\\
\end{eqnarray*}
If $A_t \equiv A$ , $B_t\equiv B$ and $a'_0(t) = b'_0(t) = 0$, we
obtain
$$\nabla_{hX'_0(t)}^{g_t}hX^{'}_0(t) = h^2(t) \Gamma_0(t) +
\frac{1}{2} dh^2 (X_0(t)). \frac{d}{dt} X_0(t) \frac{d}{dt} X_0(t)
=
h^2(t)\Gamma_0(t)+\frac{1}{2}\frac{d}{dt}h^2(t)\frac{d}{dt}X_0(t),$$
$$\nabla^{g_t}_{h\frac{d}{dt}X_1(t)}h\frac{d}{dt}X_1(t) =
\nabla_{h\frac{d_*}{dt}X_1(t)}^{g_t}h\frac{d_*}{dt}X_1(t) =
\nabla_{h A.X'_0(t)}^{g_t}h A.X'_0(t)$$
$$ \hskip 3cm = h^2
\nabla^{g_t}_{AX^{'}_0(t)} AX^{'}_0(t) + h A
  X^{'}_0(t).(h(X_0(t)))AX^{'}_0(t)$$
$$ \hskip 3cm = h^2 \nabla^{g_t}_{X^{'}_1(t)}X^{'}_1(t) + h \;\;dh(X_0(t)).
AX^{'}_0(t)AX^{'}_0(t)$$
$$\hskip 3cm  = h^2(t)\Gamma_1(t)+\frac{1}{2}dh^2(X_0(t)).\frac{d}{dt}X_1(t) \frac{d}{dt}X_1(t)
$$
and
$$\nabla
_{h\frac{d}{dt}X_2(t)}^{g_t}h\frac{d}{dt}X_2(t)=\nabla
_{h\frac{d_*}{dt}X_2(t)}^{g_t}h\frac{d_*}{dt}X_2(t)
=h^2(t)\Gamma_{2}(t)+\frac{1}{2}dh^2(X_0(t)).\frac{d}{dt}X_2(t)
\frac{d}{dt}X_2(t).$$ 
In the general case, the above equalities show that the two vectors
$$\nabla_{h\frac{d_*}{dt}X_1(t)}^{g_t}h\frac{d_*}{dt}X_1(t) \mbox{ and } \nabla
_{h\frac{d_*}{dt}X_2(t)}^{g_t}h\frac{d_*}{dt}X_2(t)$$ 
transform in the same manner when we multiply the speed vectors by a given
function $h$. If, in addition, $h$ is constant on a neighborhood of $X_0(t)$ or if $h$ is constant on $X_0(t)$ and $A_t\equiv Id_{\mathbb{R}^3}\equiv B_t$, then the two expressions of acceleration
vector are multiplied by $h_0^2$. These properties imply a sort of
tensoriality of the speed vector and the acceleration vector in
the following sense:\\
Suppose that $\chi (X)$ is a vector field on $B(O,T)$ having the
integral curves $X_0(t)$ (i.e. $X'_0(t)=\chi (X_0(t))$) for $t_0
\leq t \leq T$. If we multiply $\chi$ by a function $h(X)$
supposed to be constant on a neighborhood of the integral curves $X_0(t)$ or if $h$ is constant on $X_0(t)$ and $A_t\equiv Id_{\mathbb{R}^3}\equiv B_t$ (i.e.
$h(X_0(t))\equiv h(x_0,y_0,z_0)= h_0$), then the two expressions
of the speed vector are multiplied by the same constant $h_0$ and
the two expressions of the acceleration vector are multiplied by
$h_0^2$.\\
When $g_t$ is flat, $A_t\equiv A$, $B_t\equiv B$ and $a_0(t)$ and $b_0(t)$ are geodesics (or
$a_0(t) = b_0(t) = 0$) then this tensoriality property is still
valid for the ordinary acceleration vector
$$\Gamma(t) = \frac{d^2}{dt^2} X(t) = X^{''}(t),$$
where we have used $(d_*)$ and the fact that $\Gamma(t)=\Gamma_*(t)$.\\
Finally, let us notice that we can define (in the general case
where $g_t = g_t(X)$), globally or locally, all geometrical
objects (i.e. depending only on the metric $g_t$) such as:
Isometry groups, gradient and Hessian of functions, divergence of
vector fields or differential forms, Laplacian of a function or a
differential form (Hodge operator), Dirac operator....

\section{Physical modeling of the expanding universe}

$\hskip 0.5cm$We begin here by showing that the presumed
invalidity of the Maxwell equation covariance regarding inertial
frames' exchanges is only apparent and that the covariance
principle is valid, with respect to such exchanges, for
electromagnetism Maxwell's equations.\\
Indeed, let us consider two inertial Euclidean frames
$\mathcal{R}_1$ and $\mathcal{R}_2$ so that the motion of the
second frame with respect to the first one is uniform. This
situation can be brought back to suppose that, if
$(x_1(t),y_1(t),z_1(t))$ is any trajectory in $\mathcal{R}_1,$
then the trajectory in $\mathcal{R}_2$ is given by
$(x_2(t),y_1(t),z_1(t))$ with $x_2(t)\;=\;x_1(t)\;-\;vt.$
Nevertheless these two frames (the study of the passage from each
to other has led to the presumed invalidity of the covariance
principle) are in fact both uniformly moving. Therefore, if $v_1$
and $v_2$ are respectively the constant speeds of these frames
with respect to a virtual frame $\mathcal{R}_0$ supposed to be
fixed (we can assume that the two speed vectors have both the same
direction as the $Ox$ axis of $\mathcal{R}_0$) and if $\varphi$ is
a function that can be supposed of the form $\varphi(x_1,t)$ in
the frame $\mathcal{R}_1,$ then the wave equation is written in
$\mathcal{R}_1$ as:
\begin{equation}\label{r9}
\square_1\varphi(x_1,t)\;:=\;\frac{\partial^2\varphi}{\partial
t^2}(x_1,t)\;-\;\frac{\partial^2\varphi}{\partial
x_1^2}(x_1,t)\;=\;f(x_1,t).
\end{equation}
Putting $\varphi(x_2,t)\;=\;\varphi(x_1-vt,t),$ the form of this
equation remains the same when written in the frame
$\mathcal{R}_2;$ that is it can be written as
\begin{equation}\label{r10}
\square_2\varphi(x_2,t)\;:=\;\frac{\partial^2\varphi}{\partial
t^2}(x_2,t)\;-\;\frac{\partial^2\varphi}{\partial
x_2^2}(x_2,t)\;=\;f(x_2,t)
\end{equation}
in the following sense:\\
When we write $\varphi(x_1,t)\;=\;\varphi(x_0-v_1t,t),$ the wave
equation being written in $\mathcal{R}_0$ under its canonical form
\begin{equation}\label{r11}
\square_0\varphi(x_0,t)\;:=\;\frac{\partial^2\varphi}{\partial
t^2}(x_0,t)\;-\;\frac{\partial^2\varphi}{\partial
x_0^2}(x_0,t)\;=\;f(x_0,t),
\end{equation}
then equation (10) is obtained from equation (11) in the same way
as equation (9) is obtained from the same equation (11). This is
done by giving to the partial derivative
$\frac{\partial\varphi}{\partial t}$ in (10) the same meaning as
we have given to $\frac{\partial\varphi}{\partial t}$ in (9).
Namely, we define the derivative $\frac{\partial\varphi}{\partial
t}(x_1,t)$ by replacing, into
$$\frac{d}{dt}\varphi(x_0,t)=\partial_1\varphi(x_0,t)x_0^{'}(t)+\partial_2\varphi(x_0,t),$$
$x_0(t)$ by $x_1(t)(=x_0(t)-v_1t)$ and $x^{'}_0(t)$ by
$x^{'}_1(t)$,supposed to be the same as $x^{'}_0(t)$, which is not
the same as taking
$$\frac{\partial \varphi}{\partial t} (x_1,t) = \frac{d}{dt}\varphi(x_0-v_1t,t)=\partial_1\varphi(x_0-v_1t,t)(x_0^{'}(t)-v_1)+\partial_2\varphi(x_0-v_1t,t).$$
The same operation is made for $\frac{\partial\varphi}{\partial
t}(x_2,t)$ by replacing, into $\frac{d}{dt}\varphi(x_0(t),t),$
$x_0(t)$ by $x_2(t)(=x_0(t)-v_2t)$ and $x^{'}_0(t)$ by
$x^{'}_2(t)$,supposed to be the same as $x^{'}_0(t)$, which is not
the same as
$$\frac{\partial \varphi}{\partial t} (x_2,t) := \frac{d}{dt}
\varphi(x_0(t) - v_2 t , t).$$
 So, if we assume that
$\mathcal{R}_1$ is at rest and $\mathcal{R}_2$ is inertial with
respect to $\mathcal{R}_1,$ we deduce (10) from (9) by using the
partial derivative $\frac{\partial\varphi}{\partial t}(x_2,t)$
obtained from
$$\frac{d}{dt}\varphi(x_1,t) = \partial_1 \varphi(x_1,t)
x^{'}_1(t) + \partial_2 \varphi(x_1,t)$$ by replacing $x_1(t)$ by
$x_2(t)(=x_1(t)-vt)$ and $x^{'}_1(t)$ by $x^{'}_2(t)$, supposed to
be the same as $x^{'}_1(t)$, which is not the same as
$$ \frac{\partial \varphi}{\partial t} (x_2,t) := \frac{d}{dt}
\varphi(x_1(t) - v t, t).$$
In other words, when taking the
derivative with respect to the second variable $t$ of the function
$\varphi(x_1-vt,t),$ i.e. $\frac{\partial\varphi}{\partial
t}(x_1-vt,t),$ we assume that the first variable is $x_1(=x_2+vt)$
and not $x_1-vt(=x_2).$ Briefly speaking, when making this
operation, we consider the first
variable as being $x_1$ instead of $x_1-vt$ (i.e. considering
the time-dependence of the variable $x_2$ as being the same as that of the variable $x_1$) and then we replace $x_1 -vt$ by $x_2$.\\

More generally, we may briefly recall some general tensoriality
properties regarding frame exchanges. Indeed, let
$\mathcal{R}_0=(O_0,\overrightarrow{i_0},\overrightarrow{j_0},\overrightarrow{k_0})$
be a virtual fixed Euclidean frame of $\mathbb{R}^3$ and let
$\mathcal{R}_1=(O_1,\overrightarrow{i_1},\overrightarrow{j_1},\overrightarrow{k_1})$
and
$\mathcal{R}_2=(O_2,\overrightarrow{i_2},\overrightarrow{j_2},\overrightarrow{k_2})$
be two other frames. We assume that $O_1$ and $O_2$ move
respectively with relative constant velocities
$\overrightarrow{v_1}$ and $\overrightarrow{v_2}$ with respect to
$\mathcal{R}_0.$ We furthermore suppose that $\mathcal{R}_1$ and
$\mathcal{R}_2$ are respectively obtained from $\mathcal{R}_0,$
for $t\geq0,$ by linear transformations $A_t$ and $B_t.$ Finally,
let $X_0(t)=(x_0(t),y_0(t),z_0(t)),$
$X_1(t)=(x_1(t),y_1(t),z_1(t))$ and
$X_2(t)=(x_2(t),y_2(t),z_2(t))$ be respectively the coordinates of
a punctual particle trajectory in $\mathbb{R}^3$ with respect to
these three frames. We then have, for $t\geq0,$
$$
    X_1(t)=A_t(X_0(t)-t\overrightarrow{v_1})=:A_t.Y_0(t),
$$
$$
    X_2(t)=B_t(X_0(t)-t\overrightarrow{v_2})=:B_t.Z_0(t)
$$
and
\begin{eqnarray*}
  X_2(t) &=& B_t(X_0(t)-t\overrightarrow{v_1}+t\overrightarrow{v_1}-t\overrightarrow{v_2})= B_t.(A_t^{-1}.X_1(t)+t\overrightarrow{v_1}-t\overrightarrow{v_2})\\
   &=& B_t\circ A_t^{-1}.X_1(t)-B_t.t(\overrightarrow{v_2}-\overrightarrow{v_1})\\
  &=& B_t\circ A_t^{-1}.X_1(t)-B_t.t\overrightarrow{v}
\end{eqnarray*}
where $\overrightarrow{v}$ is the relative velocity.\\
Now, we recall the definition of the two derivative notions
$\frac{d_1}{dt}$ and $\frac{d_*}{dt}$ previously introduced:
\begin{eqnarray*}
  \frac{d_1}{dt}X_1(t) &:=& \frac{d}{dt}X_1(t)-A_t^{'}(X_0(t)-t\overrightarrow{v_1})\\
 &=& \frac{d}{dt} X_1(t) - A^{'}_t.Y_0(t)\\
 &=& \frac{d}{dt}X_1(t)-A_t^{'}\circ A_t^{-1}.X_1(t)
\end{eqnarray*}
and
$$\frac{d_*}{dt}X_1(t):=\frac{d}{dt}X_1(t)+A_t.\overrightarrow{v_1}-A_t^{'}(X_0(t)-t\overrightarrow{v_1}).$$
On the other hand, we have
$$\frac{d}{dt}X_1(t)=A_t\left(\frac{d}{dt}X_0(t)-\overrightarrow{v_1}\right)+A_t^{'}(X_0(t)-t\overrightarrow{v_1})$$
which is
$$\frac{d}{dt}X_1(t)=A_t.\frac{d}{dt}Y_0(t)+A_t^{'}.Y_0(t).$$
Therefore, we obtain
$$
   \hskip 1.5cm  \frac{d_1}{dt}X_1(t)=A_t.\frac{d}{dt}Y_0(t)=A_t.\frac{d_1}{dt}Y_0(t)
    \;\;\;\;\;\;\;\;\;\;\;\;\;\;\;\;\;\;\;\hskip 1cm (d_1^{'})
$$
and
\begin{eqnarray*}
  \frac{d_*}{dt}X_1(t) &=& A_t.\frac{d}{dt}Y_0(t)+A_t.\overrightarrow{v_1}\\
 &=& A_t.\frac{d}{dt}X_0(t)=A_t.\frac{d_*}{dt}X_0(t)
\end{eqnarray*}
since $\frac{d_1}{dt}$ and $\frac{d_*}{dt}$ are the same as
$\frac{d}{dt}$ for the $\mathcal{R}_0-$coordinates.\\
We notice that we have
$$\frac{d_2}{dt}X_2(t):=B_t.\frac{d}{dt}Z_0(t)\;\;\; \mbox{for}
\;\;\; Z_0(t)=X_0(t) - tv_2=Y_0(t)-t\overrightarrow{v}
\;\;\;\;\;\; (d_2^{'})$$ and
$$\frac{d_*}{dt}X_2(t):=B_t.\frac{d}{dt}X_0(t)$$
which yields
$$
\frac{d_*}{dt}X_2(t)=B_t\circ A_t^{-1}.\frac{d_*}{dt}X_1(t).
$$
Then, when $B_t\equiv A_t,$ we get
$$\frac{d_*}{dt}X_2(t)=\frac{d_*}{dt}X_1(t).$$
Moreover, when $\overrightarrow{v_1}=\overrightarrow{v_2}$ (i.e.
$\overrightarrow{v}=0$), we have (using ($d_2^{'}$) and
($d_1^{'}$))
$$
\frac{d_2}{dt}X_2(t)=B_t\circ A_t^{-1}.\frac{d_1}{dt}X_1(t)
$$
and if we furthermore have $B_t\equiv A_t,$ then
$$\frac{d_2}{dt}X_2(t)=\frac{d_1}{dt}X_1(t),$$
which is valid, in particular, when $A_t\equiv\mbox{Id}_{\mathbb{R}^3}$ and $B_t\equiv\mbox{Id}_{\mathbb{R}^3}$.\\
These derivative notions take into account, in a most natural way,
the relative speeds and rotations of the moving frames. The first
one coincides, for $A_t\equiv A,$ with the classical derivation.
The second one becomes, for $A_t\equiv A$,
$$\frac{d_*}{dt}X_1(t)=\frac{d}{dt}X_1(t)\;+\;A.\overrightarrow{v_1}.$$
The previous relations show some tensoriality properties for the
speed vector regarding frame
exchanges.\\

In the particular case where $A_t\equiv Id_{\mathbb{R}^3},$ we
have
$$\frac{d_1}{dt}X_1(t)=X_1'(t)=\frac{d_1}{dt}Y_0(t)=Y_0'(t)=X_0'(t)-\overrightarrow{v_1} , $$
which is the classical Galilean derivation, and
$$\frac{d_*}{dt}X_1(t)=X_0'(t).$$
In the light of the preceding, we can clarify more precisely the
Maxwell's equation covariance problem. Indeed, after replacing
$X_0(t),$ $X_1(t)$ and $X_2(t)$ by $x_0(t),$ $x_1(t)$ and $x_2(t)$
and the speed vectors $\overrightarrow{v_1},$
$\overrightarrow{v_2}$ and $\overrightarrow{v},$ simply by the
scalar speeds $v_1,$ $v_2$ and $v,$ we can assert that this
covariance validity needs only to make a slight modification of
the spatial variable notion $(x\longrightarrow x-vt)$ and also of
the notion of the derivation with respect to the time variable
$\left(\frac{d}{dt} \longrightarrow\frac{d_*}{dt}\right).$ These
modifications constitute the two natural procedures for including
the frames' motions into the general motion within the universe,
unlike the relativistic spacetime conception which leads to the
deterioration of the natural relationship between space and time
since, for us, distances are essentially proportional to time.\\

\subsection*{Canonical form of the Maxwell equation}
Now, let us consider the classical time derivative
$$\frac{d}{dt}\varphi(x_1,t)=\frac{d}{dt}\varphi(x_0-v_1t,t)=\partial_1\varphi(x_0-v_1t,t)(x_0^{'}(t)-v_1)+\partial_2\varphi(x_0-v_1t,t).$$
Then, taking the $\frac{d_*}{dt}$ derivative instead of
$\frac{d}{dt},$ we obtain
\begin{eqnarray*}
   \frac{\partial_*\varphi}{\partial t}(x_1,t)&=& \frac{\partial_*\varphi}{\partial t}(x_0-v_1t,t) \\
   &:=& \partial_1\varphi(x_0-v_1t,t)\frac{d_*}{dt}x_1(t)+\partial_2\varphi(x_0-v_1t,t) \\
   &=& \partial_1\varphi(x_0-v_1t,t)x_0^{'}(t)+\partial_2\varphi(x_0-v_1t,t) \\
  &=&\frac{\partial\varphi}{\partial t}(x_1,t)\;=\;\frac{\partial\varphi}{\partial t}(x_0-v_1t,t)
\end{eqnarray*}
where the latter derivative has the meaning previously specified
as here we have $ \frac{d_*}{dt} x_1(t) = x^{'}_0(t)$ since $A_t
\equiv Id_{\mathbb{R}^3}$. It is made up by considering the time
dependence of the first variable $x_1(t)=x_0(t)-v_1t$ as being the
same as that of the variable $x_0(t)(=x_1(t)+v_1t),$ i.e. by
neglecting the term $v_1t$ which comes from the relative motion of
the frame $\mathcal{R}_1.$ In other words, we have to take the
derivative of the trajectory seen by $\mathcal{R}_1$ (i.e. $x_1(t)$) as
being the derivative of this same trajectory when it is seen by
$\mathcal{R}_0$ or by any other fixed frame $\mathcal{R}$ that is isometric to $\mathcal{R}_0$;
in particular by $\mathcal{R}_1$ provided that this latter is considered as
being at rest with respect to $\mathcal{R}_0$. In the same order of ideas we
can write
$$ \frac{\partial_*^2\varphi}{\partial t^2}(x_1,t)\;=\; \frac{\partial^2\varphi}{\partial
t^2}(x_0-v_1t,t).$$ We can now maintain that the general intrinsic
canonical form of the Maxwell equation is
$$
\square_*\varphi(x,t)\;:=\;\frac{\partial^2_*\varphi}{\partial
t^2}(x,t)\;-\;\frac{\partial^2\varphi}{\partial x^2}(x,t)\;=\;0
$$
which reduces to the classical Maxwell equation
$$
\square\varphi(x,t)\;=\;\frac{\partial^2\varphi}{\partial
t^2}(x,t)\;-\;\frac{\partial^2\varphi}{\partial x^2}(x,t)\;=\;0
$$
when we consider the rest frame with respect to the motion of the
origin. If we consider two other frames $\mathcal{R}_1$ and
$\mathcal{R}_2,$ the canonical equation becomes respectively
$$
\square_*\varphi(x_1,t)\;=\;\frac{\partial^2_*\varphi}{\partial
t^2}(x_1,t)\;-\;\frac{\partial^2\varphi}{\partial
x_1^2}(x_1,t)\;=\;0
$$
and
$$
\square_*\varphi(x_2,t)\;=\;\frac{\partial^2_*\varphi}{\partial
t^2}(x_2,t)\;-\;\frac{\partial^2\varphi}{\partial
x_2^2}(x_2,t)\;=\;0
$$
which both reduce to the classical one when considering the
respective
rest frames regarding the respective origins' motions.\\
However, when we consider the $\mathcal{R}_1$ to $\mathcal{R}_2$
frame exchange, the $x_2$ variable in the equation for $x_2$ is to
be considered as $x_2=x_1-vt,$ where $v$ is the relative speed of
$\mathcal{R}_2$ with respect to $\mathcal{R}_1,$ and this equation
does not reduce to the classical form. The left hand side of this
equation implies the meaning (specified above)
$$\frac{\partial^2_*\varphi}{\partial
t^2}(x_1-vt,t)\;-\;\frac{\partial^2\varphi}{\partial
x_1^2}(x_1-vt,t).$$ This means that, in the derivative of
$\varphi(x_1-vt,t)$ with respect to time, the first term is
obtained by considering the time dependence of the first variable
as coming only from the dependence on time of the variable
$x_1(t).$ The time dependence of $x_2(t)=x_1(t)-vt$ by the
intermediate of $vt$ is to be neglected, which is very natural
since the motion of the considered frame with respect to any other
one can not be a canonical feature of the wave propagation or more
generally of any real physical motion. If $v=0,$ both equations
for $x_1$ and $x_2$ are identical and reduce to
the classical one when written in any rest frame.\\

Obviously both operators $\square_1$ and $\square_2,$ as they have
been introduced using the operator $\square_0,$ are identical.
Then, both of them reduce, when written each in its appropriate
rest frame, to the classical operator $\square.$ The study of
these operators aimed only to prove that we can not privilege one
inertial frame at the expense of all others. All such frames have
the same right for expressing a physical law. Consequently, we can
not choose one of them and deduce the non validity of the
covariance principle relying on the only reason that the same
physical law is expressed differently in another inertial frame.
Therefore, only the operator $\square_*$ is canonical and
consequently it is the only one that can be used canonically in
order to traduce a physical law in term of local coordinates.\\

\subsection*{Canonical constancy of the light speed}
We can now
conclude that the previous framework is the proper context that
permits a correct interpretation of the principle of invariability
of the light speed in vacuum with respect to all inertial frames.
Indeed, if in an inertial frame $\mathcal{R}_0$ the trajectory of
a light beam is $x_0(t)=ct,$ then when we consider another frame
$\mathcal{R},$ moving uniformly following the $Ox$ axis of
$\mathcal{R}_0$ with a constant speed $v$ (fig.5), we have
$$x(t)\;=\;x_0(t)\;-\;vt\;=\;ct\;-\;vt$$
where $x(t)$ is the expression of the same trajectory in the frame
$\mathcal{R}$ since here we have $x(t)=A_t(x_0(t)-vt)$ and
$A_t=Id_{\mathbb{R}^3}.$ So we have
$$\frac{d_*}{dt}x(t)\;=\;\frac{d}{dt}x_0(t)\;=\;c \;(= 1), $$
which is the canonical speed of light, and
$$\frac{d_1}{dt}x(t)\;=\;\frac{d}{dt}x(t)\;=\;c-v$$
which conforms with the Galileo-Newtonian notion of speed.\\
This again shows the invalidity of the special relativity second
postulate when it is interpreted as saying that the speed of light
does not depend on the speed of the moving frame that is used to
measure it. Naturally the fundamental principle of the light speed
independence of the
moving source speed is perfectly valid.\\

We also can sum up the preceding by saying that when we write the
Maxwell equation in an arbitrary Galilean referential $\mathcal{R}$ under
the form
$$ \square \varphi(x,t) = \frac{\partial^2 \varphi}{\partial t^2}
(x,t) - \frac{\partial^2 \varphi}{\partial x^2} (x,t) = 0$$we
actually are considering implicitly that this frame is fixed and
that the speed of light in this frame is 1. The Maxwell equation
has to be written then as
$$ \square_* \varphi(x_1,t) = \frac{\partial^2_* \varphi}{\partial t^2}
(x_1,t) - \frac{\partial^2 \varphi}{\partial x_1^2} (x_1,t) = 0$$
in any other arbitrary Galilean referential $\mathcal{R}_1$ that moves with
an arbitrary given speed $v_1$ with respect to $\mathcal{R}$ and the speed
of light relatively to this frame is then $1-v_1$ $(\frac{d_*}{dt}x_1(t)=\frac{d}{dt}x_1(t)+v_1)$.\\
If we write, for example, the Maxwell equation with respect to the
proper referential of the Sun under the above first form, we
actually are considering that the speed of the light with respect
to the Sun is 1. The Maxwell equation in the proper referential of
the Earth has to be written under the above second form and the
speed of the light, emitted by the sun, with respect to this
referential is then
approximately $1-10^{-4}$.\\\\
Besides, we have to notice that there is, for us, a real physical
difference between the real motion (with respect to an initial
referential frame) of the inertial frame and the real motion of a
body or a particle. Actually, when we assume that the particle
(for example) is accelerating with respect to the initial
referential then it is accelerating with respect to the inertial
frame and the particle is (physically) radiating while if we
assume that the particle is at rest (or animated with a uniform
motion) with respect to the initial referential then whatever is
the motion of the moving frame it is obvious that the particle
is not radiating.\\\\
\textbf{Remark:} When we attest that the speed of light does
depend on the inertial frame used for measuring it, we simply mean
that:\\
When supposing that a light source is located at a supposed
constant distance $d$ on the $x$ axis of a Euclidean frame $R_1 =
(O_1, e_1, e_2, e_3)$ and that another frame $R_2 = (O_2, e_1,
e_2, e_3)$ which coincides with $R_1$ at time $t=0$, is moving
along the $x$ axis with a uniform speed $v >0$ (with respect to
the frame $R_1$) towards the source, then we attest that the light
that is emitted by the source at a time $t > 0$ reaches the
observer that is located at $O_1$ a little later than the observer
that is located at $O_2$. Similarly, if we suppose that $R_1$ is
fixed and the two sources are located, at time $t =0$, at the same
distance $d$ from $O_1$ and if we suppose that they both are
moving away from $O_1$, the first one with a uniform speed $v_1$
and the second one with a uniform speed $v_2 > v_1$ with respect
to $R_1$, then the light that is emitted by the first source at a
time $t
>0$ reaches the observer $O_1$ before the light that is emitted at
the same time by the second source.\\\\
In order to illustrate the significance of the error that is
committed when we hastily write the change of variables between
two Galilean referentials, we shall consider the following example
extracted from the excellent book on the Relativity theory
entitled: Relativit\'{e} - Fondements et appliquations ([7]). In
chapter 1 of this reference, we consider a light source located at
the origin $O'$ of a Galilean referential $R'= O' x' y' z'$, of
speed $v = \beta c$ (where $\beta = \frac{v}{c}$) with respect to
another Galilean referential $R = O x y z$. We call $E_1$ the
event "emission of the light at $O'$ at time $t=t'=0$", $E_2$ the event "reflexion
on a mirror $M'$", located on the $O'y'$ axis at a distance $l$
from $O'$ and $E_3$ the event "detection of the light at $O'$"
(fig. 12). When writing the coordinates of $E_1, E_2$ and $E_3$ in
$R'$ as being respectively
$$ E_{1_{R'}} \left\{%
\begin{array}{ll}
    x' = 0 \\
    y' = 0 \\
    z' = 0 \\
    ct' = 0 \\
\end{array}%
\right. \;\;\; E_{2_{R'}} \left\{%
\begin{array}{ll}
    x' = 0 \\
    y' = l \\
    z' = 0 \\
    ct' = l \\
\end{array}%
\right. \;\;\; E_{3_{R'}} \left\{%
\begin{array}{ll}
    x' = 0 \\
    y' = 0 \\
    z' = 0 \\
    ct' = 2l \\
\end{array}%
\right.$$ we have hastily decided that $ct' = l$. But this is only
the case of an exceptional photon and this fact does not allow us
to describe an arbitrary Galilean change of variables in this way.
In the referential $R'$, the distance travelled by an arbitrary
photon, during the time $t'$, is not equal to $l$ (i.e. $ct' \neq
l$), as we have showed in the train, the mirror and the two observers experiment.\\
Likewise, when we decide that, according to the Galilean
transformation, the events $E_1, E_2$ and $E_3$ are written, in
$R$ as being
$$ E_{1_{R}} \left\{%
\begin{array}{ll}
    x = 0 \\
    y = 0 \\
    z = 0 \\
    ct = 0 \\
\end{array}%
\right. \;\;\; E_{2_{R}} \left\{%
\begin{array}{ll}
    x = \beta l \\
    y = l \\
    z = 0 \\
    ct = ct' = l \\
\end{array}%
\right. \;\;\; E_{3_{R}} \left\{%
\begin{array}{ll}
    x = 2 \beta l \\
    y = 0 \\
    z = 0 \\
    ct = 2l \\
\end{array}%
\right.$$
we commit the same error by considering that $ct = l$ and
$ x = \beta l = v \frac{l}{c}$ as, in general, $ct' = ct \neq l$
and $\frac{l}{c} \neq t$. Thus, the fact of drawing the conclusion
that the distance between the events $E_1$ and $E_2$ is not
invariant by a Galilean change of referentials relying on the
equality $d' = l$ (where $d'$ is assumed of being the distance travelled by the light as measured by $R'$), is not legitimate as the distance travelled by
each photon is different from the distances travelled by all other
ones and the referential $R'$ can not be adequately used in order
to determine the "distance travelled by the light". Actually, this
distance $d$ can only be measured by $R$ and it verifies $d = (l^2 + \beta^2 d^2)^{\frac{1}{2}}$ which gives, for $v\ll c$, $d\simeq l (1 +
\beta^2)^{\frac{1}{2}} \simeq
\gamma l$.\\
In the same manner, when we use the relativistic transformation
$$ x = \gamma (x' + \beta c t')$$
$$y = y' \hskip 1.8cm$$
$$ z = z' \hskip 1.8cm$$
$$ ct = \gamma (ct' + \beta x'),$$ we write the coordinates of the
events $E_1, E_2$ and $E_3$ in the frame $R$ under the form
$$ E_{1_{R}} \left\{%
\begin{array}{ll}
    0 \\
    0 \\
    0 \\
    0 \\
\end{array}%
\right. \;\;\; E_{2_{R}} \left\{%
\begin{array}{ll}
   \gamma \beta l \\
    l \\
    0 \\
   \gamma l \\
\end{array}%
\right. \;\;\; E_{3_{R}} \left\{%
\begin{array}{ll}
    2 \gamma \beta l \\
    0 \\
    0 \\
    2 \gamma l \\
\end{array}%
\right.$$and so we commit the same error by declaring that $ct' =
l$.\\\\
We can also add to the preceding arguments that the fact of
considering negative times ($t<0$) is fundamentally incompatible
with the confirmed theory of the universe expansion which
pre-assumes the existence of an original time or an origin for
time labeled by $t=0$, starting from which (or more precisely just after
which) the time progresses homogeneously increasing ($t>0$) simultaneously with the expansion process.

\subsection*{Alternative interpretation of Michelson-Morley
experiment} In order to prove the deficiency of the results
obtained from the Michelson-Morley experiment, we will describe
first an experiment that can be used also for showing the
interpretation deficiency of the emitter in the middle of a track
(and the two mirrors on both sides) experiment. It also suggests
the way for analyzing other experiments such as the emitter in a
plane heading toward a given
observer.\\

So, we consider an emitter of light that is at rest with respect
to the earth and sending light rays parallel to the earth movement
(supposed to be uniform) toward a mirror located at a distance $L$
away from it. Let $0<v<1$ and $c=1$ be respectively the speeds of
the earth and the light with respect to a virtual fixed frame
coinciding at $t=0$ with the emitter and having its $Ox$ axis in
the
movement direction.\\
Let $t_1$ and $t_2$ be respectively the times taken by the beam
(or more precisely by a photon) to reach the mirror and then to
return to the emitter (fig.4$^{''}$). We obviously have, by
measuring both path lengths according to the fixed frame:
$$L\;+\;t_1v\;=\;t_1$$
and
$$L\;-\;t_2v\;=\;t_2.$$
We then have
$$t_1\;=\;\frac{L}{1-v}\qquad\qquad\mbox{and}\qquad\qquad
t_2\;=\;\frac{L}{1+v}$$ which yield
$$t_1\;+\;t_2\;=\frac{2L}{1-v^2}.$$
If we assume, according to the second part of the special
relativity second postulate, that the speed of light with respect
to the frame for which the emitter is at rest is given by $1,$ we
get
$$t^{'}_1\;+\;t^{'}_2\;=\;2L$$
where
$$t^{'}_i\;=\;\gamma(t_i\;-\;xv)\qquad\mbox{for}\qquad
i=1,2\qquad\mbox{and}\qquad \gamma\;=\;\frac{1}{\sqrt{1-v^2}}.$$

We will now show that this equality yields a contradiction. Indeed
\begin{eqnarray*}
2L =  t^{'}_1\;+\;t^{'}_2 &=& \gamma\left[t_1\;-\;(L+t_1v)v\right]+ \gamma [t_2 - (-t_2) v] \\
  &{}&((-t_2)\;\mbox{is the algebraic distance travelled by the} \\
 &{}& \mbox{light in the fixed frame})\\
  &=& \gamma(t_1+t_2-Lv-t_1v^2+t_2v) \\
  &=&\gamma\left(\frac{2L}{1-v^2}-Lv-\frac{L}{1-v}v^2+\frac{L}{1+v}v\right)
\end{eqnarray*}
and then
$$\gamma\left(\frac{2}{1-v^2}\;-\;v\;-\;\frac{v^2}{1-v}\;+\;\frac{v}{1+v}\right)\;=\;2$$
that is
$$\frac{2-v+v^3-v^2-v^3+v-v^2}{1-v^2}\;=\;2\sqrt{1-v^2}$$
or
$$\frac{2-2v^2}{1-v^2}\;=\;2\sqrt{1-v^2}$$
which yields $1=\sqrt{1-v^2}$ and then $v=0,$ which is absurd.\\

\noindent Otherwise, if we use the classical Galilean
transformation we have
$$y(t)\;=\;x(t)\;-\;vt\;=\;t\;-\;vt\qquad\qquad\;\;\;\;\quad\quad\mbox{for}\qquad 0<t<t_1\;\;\;\;\;\;\;$$
and (starting from time $t_1$)
$$y(t)\;=\;x(t)-v(t-t_1)\;=\;t_1\;-\;vt_1\;-\;(t-t_1)\;-\;v(t-t_1)\qquad\mbox{for}\qquad t_1<t<t_1+t_2,$$
where $x(t)$ is the abscissa with respect to the frame of the emitter when it was located at time $t_1$ and supposed as being fixed there, which gives
$$y(t_1)\;=\;t_1\;-\;vt_1$$
and
$$\qquad \qquad y(t_1+t_2)\;=\;t_1\;-\;vt_1\;-\;t_2\;-\;vt_2.$$
Now, $y(t_1+t_2)=0$  gives
$$t_1\;-\;t_2\;=\;v(t_1+t_2)$$
and
$$v\;=\;\frac{t_1-t_2}{t_1+t_2}$$
as it must be (the speed of the moving frame with respect to the fixed one is equal to distance divided by time).\\
Furthermore, we have
$$y^{'}(t)\;=\;1-v\qquad\quad\quad\mbox{for}\qquad 0<t<t_1\qquad$$
and
$$y^{'}(t)\;=\;-(1+v)\qquad\;\;\mbox{for}\qquad t_1<t<t_1+t_2.$$
Then we get
\begin{eqnarray*}
  y^{'}(t)t_1\;+\;(-y^{'}(t)t_2)&=&(1-v)t_1+(1+v)t_2 \\
  &=&L+L\;=\;2L
\end{eqnarray*}
which agrees with the Galileo-Newtonian notions.\\

Now, in the light of the two previous experiments, we can show
that the interpretations of experiments of type Michelson-Morley
are erroneous. Indeed, we assume (for simplicity) that the
apparatus is made up (as in [7]) with two mirrors $M_1$ and $M_2$
and a half-transparent strip $L_s$ which divides the incident beam
of light into two parts of equal intensities and that we realize
the interference of the waves originated from the image $S_1$,
given by $L_s$ and $M_1$ and the image $S_2$, given by $L_s$ and
$M_2$
(which is slightly slanted) (fig.13).\\
For the experimenter, the variation of the illumination at the
point $P$, where is lying the detector, depends on the difference
$\tau$ between the times taken by the waves originated in $S$ and
detected at $P$. More precisely, the interferential pattern
depends on the phase difference $2 \pi \nu \tau$, where $\nu$ is
the
frequency of the monochromatic light emitted by $S$.\\
Thus, if $\tau_1$ is the time taken by the light to go from $I$ to
$I_1$ and then return to $I$ and $\tau_2$ is the time taken by the
light to go from $I$ to $I_2$ and then return to $I$, we have
$\tau = \tau_2 - \tau_1$ as the distances $SI$ and $IP$ are
common. Let us now express $\tau_1$ and $\tau_2$ by means of the
length $I I_1 = I I_2 = l$, the relative speed
$\overrightarrow{v_{e}}$ of the laboratory, i.e. the speed of the
Earth with respect to a Galilean referential $R$ considered as
being fixed, and the speed $v = c$ of the light in $R$. The
Galilean relation relating the speeds $\overrightarrow{v}$,
$\overrightarrow{v_{e}}$ and $\overrightarrow{v'}$, where
$\overrightarrow{v'}$ is the speed of light with respect to the
referential $R'$ fixed to the Earth, is written as
$\overrightarrow{v} = \overrightarrow{v'}+
\overrightarrow{v_{e}}$. Now, as we have showed above, we have, in
$R$,
$$ l + t_1 v_e = vt_1 = ct_1 \;\;\; \mbox{and} \;\;\; l - t_2 v_e =
vt_2 = ct_2$$ which implies
$$ t_1 = \frac{l}{c - v_e} \;\;\; \mbox{ and } \;\;\; t_2 =
\frac{l}{c +v_e}$$and then
$$ \tau_2 = \frac{l}{c-v_e} + \frac{l}{c+v_e} =
\frac{2lc}{c^2-v_e^2} = \frac{2l}{c} \frac{1}{1-\beta_e^2} =
\frac{2l}{c} \gamma_e^2.$$ Likewise, by applying the Galilean
relation, we obtain, in $R'$,
$$ \tau^{'}_2 = \frac{l}{c-v_e} + \frac{l}{c+v_e} = \frac{2l}{c}
\gamma_e^2= \tau_2$$ as the speed of light in $R'$ is $c -v_e$ in
the going journey and $c+v_e$ in the return one.\\
In other respects, we have in $R$
$$\tau_1 = \frac{2d_1}{c} \simeq \frac{2}{c} \sqrt{l^2+\frac{v_e^2l^2}{c^2}}\simeq 2\gamma_e\frac{l}{c}$$
and then
$$ \tau = \tau_2 - \tau_1 = \frac{2l}{c} \gamma_e^2 - \frac{2l}{c}\gamma_e.$$ 
But, when we measure $\tau_1$
in $R'$ by using the diagram of figure 13, we are asserting that
the "velocity of light" in $R'$ is $\overrightarrow{c} -
\overrightarrow{v_{e}}$ which actually is the velocity of the
photon that stays continually at the vertical above $I$ and, as we
showed above, this is not the case in this experiment and so, we
can not maintain that
$$ || \overrightarrow{c} - \overrightarrow{v_e}|| = (c^2 + v_e^2 -
2 v_e c\; cos \theta)^{\frac{1}{2}} = (c^2
-v_e^2)^{\frac{1}{2}},$$where $\theta$ is defined by $cos \theta =
\frac{v_e}{c}$. Therefore, we can not push the reasoning forward
and maintain that
$$ \tau_1 = \frac{2l}{(c^2
-v_e^2)^{\frac{1}{2}}} = \frac{2l}{c} \gamma_e\quad\mbox{and}\quad \tau=\frac{2l}{c}\gamma_e(\gamma_e-1)\simeq \frac{l}{c}\beta^2_e$$
in order to
conclude that the difference of phase is
$$ \varphi = \varphi_2 - \varphi_1 = 2 \pi \nu \tau \simeq
\frac{2 \pi}{\lambda_0} l \beta_e^2.$$ 
Nevertheless, we can
correctly maintain that, measured in $R$, we have
$$ \tau \simeq \frac{2l}{c} \gamma_e^2 - \frac{2l}{c}\gamma_e.$$ 
So, when the apparatus is turned by
$90^\circ$, the roles played by the mirrors $M_1$ and $M_2$ are
reversed but, in both cases, $|\tau| = |\tau_2 - \tau_1 |$ is
unchanged. Consequently, in both cases, the difference between the
distances that are travelled by both of the beams of light is the
same; which explains the invariance of the interference pattern.\\\\
\textbf{Remark:} Although the proper time $\tau$ plays a
primordial role in the relativity theory, it is, in the framework
of our model, only a parameter that is related to the universal
time $t$ by the intermediate of the relation of change of
variable: $\frac{d \tau}{dt} = \frac{1}{\gamma}$, where  $\gamma =
\frac{1}{\sqrt{1-\frac{v^2}{c^2}}}$ is the  Lorentz factor, which
appeared in the work of W. Kaufmann as being a proportionality
factor such that the quantity $\gamma m_0$ (where $m_0$ is the
rest mass of a material particle) was qualified by him as being
the apparent mass of a material particle into movement. This
factor also appears in the above calculation of distances and it
will play an important role in our theory (as well as in the
relativity theory), notably for the definition of the global
energy and the momentum of a moving particle.\\

\subsection*{Geometrization of the physical universe}
 So, we propose in this paper to conserve the
pre-relativistic (Galileo-Newtonian) conception of space and time
and to consider the half-cone $C= \{(x,y,z,t) \in \mathbb{R}^4;\;
x^2+y^2+z^2 \leq t^2\;\mbox{and}\;t\geq0\}$ as being the world of trajectories in
$\mathbb R^4$, with some restrictions imposed on the real ones in
order to conform with the causality principle. Then, according to
the preceding properties, we can deduce that all mechanical and
physical laws, which are based on equalities implying speed
vectors, acceleration vectors and vector fields (or more generally
tensor fields), are invariant under frame exchanges carried out in
$(\mathbb R^3, g_t)$ implicating relative constant speeds and
constant isometries ($a'_0(t) = \vec {V_1}, b'_0(t) = \vec{V_2},
A, B \in O(g_{t})$ for any $t$). We can mention for example: the
fundamental laws of Mechanics, Maxwell's equation, the
conservation of energy principle,
the least action principle....\\\\
Besides, we have already proved, in the general case of a variable
metric (with variable curvature depending on time) $g_t$, some
properties of tensoriality with respect to frame exchanges and the
covariance of physical laws with respect to some of them, notably
those that involve isometries (for $g_t$) including all
isometries and not only a sub-group of their global group.\\\\
We then propose to go forward in the direction of the physical
reality of the universe taking into account the mass distribution,
the gravitational phenomenon and other manifestations of matter
such as electromagnetism and different forms of energy (excepting
quantum phenomena and energy singularities which are, within our
new framework, much easier to deal with).\\\\
Thus, let us first consider that an electrically neutral material
particle of (nearly constant) inertial mass $m_I$ is in a state of
free fall within a uniform gravitational field (locally in
$\mathbb R^3$) defined by $\vec{g}=(0,0,-g)$ in a fixed Euclidean
frame. When this particle is observed with a Euclidean frame
($O,e_1,e_2,e_3$) which follows itself a similar trajectory of free fall
such that $ e_3=(0,0,1) $, then (as Einstein showed it) the
particle seems as being at rest in this new frame. Likewise, if
this particle had an initial horizontal speed $\overrightarrow
{V_0}$ (in the fixed frame) at $t=t_0$, then its motion seems, in
the moving frame, as if it was uniform whereas it is in fact of
parabolic shape in the fixed frame. The moving frame having
uniform acceleration $\overrightarrow {\Gamma}=\overrightarrow
{g}$, in the fixed frame, plays then the role of an inertial frame
regarding the two Newtonian inertia principles. This same property
is valid for any other frame carrying out a free fall motion like
our previous moving frame. We call such a frame $\overrightarrow
{g}$-inertial frame. There exists (locally in $\mathbb R^3$) a
dynamical metric tensor $g_{ab}$ such that $\nabla
_{x'(t)}^{g_{ab}}x'(t)=0$ for any free fall movement described in
the fixed frame by $x(t)$, i.e. such that all trajectories
corresponding to a free fall movement are geodesics for this
metric (in other words, there exists theoretically a universal
metric for which all natural motion are geodesics). Indeed, in
favor of the presupposed symmetries of our (isolated) system and
using the coordinates in the fixed frame, we can assume that we
have
$$ g_{ij}=dx_{1}^2+dx_{2}^2+b(t,x_3,g)dx_{3}^2. $$Actually, we
have, using this metric denoted by $g_t(x)$:
$$||x||^{2}_{g_t(x)}=x_1^2+x_2^2+b(t,x_3,g)x_{3}^2$$and, if we
take $x=x_3e_3$, we
obtain$$||x_3e_3||^{2}_{g_t(x)}=b(t,x_3,g)x_{3}^2.$$Now the free
fall trajectory is described by
$$x_3(t)=a_3-\frac{1}{2}gt^2\;\mbox{ with
}\;x_3(0)=a_3,\;x_3(t_0)=0\;\mbox{ and
}\;a_3=\frac{1}{2}gt_0^2.$$So we get, for $g>0$:
$$t=\sqrt{\frac{2}{g}(a_3-x_3(t))}\;\;\;\;\;\;\;\;\;\mbox{ and }\;\;\;\;\;\;\;\;\;x'_3(t)=-gt.$$
Let us write the following equivalences:
$$\nabla_{x'_3(t)e_3}^{g_t}x'_3(t)e_3 =0\Leftrightarrow g^2t\nabla_{e_3}^{g_t}te_3=0\Leftrightarrow
$$$$g^2t^2\nabla_{e_3}^{g_t}e_3+g^2t\frac{d}{dx_3}\sqrt{\frac{2}{g}(a_3-x_3(t))}e_3=0\Leftrightarrow$$
$$t\frac{\frac{d}{dx_3}b(t,x_3,g)}{b(t,x_3,g)}=-\frac{d}{dx_3}\sqrt{\frac{2}{g}(a_3-x_3(t))}\Leftrightarrow$$
$$\frac{\frac{d}{dx_3}b(t,x_3,g)}{b(t,x_3,g)}=-
\frac{\frac{d}{dx_3}\sqrt{\frac{2}{g}(a_3-x_3(t))}}{\sqrt{\frac{2}{g}(a_3-x_3(t))}}\Leftrightarrow$$
$$b(t,x_3,g)=\frac{k(t)}{\sqrt{\frac{2}{g}(a_3-x_3(t))}}=\frac{k(t)}{t}.$$
For \emph{x}(\emph{t}) = $x_3(t)e_3$, we have
$$||x_3(t)e_3||^2_{g_{ij}(t)} = b(t,x_3,g)x_3^2(t) = x_3^2(t)
||e_3||_{g_{ij}(t)}$$ and then
$$b(t,x_3,g) = ||e_3||_{g_{ij}(t)} = \frac{k(t)}{t}.$$
Now
$$ ||x^{'}(t)||_{g_{ij}(t)} = ||x^{'}_3(t)e_3||_{g_{ij}(t)} = ||g t e_3||_{g_{ij}(t)} = g t
||e_3||_{g_{ij}(t)} = c(t)$$which yields
$$||e_3||_{g_{ij}(t)} = \frac{c(t)}{gt} = b(t,x_3,g)$$
Therefore, we obtain
$$g_{ij}(t) = dx_1^2 + dx_2^2 + \frac{c(t)}{gt} dx_3^2 = dx_1^2+
dx_2^2 + \frac{c(t)}{\sqrt{2g(a_3 - x_3(t))}}dx_3^2$$
This metric
obviously depends on the chosen level $x_3(0) = a_3$.\\
When the gravitational field $\overrightarrow {g}$ is central and
has a constant (Euclidean) norm $g$ with a fixed center $C$, then
we can integrate (locally) this gravity within a metric
$g_{ab}(t,x)$ by means of its canonical connection
$\nabla^{g_{ab}}$ by defining all its geodesics by
$\nabla_{x'(t)}^{g_{ab}}x'(t)=0$, where $x(t)$ denotes the
coordinates of these curves in a Euclidean frame originated in $C$
or any other fixed frame. In favor of symmetries (intuitive local
homogeneity and isotropy principles), we can determine the metric
$g_{ab}$ using the normal Riemannian coordinates around $C$ and
then transforming them into spherical coordinates in order to
obtain
$$g_{ab}(t,r)=b(t,r,g)dr^2+r^2d\sigma ^2$$where $b(t,r,g)$
decreases when $t$ or $g$ increases.\\
Thus, distances to the center and volumes are "inversely
proportional" to the intensity of gravity. When the center of
gravity $C(t)$ is mobile, then the free fall motion does not occur
following a straight line heading towards the center, but
following a curve $x(t)$ whose tangent vector is always pointing
towards the moving center. However, a central gravitational field
is never uniform; it always depends on the distance to the center
(and then on the parametrization time of free trajectories). If
\emph{C}(\emph{t}) is moving, then the field is radially constant
with respect to a frame that is fixed at the center
\emph{C}(\emph{t}). Nevertheless, if \emph{C} is fixed, we can
consider the gravitational field as being uniform for sufficiently
distant neighboring objects. If we now assume that $\|C'(t)\|<< 1$
and $g_t<<1$, we recover approximately the Euclidean metric. This
is the case when we are located (locally) at a reasonable distance
away from the earth surface. But, this is not, in general, the
situation that corresponds to the physical reality. Indeed,
although we can imagine an inertial frame so that the relative
speed of a given star, for instance, is sufficiently small, the
vector $\overrightarrow {g}$ (or rather $\overrightarrow {g_t}$)
depends strongly on the Euclidean distance of the body in movement
to the center of this star. Nevertheless, in a supposed isolated
system, we can use empirical methods to determine the geodesics
(trajectories of bodies corresponding to free fall motions) and
study their speed vectors and then their relative acceleration
vectors (by determining their relative deviations) and then deduce
the Christoffel symbols along these curves, associated with the
metric $(g_t)_{ab} $. We can also deduce the curvature tensor
$R^{a}_{bcd} (t) $ using the infinitesimal deviation equations
([4],p.3.3.18). We may use, in addition, the identity
$\bigtriangledown^{(g_t)_{ab}}(g_t)_{ab}=0$ as well as the
possible symmetries.\\

\subsection*{The physical metric}
These same considerations hold when we aim to determine (even
locally) the global metric $g_{ab} (t,x)$ of the expanding
universe $U(t)=(B(O,t),g_{ab}(t,x))_{t > 0}$, taking into account
the other phenomena (electromagnetism, radioactivity, energy
singularities and quantum phenomena) in order to integrate them
into the metric. This leads us to an Einstein's equation type
whose resolution gives an approximate metric $g_{ab}(t,x)$ which
could specify the trajectories associated with free motions, i.e.
the geodesics of the dynamical universe
($\nabla_{x'(t)}^{g_{ab}}x'(t)$=0). The metric
$g_{ab}$(\emph{t},\emph{x}) will
be called the physical metric.\\\\
Indeed, let us denote (according to the weak equivalence
principle) $m_I=m_g$ the inertial or gravitational mass of a
particle when it is located within a generalized gravitational field
$\overrightarrow {g_t}$, induced locally by a matter-energy
distribution, that is expressed by the metric $g_{ab}(t,x)$. Under
these conditions we have, for any trajectory $x(t)$, $$\vec{g_t}
=\vec{ \Gamma_t}=\bigtriangledown^{g_e}_{x'(t)} x'(t)=x''(t)$$when
using any virtual fixed Euclidean frame and
$$\bigtriangledown^{g_{ab}}_{x'(t)} x'(t)=0$$for any given free
fall motion as we will show in the next section. So, the
$g_t$-inertial mass $m_I(t)$ depends on time by means of the
metric $(g_t)_{ab} $ which reflects all sorts of energy forms and
energy fluctuations.\\\\
If ${\cal {R}}_0$ is a fixed Euclidean frame, ${\cal {R}}_1$ is an
identical frame to ${\cal {R}}_0$ whose origin describes the curve
$a(t)$ with respect to ${\cal {R}}_0$, and if $x(t)$ is any
trajectory (in ${\cal {R}}_0$), let us define
$$\widetilde{\Gamma}_0(x(t)):=\nabla_{x'(t)}^{g_t}x'(t)$$and
$$\widetilde{\Gamma}_{01}(x(t)):=\nabla_{x'(t)}^{g_t}x'(t)-
\nabla_{a'(t)}^{g_t}a'(t)=\widetilde{\Gamma}_0(x(t))-\widetilde{\Gamma}_0(a(t)).$$
$\widetilde{\Gamma}_{01}(t) := \widetilde{\Gamma}_{01}(x(t))$ is
in fact the dynamical acceleration of the trajectory $x(t)$ with
respect to the frame ${\cal {R}}_1$ which has its proper dynamical
acceleration $\widetilde{\Gamma}_0(a(t))$ =:
$\widetilde{\Gamma}_0(t)$. Then we have
$$\widetilde{\Gamma}_{01}(t)=\widetilde{\Gamma}_{0}(x(t)) \mbox { if and only
if the frame} {\cal {R}}_1 \mbox{ is } g_t-\mbox{ inertial (i.e.}
\nabla_{a'(t)}^{g_t}a'(t)=0).$$ According to these notions and
notations, we can state
$$x(t)\mbox{ is a geodesic for the dynamical metric}
\Leftrightarrow \widetilde{\Gamma}_{0}(x(t))=0,$$and then we have,
for all such ${\cal {R}}_1$:
$$ \widetilde{\Gamma}_{01}(t)=0 \Leftrightarrow
a(t)\mbox{ is a geodesic }.$$
This constitutes a generalization of
the first principle of inertia which states that\\ "a movement is
uniform in an inertial frame if and only if it is uniform in any
other inertial
frame".\\\\
Notice that $\nabla_{x'(t)-a'(t)}^{g_t}x'(t)-a'(t)$ is not
necessarily null when
$\nabla_{x'(t)}^{g_t}x'(t)=\nabla_{a'(t)}^{g_t}a'(t)=0$ unlike the
property $\nabla_{x'(t)-a'(t)}^{g_e}x'(t)-a'(t)=0$ when
$x''(t)=a''(t)=0.$\\\\
In the light of the universe physical realities, the fundamental
principle of Mach adopted by Einstein (Matter=curvature) and the
general principle of modeling (Physics=Geometry), we have to deal
with all aspects of matter-energy. This leads us to the
symmetrical (0,2)-tensor, defined on $B(O,t) \subset
\mathbb{R}^3$, and modified in a sense that will be specified
below, which is the (mass-energy) tensor $T_{ab}^{*}(t)$ whose
variable independent elements are only six.\\\\
We consider then the physical universe identified, at every time
$t > 0$, to$$U(t)=(B(O,t) , g_{ab}(t,x))$$where $B(O,t)$ is the
Euclidean ball of radius $t$ and $g_{ab}(t,x)$ is the
(regularized) Riemannian metric associated to the physical
universe at time $t$. Recall that this metric is variable with the
position and time and its curvature is, in general, positive and
is itself variable with the position and time. This curvature is
due to the distribution of matter-energy at any time $t$ and then
essentially to the entire gravitational field and other
manifestations and effects of matter.\\
The half-cone of space and time is constituted of all sections
$t=const.$ ($t\geq 0$), which are the ball-hypersurfaces of
$C\subset\mathbb R^4$ (except for \emph{t} = 0). They are
orthogonal to the time axis for the Minkowski metric on $C$. Each
of these balls, $B(I,t_0)$ of center $I(0,0,0,t_0)$ and Euclidean
radius $t_0$, is provided with the Riemannian metric $g_{t_0}(x)$
which depends on the position $x$ and has a variable curvature.
The Einstein's equation is written (within our new context) as
$$ G_{ab}(t):= R_{ab}(t)- \frac {1}{2}R(t) g_{ab}(t)=:T_{ab}^{*}(t)\hskip 2cm    ({\cal {E}})   $$
where $R_{ab}(t)$ is the Ricci (0,2)-tensor associated with the
metric $g_{ab}(t)$ in $U(t) \subset \mathbb{R}^3$, $R(t)$ is the
scalar curvature of $U(t)$ and $ T_{ab}^{*}(t)$ is the symmetrical
matter-energy tensor. This tensor depends naturally on the density
$\rho(t)$ of the mass distribution, the ambient gravitational
field, the pressure $P(t)$ resulting from phenomena typically
associated to perfect fluids, the global electrical field $
\vec{E(t)}$ and the global magnetic field $\vec{B(t)}$ and some
other energy manifestations. All these tensorial objects depend on
position and time according to whether space has, locally and at
fixed time, a matter dominance or a radiational one. We notice
that we have
$$\bigtriangledown^a T_{ab}^{*}(t)=0$$(according to the second Bianchi
identity) where $\bigtriangledown=\bigtriangledown (g_{ab}(t))$ is
the Levi-Civita connection associated with the Riemannian metric
$g_{ab}(t,x)$, insuring the validity of the conservation of energy
law. Moreover, the trajectories of bodies which are only under the
action of natural forces are substantially and theoretically
geodesics for this metric. The crucial task is the approximate
(local) resolution of Einstein equation on the base of a bank of
dynamic data as precise as possible. This will be achieved in section 10.\\\\
On the other hand, we notice that the three bodies' problem (or
more generally any $n$ bodies' problem), whether they constitute
or not an isolated system, has to be treated within the above
setting. If $(g_t)_{ab} $ is the (local) ambient metric and if
$x_i(t) (i=1,2,3)$ describes the trajectory of any of these bodies
in any virtual fixed frame, then we have
$$\bigtriangledown^{(g_t)_{ab}}_{x_i'(t)} x_i'(t)=0 \quad
\mbox{for } i=1,2,3.$$If we assume that these three bodies form an
isolated system, then we can conversely use these equations
together with
all data of the problem to (locally) recover the metric $(g_t)_{ab}$.\\\\
More generally, we can locally determine the metric $(g_t)_{ab}$
using empirical methods. If $X(t)$ denotes the trajectory of a
particle or a body moving only under the natural forces (i.e.
forces originated by non singular energy phenomena. Such a motion
will be qualified as a free motion) with respect to any fixed
frame, then we have
$$\nabla _{X'(t)}^{g_t}X'(t)=0\;\mbox{ at every time }\;t.$$
So, the empirical or numerical determination of sufficiently many
geodesics $X(t)$ permits to determine approximately the
Christoffel symbols and the metric $g_t$ at each point of $X(t)$.
Therefore the metric $g_t$ is determined either by means of its
geodesics or by determining the tensor $T_{ab}^{*}$ and then
resolving the simplified Einstein's equation.\\
We notice that, when $T_{ab}^{*}=0$ on a region $D\subset B(O,t),$
then this means that, according to our definition of this tensor,
the region \emph{D} is not only devoid of Matter, but also it is
apart from any energy influence of matter and so we have $g_{ab}$
= $g_e$ on \emph{D} as well as $R_{ab}$ = 0 and $R$ = 0. In the
particular cases (perfect fluids, electromagnetic fields, scalar
fields of Klein-Gordon), the asymptotical cases and
quasi-newtonian cases, the resolution of equation $({\cal {E}})$
is much easier, within our
model, than within the framework of the standard general relativity. Some of these issues will be reviewed below.\\\\
\textbf{Remark:} Our tensor $T^*_{ab}$ and our associated physical
metric $g_t$ incorporate, within their definition all
matter-energy forms and effects which is not the case of the
Einstein's tensor and its associated spacetime metric. The
Einstein's tensors and metric reflect only the gravitational field
that is caused by a given mass in the presence of a matter field
(such as a perfect fluid with or without pressure) and possibly an
electromagnetic field or in the vacuum. The vacuum Einstein's
equation
$$ R_{ab} - \frac{1}{2} R \; g_{ab} = 0$$ is characterized by
$\rho$ =0, $T_{ab}$=0, $R_{ab}$=0 and $R$=0 and does not imply
that the spacetime metric is the flat one.\\
Moreover, when we introduce the cosmological constant $\Lambda$,
the vacuum Einstein's equation becomes
$$ R_{ab} = \frac{1}{2} R\; g_{ab} - \Lambda \; g_{ab}$$ and then
we have $R= 4 \Lambda$ and if $\Lambda \neq$ 0 we have $R \neq$ 0
which implies that the vacuum geodesics are not the same as those
of the flat spacetime. Thus, in the light of our model, $\Lambda$
reflects the influence of the cosmic matter on the regions where
there is no matter (even though there are the intergalactic
gravity, the cosmic radiations and the neutrinos) and logically it
must depend on time and probably on space regions. For us a space
region is in a state of absolute vacuum (i.e. $T^*_{ab}$ =0) if
and only if it is (nearly) apart of all matter manifestations as
well as of their effects.\\\\
Finally, we mention that if $P_0$ is an isotropic observer (i.e.
moving approximately in the expansion direction at the light
speed), the horizon particle for him is roughly the union of all
balls of center $P_t$ that are located on the isotropic line
$(OP_0)$ and therefore this horizon is the half-space located
beyond the perpendicular plane at $P_0$ to this line
(fig.6).Clearly, this horizon is purely theoretical because of the
singularities (essentially the black holes) that are scattered in
the real physical universe. An ordinary observer can only see (or
interact with) a little part of the expanding universe.\\

\subsection*{ Comparison with the relativistic invariance of the
speed of light}

Within the special relativity framework, the speed of light is
independent of the Galilean observer which generally is
represented into the relativistic spacetime ($\mathbb{R}^4,\eta$)
(where $\eta$ is the Minkowsky metric
$-dt^2+dx_1^2+dx_2^2+dx_3^2$) by a timelike straight line $D$.
Considering a normal parametrization $c: I \subset \mathbb{R}
\longrightarrow \mathbb{R}^4$ of $D$ (i.e. $\eta (c^{'}(t),
c^{'}(t))$ = -1) and  an arbitrary parametrization $\widetilde{c}:
J \subset \mathbb{R} \longrightarrow \mathbb{R}^4$ of another
Galilean observer $\widetilde{D}$, we can write
$$ {\widetilde{c}}\;^{'}(t) = \overrightarrow{k} + \alpha c^{'}(t)$$
for $\overrightarrow{k} \in c^{'}(t)^{\perp}$ and $\alpha \in
\mathbb{R}$, where $c^{'}(t)^{\perp}$ is the orthogonal hyperplane
(for the metric $\eta$) of the vector $c^{'}(t)$ which also is the
set of simultaneous points of $c(t) \in D$ at time $t$. We then
define the relative speed vector of the observer $\widetilde{D}$
with respect to $D$ by
$$ {\overrightarrow{v}}_{\widetilde{D}/D} =
\frac{\overrightarrow{k}}{\alpha}$$and if $\widetilde{c}$ is a
normal parametrization with respect to $D$ of $\widetilde{D}$,
then we have $\alpha$ =1 and
${\overrightarrow{v}}_{\widetilde{D}/D} =
\overrightarrow{k}$.\\
One then proves that the relative speed of light with respect to
an arbitrary Galilean observer $D$ is $c$ = 1.\\This is
established by considering a lightlike line $L$ parametrized by
$\widetilde{c}(t) = M + tl$ with $\eta(l,l)$ =0 and $M$ is an
arbitrary point of the timelike cone and taking $A = c(t) \in D$
and $B = \widetilde{c}(t)$ the point of $L$ that is simultaneous
to $c(t)$ for $D$ and writing
$$ {\widetilde{c}}\;^{'}(t) = l = \overrightarrow{k} + \alpha
c^{'}(t)$$where $\alpha \in \mathbb{R}$ can be considered as being
positive and $\overrightarrow{k} \in c^{'}(t)^{\perp}$ and finally
by considering the relative speed vector as being
$$ \overrightarrow{v}_{L/D} =
\frac{\overrightarrow{k}}{\alpha}$$that is assumed to represent
the speed vector of light with respect to the observer $D$.\\
Thus, the relation $\eta(l,l) =
\eta(\overrightarrow{k},\overrightarrow{k}) + \alpha^2
\eta(c^{'}(t),c^{'}(t)) =
\eta(\overrightarrow{k},\overrightarrow{k}) - \alpha^2 = 0$
implies
$$\alpha = \sqrt{\eta(\overrightarrow{k},\overrightarrow{k})} \;\;\;\; \mbox{
and} \;\;\;\; \parallel \overrightarrow{v}_{L/D} \parallel_\eta =
\sqrt{\eta
(\frac{\overrightarrow{k}}{\alpha},\frac{\overrightarrow{k}}{\alpha})}
=
\frac{\sqrt{\eta(\overrightarrow{k},\overrightarrow{k})}}{\alpha}
= 1.$$ 
For us, this reasoning is only valid for a stationary
observer i.e. when $D$ identifies with the time axis where we have
$c(t) = N + t(1,0,0,0) \hskip 0.1cm$, $\hskip 0.1cm$ $c^{'}(t) =
(1,0,0,0) \hskip 0.1cm$, $\hskip 0.1cm$ $c^{'}(t) ^{\perp}$ is a
hyperplane that is parallel to $(O,x_1,x_2,x_3)$ for $x_1,x_2,x_3
\in \mathbb{R}$ and $\widetilde{c}(t) = M +
t(1,\frac{1}{\sqrt{3}},\frac{1}{\sqrt{3}},\frac{1}{\sqrt{3}})$
satisfies $\eta
({\widetilde{c}}\;^{'}(t),{\widetilde{c}}\;^{'}(t)) =0$.\\
When we consider an arbitrary Galilean observer, $c^{'}(t)
^{\perp}$ is not necessarily (within the relativistic framework)
the spacelike hyperplane previously considered).\\
The previously established relation for the Galilean observer $D$
and $\widetilde{D}$ which led to the above definition of the
relative speed $\overrightarrow{v}_{ \widetilde{D}/D } =
\overrightarrow{k}$ (within the special relativity framework) does
not have for us any physical significance. This definition, as
well as the relativistic spacetime notion, is only conceived in
order to justify the erroneous aspect of the special relativity
second postulate which stipulates that the speed of light is the
same for all Galilean observers. Actually, this aspect of this
postulate has been adopted by Einstein only in order to conform
with the Galilean relativistic principle which stipulates that all
physical laws (and particularly the Maxwell's laws) have to be the
same for all inertial observers. But we have just now proven that
this was useless by giving a physical content to the notion of
Maxwell's equation
canonicity.\\

For us (as it will be shown in section 5), the real physical
universe actually is a part of $\mathbb{R}^3$ that is modeled, at
every time $t$, by the Euclidean ball $B(O,R(t))$ of
$\mathbb{R}^3$ provided with the physical metric $g_t$ (that
reflects the physical consistency of the universe) and the measure
of any observer speed or any trajectory speed into $B(O,R(t))$,
that is parametrized into the space-time semi-cone
$$ C = \{ (x,y,z,t) ; x^2+y^2+z^2 \leq R^2(t) \sim t^2, t> 0 \}
\simeq \bigcup_{t > 0} B(O,t) \times \{t\}$$by $c(t) =
(t,x_1(t),x_2(t),x_3(t))$ and into $B(O,R(t))$ by $X(t) =
(x_1(t),x_2(t),x_3(t))$, is given by $||X^{'}(t)||_{g_t}$. So,
this speed is measured within $B(O,R(t))$ using the physical
metric $g_t$ and not within the semi-cone $C$ of $\mathbb{R}^4$
and if we denote by $h_t := dt^2 - g_t$ the metric on the
semi-cone that is associated with $g_t$ at time $t$, we have
$$ 0 < ||c^{'}(t)||_{h_t} < 1 \;\;\; \mbox{ and } \;\;\; ||X^{'}(t)||_{g_t} <
1$$for any observer, meanwhile
$$ ||c^{'}(t)||_{h_t} = 0 \;\;\; \mbox{ and } \;\;\;
||X^{'}(t)||_{g_t} = 1$$for the trajectories that characterize the
light rays (which actually are also characterized by $\nabla_{X^{'}(t)}^{g_t} X^{'}(t) = 0$).\\
Thus, when a part of the universe can be assimilated to the
absolute vacuum, we then have (within this part) $g_t = g_e$ and
$h_t = dt^2 - g_e = - \eta$ and we can consider two Galilean
observers $D_1$ and $D_2$ that will have a relative speed with
respect to a stationary observer $D_0$ (represented by the time
axis or any vertical straight line) for each of them. So, if
$c_1(t)$ and $c_2(t)$ are the normal parametrizations of $D_1$ and
$D_2$, then we have
$$ c^{'}_1(t) = \overrightarrow{k_1} + (a,0,0,0)$$
$$ c^{'}_2(t) = \overrightarrow{k_2} + (b,0,0,0)$$where
$\overrightarrow{k_1}$ and $\overrightarrow{k_2}$ are the parallel
to the time-axis projections on $B(O,R(t))$ (which is an
orthogonal projection) of $c^{'}_1(t)$ and $c^{'}_2(t)$ and the
relative speed of $D_2$ with respect to $D_1$ is
$$\overrightarrow{v}_{D_2 / D_1} =: \overrightarrow{k_r} =
\overrightarrow{k_2} - \overrightarrow{k_1}.$$ 
Similarly, the speed of light with respect to $D_0$ is in that case $||
\overrightarrow{k}||_{g_e}$ = 1 where $\widetilde{c}(t)$
designates the light trajectory and ${\widetilde{c}}\;^{'}(t) = l$
is written as $l = \overrightarrow{k} + (1,0,0,0)$ with $h(l,l) :=
(dt^2 - g_e)(l,l)$ =0. Moreover, the relative speed of light with
respect to the observer $D_1$ is
$\overrightarrow{k}-\overrightarrow{k_1}$. Thus, if $D_0, D_1$ and
$L$ are located in the same plane and if $v_{D_1 / D_0} = v$,
then the relative speed of light with respect to $D_1$ is $1-v$.\\
Concerning the real physical universe $(B(O,R(t),g_t)$, we can
only define the instantaneous relative speed vector of the light
with respect to an observer $D_1$ at the intersection point of
$D_1$ with the light trajectory $L$ by
$\overrightarrow{k}-\overrightarrow{k_1}$ having the physical
magnitude $||\overrightarrow{k}-\overrightarrow{k_1}||_{g_t}$
where $\overrightarrow{k}$ and $\overrightarrow{k_1}$ are the
sapcelike associated vectors with $L$ and $D$. this magnitude
reduces to $||\overrightarrow{k} - \overrightarrow{k_1}||_{g_e}$
within the absolute vacuum.

\subsection*{Simultaneity}
On the other hand, we shall illustrate, with the help of a classical example (c.f. [2]), that simultaneity has a universal character. Actually, let us assume that, in a supposed fixed inertial frame $R$, a particle is produced with a constant speed $v$ and that, when it is located at the point $x=0$, it decays into two photons $p_1$ and $p_2$ that propagate along the $Ox$ axis in opposite directions. If detectors $D_1$ and $D_2$ are located at points $x=-L$ and $x=L,$ then $p_1$ and $p_2$ hit $D_1$ and $D_2$ at the same time $t=\frac{L}{c}$ because the speed of these photons in $R$ is $c$. When we analyze this event by means of the frame $R'$ for which the particle is at rest using the relativistic transformation formulas we find that $p_2$ hits $D_2$ before that $p_1$ hits $D_1$. This leads to the fact that two events which are simultaneous in the frame $R$ are not simultaneous in $R'$.\\
Our interpretation is completely different. For us, when these photons are emitted, each of them heads along its own direction with the speed $c$ (in $R$) independently of the particle that gave birth to them and of the frame $R'$. When one insists to use $R'$, which has the speed $v$ with respect to $R$, and the galilean transformation one obtains the following result:\\
In $R'$, the speed of $p_2$ is $c-v$ and $p_2$ travels the distance $L'-vt_2-0=L-vt_2$, where $t_2$ is the time taken (in $R'$) by this photon before hitting $D_2$; which gives $t_2=\frac{L-vt_2}{c-v}$ and then $ct_2=L$. Likewise, the speed of $p_1$ in $R'$ being $c+v$ and the distance made by $p_1$ being $L'+vt_1-0=L+vt_1$, where $t_1$ is the time taken (in $R'$) by this photon before hitting $D_1$, then we have $t_1=\frac{L+vt_1}{c+v}$ which implies $ct_1=L$. Thus, we obtain $ct_2=ct_1$ and $t_1=t_2=\frac{L}{c}=t$.

\subsection*{Final remarks}
We mention that the relativistic phenomenon of the length contraction is logically and physically unconceivable. Indeed, if we consider two identical metallic sticks $s$ and $s'$ having the same length $l$ and if the stick $s$ is fixed along the $Ox$ axis of a supposed fixed frame $R$ and if the stick $s'$ is moving uniformly along $Ox$, then the use of the relativistic formulas relating the coordinates written in $R$ to those written in the frame $R'$, for which the stick $s'$ is at rest, leads to the following physical aberration:\\
When we use the frame $R$ for analyzing the situation, we would see that at the time when the origin of the stick $s'$ coincides with that of $s$ the extremity of $s'$ does not coincide with that of $s$ and at the time when the extremity of $s'$ coincides with that of $s$ then the origin of $s'$ does not coincide with that of $s$ (c.f. [7]). This situation which could be theoretically accepted is not physically conceivable.\\
The preceding shows, one more time, that the hypothesis of the invariability of the speed of light with respect to all Galilean frames and the introduction of the relativistic notion of spacetime lead to physically unrealizable situations even though they appeared to have succeeded  to resolve some illposed problems without forgetting the theoretical complications originated in the consideration of the most general Lorentzian spacetime when we are studying the general relativity which is the authentic theory that thoroughly explains the laws of the universe expansion.\\
Finally, we notice that, for us, the paradox of Langevin's twins has no any relation to the reality as the relativistic proper time is just a practical and useful parameter that has no any real significance when compared to the absolute universal time.

\section{Matter-Energy, black holes and inertial mass}

$\hskip 0.5cm$We begin this section by mentioning that the whole
of this study is based on all main, seriously confirmed principles
of Mechanics and Physics, established by Newton, Einstein, Hubble
and many others, that coincide in all special cases with the
codified, measured and verified laws by a large number of
physicists as Maxwell, Lagrange, Hamilton, Shr\"{o}dinger, Bohr
and Planck and perfectly modelled by means of the work of Gauss,
Euler and
Riemann between many others.\\\\
Our starting line here is the following simplified and modified
Einstein's equation
\begin {equation}\label{r12}
 G_{ab}:=R_{ab}-\frac{1}{2}Rg_{ab}=:T^{*}_{ab}
\end {equation}
where all symmetrical tensors used here are defined on the ball
$B_{e}(O,R(t)) \subset \mathbb{R}^3$ which will be considered as
being the Euclidean ball $B_{e}(O,t)$ for sake of simplicity. This
reduces in fact to assume that the universe expansion speed is the
same as the electromagnetic waves' speed which is assumed of being
always the same as the light speed in the absolute vacuum (i.e.
$c=1$), although this is incorrect for sufficiently small $t$ (the general case will be discussed at the end of this article).\\
Recall that all tensors here describe physical and geometrical
phenomena that are inherent in the real physical universe at time
$t$. The distribution of matter located in a region of the
universe determines, at every time $t$, a distribution $m_t(X)$ of
inertial mass which gives birth to a global gravitational field
and other force fields. This inertial mass is subject to many
different evolutive energy transformations (electromagnetic,
thermonuclear and radioactive).\\In addition to the inertial
static aspect of matter (characterized by the inertial mass) and
to this evolutive aspect, we must add the dynamical aspect, i.e.
the creation and the transformation of kinetic energy by the
movement of matter. We notice that here the word energy must be
interpreted with a wide, total and unified
sense.\\
The physical metric $(g_t)_{ab}$ or $g_t$ (depending on position
and time) reflects, at every time $t$, the permanent perturbation
of the geometrical flat space ($B_e(O,t),g_e$). This perturbation
is created, according to Mach-Einstein principle, by the matter -
energy distribution; it is expressed by the creation of spatial
curvature which is taken into account in the equation (12) through
the Ricci tensor $R_{ab}(t)$ and the non negative scalar curvature
$R(t)$. So the metric $g_t$ measures in fact the effect of the
matter contained into the space rather than the geometric volume
of this space which is conventionally measured by the Euclidean
metric $g_e$. However, we notice that these tensors, contrary to
Riemannian tensors, can admit singularities that are essentially
related to the phenomenon of collapsing of matter generating black
holes. The equation (12) is written as
\begin {equation}\label{r13}
Rg_{ab}=2(R_{ab}-T^{*}_{ab})=:G^{*}_{ab}.
\end {equation}
It contains, at every time $t$, all geometrical, physical and
cosmological features of the universe.\\
If we assume that $G^{*}_{ab}$ vanish on a ball $B:=B(I,r)\subset
B(O,t)$, then $$R_{ab} = 0 \mbox{ is equivalent to } T^*_{ab} = 0
\mbox{ on } B,$$ but, according to the definition of the tensor
$T^*_{ab},$ the relation $T_{ab}^{*} = 0$ is equivalent to
$g_{ab}$ = $g_e$ on \emph{B} and then we have $R=0.$ If $R \not=0$
on a neighborhood of $I$ in $B$, we have
$g_{ab}=\frac{1}{R}G^{*}_{ab}=0$ on this neighborhood and then $R
= 0$, which is absurd. Likewise, $R$ can not vanish at any
isolated point in $B$ (at $I$ for example) and even on any curve
of empty interior in $B$ (which passes through $I$ for example)
since, out of this point or this curve, we would have $R \not= 0$,
which also is impossible. Therefore if $G^{*}_{ab}=0$ on $B\setminus I$, $g_{ab}$
can not be neither a vanishing nor a non vanishing (0,2)-tensor of class $C^2$ on $B$ with
$R \not= 0$. However, $g_{ab}$ can then be considered as being a
distribution whose support is included in $I$ (for symmetry
reason) and having the form $k \delta _{I}g_e$ where
$$\delta _{I}g_e(X(I),Y(I))=g_e(X(I),Y(I))$$for any two tangent
vectors $X(I)$ and $Y(I)$ at $I$ and$$\delta
_{I}g_e(X(P),Y(P))=0$$ for $P\in B$, $P\not=I$ and any two tangent
vectors $X(P)$ and $Y(P)$ at $P$.\\
Actually, this situation corresponds to the fact of a formation of
a black hole created by a complete concentric collapsing of a
material agglomeration having a very large volume density of its
inertial mass, which is expressed by a very large intensity of
central self gravity. So $g_{ab}$ can not be null and therefore
$supp (g_{ab})$ is reduced to the center of mass $I$ and we have
(as we will show it below):
$$g_{ab} \simeq (V_e(B) - E)\delta_I g_e \hskip 0.5cm , \hskip 1cm dv_g \simeq (V_e(B) -
E)\delta_I$$
$$dv_g(I) \simeq V_e(B) - E \hskip 0.5cm \mbox { and } \hskip 0.5cm E_t(X) \simeq E\delta_I$$
where $\delta_{I}$ is the Dirac mass at \emph{I} and \emph{E} is
the equivalent mass energy to the global inertial mass \emph{m} of
the agglomeration just before the last phase of the collapsing.
The point $I$ is the center of a central gravity which absorbs all
particles reaching $B$. Thus the black hole $B$ constitutes, in
some meaning, a region of total absorption of matter as well as of
electromagnetic waves (a region of no escape). In other words, $B$
is the Schwarzchild ball that is characterized by the fact that
(roughly speaking) no signal can be emitted from it. The total
potential mass energy concentrated in $I$ of a black hole $B$ is,
approximately, $E \sim m$. At $I$, we can consider the
matter-energy volume density and the curvature as being infinite.
$I$ is a singularity of the physical space $B(O,t)$.\\Moreover we
neither can have $g_{ab}$ of class $C^2$ with $R \not= 0$ such as
one of the eigenvalues $\lambda _i(t,X)$ ($i=1,2,3$) of
$G^{*}_{ab}$, being null (resp. two among them being null) on a
ball $B(I,r)$ since we then would have $vol (B,g_{ab})=0$ and
$g_{ab}$ would be, using a $g_e$-orthonormal basis of
eigenvectors, of a diagonal form like
$$\frac{\lambda_1(t,X)}{R}dx_1^2+ \frac{\lambda _2(t,X)}{R}dx_2^2 \;\;\;\;\;\; \mbox{(resp. } \frac {\lambda _1(t,X)}{R}dx_1^2 \mbox{)};$$ which
constitutes a physical phenomenon that is incompatible with the
(intuitive) local isotropy and homogeneity principles. So, each of
these cases implies the vanishing of all eigenvalues on $B$ and
leads again to the identity $G^{*}_{ab}=0$ on $B$ and to the same absurdity. We are again in the situation of a static Schwarzchild black
hole or other types of black holes.\\\\
Coming back to the universe $U(t),$ let us consider a material
agglomeration filling a connected region included in $B(O,t).$ Let
\emph{m}(\emph{t}) $\sim$ \emph{e}(\emph{t}) be its global
inertial mass or equivalently its global potential matter energy.
Now we propose to distinguish between two kinds of gravity
inherent in this agglomeration of matter: the internal gravity
(responsible, together with other interaction forces, of the
cohesion or the non dispersion of the agglomeration) defined
within this region and the external gravity defined around this
region. For the universe at $t=0$, having the total inertial mass
$M_0$ converted into the total energy $E_0$ (supposed to be finite
as it is generally admitted by physicists), the totality of
gravity before the Big Bang is internal. For the center $I$ of a
static black hole $B(I,r)$, internal gravity is concentrated in
$I$ and external one is defined, in extremal form, within $B$ and
extended around $B$ as classical newtonian gravity. For an
isolated material system such as, for instance, a galaxy with its
significant gravitational extent, having a global inertial mass
$m$, the global gravity of this system is essentially internal,
whereas if we consider any star of this galaxy, then we have to
distinguish between its internal gravity and its external one
within this system. This is still true for all scale of material
formations. We will show later on that internal gravity is
strongly related to the binding energy and binding
forces.\\Likewise, we must distinguish, in the dynamical universe,
between kinetic energy and potential mass energy of a material
system when it is moving, along a trajectory $X(t)$, with the speed $v(t):=\|X'(t)\|_{g_t}\leq\|X'(t)\|_{g_e}<1$. The first one is
in fact equal to
$$E_k(t)=\frac{1}{2}m_1(t)v^2(t)$$where $m_1(t)$ is its reduced inertial mass
at time $t$ that will be defined below. For the universe at $t=0$,
the kinetic energy is null and the potential mass energy is $E_0
\sim M_0$. For an isolated material system such as a galaxy moving
with a speed $v$, we have $E_k=\frac{1}{2}m_1(t) v^2$ and the
total energy is
$$E(t)=m(t)c^2+\frac{1}{2}m_1(t)v^2(t)$$
where $m(t) = \gamma(t) m_0 :=
(1-\frac{v^2(t)}{c^2})^{-\frac{1}{2}} m_0$ has been qualified by
W. Kaufmann as being the apparent mass of the moving particle.\\
Recall that, for any freely moving particle, we have (in a virtual
fixed frame) $\widetilde{\Gamma}(t)=\nabla
_{X'(t)}^{g_{ab}}X'(t)=0,$ whereas
$\Gamma(t)=\nabla_{X'(t)}^{g_{e}}X'(t)=X''(t)=0$ if and only if
$X(t)$ is the coordinate vector of any constant velocity trajectory in any inertial frame.\\\\

\subsection*{Mass and Energy distributions}

We now assume that the inertial mass distribution of matter (into
movement) in the universe $B(O,t)$ is given by $m_t(X)$ to which
we associate the measure $dm_t=:\rho_t$. We denote by
$g_t=g_{ab}(t,X)$ the Riemannian metric on $B(O,t)$ which
characterizes the real physical universe (reflecting all effects of
this distribution as well as of other forms of energy) and by
$$\mu _t:=dv_{g_t}=:v_t(X)dX$$ the measure of density $v_t(X)$
with respect to the Lebesgue measure $dX$ on $B(O,t)$. The global
inertial mass of the universe at every time $t$ is given by
$$M(t)=\int _{B(O,t)}\rho_{t}:=\int _{B(O,t)}m_t(X)dX$$
On the other hand, we denote by $E(t,X)=E_t(X)$ the distribution
of the generalized potential energy which includes, by definition,
all manifestations and effects of matter (material distribution
$m_t(X)$, pure energy of black holes, gravity, electromagnetism
and interaction forces). Then we denote by $\nu_t=E_t(X)dX$ the
measure associated with this distribution. On the base of all
preceding data, we can state
$$\nu _t=dX-\mu _t.$$ This equality expresses the fact that $\nu_t$
measures the failure of the real physical volume of a domain in
$U(t)$, containing matter-energy distribution, to be equal to the
spatial Euclidean volume of this domain when it is supposed to be
empty. This equality can also be written as
$$\mu
_t=dX-\nu _t\;\mbox{ or }\;v_t(X)=1-E_t(X)$$which, in that way,
expresses that $\mu_t$ measures in fact the real physical volume
of this domain taking into account the modification of Euclidean
distances imposed by the metric $g_t$ which itself reflects the
existence of the matter-energy in that domain. We have naturally
$$\rho_t \leq \nu_t \leq dX \hskip 0.2cm\mbox { or } \hskip 0.2cm m_t(X) \leq E_t(X) \leq
1.$$All of this is confirmed by our explicit calculation of the
metrics corresponding to both uniform and central gravitation
studied in section 4. Moreover these relations justify the metric $g_{ab}$ characterization previously established for black holes.\\
The equivalence principle and the law of energy conservation give
\begin{eqnarray*}
  E(t)&:=& \int_{B(O,t)}E_{t}(X)dX := \int _{B(O,t)\backslash {\bigcup _{\alpha \in A}}B_{\alpha}} E_t(X)dX + \sum_{\alpha \in A} e_\alpha \\
   &=&E(0)=:E_0\sim M_0
\end{eqnarray*}
where $e_\alpha$ denotes the $B_\alpha$black hole energy for
$\alpha \in A.$\\Let us consider, for a while, the space and time
half-cone
$$C=\{(x,y,z,t)\in \mathbb R^4;\;x^2+y^2+z^2 \leq t^2, t\geq 0\}=\bigcup _{t\geq 0} B(O,t)
\times\{ t \}
$$equipped with the metric $\eta$ given
by$$\eta(t,X)=dt^2-g_e(X)$$ which is induced by the Minkowski
metric defined on the virtual classical space $\mathbb R^4$.
Within this half-cone occurs the dynamic creation of real
geometrical temporal space, expanding permanently, $B(O,t)$, and
making up the real physical space $U(t)=(B(O,t),g_t)$ always in
expansion. We notice that, in the interior of $C$, $\eta$ is a
Riemannian
metric.\\
We consider then the generalized potential energy function
$E(t,X)=E_t(X)$, for $X \in B(O,t)$ and we assume that $E$ is
continuous on $C$ and that all its partial derivatives of order
$\leq 2$ exist and are continuous and bounded on
$C^{*}=C\backslash \{O\}$. The global potential energy of the
universe at time $t$ is then written
$$E(t)=\int _{B(O,t)}E(t,X)dX=\int _{0}^{t}dr\int _{S(O,r)}E(r,X)d\sigma
_r\equiv E_0$$then
$$E'(t)= \int _{S(O,t)}E(t,X)d\sigma_t=:S(t)=0,$$for any $t>0$.
So, the function of time $S$ is given by $E_0\delta_
{\mathbb R^+}$ and we have $$E_t|_{S(O,t)}=0 \mbox { for }  t>0.$$\\
Notice that, in the interior of a black hole $B(I,r)\subset
B(O,t)$, when it exists, we have
$$E_t(X)=e(I)\delta _I=m(I)\delta _I,$$
where $e(I)$ is the potential mass energy which is equivalent to
the inertial mass $m(I)$. Thus $e(I)$ is a part of the initial
energy $E_0$ of the original universe that has been reconcentrated
at a given time (after the Big Bang) at $I$; whereas $E(t,X)$ =
\emph{e}(\emph{I})$\delta_{I}$ denotes the generalized potential
mass energy function on $B(I,r)$. Notice also that the function
$E_t(X)$ is null in any region of the universe that can be
considered as deprived of matter and of its effects. Finally we
mention that, although the function $E(t,X)$ is far from being of
class $C^2$, we however can reasonably approach it by such a
function (idealizing in such a way the universe) or consider it,
as well as their partial derivatives of order $\leq 2$, in the
distributional sense.\\

\subsection*{Matter-Energy equation}
We now consider a part $C_1$ of $C$ located between $t=t_1$ and
$t=t_2$ for $t_1<t_2$. We have $\partial C_1=B(O,t_1)\bigcup
B(O,t_2)\bigcup S_1 $ where $S_1$ is the lateral boundary of
$C_1$. We then consider the force field $F_{\eta}$ defined on
$C^{*}$ and deriving from the global potential function $E(t,X)$,
i.e.
$$F_{\eta}(t,X):=-\nabla
^{\eta}E(t,X):=-grad_{\eta}E(t,X).$$So, if $u(t)=(t,X(t))$ is a
given trajectory in $C$ and if $F_{\eta}(u(t))=0$, then we have
$$\eta(F_{\eta}(u(t)),u'(t))=0 \Leftrightarrow \eta(\nabla ^{\eta}E(u(t)),u'(t))=0$$$$\Leftrightarrow
dE(u(t)).u'(t)=0 \Leftrightarrow \frac{d}{dt} E(u(t)) = 0
\Leftrightarrow E(u(t))=const.$$Likewise, the identity
$$||F_{\eta}(u(t)||_{\eta}=||\nabla ^{\eta}E(u(t))||_{\eta}=0$$
is equivalent to
$$dE(u(t)).\nabla ^{\eta}E(u(t))=0$$and to
$$\frac{\partial E}{\partial t}(u(t))^2 =|\nabla ^{g_e}E_t(X(t))|^2=\sum _{i=1}^{3}
\frac {\partial E}{\partial x_i}(u(t))^2.$$On the other hand we
have
$$||u'(t)||_{\eta}=0 \Leftrightarrow ||(1,X'(t))||_{\eta}=0\Leftrightarrow |X'(t)|=1,$$
which means that the Euclidean speed is equal to 1 and then we get
$$|X(t)-X(t_0)|=t-t_0 \mbox{ for } t \geq t_0 \geq 0$$and the
trajectory in $C$ is reduced to a ray of a light cone surface.\\
Let us denote by $d\eta$ the measure associated with $\eta$ in $C$
and by $\Delta_{\eta}$ the Laplace-Beltrami operator on $C$
($d\eta$ is a measure of density $f(t,X)\geq 0$ with respect to
the Lebesgue measure on $C$ with $f(t,X)=0$ on $\partial C\setminus O$).
According to Stokes theorem, we have (where $\overrightarrow{n}$ is the normal vector to
$S_1$ that permits the application of Stockes' theorem as explained in $[4]$, p. $434$):
\begin{eqnarray*}
&{}& \int _{C_1}\Delta _{\eta} E(t,X)d\eta=\int _{C_1} div_{\eta}(\nabla^{\eta}E(t,X))d\eta \\
&{}& = \int _{B(O,t_2)} \eta
(\nabla^{\eta}E(t_2,X),\frac{\partial}{\partial
t})f(t_2,X)\,dX \\
&{}& -\int _{B(O,t_1)} \eta
(\nabla^{\eta}E(t_1,X),\frac{\partial}{\partial t})f(t_1,X)dX+\int
_{S_1}\eta
(\nabla^{\eta}E(t,X),\vec{n})d\eta \\
&{}& = \int_{B(O,t_2)} \frac{\partial E}{\partial t}
(t_2,X)f(t_2,X)dX - \int_{B(O,t_1)} \frac{\partial E}{\partial
t}(t_1,X)f(t_1,X)dX,
\end{eqnarray*}
since $E(t,X)=0$ on $S_1$, $E(t,X) \cong 0$ on a neighborhood of
$S_1$ and $\nabla^{\eta} E(t,X) = 0$ on $S_1$.\\
Thus we obtain:
\begin{eqnarray*}
&{}& \int _{C_1}\Delta _{\eta} E(t,X)d\eta=\int _{t_1}^{t_2}dt\int _{B(O,t)}\Delta _{\eta} E(t,X)f(t,X)dX \\
&{}& =\int _{0}^{t_2}dt\int _{B(O,t)}\Delta _{\eta}
E(t,X)f(t,X)dX-
\int _{0}^{t_1}dt\int _{B(O,t)}\Delta _{\eta} E(t,X)f(t,X)dX \\
&{}& =\int _{B(O,t_2)}\frac {\partial E}{\partial
t}(t_2,X)f(t_2,X)dX- \int _{B(O,t_1)}\frac {\partial E}{\partial
t}(t_1,X)f(t_1,X)dX=F(t_2)-F(t_1)
\end{eqnarray*}
where
$$F(t):=\int
_{B(O,t)}\frac {\partial E}{\partial t}(t,X)f(t,X)dX$$Therefore
$$\int _{B(O,t_2)}\Delta _{\eta} E(t_2,X)f(t_2,X)dX=F'(t_2)$$and
$$\int _{B(O,t_1)}\Delta _{\eta} E(t_1,X)f(t_1,X)dX=F'(t_1)$$Then
we have
$$\int _{B(O,t)}\Delta _{\eta}
E(t,X)f(t,X)dX=F'(t)$$for $t>0$. This equality can be written as
$$\int _{0}^{t}dr\int _{S(O,r)}\Delta _{\eta} E(r,X)f(r,X)d\sigma _r=
\frac{d}{dt}\int _{B(O,t)} \frac {\partial E}{\partial
t}(t,X)f(t,X)dX$$which implies
$$\int _{S(O,t)}\Delta _{\eta} E(t,X)f(t,X)d\sigma _t=
\frac{d^2}{dt^2}\int _{B(O,t)} \frac {\partial E}{\partial
t}(t,X)f(t,X)dX$$$$= F^{''}(t) = \frac{d^2}{dt^2}\int
_{0}^{t}dr\int _{S(O,r)}\frac {\partial E}{\partial
r}(r,X)f(r,X)d\sigma _r$$
$$=\frac{d}{dt}\int _{S(O,t)}\frac {\partial E}{\partial t}(t,X)f(t,X)d\sigma
_t=0$$for $t>0$, since $f(t,X) \equiv 0$ on $S(O,t)$.\\
So, we have
$$F'(t) = \int
_{B(O,t)}\Delta _{\eta} E(t,X)f(t,X)dX=const.$$which implies
$$\int _{C_1}\Delta _{\eta} E(t,X)d\eta=a(t_2-t_1)$$
with $a=F'(t)$ and then we get, for $t>0$ and \emph{C}(\emph{t}) = $\{$
(\emph{x},\emph{y},\emph{z},\emph{r}) $\in \mathbb{R}^4$;
$x^2+y^2+z^2 \leq r^2$, 0 $\leq r \leq t \}$:
$$\int _{C(t)}\Delta _{\eta} E(t,X)d\eta=at.$$Using the change of
variables

$$(t,X)\rightarrow (\lambda s,\lambda
X)\;\;\mbox { for }\;\; \lambda >0 \;\;\mbox { in } C(t),$$we
obtain
$$\int _{C(s)}\Delta _{\eta} E(\lambda s,\lambda X)\lambda ^4 d\eta=a\lambda
s$$or
$$\int _{C(s)}\Delta _{\eta} E(\lambda (s, X))
d\eta=\frac{a}{\lambda^3}s.$$Now, we have
$$\int _{C(s)}\Delta _{\eta} E(s,X)d\eta=as$$and then
$$\int _{C(s)}\Delta _{\eta} E(s,X)d\eta=\lambda ^3\int _{C(s)}\Delta _{\eta} E(\lambda (s,
X))d\eta$$which implies ($\lambda$ being arbitrary)
$$\int _{C(s)}\Delta _{\eta} E(s,X)d\eta=as=0$$and then$$a=0,\;\;F(t)=const.\hskip 1cm \mbox {
for }\;\;t>0$$So, we finally have$$\int _{C_1}\Delta _{\eta}
E(t,X)d\eta=0\;\mbox{ for any }\;t_1\;\mbox { and }\;t_2 \; \mbox
{ satisfying } 0<t_1<t_2.$$Let us now show that
$$\Delta _{\eta}
E(t,X)=0\;\mbox{ on }\; C^*.$$Indeed, if we suppose that $\Delta
_{\eta} E(t_0,X_0)>0$, for example, then we would have $\Delta
_{\eta} E(t,X)>0$ on a neighborhood $B$ of $(t_0,X_0)$ in
$\{t_0\}\times B(O,t_0)$. Let us then consider the union $C'$ of
all future causal half-cones having their vertices on $B$.\\
Now, by reasoning in the same way as above on a part $C'_1$ of
$C'$ located between $t_1$ and $t_2$ such that $t_0<t_1<t_2$, we
can show that
$$\int_{C_1'}\Delta_{\eta} E(t,X)\,d\eta=0,\qquad \mbox{ for any $t_1$ and $t_2$ sufficiently close to $t_0$}
$$ and then
$$\int_{B}\Delta _{\eta} E_{/B}(t_0,X)f(t_0,X)dX=0$$
(the contribution of each half-cone being null since $E(t,X)$ can be considered as being constant on every half-cone surface for such $t_1$ and $t_2$) which is
contradictory since $\Delta _{\eta} E(t_0,X)$ is supposed to be
continuous and positive on \emph{B}. Therefore we have
\begin{equation}\label{r14}
\Delta _{\eta}E(t,X)=0\mbox{ on }C^*
\end{equation}
Thus for $t>0$, we obtain $$\hskip 4cm \frac{\partial ^2}{\partial
t^2}E(t,X)-\Delta E(t,X)=0 \hskip 3cm (E^*)$$ 
This same result can be obtained, outside of the black holes or other singularities,
supposing some regularity of the distribution $E(t,X)$ on a region
of $C$ resulting from excluding small neighborhoods of black holes
evolving with the expansion. If $B(I_{\alpha}(t),r_\alpha(t))$ is such a
black hole included in $B(O,t)$, for $t\in[t_1,t_2]$, and if we
assume that $\sum _{\alpha \in A}e_{\alpha}(t)$ is negligible
compared to $E_0(t)$ or, simply, is constant on $[t_1,t_2]$, we
can apply the conservation of energy principle on this region of
$C$ to obtain the equation ($E^*$) on $B(O,t)\backslash \bigcup
_{\alpha \in A}B_{\alpha}$.\\\\
We now consider the dynamical real universe $(U(t))_{t > 0}
=(B(O,t),g_t)_{t > 0}$ and the half-cone $C$ equipped with the
Riemannian metric $h$ defined on the interior of $C$ (and the non negative Lorentzian metric on $C^*$) by
$$h(t,X)=dt^2-g_t(X)=dt^2-g(t,X)$$
where $g_t$ is the Riemannian metric
$g_{ab}$ defined on $B(O,t)$ by equation ($\cal {E}$). We denote
by $d\beta$ the measure associated to $h$ in $C^*$ and we apply
Stokes theorem on any part $C_1$ of $C^*$ (defined in the same way
as before), then we can follow the same steps as above and use the
light half-cones within $C$ whose rays are adapted to the metrics
$g_t$ and $h$ (i.e. null geodesics for \emph{h}). After that, we can
replace into the preceding results $d\eta$ by $d\beta$, $dX$ by
$d\mu$, $\Delta_{\eta}$ by $\Delta_{h}$ and $F_{\eta}(t,X_t)$ by
$F_h(t,X_t):=-\nabla^{h}E(t,X_t):=-grad_{h}E(t,X_t)$, for $X_t \in
B(O,t)$, to obtain with the same hypotheses as above:
$$\int _{B(O,t)}\Delta_{h}E(t,X)k(t,X)dX=0 \hskip 1cm \mbox { for }t>0,$$with
$k(t,X) \geq $0 on \emph{C} and $k(t,X)=0$ on $\partial C\setminus O$,
$$\int _{C_1}\Delta_{h}E(t,X)d\beta=0 \hskip 1cm \mbox { for }0<t_1<t_2,$$
$$\Delta_{h}E(t,X)=0 \hskip 1cm \mbox{ on } C^*$$and
\begin{equation}\label{15}
\frac{\partial^2}{\partial t^2}E(t,X)-\Delta_{g_t}E(t,X)=0
\end{equation}
on $B(O,t)$ with $E(t,X)|_{S(O,t)}=0$ for any $t>0$.\\
This equation approximately coincides with $(E^*)$, for $t\gg1$, since then the metric $g_t$ can be approximated by $g_e$ when we deal with macroscopic global results.\\\\

\noindent \textbf{Remark}: The Lorentz factor $\gamma=(1-v^2)^{-\frac{1}{2}}$ is naturally introduced by the evolution metric $h=dt^2-g_t$ on $C$. Indeed, this relation implies $\tau^2=t^2-d^2$, where $\tau$ is the interval between two events that respect the causality principle within $C$ and $d$ is the spatial distance when measured by $g_t$. The later relation implies 
$$\frac{\tau^2}{t^2}=1-v^2=\frac{1}{\gamma^2}\quad\mbox{or}\quad t=\gamma\tau.$$

\subsection*{Universal time and proper time}

Reconsider the experiment of the slowing down of the mobile watches (see [7], p. 43); we have then considered that the period $T_p$ given by the airborn watches and indicated by the watch on the ground $T$ satisfy the relation
$$T=\gamma T_p\qquad\mbox{where}\qquad\gamma=(1-v^2)^{-\frac{1}{2}}\simeq 1+\frac{v^2}{2}$$
because the relative speed with respect to the Earth $v\ll 1$ for $c=1$.\\
Now, according to wether the travel (which takes place at a distance $R$ to the rotation axis of the Earth) is directed to the West or to the East the relative speeds to be considered are
$$v_w=v+\Omega R\qquad\mbox{and}\qquad v_e=v-\Omega R$$
where $\Omega\simeq 7.3\times 10^{-5}$ rad s$^{-1}$ is the rotation speed of the Earth with respect to its rotation axis.\\
It have been then considered that the relative difference of the period calculated for the aeroplane moving toward the West is
$$\frac{T_w-T_p}{T_p}=\gamma_w-1=\frac{(v+\Omega R)^2}{2}\simeq v(v+2\Omega R)$$
meanwhile, for the aeroplane moving toward the East is
$$\frac{T_e-T_p}{T_p}=\gamma_e-1=\frac{(v-\Omega R)^2}{2}\simeq v(v-2\Omega R).$$
It was then observed that the deviations, which are $275$ ns for the displacement toward the West and $40$ ns for the displacement toward the East confirm the theoretical predictions of the special relativity.\\
In the framework of our model, we obtain the same results but in a really physical manner. Indeed, our unit of time I.S. is based on the speed of light in the absolute vacuum where the metric is $g_e$, which corresponds to the energy density $\rho=0$, neglecting the cosmological constant $\Lambda$, and on the metric $-\eta=dt^2-g_e$ in the semi-cone of space and time. At the Earth surface, the correspondent metrics will be $g$ and $h=dt^2-g$, which are different from the metrics $g_1$ and $h_1=dt^2-g_1$ at the altitude of the two aeroplanes. The difference of the $2\times 3$ metrics is due essentially to the differences of the gravitational and magnetic fields.\\
We recall that, in our model, the metrics take into account both of these fields and contract the distances:
$$\|X'(t)\|_g\leq\|X'(t)\|_{g_1}\leq\|X'(t)\|_{g_e}.$$ 
Moreover, the universal time for us is intrinsically related to the speed of light in the vacuum $c=1$ with respect to a referential that is supposed to be fixed and related to the proper time $\tau$ (which depends on the metric and on the speed $v$ of the referential) by
$$\frac{\tau}{t}=\frac{1}{\gamma}=(1-v^2)^{-\frac{1}{2}}.$$
Therefore, in the above experiment, we should have to compare the two proper times $\tau$ and $\tau_1$ (relative to $g$ and $g_1$) to the universal time $t$ before comparing them to each other; which gives the same results. Our watches on the Earth, which are set, on one hand, on the speed of light in the vacuum using the metric $g_e$, and on the other hand, on the metric $g$ that is different from $g_e$ as well as from $g_1$. The quotient $\displaystyle \frac{\tau_1}{\tau}=\frac{\tau_1/t}{\tau/t}$ give us more physically realist results for the three used watches.\\
The relativity is a notion related to the proper time $\tau$ and not to the universal time $t$. We have not the  right to define the notion of time with respect to a privileged metric other than $g_e$ nor to set our watches on the Earth surface or on the surface of another planet or near a black hole.\\
Only the time for an observer supposed to be fixed inside the absolute vacuum has an intrinsic character. Even the proper time $\tau$ for a mobile observer lying in the vacuum, where the metric is $g_e$, do not have the right of defining any intrinsic unit of time. Only for a fixed observer in the vacuum we have $\tau=t$. For a black hole of center $I$, $B(I,r)$, and energy $e$ we have $\tau=t$ for every referential moving in $B\setminus I$ because then we have
$$v_{g_t}:=\|X'(t)\|_{g_t}=0,$$
meanwhile for a referential achieving $I$, for $t=t_0$, we have
$$\frac{\tau}{t}=\sqrt{1-e^2}=\sqrt{1-\|X'(t)\|_{g_t}}$$
when we consider the total energy of the universe $E_0$ as being the unit of energy.

\subsection*{Force field, Acceleration and Geodesics}
Moreover, let us define
$$F_{g_t}(X):=-\nabla ^{g_t}E_t(X)$$and
$$F_{g_e}(X):=-\nabla ^{g_e}E_t(X)$$for $X\in B(O,t)$ and then
consider a trajectory $X(t)$ in the dynamical universe $(U(t))_{t
> 0}$. Then
$$\frac{d}{dt}E(u(t))=\frac{d}{dt}E(t,X(t))=\frac {\partial E}{\partial t}(t,X(t))+
\frac{d}{ds}E_t(X(s))|_{s=t}$$
$$=\frac {\partial E}{\partial t}(t,X(t))+
dE_t(X(s)).X'(s)|_{s=t}=\frac{\partial}{\partial
t}E(t,X(t))+dE_t(X(t)).X'(t)$$Moreover
$$\eta(\nabla^{\eta}E(u(t)),u'(t))=\frac{d}{dt}E(u(t))=h(\nabla^{h}E(u(t)),u'(t))$$
$$=\frac {\partial E}{\partial t}(u(t))-g_t(\nabla ^{g_t}E_t(X(t)),X'(t))=
\frac {\partial E}{\partial t}(u(t))-g_e(\nabla
^{g_e}E_t(X(t)),X'(t)).$$Therefore
$$g_t(F_{g_t}(X(t)),X'(t))=g_e(F_{g_e}(X(t)),X'(t))$$
$$=\frac{d}{dt}E(t,X(t))-\frac {\partial E}{\partial t}(t,X(t))=dE_t(X(t)).X'(t)=
\frac{d}{ds}E_t(X(s))|_{s=t};$$which implies, according to the
generalized fundamental law of Dynamics,
$$F_{g_e}(X(t))=-\nabla^{g_e}E_t(X(t)) = \nabla^{g_e}_{X^{'}(t)} X^{'}(t) =\Gamma(t)=X''(t)$$and
$$F_{g_t}(X(t))=-\nabla^{g_t}E_t(X(t))=\nabla_{X'(t)}^{g_t}X'(t)=\widetilde{\Gamma}(t),$$
the last two identical vectors being 0 for free motion.\\
Thus, $F_{g_{e}}$(\emph{X}) is the global natural forces' vector
field. The force acting on a punctual material particle of mass
$m_{t}$ located at \emph{X} actually is $m_{t}
F_{g_{e}}$(\emph{X}) $\equiv$ -$m_{t} \nabla^{g_{e}}
E_{t}$(\emph{X}). For a moving punctual material particle, the force is
\emph{m}(\emph{X}(\emph{t}))$F_{g_{e}}$(\emph{X}(\emph{t})) =
\emph{m}(\emph{t})$\Gamma$(\emph{t}).\\\\
In particular,
$$X(t) \mbox{ is a geodesic for }g_e, \mbox{ with
}||X'(t)||_{g_e}=1 $$$$\mbox{if and only if }u(t) \mbox { is a
geodesic for }\eta \mbox{ with }||u'(t)||_{\eta}=0$$ and then
$u(t)$ is a ray of a light cone surface in $(C, \eta)$,
$E(u(t))=const.$ and
$$F_{g_e}(X(t))=-\nabla^{g_e}E_t(X(t))=\Gamma(t)=X''(t)=0.$$
Likewise,
$$X(t) \mbox{ is a geodesic for }g_t, \mbox{ with
}||X'(t)||_{g_t}=1 $$$$\mbox{if and only if }u(t) \mbox { is a
geodesic for }h \mbox{ with }||u'(t)||_{h}=0$$ and then $u(t)$ is
a light ray in $(C,h)$ and we have $E(u(t)) = const.$ and
$$F_{g_t}(X(t))=-\nabla^{g_t}E_t(X(t))=\nabla _{X'(t)}^{g_t}X'(t)=\widetilde{\Gamma}(t)=0.$$
Recall that, the
statement$$F_{g_t}(X(t))=\widetilde{\Gamma}(t)=0$$for any free
motion in $(U(t))_{t > 0}$ constitutes a generalization of the
two fundamental Newtonian laws of inertia.\\\\
Moreover, we notice that, if we assume the virtual existence
(purely theoretical) of the space $\mathbb R^3$ and the half-cone
of space and time before their real, physical and temporal
existence starting with the Big Bang, which are modeled much later
by Euclid and Descartes and afterwards by Galileo and Newton, then
we can consider the original universe as being $E_0\delta_{\mathbb
R^3}$ and $E_0\delta_{C}$. Otherwise, after the Big Bang, we have,
in the half-cone of space and time, $E(t,X)=0$ on $\partial C\setminus O$
and $E(0,O)=E_0$. Likewise, we have $\Delta _h E(t,X)=0$ on $ C^*$
and $\Delta _h E=E_0\Delta _h \delta_{C}$ where we have denoted,
for a class $C^2$ function $\varphi(t,X)$ with compact support in
$C$:
$$<\Delta _h \delta_{C},\varphi(t,X)>=\lim_{t\longrightarrow 0}\Delta _h\varphi(t,O).$$

\subsection*{Conclusions}
According to the preceding results, we conclude that our real,
geometrical, physical and dynamical universe, modeled by
$(U(t))_{t > 0} =(B(O,t),g_t)_{t > 0}$ is
characterized by each of the equivalent following notions:\\
a) The material distribution of (the density of) the inertial mass
$m_t(X)$ together with the $e_\alpha(t) \delta_{I(t)}$
corresponding to
the black holes scattered into \emph{B}(\emph{O},\emph{t}).\\
b) The regularized scalar field of the total potential energy
represented by the distribution of generalized potential energy $E_t(X)$\\
c) The Riemannian regularized metric $g_t$ on $B(O,t)$.\\
d) The modified (matter-energy) tensor
$T^*_{ab}$.\\
e) The physical measure $\mu_t$ on $B(O,t)$ given by $\mu
_t=dv_{g_t}=v_t(X)dX$ where $dX$ is the Lebesgue measure on
$B(O,t)$.\\
f) The measure $\nu _t=E_t(X)dX$ of density $E_t(X)$ with respect
to the Lebesgue measure on $B(O,t)$.\\
g) The vector field $\nabla ^h E(t,X)$ on $C$, where $h=dt^2-g_t$
is the Riemannian metric on the half-cone of space and time.\\
h) The total and global vector field $\nabla ^{g_e}E_t(X)$ which
is equal to the total and global force vector field $F_{g_e}(X)$
and is given by $-\nabla ^{g_e}E_t(X)=F_{g_e}(X)$(=
\emph{$\Gamma$}$_{t}=X''(t)$, for
any material particle's free motion into $(U(t))_{t > 0}$).\\
k) The set of all dynamical geodesics $X(t)$ for the metric $g_t$
evolving with time (i.e. satisfying $\widetilde{\Gamma}(t)=\nabla
_{X'(t)}^{g_t}X'(t)=F_{g_t}(X(t))=-\nabla ^{g_t}E_t(X(t)) = 0)$.\\\\
Let us now notice that our model is definitely consistent in the
sense that, on one hand, it proves, a posteriori, the legitimacy
of all mechanical and physical principles which are discovered and
stated by all great scientists of the humanity, and on the other
hand we have\\
$E_t(X)\equiv 0$ on a domain $D$ of the dynamical universe between
$t=t_1$ and $t=t_2\Leftrightarrow  \nu_t=0 \Leftrightarrow
T^*_{ab}\equiv 0$ on the domains $D_t$ in $B(O,t)$ corresponding
to $D\Leftrightarrow g_t=g_e$ in $D_t\Leftrightarrow \mu_t =dX$ on
$D_t\Leftrightarrow h=\eta$ on the domain of $C$ that corresponds
to $D\Leftrightarrow \Gamma_t=X''(t)\equiv 0$ for free motion in
$D\Leftrightarrow$ All trajectories corresponding to free motion
in $D$ are geodesics constituted of straight lines.\\\\
\subsection*{Remark} The property, previously noticed, of our physical
metric $g_t$ shows that our space - time provided with our metric
$h_t$ = $dt^2 - g_t$ satisfies the three (slightly modified)
postulates of the "metric theories" which stipulate that:\\

(\emph{i}) the space - time is provided with a metric,\\

(\emph{ii}) the free falling bodies trajectories are geodesics.\\

(\emph{iii}) In any local referential frame, the non gravitational
Physics' laws are the same as classical Physics' laws (without
special relativity
considerations).\\\\
This shows that our gravitational model satisfies the Einstein's
equivalence
principle.\\\\
We end this section by noticing that, in our model, the
geometrical space does not exist around the point of concentration
of the original energy $E_0$ before the Big Bang, whereas the
geometrical space $B(I,r)$ exists actually around the point of
concentration $I$ of the energy of a black hole, but this space is
equipped with a null metric outside \emph{I}. Conversely, when a
domain of the geometrical space of the universe $B(O,t)$ does not
contain matter nor its effects, then this part of space has a very
existence and is equipped with the Euclidean metric $g_e$. The
gravitational field in this case is vanishing, but around the
center $I$ of a black hole it is omnipresent.\\\\

\section{Energy, Pseudo-waves and Frequencies}

$\hskip 0.5cm$ Let us now consider the wave equation (which is the
matter - energy equation):
$$\Box E(t,X)=\frac {\partial^2E}{\partial t^2}(t,X)-\Delta
E(t,X)=0 \hskip 4cm(E^*)$$ and (using the separation of variables
method) let us determine, for any $t>0$, the solutions on $B(O,t)$
that satisfies $E(t,X)_{|S(O,t)}=0$. So we consider the functions
of the form
$$E_0(t,X)=f_0(t)F_0(X)\;\mbox{ for a fixed }\;t_0 > 0$$satisfying
$$\Box f_0(t)F_0(X)=0 \hskip 0.5cm\mbox { with} \hskip 0.5cm F_0(X)|_{S(O,t_0)} = 0 \hskip 3cm(E_0)$$
It is well known that solutions $f_0$ and $F_0$ of equation
$(E_0)$ are obtained from the increasing sequence $\lambda_i(t_0)$
of eigenvalues of the Laplace-Beltrami operator $-\Delta$ and the
sequence of eigenfunctions $\varphi _{t_0,i}(X)$ that are
associated with the Dirichlet problem on the ball $B(O,t_0)$
provided with the metric $g_e.$ The corresponding solutions,
$f_{0,i}(t)\varphi _{t_0,i}(X)$, are defined by
$$\Delta \varphi_{t_0,i}(X)=-\lambda _i(t_0)\varphi_{t_0,i}(X)$$and
$$f''_{0,i}(t)+\lambda _i(t_0)f_{0,i}(t)=0.$$
We pick out one of these solutions which will be denoted by
$$f_0(t)\varphi_{t_0}(X);$$and then we get, for $0\leq t\leq t_0$:
$$f''_{0}(t)+\lambda (t_0)f_{0}(t)=0.$$
Now, we consider the isometry that identifies $(B(O,t_0),g_e)$ with $(B(O,1),t_0^2g_e)$ given by $X\rightarrow\frac{X}{t_0}$. Then if $\mu_0$ and $\mu(t_0)$ are the eigenvalues having the same rank as $\lambda(t_0)$ with respect to the Dirichlet problem respectively on $(B(O,1),g_e)$ and $(B(O,1),t_0^2g_e)$, we have $\lambda(t_0)=\frac{\mu_0}{t_0^2}$ $(\lambda(t_0)=t_0^2\mu(t_0),\;\nabla_{g_e}\varphi(\frac{x}{t_0})=\frac{1}{t_0}\nabla_{g_e}\varphi(X)$ and $\Delta_{t_0^2g_e}\varphi=\frac{1}{t_0^2}\Delta_{g_e}\varphi)$ and $f_0$ is the solution of the equation
$$f''_{0}(t)+\frac{\mu_0}{t_0^2}f_{0}(t)=0, \;\;\;\;0\leq t\leq t_0\;(t_0>0).$$
The solution of this equation is obviously the periodic function
$$
f_0(t)= f_0(0)\cos \frac{\sqrt \mu_0}{t_0}t+\frac{t_0}{\sqrt
\mu}f'_0(0)\sin \frac{\sqrt \mu_0}{t_0}t.
$$
The solution of $(E_0)$ corresponding to the eigenvalue $\lambda
(t_0)=\frac{\mu_0}{t_0^2}$ is defined as
\begin{equation}\label{r16}
E_{\mu_0}(t,X)=(f_0(0)\cos \frac{\sqrt \mu_0}{t_0}t+\frac{t_0}{\sqrt
\mu_0}f'_0(0)\sin \frac{\sqrt \mu_0}{t_0}t)\varphi_{t_0}(X)
\end{equation}
for $\;X \in B(O,t_0),\;t_0>0$ and $\;0\leq t\leq
t_0$.\\
Similarly, if $h_{\mu_0}(t)\psi_{\mu_0}(X)$ is the solution of the Dirichlet problem on the unit ball $(B(O,1),g_e)$ associated to the eigenvalue $\mu_0$, then we have, for $t\leq t_0$ and $X\in B(O,t_0)$:
$$f_{\mu_0}(t)=h_{\mu_0}\left(\frac{t}{t_0}\right)\quad\mbox{and}\quad\varphi_{t_0,\mu_0}(X)=\psi_{\mu_0}\left(\frac{X}{t_0}\right).$$
Indeed, the equation
$$h''_{\mu_0}(t)+\mu_0h_{\mu_0}(t)=0$$
is equivalent to the equation
$$f''_{\mu_0}(t)+\frac{\mu_0}{t_0^2}f_{\mu_0}(t)=0,$$
and consequently we have
$$E_{\mu_0}(t,X)=\left(h_{\mu_0}(0)\cos\frac{\sqrt \mu_0}{t_0}t+\frac{1}{\sqrt
\mu_0}h'_{\mu_0}(0)\sin \frac{\sqrt \mu_0}{t_0}t\right)\psi_{\mu_0}\left(\frac{X}{t_0}\right).$$
$E_{\mu_0}(t,X)$ is then a
periodic function of period $T_0= 2\pi\frac{t_0}{\sqrt
\mu_0}$ and frequency $f(t_0)=\frac{1}{2\pi}\frac{\sqrt{\mu_0}}{t_0}$.\\\\
\textbf{Remark}: We could also consider $E(t,X) = k_0(t)G_0(X)$ as
a solution of the equation (15). One solution corresponding to an
eigenvalue $\alpha(t_{0})$ associated to the Laplace-Beltrami
operator $-\Delta_{g_{t_0}}$ on the Riemannian manifold
(\emph{B}(\emph{O},$t_0$),$g_{t_0}$) would be then of the form
$$E(t,X) = k_0(t) \theta_{t_0}(X)$$
with $$k_0^{''}(t) + \alpha(t_0)k_0(t) = 0$$ and
$$\Delta_{g_{t_0}} \theta_{t_0}(X) = - \alpha(t_0)
\theta_{t_0}(X).$$
However, in that case, we can not reduce the
considered problem to the study of the Dirichlet problem on the
space $(B(O,1),t_0^2g_1)$ unless the function $X\longrightarrow
\frac{X}{t_0}$ would be an isometry from $(B(O,t_0),g_{t_0})$ on
$(B(O,1),t_0^2g_1),$ which is,
in the most favorable cases, a crude approximation.\\\\
In fact, the above solutions $E_\mu$ can not be assimilated to the
solutions of our matter-energy equation on the dynamical universe
$(U(t))_{t > 0},$ i.e. equation $(E^*),$ but only on an interval
of time, $t_1\leq t<t_0$ with $t_0>t_1$ and $t_1$ sufficiently
(and relatively) close to $t_0$ in such a manner that the
eigenfunctions $\varphi _t(X)$ and eigenvalues $\lambda (t)$
corresponding to $t\in [t_1,t_0]$ can be respectively considered
as being reasonable approximations of $\varphi_{t_0}(X)$ and
$\lambda (t_0)$. Moreover, in order to obtain a reasonable
periodic approximation of equation $(E^*)$ on $(U(t))_{t_1\leq t
\leq t_0}$, it is necessary that the period $T_0=
2\pi\frac{t_0}{\sqrt \mu}$ be significantly less than
$t_0-t_1$, and then $\mu$ be significantly greater than $4\pi^2(\frac{t_0}{t_0-t_1})^2$.\\
By giving to $t_0$ a large number of convenable values $t_i$, we
obtain approximate periodic solutions (of periods
$T_i=2\pi\frac{t_i}{\sqrt \mu}$) to our problem on juxtaposed
rings of $B(O,t)$ for \emph{t} = sup$_{i} t_{i}$.\\\\
By replacing in (16) $t$ by $t_0$ and rewriting it as a function
of the variable $t$ in place of $t_0$, we obtain the solution
$E_{\mu}$ of $(E^*)$ defined on $B(O,t)$, for $t>0$, by
$$E_{\mu}(t,X)=f_{\mu}(t)\varphi_{t,\mu}(X)=(f_{\mu}(0)\cos \sqrt \mu+\frac{t}{
\sqrt{\mu}}f'_{\mu}(0)\sin \sqrt
\mu)\varphi_{\mu}(X)$$
where $\varphi_{\mu}(X)=\psi_{\mu}(\frac{X}{t})$, $f_{\mu}(0)=h_{\mu}(0)$ depends
on $\mu$ and $f'_{\mu}(0)$ depends on $\mu $ and $t$. This
solution can be approximated, on appropriate rings
$B(O,t)\backslash B(O,t')$, by periodic functions of periods
$T(t)=2\pi\frac{t}{\sqrt \mu}$ and frequencies
$f(t)=\frac{1}{2\pi}\frac{\sqrt \mu}{t}$.\\Therefore $E_{\mu}$ is
a pseudo-wave (that will be incorrectly called wave) of
pseudo-period $T(t)=2\pi\frac{t}{\sqrt \mu}$ and pseudo-frequency
$f(t)=\frac{1}{2\pi}\frac{\sqrt \mu}{t}$
respectively (both depending on time $t$).\\
Thus, in order to assimilate these solutions to waves on
significant intervals of time, we put $t=e ^{\alpha}$, $\mu=4\pi^2
e^{2\beta}$ and then noticing that we have $T=e^{\alpha-\beta}$,
we have to take $e^{\alpha-\beta}<< e^{\alpha}$. So we must take
$\beta>>0$ and therefore the
eigenvalue $\mu>>0$.\\
When $X(t)$ is a trajectory of a free motion on a given interval
of time, i.e.
$$\widetilde{\Gamma}(t)=\nabla_{X'(t)}^{g_t}X'(t)=0\;\mbox{ for }\; t\in I,$$
the wave $E_{\mu}(t,X(t))$ corresponding to an eigenvalue $\mu$,
satisfies the identity
$$-\nabla
^{g_t}E_{\mu}(t,X(t))=F_{g_t}(X(t))=\widetilde{\Gamma}(t)=0.$$
Now, when the metric is Euclidean, the Newton's inertia principle
states that the trajectory \emph{X}(\emph{t}) of an arbitrary
particle is uniform (i.e. \emph{X}(\emph{t}) is a geodesic and the
speed $\parallel X^{'}$(\emph{t})$\parallel_{g_e}$ = \emph{v} is
constant) if and only if the force field acting on the particle,
$$F_{g_e}(t) = -\nabla^{g_e}E(t,
X(t)) = \nabla^{g_e}_{X^{'}(t)} X^{'}(t) = X^{''}(t),$$ is null
and then the particle energy is conserved along this trajectory,
i.e.
$$ E(t,X(t)) = \mbox{ const }.$$
In the case of our physical metric $g_t$, this same principle is
generalized in the following way:\\
The particle trajectory \emph{X}(\emph{t}) is a geodesic with
respect to the metric $g_t$ (with $\parallel
X^{'}$(\emph{t})$\parallel_{g_t}$ = const.) if and only if
$$-\nabla^{g_t} E(t,X(t)) = F_{g_t}(t) =
\nabla^{g_t}_{X^{'}(t)}X^{'}(t) = \widetilde{\Gamma}(t) = 0$$and
then the particle punctual energy is conserved along this
geodesic, i.e.
$$ E(t, X(t)) = \mbox{ const }.$$
In our setting, we then have
$$E_\mu(t,X(t)) = f_\mu(t) \varphi_{t,\mu} (X(t)) = e(\mu)$$
(where \emph{e}($\mu$) is a constant depending only on $\mu$) and
$$\triangle E_\mu (t,X(t)) = f_\mu(t) \frac{\mu}{t^2}
\varphi_{t,\mu} (X(t)) = \frac{\mu}{t^2} e(\mu).$$
For original particles that propagate with the speed $1$ (or $v(t)$), we have $X(t)\in S(O,t)$ (or $X(t)\in S(O,R(t))$) and
$E_\mu$(\emph{t}, \emph{X}(\emph{t})) = 0; these are the original
electromagnetic waves that made up the semi - cone of space and
time. For a material particle that propagates along a geodesic
(with respect to $g_t$) with a speed \emph{v} = $\parallel
X^{'}$(\emph{t})$\parallel_{g_t} <$ 1, we have
$$ E_\mu (t, X(t)) = e_0(\mu) > 0.$$
We notice that the photon energy other then the original ones
equally are positif.\\

\subsection*{Planck-Einstein Energy}
Moreover, by adapting the undulatory principle of Planck-Einstein
to our setting, we conclude that, for any material or immaterial
point $X$, moving freely in $(U(t))_{t > 0},$ we have
$$E_{\mu}(t,X(t))=h_{\mu}(t)f_{\mu}(t)=h_{\mu}(t)\frac{1}{2\pi}
{\frac{\sqrt \mu}{t}}=:\overline{h}_{\mu}(t)
{\frac{\sqrt\mu}{t}}$$where $f_\mu$(\emph{t}) denotes here the
frequency and $\overline{h}_{\mu}(t)$ replaces, in some way, the
Planck constant. This equality implies
$$\overline{h}_{\mu}(t)=tc(\mu)$$where $c(\mu)=\frac{e(\mu)}{\sqrt{\mu}}$ is a constant
depending only on $\mu;$ so
$$E_{\mu}(t,X(t))=c(\mu)\sqrt {\mu} =e(\mu)$$
\textbf{Remarks}: $1^\circ$) We notice that, as we will see in the
next section, this constant indeed depends on time when being
considered on a large scale of time. This is due to the perpetual cooling of
the cosmos. Moreover, we shall prove in Section $12$ that if $\rho(t)$ is the energy mean density of the Universe at time $t$ and $f(t)$ is the mean frequency of the cosmic matter-energy, then $E(t,X(t))=:e(t)=h(t)f(t)=\rho(t)\propto\frac{1}{t^3}$ and consequently $f(t)\propto\frac{1}{t^4}$.\\
$2^\circ$) Contrary to the Planck constant (which is generally
considered as a universal constant), our constant
$\overline{h}_{\mu}(t)$ is proportional to time.\\\\
In other respects, let us consider a material agglomeration that
is filling up a domain $D_{t}$ in $B(O,t).$ This domain is
subdivided into some subdomains $D_{t,n}$, on each of them is
defined an energy distribution $E_{n}(X_{t})$ which coincides, for
each domain $D_{t,n}$ that is filled up by a fundamental material
particle, with a constant material distribution
$m_n(X_t)\;=:\;m_n$ (the fundamental particles will be classified
in section 8). The energy distributions $E_n(X_t)$ defined on all
other subdomains are equally assumed to be constant. We then have
$$D_t=\bigcup_{1 \leq n \leq N}D_{t,n}$$
with  
$$E_n(X_t) = e_n( \sim m_n \mbox { for } X_t \in
D_{t,n}\; \mbox{when $D_{t,n}$ corresponds to a material particle}).$$
So, the energy of the domain $D_{t}$ is
$$E(D_t)=\sum _{n}vol(D_{t,n})e_n=:\sum
_{n}V_n(t)e_n,$$ 
where vol(\emph{D$_{t,n}$}) =: \emph{V$_n(t)$}
denotes here the Euclidean volume of the subdomain \emph{D$_{t,n}$}.\\
Obviously, even if $D_t$ constitutes an isolated system such that
$vol(D_t)$ remains constant, $V_n(t)$ evolves with time. This is
due to several evolutive dynamical phenomena: Transformation
matter-pure energy, radiations of all sorts, disintegration,
collision, fission, fusion and chemical, nuclear
and thermic interaction.\\
In the dynamical universe, we have to take into account the
kinetic energy of matter into movement. However, we can speak
about inertial mass, potential energy, kinetic energy and quantity
of movement (i.e. $mv$) only when considering a material point or
a material domain moving with a speed inferior to 1.\\
The kinetic energy of a material punctual particle following a
trajectory $X(t)$ is in fact $\frac{1}{2}m_1(X(t))X'(t)^2$ where
$m_1(X(t)) = \gamma_1(X(t))m_0$, $m_0$ is the rest mass of the
particle and $\gamma_1(X(t))$ is a factor that decreases from 1 to
0 when the particle speed increases from 0 to 1, which expresses
the loss of mass undergone by the particle caused by the
radiations produced by the acceleration. This factor can be
determined theoretically or experimentally for all sortes of
particles. The inertial mass or the potential mass energy of a
material domain $D_t$, at the time $t$, is given
by$$\rho_t(D_t)=\int_{D_t}m_1(X_t)dX_t.$$ Its kinetic energy is
given by
$$\frac{1}{2}\int_{D_t}m_1(X_t)v(X_t)^2dX_t.$$
When we are dealing with a domain devoid of matter and crossed
everywhere by radiations as electromagnetic waves (visible or
invisible light, X rays, $\gamma$ rays) we can not speak neither
about inertial mass nor about potential energy, kinetic energy or
quantity of movement. We can only speak about the energy of
propagated waves, beams and photons. Nevertheless, we can define,
within our model, the linear momentum $\overrightarrow{p}$ of an
electromagnetic wave or photon as being $\overrightarrow{p(t)}$ =
$\frac{1}{c^2}$ \emph{E}(\emph{t}, \emph{X}(\emph{t}))
$\overrightarrow{X'(t)}$ and we then have $p(t) c =E(t)$ or $p c
=E$ which agrees with the special relativity notion (here $ ||
\overrightarrow{X^{'}(t)}||_{g_e}=c $). Thus the kinetic energy of
a material point $X$ with $m_0(X)=m_0$, making up a trajectory
$X(t)$ is $\frac{1}{2} m_1(X(t)) X'(t)^2$ and its total energy is
$\emph{m}(t) c^2+\frac{1}{2} m_1(t) X'(t)^2)$ whenever its speed
is $<c$ with $m(t)=\gamma(t)m_0$ when dealing with equation $(15)$ (relative to the metric $g_t$ in place of $g_e$), we have to replace the Euclidean speed by $\|X'(t)\|_{g_t}$ and then we obtain $\|X'(t)\|_{g_t}\simeq c$ and $p(t)c\simeq E(t)$ for electromagnetic waves  unless inside black holes $B\setminus I$ where $\|X'(t)\|_{g_t}=0$.\\
For a material punctual particle having $m_0$ as rest mass, the
quantity $m_1(t) = \gamma_1(t) m_0$ will be called the reduced
mass and $m(t) = \gamma (t) m_0 =
(1-\frac{v^2(t)}{c^2})^{-\frac{1}{2}} m_0$ is the apparent mass of
the particle into movement. However, for a black hole \emph{B}
having an initial mass energy \emph{E} and moving with any speed
\emph{v}, we can consider its total energy as being
$\frac{E}{c^2}(c^2+\frac{1}{2}v^2)$ after the famous Einstein's
relation $E= mc^2$. Nevertheless, the distribution $E(t,X)$ is
entirely determined by the distribution $m_t(X)$ together with
black holes' energies. That is also the case of the total force
field $F_{g_e}(X)=- \nabla^{g_e}E_t(X)$ and the acceleration
vector field $\Gamma(X(t))=X''(t)$ for any free motion. Therefore
the metric $g_t$ which is intrinsically related to the
distribution $E_t(X)$ and which satisfies
$\widetilde{\Gamma}(X(t))=\nabla _{X'(t)}^{g_t}X'(t)=0$ (for free
motions), takes into account all manifestations and effects of
matter, including electromagnetic fields, interaction forces and
the resulting binding forces, and not only the gravitational
one.\\
Notice that, the kinetic energy of a system is not necessarily
conserved neither globally nor locally as it is shown, for
example, by the transformation of a part of the kinetic energy of
a system into heat during a collision, for example. Conversely,
the principle of the global momentum conservation for an isolated
system such as the whole universe is valid. We then have at any
time $t$:
$$\int_{B(O,t)} \frac{1}{c^2} E_t(X_t) \emph{\textbf{v}} (X_t) dX_t =
\textbf{0}.$$where \emph{\textbf{v}}($X_t$) denotes here the
velocity vector.\\
In particular the center of gravity of the universe is fixed. This
relation is written, for $B(O,t)=\bigcup_{n} D_n(t)$, where each
$D_n(t)$ is characterized by its mass density $m_n:=\frac{e_n}{c^2}$, as
$$\sum_{n}\int _{D_n(t)}m_n \emph{\textbf{v}} (X_t)dX_t = \textbf{0}.$$
When, for an isolated domain $D(t)=\bigcup_{n}D_{n}(t)$ we assign
to the center of mass $G_n(t)$ of each $D_n(t)$ the resultant
speed vector \emph{\textbf{v}}$_n$(\emph{t}), we obtain
$$\sum_nV(D_n(t)) \frac{E_{t,n}}{c^2}(G_n(t)) \emph{\textbf{v}}_n(t) = \emph{\textbf{a}}.$$or
$$\sum_nV_n(t)m_n\emph{\textbf{v}}_n(t) =
\emph{\textbf{a}}.$$where \emph{\textbf{a}} is a constant
vector.\\
When a particle of mass $m$ is subject to a constant exterior force
and is filling a domain $D=\bigcup_{n}D_{n}(t)$ and moving at a
speed vector \emph{\textbf{v}}(\emph{t}), we have
$$\sum_nV_n(t)m_n \emph{\textbf{v}}_n (t) =
m \emph{\textbf{v}}(t).$$ For an atom, for example, of mass $m$
and speed \emph{\textbf{v}} having $k_1$ electrons and $k_2$
quarks filling respectively the volumes $V_{1,i}$ and $V_{2,j}$
and having respectively the speeds \emph{\textbf{v}}$_{1,i}$ and
\emph{\textbf{v}}$_{2,j}$, we have
$$\sum _{i=1}^{k_1}V_{1,i}m_{1,i}\emph{\textbf{v}}_{1,i}(t)+
\sum _{j=1}^{k_2}V_{2,j}m_{2,j}\emph{\textbf{v}}_{2,j}(t) =
m\emph{\textbf{v}}.$$ The solar system equally obeys the same scheme.\\
Indeed, let us assume that the solar system with its \emph{N}
planets is isolated (which is not the case as it is inside the
milky way) and designate the absolute speed vector of the system
gravity center (i.e. with respect to a virtual fixed frame) by
\textbf{\emph{v}}. Likewise, let us designate the absolute speed
vector of the sun (resp. of the \emph{N} planets) by
$\emph{\textbf{v}}_0$ (resp. by $\emph{\textbf{v}}_i$, \emph{i} =
1,2,...\emph{N}). All trajectories are geodesic with respect to
the cosmological metric $g_t$ (i.e. $\nabla_{\emph{\textbf{v}}_i
(t)}^{g_t} \emph{\textbf{v}}_i$(\emph{t}) = \textbf{\emph{0}}) and
we have
$$|| \emph{\textbf{v}}_i(t)||_{g_t} = v_i \;\;\; \mbox {where } v_i \;\;\; \mbox {is constant
for } i = 0,1,...N.$$So, we have
$$\sum_{i=0}^{N} m_i \emph{\textbf{v}}_i = m_0 \emph{\textbf{v}}_0 + \sum_{i=1}^{N} m_i \emph{\textbf{v}}_i  =
(\sum_{i=0}^{N}m_i) \emph{\textbf{v}} = m_0 \emph{\textbf{v}} +
(\sum_{i=1}^{N} m_i)\emph{\textbf{v}}.$$ Now, for \emph{i} =
1,..., \emph{N}, we have
$$\emph{\textbf{v}}_i = \emph{\textbf{v}} + \emph{\textbf{u}}_i$$where $\emph{\textbf{u}}_{i}$ is the relative speed vector with
respect to the sun. Then
$$m_0 \emph{\textbf{v}}_0 + \sum_{i=1}^{N}m_i(\emph{\textbf{v}} + \emph{\textbf{u}}_i) = m_0 \emph{\textbf{v}} +
(\sum_{i=1}^{N}m_i)\emph{\textbf{v}}.$$ But the gravity center of
the system is nearly the same as that of the sun and then we have
$\emph{\textbf{v}}_0 \simeq \emph{\textbf{v}}$. Consequently, we
get
$$\sum_{i=1}^{N} m_i \emph{\textbf{u}}_i = \emph{\textbf{0}}.$$
Notice also that, apart from the whole universe, null other system
is durably isolated (including galaxies, black holes and naturally
all systems with planetary scale). However, it is the distribution
(essentially local distribution) $E_t(X)$ that governs the free
motion in all local or microlocal system. Thus, at the atom level,
for instance, the free movement of electrons along their
respective orbits or rather the trajectory of a material point of
each electron, is governed by the acceleration
$\Gamma(t)=-\nabla^{g_e}E_t(X(t))$ and satisfy the identity
$\widetilde{\Gamma}(t)=\nabla _{X'(t)}^{g_t}X'(t)=0.$ However,
after an exterior energetic support (thermic or electromagnetic,
for example), the electron is submitted to possibly many energy
transformations as a change of its orbital energy level or a pure
separation from its original atom with a well determined kinetic
energy. We notice that, in the inverse
case, the energy conservation law is insured by emission of photons.\\
This is still valid for a nucleus which is, in addition, submitted
to nuclear interaction forces that (under external stimulation or
by a natural process) lead to matter-energy transformations such
as: disintegration, fission, fusion, excitation and radiations and
to chemical reactions leading to a transfer or liberation of
energy that are governed by the energy conservation principle, the
best energetic
equilibrium rule and the mechanical principle of the least action and also by the Pauli exclusion principle.\\
We finally mention that each solution $E(t,X(t))$ of equation
($E^*$), propagating along $X(t),$ is of the form $E
_{\mu_0}(X(t))=f_{\mu_0}(t)\varphi_{t,\mu_0}(X(t))$ and can not be
equal to a linear combination of such solutions as we can verify
this with the phenomenon of the diffraction of light. A
monochromatic light ray can not be diffracted into multiple rays
that will have different
energy than the incident ray.\\\\
Equality $f_{\mu}(t)=\frac{1}{2\pi}\frac{\sqrt \mu}{t}$ shows that
the undulatory frequency of the matter-energy increases with $\mu$
and decreases with $t$. So, just after the Big Bang (for $t$
sufficiently small) all propagations have an undulatory character
which is as pronounced as $\mu$ is larger. For $t$ sufficiently
large, only the very large eigenvalues lead to waves that have a
significant undulatory character. However, we have
$v(t)=\lambda(t)f_{\mu}(t)$ where $\lambda(t)$ is the wavelength
and $v(t)\leq 1$. Then, for $t<<1$, we have $f_{\mu}(t)\gg1$ for
all $\mu$ and so we have $\lambda(t)\ll1$. Thus we can conceive
intuitively that, when $t=0$ (that is, before the Big Bang), we have $\lambda=0$ and then there
is neither material nor non material propagation. Moreover when
$v_e(t)$ is the Euclidean speed of propagation of a wave starting at an
infinitesimal time $t\sim0$, we have
$$v_e(t)=\frac{1}{2\pi}\frac{\sqrt{\lambda^2(t)\mu}}{t}$$and then,
for finite $t\gg0$, when $v_e\sim 1,$ we have
$$\frac{\lambda^2(t)\mu}{t^2}\leq 4\pi^2.$$
So, for the wave of speed $v_e=1$ (light, X rays, $\gamma $ rays),
we have $\lambda^2(t)\mu=4\pi^2t^2$ and
$\displaystyle \lambda(t)=\frac{2\pi t}{\sqrt \mu}$. Nevertheless, for the first original photons we have $\displaystyle E(t,X(t))=h_\mu(t)f_\mu(t)=0$, $v_e(t)\neq0$ and $\displaystyle \lim_{t\rightarrow0}\lambda(t)=0$ beside of $\displaystyle \lim_{t\rightarrow0}h_\mu(t)=0$ and $\displaystyle \lim_{t\rightarrow0}f_\mu(t)=+\infty$.\\
In order to come back to our starting point, we notice that these
latter waves are those that create, since the Big Bang until now,
the geometrical space whose expansion occurs at an increasing speed \emph{v(t)}
which is presently very near to 1 and theoretically must tend to 1. The
material corpuscular universe is expanding at a speed inferior to
1 and its acceleration is submitted to the possibility of the
material perception, which needs to be determined more and more
precisely, although it theoretically must be nearly null.\\\\
\textbf{Remark}: If we have considered a ball of matter-energy of mass-energy $M_0\sim E_0$ with a very large density of energy (such as
an ultradense neutron star) instead of the potential energy $E_0
\sim M_0$ concentrated at a single point $O$ (as we have represented
the potential energy of a black hole as
$e(I)\delta_I=m(I)\delta_I$), then we would have not altered our
mathematical model in virtue of
the enormous extent of the virtual or the geometrical space. Thus, this fact allows us to
avoid appealing to the infinity notion (Infinite density, infinite curvature, infinite pressure and infinite temperature). \\\\

\section{Some repercussions on modern Physics}
\subsection*{Temperature and pressure}
$\hskip 0.5cm$The major absent factor along our study until now is
the temperature (intrinsically related to the pressure) factor.
However, temperature is an inherent characteristic in the
expansion operation: The universe is permanently expanding and
cooling. Moreover, temperature is indissociable from all energy
forms: heat, radiations (via thermal spectrum),chemical and
nuclear reactions, internal energy of stars (via fusion and
internal fluctuation pressure) and particularly, temperature is
associated with the average kinetic energy of molecules in thermal
equilibrium through the formula
$$<E_k> = \frac{3}{2}kT.$$
So we can briefly say that: temperature intervenes into and
fashions equilibrium states of all systems' energy. We then start
by specifying that the relation used in the preceding sections, in
order to describe the free trajectories $X(t)$ (i.e.
$\nabla_{X'(t)}^{g_t}X'(t)$ = 0) for punctual material particles
and photons as
$$E_\mu(t,X(t)) = e(\mu),$$
is only valid on small time intervals where we can consider the
temperature as being constant. Indeed, although the metric $g_t$
takes implicitly into account the surrounding temperature factor,
we have to recognize that the preceding formula must be written as
$$E_\mu(t,T(t),X(t)) = e(\mu,T(t)).$$
Beside of the dependence of the energy $E_\mu(t,X(t))\;=\;
h_\mu(t)f_\mu(t)$ on $\mu,$ we must add necessarily its dependence
on $T(t)$ through the dependence of the metric itself on $T(t).$
The energy $E(t, X(t))$ which is conveyed to us by radiations from
long time ago and long distance away is attenuated not only
because of collisions, but above all under the influence of the
undulatory universe cooling.The decrease of frequencies (i.e. the
increase of wavelengths) is counterbalanced by the increase of
$h(t).$ On the other hand, if the waves are propagating along
$X(t)$ with a constant energy $E(t,X(t)),$ then we can not have
$$\int_{B(O,t)} E_t dX_t = \mbox{const.}$$
for each $t\gg0$.\\
If we adapt the perfect gas model to the whole universe, we can
assume that we have permanently
$$P(t)V(t) = K(t)T(t);$$
which implies, for large enough positive \emph{t}:
$$P(t)t^3 = K'(t)T(t).$$
For \emph{t} $\ll$ 1, the situation could not be the same because
then vol($B_{e}(O,R(t)))$ could not be proportional to
\emph{t$^3$}. This is due to the fact that the speed of
radiations' propagation that verifies $\nabla^{g_t}_{X'(t)}X'(t)$
= 0 could be originally less than 1 which is the speed of light in
the vacuum. Indeed the metric \emph{g$_t$} contracts the distances
significantly at the early expansion because of the largeness of
the interaction forces' magnitude (gravitational and other
interaction forces) at the origin as well as the largeness of the
energy density and the pressure and
temperature intensities.\\
We notice that, we can avoid using the notion of a very large
energy that is concentrated in a point at the origin of time
$t\;=\;0$ with infinite pressure and temperature and conceive
starting our study by considering the quasi-original universe as
being reduced to a small ball of Euclidean radius $r_0$ at a small time $t_0\;>\;0.$ The universe is then considered at
time $t_0$ as being a small ball, within it radiations are
characterized by very large (but finite) pressure and temperature.
This situation evolves with the time's progress toward a state
qualified as a soup of quarks and leptons before the formation of
hadrons followed by nucleons, atoms and galaxies that marks the
passage from a radiations' dominated state to a matter dominated
state. Temperature had certainly an important influence during
this evolution which led to the current situation characterized by
an approximate average temperature of $2,74K.$ However, a large
number of technical (theoretical or experimental) means are
available for us to go forward in our investigations in order to
discover more and more thoroughly the original states of our
universe and the laws
that govern its evolution.\\
Let us show now, using some examples, that we can recover some
confirmed results in modern Physics without using neither the
erroneous part of the second postulate of
special relativity nor the uncertainty principle.\\\\
\subsection*{Remarks on relativistic formulas}

In the following, we will notice some remarks and establish some
properties and results based on the refutation of the erroneous
interpretation of the special relativity second postulate and the
spacetime relativistic notion as well as on the canonicity of the
Maxwell equation and the principle of the speed of light constancy
that are established by using the derivations $\frac{d_1}{dt}$ and
$\frac{d_*}{dt}$ which take into account the moving frame speed
(see sections: 2,3
and 4.)\\

The speed actually is a continuous variable on $[0,1[,$ then we
have opted to do not make a sharp distinction, according to their
speed, between relativistic and non relativistic particles
concerning either their energy or their momentum. What is the
critical speed starting from which we can use the relativistic
formulas:\\\\
$$p=\frac{mv}{\sqrt{1-\frac{v^2}{c^2}}}=:\gamma mv , E = \sqrt{p^2c^2+(mc^2)^2}, E =
\gamma m c^2 \mbox{ and } v = \frac{pc^2}{E}?$$\\
These formulas can not coincide with classical ones that give the
particle energy at any non vanishing speed:
$$E = m c^2 + \frac{1}{2}mv^2 = \gamma m c^2 \Longleftrightarrow \gamma = 1
+ \frac{v^2}{2c^2} \Longleftrightarrow \frac{1}{1-\frac{v^2}{c^2}}
= 1+ \frac{v^2}{c^2}
+\frac{v^4}{4c^4}$$\\
$$\Longleftrightarrow 1 = 1 - \frac{v^4}{c^4} + \frac{v^4}{4c^4} - \frac{v^6}{4c^6}
\Longleftrightarrow 0 = -\frac{3v^4}{4c^4} - \frac{v^6}{4c^6}.$$
Moreover, we have in classical Physics (for \emph{v} $\ll$ 1):
$$p = mv, \hskip 0.15cm E_k = \frac{1}{2}mv^2 = \frac{1}{2}pv
\hskip 0.15cm and \hskip 0.15cm E = m c^2 + \frac{1}{2}mv^2$$\\
whereas, in the relativistic framework, we have
$$p c = \frac {h c}{\lambda} = hf = E$$
for photons and \emph{v} = $\frac {p c^2}{E}$ for material
particles. For these last particles, the two notions coincide only
for \emph{v} = 0. Indeed, if $v>0$ then
$$v \;=\; \frac{p c^2}{E} = \frac{mvc^2}{E} \quad\mbox{ implies }\quad E \;=\;mc^2,$$
which is contradictory ($E\;=\;m c^2$ $\Longrightarrow$ $v\;=\;0$).\\

\noindent{\bf Remark}: We briefly notice that all results and formulas that have been
established by using the erroneous part of the second postulate
can be established more precisely in a consistent
manner. However, using relativistic formulas leads to very useful approximate results.\\

\subsection*{The famous $E = mc^2$ statement}

Within our model, we have for material or immaterial points:
$$E_\mu(t, T(t), X(t)) = h_\mu(t, T(t))f_\mu(t, T(t)) = e(\mu,
T(t))$$ and, for any fundamental material particle with mass
$m(t)$ that occupies a domain of volume $V(D_t)\;=:\;V(t)$ at time
$t$ and for all speeds $v(t)\leq v_e(t)<1$ we have
\begin{eqnarray*}
  E(t) &=& \int_{D_t}E_\mu(t, T(t), X(t))dX_t = \int_{D_t}
  h_\mu(t,T(t)) f_\mu(t,T(t)) dX_t \\
  &=& h_\mu(t, T(t))f_\mu(t, T(t))V(t) = e(\mu, T(t))V(t) \\
  p(t) &=&  \int_{D_t}m(X_t)v(X_t)dX_t = m(t)v(t)\\
  E_k(t) &=& \frac{1}{2} \int_{D_t}m_1(X_t)v^2(X_t)dX_t =
\frac{1}{2}m_1(t)v^2(t)
\end{eqnarray*}
and
$$E(t) = m(t)c^2 + \frac{1}{2}m_1(t)v^2(t) = m(t)(c^2 +
\frac{1}{2} \frac{m_1(t)}{m(t)} v^2(t)) =: \rho(t)V(t)(c^2 +
\frac{1}{2} \frac{m_1(t)}{m(t)} v^2(t)),$$
where $\rho(t)$ is the density of mass of the material particle.\\ 
So we have
$$h_\mu(t,T(t)) f_\mu(t,T(t)) V(t) = \rho(t) V(t)
(c^2+\frac{1}{2} \frac{m_1(t)}{m(t)} v^2(t))$$and then
$$E(t,X(t)) = h(t)f(t) = \rho(t)(c^2 + \frac{1}{2} \frac{m_1(t)}{m(t)} v^2(t))$$
where we have denoted $E_\mu(t,T(t),X(t))$ by $E(t,X(t))$,
$h_\mu(t,T(t))$ by $h(t)$ and $f_\mu(t,T(t))$ by $f(t).$\\
For $v\;=\;0$ we get $E(t_0, X(t_0))\;=\;
\rho(t_0)c^2$ and so:\\
The punctual energy of matter at rest is equal to its energy
density, and then we recover the famous Einstein's statement
$$ E_0 \;=\; m_0 c^2$$
Thus, according to our model, the total energy of a material particle is
$$E(t) = m(t)(c^2 + \frac{1}{2} \frac{m_1(t)}{m(t)} v^2(t))$$
where $m(t)$ is the initial (at rest) mass $m(0)\;=\;m_0$
multiplied by a factor $\gamma(t)$ depending on the speed and time
$$m(t) = \gamma(t)m_0.$$
This factor has been experimentally determined (several years
before the special relativity theory) by the physician W. Kaufmann
who qualifies the expression $\gamma m$ as being the apparent
mass. The $\gamma(t)$ factor actually is the Lorentz factor:
$$\gamma(t) = \gamma = \frac{1}{\sqrt{1-\beta^2}} = (1 -
\frac{v^2(t)}{c^2})^{-\frac{1}{2}}$$ for $\beta =
\frac{v(t)}{c}$.\\
Thus, writing $m(t)\;=\;\gamma(t)m_0$ and $m_1(t) = \gamma_1(t)
m_0$, one obtains
\begin{equation} \label {r17}
E(t) \;=\; \gamma(t)m_0\left(c^2 + \frac{1}{2}
\frac{\gamma_1(t)}{\gamma(t)} v^2 \right) \simeq \gamma(t) m_0
c^2,
\end{equation}
as, for small speeds as well as for large speeds, the term
$\frac{1}{2} \frac{\gamma_1(t)}{\gamma(t)} v^2$ is negligible
compared with $c^2$, and
\begin{equation} \label {r18}
p(t) \;=\; \gamma(t)m_0v
\end{equation}
and for $v(t) \neq$ 0
$$\frac{E(t)}{p(t)} \;=\; \frac{c^2 + \frac{1}{2} \frac{\gamma_1(t)}{\gamma(t)} v^2}{v}.$$
So, for $v(t)\equiv0,$ we get $p(t)\equiv0$ and
$E(t)\;=\;\gamma(t)m_0 c^2,$ which yields, together with
$E(t)\equiv
m_0 c^2,$ $\;\;\;$ $\gamma(t)\equiv\gamma(0)=1=\gamma_0.$\\
For $v\sim c,$ we have $E(t) \simeq p(t) c$ and for $v\sim0,$ we
have
$E(t) \simeq m_0 c^2$ and $p(t)\sim0.$\\
For $v\ll1$, we get $\gamma\simeq1$, $m(t)\simeq m_0$, $\gamma_1\simeq 1$, $m_1\simeq m_0$ and $E(t)\simeq m_0c^2+\frac{1}{2}m_0v^2$.\\

\noindent Differentiating the approximate equality (17) and equality (18)
with respect to
$t,$ we obtain\\
\begin{equation} \label {r19}
E^{'}(t)\;=\;m_0\gamma^{'}(t) c^2
\end{equation}
\begin{equation} \label {r20}
p^{'}(t)\;=\;m_0\gamma^{'}(t)v\;+\; m_0\gamma(t)v^{'}
\end{equation}
and by differentiating them with respect to the speed we get\\
\begin{equation} \label {r21}
\frac{dE}{dv} \;=\;m_0\frac{d\gamma}{dv} c^2
\end{equation}
\begin{equation} \label {r22}
\frac{dp}{dv} \;=\;m_0\frac{d\gamma}{dv}v\;+\;m_0\gamma(t)
\end{equation}\\\\
We notice that the approximate equation (17) and equation (18) are
consistent as they imply
$$ p^{'}(t) = \frac{E^{'}(t)}{c^2} v+ m_0 \gamma(t) v^{'}$$
and then
$$ \frac{dp}{dv} v^{'} = \frac{1}{c^2} \frac{dE}{dv} v^{'} v + m_0
\gamma(t) v^{'}$$ Similarly, equations (21) and (22) imply
$$ \frac{dp}{dv} = \frac{1}{c^2} \frac{dE}{dv}v + m_0 \gamma(t)$$
which is equivalent to the above equation.\\
The approximate relation (17) and relation (18) actually are the famous
Einstein's equations for the energy and momentum.\\\\
\textbf{Remark:} The differentiation of the exact relations (17)
and (18) can give some indications on the factor $\gamma_1$.\\\\
We notice also that a particle with mass $m(t)$ can not
achieve a final speed $v = 1$ and having a final mass $m_f>0$ as,
$$ \lim_{v \rightarrow 1} \gamma(v) = + \infty,$$
which means that the energy needed for such a particle to achieve
this speed would be infinite.\\

\subsection*{Momentum, Kinetic energy and Mass}
Within the framework of our model, we have privileged (as Einstein
has originally done) the notion of the mass at rest $m_0$ of
particles. However, we have adopted, for particles having
significant speed, the notion of speed and time-dependent mass
under the form
$$m(t)=\gamma(t)m_0$$
where $\gamma(t) = (1+\frac{v^2(t)}{c^2})^{-\frac{1}{2}}$. We also
have adopted the expression
$$E_k(t)=\frac{1}{2}m_1(t)v^2(t)=\frac{1}{2}\gamma_1(t)m_0v^2(t)$$
for the kinetic energy and the expression
$$\overrightarrow{p}(t)=m(t)\overrightarrow{v(t)} = \gamma(t)m_0\overrightarrow{v(t)}$$
for the momentum of a moving particle.\\
The momentum of any particle is classically defined as
$$\overrightarrow{p}=m\overrightarrow{v}\quad\quad\hbox{with}\quad\frac{d\overrightarrow{p}}{dt}=\overrightarrow{F}$$
where $\overrightarrow{F}=m\overrightarrow{\Gamma}$ is the force
that acts on the particle, whereas the classical definition of the
kinetic energy is
$$E_k=\frac{1}{2}mv^2.$$
Einstein has earlier showed that these definitions are erronuous for particles moving with high speeds. Indeed, a simple example ([2], p.$112$) shows that the classical definition of kinetic energy leads to the relation $v=\sqrt{\frac{2E_k}{m}}$ which contradicts the fundamental law of special relativity (and Physics in general) asserting that the speed of any material particle cannot exceed $c=1.$\\

\noindent However, the second example ($[2], p.113$), which has been used for showing the non conservation of momentum during the collision of two particles $A$ and $B$ having the same mass $m$ and opposite speeds $\overrightarrow{v}$ and $-\overrightarrow{v}$ with respect to a given referential frame $S^{'},$ do not permit to draw the same consequences that are established by using the relativistic formulas of the frame exchange. For us, the momentum is naturally conserved when using the rest frame $S$ for either $A$ or $B.$
This is clearly showed by the diagrams of figure 11.\\\\

In other respects, it is clear that ([2],p.112) the classical
formulas $p = mv$ and $F = \frac{dp}{dt}$ contradict the
fundamental principle which stipulates that the speed of a body of
a non
vanishing mass can not exceed the speed of light.\\
Likewise, the relation $F=\frac{dp}{dt}$, for $F \neq 0$, leads,
within our framework, to a contradiction of the same nature as
previously. Indeed, let us consider (for instance) an electron
having a rest mass $m_0$ which is accelerated through an
electrical field $E$ such as the exerted electrical force on the
electron is a non vanishing constant.\\
When we write
$$F=m(t)\Gamma(t)=\gamma(t)m_0\Gamma(t)$$
and
$$p=m(t)v(t)=\gamma(t)m_0v(t),$$
for $m(t)\neq0$ and then for $\gamma(t)\neq0$ and $v(t)<1,$ we get
$$F=\frac{dp}{dt}\;\Leftrightarrow\;\gamma(t)m_0\Gamma(t)=\frac{d}{dt}(\gamma(t)m_0v(t))\;\Leftrightarrow\;\frac{F}{m_0}=\frac{d}{dt}(\gamma(t)v(t)).$$
As $F$ is assumed to be constant, we obtain
$$\gamma(t)v(t)=\frac{F}{m_0}t+C=\frac{F}{m_0}t+\gamma(\tau)v(\tau)-\frac{F}{m_0}\tau$$
for a $\tau>0.$\\
Therefore, we have
$$\frac{d}{dt}(\gamma(t)v(t))=\gamma^{'}(t)v(t)+\gamma(t)\Gamma(t)=\frac{F}{m_0}=\gamma(t)\Gamma(t)$$
which implies
$$\gamma^{'}(t)=0\quad(\hbox{since}\;v(t)\neq0)$$
and
$$\gamma(t)=\gamma=\hbox{const.}\quad\hbox{and}\quad m(t)=\gamma m_0=\hbox{const}. \hskip0.5cm $$
which is impossible since accelerated particles can not have constant mass.\\
Likewise, the above relation implies
\begin{eqnarray*}
  v(t) &=& \frac{F}{m_0\gamma}t+\frac{C}{\gamma} \\
   &=&  \frac{F}{m}t+\frac{C}{\gamma}=\Gamma t+\frac{C}{\gamma}
\end{eqnarray*}
for strictly positive constant $\Gamma,$ which is equally impossible since the speed of any moving particle of mass different from zero can not exceed $1.$\\
Therefore the relation $\frac{dp}{dt}=F$ can only be approximately
correct for minimal speeds where $\gamma^{'}(t)\sim
0,\;\gamma(t)\sim1$ and $m(t)\sim m_0$, provided that $m(t)\neq0.$
In this situation, we may write
$$\frac{dp}{dt}=\frac{d}{dt}m_0v(t)=m_0\Gamma(t)=F(t)$$ and, for
$v=\hbox{const.},$ we have
$$\frac{dp}{dt}=\frac{d}{dt}m_0v=0.$$
Recall that, within the framework of our model, we have
\begin{eqnarray*}
  p(t) &=& m(t)v(t)=\gamma(t)m_0v(t) \\
  E_k(t) &=& \frac{1}{2}m_1(t)v^2(t)=\frac{1}{2}\gamma_1(t)m_0v^2(t) \\
  E(t) &=& m(t)(c^2+\frac{1}{2} \frac{m_1(t)}{m(t)} v^2(t))=\gamma(t)m_0(c^2+\frac{1}{2} \frac{\gamma_1(t)}{\gamma(t)} v^2(t))
\end{eqnarray*}
for $m(t)\neq0$ and $v(t)\leq v_e(t)<1.$\\
These formulas conform with the two fundamental conservation laws (of energy and momentum).\\
Actually, the momentum conservation principle is clearly
expressed, within the framework of our model, by
$$\nabla_{X^{'}(t)}^{g_t} \overrightarrow{p(t)} =
\nabla_{X^{'}(t)}^{g_t} \gamma(t) m_0 X^{'}(t) = m_0
\nabla_{X^{'}(t)}^{g_t} \gamma(t) X^{'}(t) = $$
$$m_0 (\gamma(t) \nabla_{X^{'}(t)}^{g_t} X^{'}(t) +
X^{'}(t).\gamma(t) X^{'}(t)) = m_0 (\gamma(t)
\nabla_{X^{'}(t)}^{g_t} X^{'}(t) + \gamma^{'} (X(t))X^{'}(t))=\overrightarrow{F}.$$
So, if \emph{X}(\emph{t}) is a geodesic, we get
$$\widetilde{\Gamma}(t) = \nabla_{X^{'}(t)}^{g_t} X^{'}(t) =
0$$and
$$\parallel X^{'}(t)\parallel_{g_t} = v = \mbox{const.}$$which
gives $\gamma$(\emph{t}) = const. and then $\gamma^{'}(t)=0$ and
$$\overrightarrow{F}=\nabla_{X^{'}(t)}^{g_t} \overrightarrow{p(t)} = 0.$$
This quantity is truthfully null for any free trajectory
\emph{X}(\emph{t}) i.e. for any geodesic with respect to the real
physical metric which takes into account all of the natural forces
acting
on the particle. Therefore, we can state that the relation
$$ \nabla^{g_t}_{X^{'}(t)} \overrightarrow{p(t)} = \overrightarrow{F}$$is more consistent
than the above classical one as it confirms that $\overrightarrow{F}(X(t))=0\Longleftrightarrow X(t)$ is a geodesic for $g_t$.\\\\
Finally, we mention that, within the relativistic framework, the
definition of the momentum by $\overrightarrow{F} =
\frac{d\overrightarrow{p}}{dt}$ leads to the relation ([2],
(4.104))
$$ \overrightarrow{\Gamma} = \frac{d \overrightarrow{v}}{dt} =
\frac{\overrightarrow{F} -
\overrightarrow{\beta}(\overrightarrow{F}.\overrightarrow{\beta})}{m
\gamma}$$which shows that, at large speed, the acceleration is not
parallel to the force whereas within our model we have, for the
force field along the trajectory $X(t)$:
$$ F_{g_t}(X(t)) = - \nabla ^{g_t} E_t(X(t)) =
\widetilde{\Gamma}(t).$$\\

\subsection*{Remarks on the quantum theory}
Concerning the uncertainty principle, it seems illogical that,
after some experiments such as the one where particles hit a
screen through two small slits slightly spaced, we conclude that
the fact of knowing the origin of the particles hitting the screen
could really alter the physical phenomenon. It is true that the
means used in order to know this origin can alter the results by
modifying the trajectories but this is only a technical and
circumstantial phenomenon which does not allow us to conclude that
our pure knowledge can transform the physical results that are
determined objectively by the very physical conditions. Beside of
that, our theoretical or practical capacity to discover any law of
Nature does not influence the objective reality of this law. The
very long history of discoveries in all domains shows the
objectivity of natural laws independently from our circumstantial
(theoretical, approximate, experimental or technical) capacity to
discover them. The use of progressive energetic levels of
particles and greater and greater frequencies (i.e. smaller and
smaller wavelengths), for instance, has permitted the realization
of important progress in the understanding of our physical
universe through the ages as well as the understanding of matter
(nucleons, quarks, hadrons, leptons) structure and also in the
refinement of our knowledge on the quantization of both energy
levels and angular as well as intrinsic atomic and nuclear
momenta. The Schr\"{o}dinger function and Quantum Statistics have
also permitted a jump into our comprehension of the universe by
giving effective methods to discover and interpret all results
that are obtained by experimentation and led to powerful
approximations for natural phenomena rules. These phenomena
contain essentially a part of uncertainties due to the multitude
of (energetic and dynamic) evolutive factors that govern all
aspects of
matter-energy: energy levels, trajectories, interactions...\\
Although these phenomena are far from being regular
(differentiable), they are continuous. An electron, for instance,
that changes its orbit (which is in permanent evolution) for a
higher energy level (when absorbing a photon) or for a lower
energy level (together with emission of photon), spends a very
small fraction of time before achieving its final state. During
this transition, continuity of trajectory and conservation of
energy are both insured.\\\\In other respects, we notice that the
solution $\psi$ to the classical Schr\"{o}dinger equation
$$-\frac{\overline{h}^2}{2m}\frac{d^2\psi}{dx^2} +
V(x)\psi(x) \;=\; E\psi(x),$$for example, which gives the
probability of finding the particle per unit of $x$ using the
distribution
$$\frac{dP}{dx} = |\psi(x)|^2,$$
is determined by experimental and predictive way.\\
This function is completely different from our function $\psi(X)$
which comes from the resolution of the energy equation $(E^{*})$
and is given by
$$E(t,X) = g(t)\psi(\frac{X}{t})$$
and determines both the energy and the frequency of a material or
immaterial point as well as any punctual material particle at
every given time $t.$\\
Moreover both functions must be distinguished from the trajectory
$X(t)$ of this particle. So, for a simple pendulum or a quantum
harmonic oscillator, for instance, the frequency of oscillation is
not the same as the material points frequency. When the pendulum
is located at the stable vertical equilibrium, the frequency of
any material point is determined by its characteristic energy
$E(t,X)= h(t)f(t);$ its potential energy and its kinetic energy
are null and its mass energy is $E =mc^2 = m.$ Nevertheless, we
can not speak neither about its period nor its frequency
$f=\frac{v}{\lambda}.$ So we can not assume that its frequency is
different from zero and try to determine, using the uncertainty
principle, its ground state energy which would be different from
zero (even if it was very small) contrary to Newtonian principles
of classical Physics. This is also valid concerning a Tennis or
any small ball, for example, being at rest into a box assumed to
be also at rest.\\
Likewise, the minimal quantized energy level of the electron of
the hydrogen atom corresponds to the orbit which is associated to
the Bohr radius
$$r = a_0 = \frac{4\pi\varepsilon_0\overline{h}^2}{me^2}$$
which is determined by the minimal energy
$$E_m = (\frac{1}{2}mv^2 -
\frac{e^2}{4\pi\varepsilon_0r})_m = (\frac{\overline{h}^2}{2\pi
r^{2}} - \frac{e^2}{4\pi\varepsilon_0r})_m$$where $\overline{h}$
is here the classical Planck constant.\\This result has nothing to
do with the uncertainty principle; it is rather due to the fact
that the minimal total energy that can have an electron inside a
hydrogen atom (the ground state energy) is finite and
characterized by both constants $a_0$ and $\overline{h}$.\\\\

Let us finally specify that the uncertainty principle which can be
written as \emph{$\Delta$x$\Delta$p$_x$} $ \geq $
$\frac{\overline{h}}{2}$ and \emph{$\Delta$E$\Delta$t} $ \geq $
$\frac{\overline{h}}{2}$, is only a legitimate consequence of using
Schr\"{o}dinger equations:
$$\psi(x) =
\frac{1}{\sqrt{2\pi}}\int_{-\infty}^{+\infty} g(k)e^{-ikx} dk$$
and
$$g(k) = \frac{1}{\sqrt{2\pi}}\int_{-\infty}^{+\infty} \psi(x)e^{ikx}
dx$$ in order to find out the probability of localizing a given
particle under some constraints in a given position. This
principle is stated after using the probability distributions
$$\frac{dP}{dx} = |\psi(\emph{x})|^2,\hskip 0.5cm \frac{dP}{dk} =
|g(k)|^2$$and their standard deviations \emph{$\sigma_x$} and
\emph{$\sigma_k$} as well as the De Broglie relation $p =
\frac{h}{\lambda}.$ Therefore this principle states simply that
this particular approach and the use of this particular method
hold within themselves the uncertainty so quantified. Nevertheless
this does not mean that the position $x$ of the particle or its
momentum $p_x,$ at a given time, can not be well defined or can
not be determined with more precision by a
more efficient theoretical or experimental process.\\\\
Indeed, none can assert that one can not perform a technical
device or a theoretical process that could be used in order to
measure the width of a slit or the size of a particle that are
much smaller then those which are reached at the present time by
means of scattered particles. We maybe could use $\gamma-$rays
having very much smaller wavelengths after inventing an
intermediate device (or process) which makes their effects
accessible to our sensitivity or our comprehension. Likewise, we
can hope performing some new processes for measuring both position
$x$ and momentum component $p_{x}$ of a given particle which could
improve the uncertainties $\Delta x$ and $\Delta p_{x}$ as well as
their product $\Delta x \Delta p_{x}$ which is limited presently
by
$\frac{\overline{h}}{2}$ when using our current process.\\
Obviously, we can use Schr\"{o}dinger equations and quantum
Statistics as effective approaches leading to qualitative and
quantitative approximations (with naturally a margin of
uncertainty) of the studied physical phenomena when real
approximate measurements are difficult to
achieve. These issues will be discussed and made more precise in the next sections.\\\\

Notice that, within the framework of our model, the notions of
wavelength and frequency are characteristics of material or
immaterial points but not of material particles even though they
are considered physically as punctual particles such as electrons
for example. This is precisely the context within which we have to
understand and explain the undulatory character of matter. We then
consider the fact of attributing a wavelength $\lambda =
\frac{h}{p}$ and a frequency $ f = \frac{v}{\lambda}$ to a
pointlike material particle as being a practical method for making
approximate calculation and they do not correspond to a real
periodic trajectory and can not be used for making exact
calculation of relativistic or non relativistic particle's energy
by means of formulae such as:
$$E = \sqrt{(pc)^2+(mc^2)^2} = \sqrt{p^2+m^2} =
\sqrt{\frac{h^2}{\lambda^2}+m^2}=
\sqrt{\frac{h^2f^2}{v^2}+m^2},$$for
example (see section 8).\\\\
Likewise, when we use the relation 
$$p = \gamma mv =
\frac{h}{\lambda}= \frac{hf}{v}$$ 
for a non relativistic punctual
material particle, we get
$$ E_T\;:=\; hf  \;=\; \gamma mv^2$$which
implies (according to relativistic formulas)
$$\gamma m\; =\; \gamma m v^2$$
and then  $v^2=1$  which is absurd. However, the momentum \emph{p}
of a photon is really:
$$p = E = hf = \frac{h}{\lambda}$$and then we have $\lambda$ =
$\frac{h}{p}$.\\Likewise, the use of the relativistic momentum
expression $p = \frac{h}{\lambda}$ for material particles leads to
a flagrant contradiction. Indeed, let us write, for the electron
of the Bohr atom model (for instance),
$$\frac{<E_k>}{<E_p>} = -
\frac{<\frac{mv^2}{2}>}{ke^2<\frac{1}{r}>} = -
\frac{<\frac{mv^2}{2}>}{ke^2<\frac{mv}{\overline{h}}>}$$where we
have used the formula ([2],p.139)
$$<\frac{1}{r}> \simeq \frac{2\pi}{\lambda} =
\frac{p}{\overline{h}}$$and the non relativistic expression of
$E_k$, which is legitimate for the energy levels of this atom. So,
we have
$$-\frac{1}{2} = - \frac{v \overline{h}}{2ke^2} = -\frac{v
\overline{h}}{2\alpha \overline{h}c} = -\frac{1}{2\alpha
c}v,$$which is contradictory as \emph{v} decreases when $r$ increases.
\\\\
On another side, when we compare our expression for the undulatory
energy $E(t) = h_\mu(t)f_\mu(t)$ to the De Broglie-Planck-Einstein
one, which we will denote by $E =h_Pf_D$ (where $h_P$ is the
Planck constant and $f_D$ is the De Broglie frequency), we get:
$$h_{\mu}(t)f_{\mu}(t) = h_Pf_D$$which gives
$$\emph{h$_P$} = \frac{h_\mu(t)f_\mu(t)}{f_D(\mu,t)} =
\frac{e(\mu,t)}{f_D(\mu,t)}$$ and$$ f_D(\mu,t) =
\frac{1}{h_P}h_\mu(t)f_\mu(t) = \frac{1}{h_P}e(\mu,t).$$\\
\textbf{Remark:} A complementary and more systematic study of the
limits of Quantum theory is furnished in the next
section.\\

\subsection*{Repercussions on some other notions}
Finally, we notice that, in the light of our model, we can
reexamine and make precise a large number of notions and factors
having an important role in modern Physics (such as: Hubble law,
radiation spectrum dependence on the powers of temperature,
reunification of all forces problem...) without using neither the
erroneous aspect of the relativity second postulate nor the
uncertainty principle. We have to take into account the dependence
on time of some notions and constants. So, we can explain, for
instance, the redshift and the blueshift phenomena by the
dependence on time (or distance) and temperature of the frequency
and not by the speed of the
source. The only effect of the speed of the source is to make the emitter less or more distant from the receiver - analyser according to a given rate. The dependance on distance of the redshift factor $z$ explains its observed very large values $(z>2)$ that seam to contradict the classical general relativity theory. These values only prove the existence of very distant pulsars, for example.\\

Likewise, we can show easily that, within the framework of our
model, the relative speed of two isotropic galaxies moving in the
direction of the expansion is proportional to their distance. This
allows us to introduce the factor $R(t)$ that characterizes the
expansion and to follow Hubble's work by putting $r = r_0R(t)$ and
$$H(t) = \frac{\frac{dR}{dt}}{R} \mbox{ with } H_0 = \left(\frac{dR}{dt}\right)_{t = t_0}$$
in order to obtain the Hubble's law
$$v = \frac{dr}{dt} = H_0r$$
Moreover, if $m$ is the total mass of a galaxie located at a sphere of sufficiently large radius $R(t)$ and $M$ is the
total mass of the ball of radius $R(t),$ we have (J. W. Rohlf,
p.$552$) the classical results
$$E_k = \frac{1}{2}m r_0^2 \left(\frac{dR}{dt}\right)^2$$
and
$$V = -\frac{4 \pi m G r_0^2 R^2 \rho}{3}$$
where $V$ is the potential energy of the galaxy and $\rho$ is the
mean density of the ball mass.\\Next, following Einstein, we
introduce the curvature parameter $K(t)$ which is, within our
model, intrinsically related to the metric $g_t$ which, itself,
reflects the universe energy distribution. We then apply the
conservation of energy law in order to obtain the Friedmann
equation
$$\left(\frac{dR}{dt}\right)^2 = \frac{8 \pi \rho G R^2}{3} - K$$
Although this equation could give useful information on the
universe evolution, we mention that our model is different from
the Einstein-de Sitter one as it is based on another conception of
the space and time on one hand and, on the other hand, we recall
that our curvature parameter $K$ depends on time and can not be
$0.$ We notice also that our model does not conform with the first
postulate of the cosmological principle which states that the
universe would (thoroughly) look the same for any observer from
any galaxy, although it conforms with the second one which states
that the relative speed of galaxies (in the sense specified above)
is proportional to their distance. However, the readjustment of
Einstein's general
relativity theory, Hubble laws and Friedmann-Einstein's equations within our setting will be achieved in sections 10, 11 and 12.\\
We obviously have to use the quantum Statistics within its
important significance and limits. An extensive reexamination of these notions and postulates and its resulting consequences requires
naturally a long and laborious collective work.\\
All the preceding study invites us to believe that Physics is an
exact science, but this science can be revealed to us only
progressively and often approximately, meanly by joining
experimentation to theory. This is essentially due to the
complexity of natural phenomena (although the laws of Nature are
essentially simple) and to the limits imposed by our technical and
practical means and tools. However, all what remains in the
Physics domain (not in the Methaphysics one) is governed by some
number of principles although the majority of them has been
discovered by experimental and theoretical means.  The theoretical
way uses essentially Mathematics which is initiated and dynamised
by Physics and Technology, although it is purely theoretical
and
intellectual.\\\\
We conclude by saying that Mathematics and Physics are
indissociable in the same way as (more generally) they are theory
and practice (which is more experimental and utilitarian). This
shows the fundamental need of imagination, philosophy and
confidence in the collective humanity reason in order to go
forward in the scientific discovery way in all domains.\\

\noindent\textbf{Remark:} Further fundamental repercussions on Modern
physics will be developed in the following sections.\\

\section{The limits of Quantum theory}
 In this section we aim to specify the proper domain of the Quantum
theory efficiency. We will show that the wave Quantum Mechanics
(based on Schr\"odinger's equations) and quantum Statistics
essentially constitute experimental and approximate tools that
result in a probabilistic and predictive approach for explaining
physical phenomena. Consequently, they can not constitute a proper
theoretical framework for instituting any intrinsic or canonical
physical law. The De Broglie wavelength, which is a canonical
feature of electromagnetism, is simply a practical object that is
useful only for approximately studying the pointlike material
particles' behavior. Moreover the uncertainty principle is only a
legitimate consequence of the Schr\"odinger probabilistic process
and can not be considered as a universal principle. More
fundamentally, we consider that the legitimacy of the wave quantum
Mechanics (which is derived from classical Mechanics), is based on
its ability to provide, in the macroscopic cases, approximate
results that coincide with those given by Newton, Lagrange and
Hamilton's Mechanics. Classical Mechanics institutes laws for
idealized physical situations (when sufficient data are known),
whereas quantum Mechanics predicts and explains experimental
observed results; the legitimacy of the latter is insured by the
Bohr correspondence principle. Indeed, we show that quantum
Statistics uniquely relies on the very physical characteristics of
both the realized experiment and the involved particles (such as
distances, symmetries, masses, charges, momenta
and spins) as well as on mathematical Logics.\\
Quantum Mechanics can and must be used in microscopic subatomic
phenomena when our present means can not result in a theoretical
formulation.

\subsection{The De Broglie Wavelength}
\normalsize{The early $20^{th}$ century was marked by three
fundamental discoveries: the photon by Einstein, the Planck
constant and the Bohr model for the hydrogen atom. The Quantum
period has begun. The quantized nature of light as well as of
energy levels was clearly proved. Contemporaneously, many
experiments and facts have shown that matter has also some wavy
nature. The success of some new notions and the partial success of
some others led De Broglie to translate the notion of wavelength
from electromagnetism to matter particles and bodies. He then
defined the wavelength of a particle as
$$\lambda\;=\;\frac{h}{p}$$
where $p$ is the relativistic momentum of the particle and $h$ is
the Planck's constant. This was a practical and useful
approximation for analyzing the energy, momentum and speed of
particles. Nevertheless, this notion, joined to relativistic
formulas, on one hand, and to the Bohr model, on the other hand,
leads to some obvious contradictions. The wavy nature of matter
has to be explained
more generally and more precisely.\\

\noindent\textbf{Remark $1.$}$\;\;$ In the previous sections, we
have proved that some interpretations of the special relativity
second postulate are false. We also proved that the frequency $f$
is a characteristic feature of a (material or immaterial) point
that is extended only to fundamental particles (quarks and
leptons). Integration of the relation $E=hf$ over the domain
occupied by a fundamental particle gives the famous relation
$$E_0\;=\;m_0c^2\;:=\;m_0$$
for at rest matter and
$$E(t)\;=\;\gamma(t)m_0 (c^2+\frac{1}{2} \frac{\gamma_1(t)}{\gamma(t)} v^2(t)) \simeq \gamma(t) m_0 c^2$$
for a material particle into movement, where $\gamma_1(t)$
decreases from 1 to 0 when the speed increases from 0 to 1 and
$\gamma(t)$ is the Lorentz factor; $\gamma(t) m_0$ is qualified by
W. Kaufmann as being the apparent mass and we call
$\gamma_1(t) m_0$ the reduced mass of the particle into movement.\\

So, when we attribute a De Broglie wavelength
$\lambda=\frac{h}{p}$ and a frequency $f=\frac{v}{\lambda}$ to a
pointlike material particle, they do not correspond to a real
periodic movement (or trajectory) and can not be used for making
exact calculation of relativistic or non relativistic particle
energy by means of formulas such as
$$p\;=\;\gamma mv\quad,\quad E\;=\;\sqrt{p^2+m^2}\;=\;\gamma
m\quad\mbox{and}\quad v\;=\;\frac{p}{E},$$
where we have taken $c=1$
for simplicity.\\
For non relativistic particles, we obtain in this
way:
$$p\;=\;\frac{h}{\lambda}\;=\;\frac{hf}{v}\;=\;\frac{E}{v}\;=\;\frac{\gamma
m}{v}$$ which yields
$$\gamma mv\;=\;\frac{\gamma m}{v}$$
and then $v^2=1,$ which is absurd.\\

Moreover the Bohr model shows clearly that, for energy levels
$E_n$ with $n$ sufficiently large, we have $f_{orb}\simeq f_{rad}$
(whereas, for lower $n,$ we have $f_{orb}\neq f_{rad}$). So, when
we consider two consecutive high levels $E_1$ and $E_2$
corresponding to frequencies $f_1$ and $f_2,$ wavelengths
$\lambda_1$ and $\lambda_2$ and speeds $v_1$ and $v_2,$ we have
$\left(\mbox{using the relation}
\;\;E=\sqrt{p^2+m^2}=\sqrt{\frac{h^2}{\lambda^2}+m^2}=\sqrt{\frac{h^2f^2}{v^2}+m^2}\right),$
\begin{eqnarray*}
  f_2 &=& \frac{\Delta E}{h}\;=\;\frac{E_2-E_1}{h}\;=\;\frac{\sqrt{p_2^2+m^2}-\sqrt{p_1^2+m^2}}{h}\\
   &=&
   \frac{\sqrt{\frac{h^2}{\lambda_2^2}+m^2}-\sqrt{\frac{h^2}{\lambda_1^2}+m^2}}{h}\;=\;\sqrt{\frac{1}{\lambda_2^2}+\frac{m^2}{h^2}}\;-\;\sqrt{\frac{1}{\lambda_1^2}+\frac{m^2}{h^2}}.
\end{eqnarray*}
Consequently, we obtain
$$\frac{v_2}{\lambda_2}\;=\;\frac{1}{\lambda_2}\sqrt{1+m^2\frac{\lambda_2^2}{h^2}}\;-\;\frac{1}{\lambda_1}\sqrt{1+m^2\frac{\lambda_1^2}{h^2}}$$
and
\begin{eqnarray*}
  v_2 &=& \sqrt{1+\frac{m^2}{p_2^2}}\;-\;\frac{\lambda_2}{\lambda_1}\sqrt{1+\frac{m^2}{p_1^2}}\\
   &=& \sqrt{1+\frac{m^2}{p_2^2}}\;-\;\frac{p_1}{p_2}\sqrt{1+\frac{m^2}{p_1^2}} \\
   &=& \sqrt{1+\frac{m^2}{p_2^2}}\;-\;\sqrt{\frac{p_1^2}{p_2^2}+\frac{m^2}{p_2^2}}\\
   &=& \sqrt{1+\frac{m^2}{p_2^2}}\;-\;\sqrt{\frac{v_1^2}{v_2^2}+\frac{m^2}{p_2^2}} \\
   &=&\sqrt{1+\frac{m^2}{p_2^2}}\;-\;\sqrt{\left(\frac{n+1}{n}\right)^2+\frac{m^2}{p_2^2}}\;<\;0
\end{eqnarray*}
which is absurd.\\
We obtain a similar contradiction when we use $f_1=\frac{\Delta
E}{h}.$
\subsubsection*{The particle in a box case}
We consider now a small ball (or particle) in a fixed box of
length $L.$ When we are looking for the ground state energy by
using the De Broglie wavelength notion $\lambda=\frac{h}{p}$ and
the Schr\"odinger equation
$$-\frac{\overline{h}^2}{2m}\frac{d^2\psi}{dx^2}\;=\;E\psi,$$
we arbitrarily exclude the case where the particle speed is $v_0=0.$\\
Now, the introduction of the speed notion implies necessarily the
introduction of the time progress notion. Let then
$E_0=\frac{h^2}{8mL^2}$ be the quantum ground state energy that
corresponds to the speed $v_0\neq 0.$ If $|v_0|=a_0$ (a positive
constant), we obtain $|p_0|=ma_0$ when using classical Physics
formulas and $|p_0|=\gamma_0 ma_0$ when using relativistic ones,
where $\gamma_0=\frac{1}{\sqrt{1-v_0^2}}.$ But $<p_0>=0$ implies
$p_0(x)=\pm|p_0|$ and $v_0(x)=\pm a_0$ which constitute a
physically and mathematically inconceivable phenomena
($v_{0}$(\emph{x}) can not pass from $-a_{0}$ to +$a_{0}$
instantaneously). So $|p_0|$ is time dependent which is
contradictory as (according to generally accepted notions)
$$|p_0|\;=\;\frac{h}{\lambda_0}\;=\;\frac{h}{2L}\;=\;\mbox{const}.$$
Therefore, we have either $v_0=0,$ which implies $E_0=p_0=0$ (in
accordance with the classical physics minimal energy) and
$\lambda_0$ has no a real existence, or all quantities $v_0,p_0,E_0$
and (if we put $\lambda_0=\frac{h}{p_0}$) $\lambda_0$ depend on
time. In that case $L$ obviously depends on time, which is absurd
unless the legitimacy of the theoretically exact measurements'
existence is thoroughly questioned.
\subsubsection*{The pendulum and quantum harmonic oscillator case}
When we consider a pendulum, we assume either $v_1=0,$ which
corresponds to the stable vertical equilibrium state and implies
$E_1=p_1=0$ in accordance with classical Newtonian Physics, or
$v_1\neq0.$ In that case the ground state energy within the wave
quantum Mechanics framework is
$$E_1\;=\;\frac{\overline{h}\omega_1}{2}\;=\;\frac{hf_1}{2},$$
which is a non vanishing constant, since the very physical nature
of the pendulum notion imposes the attribution of a frequency
$f_1=\frac{1}{T_1}$ to the theoretically periodic movement of the
pendulum.\\
Now, when we incorrectly identify the De Broglie wavelength
$\lambda_1=\frac{h}{p_1}$ with the wavelength
$\lambda_1=\frac{v_1}{f_1}$ of the periodic movement (where $p_1$
and $v_1$ are respectively the mean scalar momentum and speed), we
obtain
$$E_1\;=\;\frac{hf_1}{2}\;=\;\frac{1}{2}\frac{h}{\lambda_1}\lambda_1
f_1\;=\;\frac{1}{2}p_1v_1$$ which is the corresponding mean
kinetic energy. The same results are valid for a quantum harmonic
oscillator. Furthermore, we obtain for the first excited state
$E_2=3E_1$ ([2], (7.121)). Now, it is physically and
mathematically undeniable that mean speed, mean momentum and mean
kinetic energy depend continuously on the initial displacement of
both pendulum and harmonic oscillator. But displacement is a
continuous variable; therefore the energy levels can not be
quantized by means of Schr\"odinger's equation.

\subsection{The uncertainty principle}
We begin this subsection by noticing that it seems strongly
illogical that, after some experiments such as the one where
particles hit a screen through small slits slightly spaced, we
conclude that the fact of knowing the origin of the particles
hitting the screen could really alter the physical phenomena. It
is true that the means used in order to know this origin can alter
the results by modifying the particles momenta and trajectories
but this is only a technical and circumstantial phenomena that
does not allow us to conclude that our pure knowledge can
transform the physical results that are determined objectively by
the real physical conditions. Beside of that, our theoretical or
practical capacity to discover any law of Nature does not
influence the objective reality of this law. The very long history
of discoveries in all domains shows the objectivity of natural
laws independently of our circumstantial (theoretical,
approximate, experimental or technical) capacity to discover them.
The use of progressively increasing energetic levels of particles
(i.e. increasing «frequencies» or decreasing «wavelengths»), for
instance, has permitted the realization of important progress in
the understanding of our physical universe through the ages as
well as the understanding of matter (nucleons, quarks, hadrons,
leptons) structure and also the refinement of our knowledge on the
quantization of both energy levels inside atoms and angular (as
well as intrinsic) atomic and
nuclear momentum.\\
The Schr\"odinger function and quantum Statistics have also
permitted a jump in our comprehension of the universe by giving
effective methods for discovering and interpreting all results
that are obtained from experimentation and led to powerful
approximations
for natural phenomena rules.\\
However, these phenomena contain essentially a part of
uncertainties due to the multitude of energetic and dynamic
evolutive factors which govern all aspects of matter-energy:
energy levels,
trajectories, interactions, transformations...\\
Although these phenomena are far from being regular
(differentiable), they are continuous. An electron, for instance,
that changes its orbit (which is permanently evolving) for a
higher energy level (when absorbing a photon) or for a lower
energy level (together with photon emission), spends an
infinitesimal time fraction before achieving its final state.
During this transition, continuity of trajectory and energy
conservation are both insured since we have
$$E_0\;\equiv\;E_e(t)\;\pm\;k(t)E_p\;\equiv\;E_e\;\pm\;E_p$$
where $E_0$ is the initial electron energy, $E_e$ its final
energy, $E_p=hf$ is the photon energy and $k(t)$ is a continuous
function that increases from $0$ to $1.$ Indeed, since photon is
fundamentally a quantum object with a fixed wavelength, its
existence is essentially related to time and distance. Its
formation (and its absorption) takes an infinitesimal fraction of
time and needs an infinitesimal extent of distance; moreover it
can not exist in a static state (i.e. independently of motion).
Then, formation and existence of photon need time, distance,
motion and speed notions. Its absorption and emission are
necessarily related to time and energy change notions.\\\\
In the following, we will give some arguments aiming to show that
the uncertainty principle, that can be written as $\Delta x\Delta
p_x\geq \frac{\overline{h}}{2}$ and $\Delta E\Delta t\geq
\frac{\overline{h}}{2},$ can not be of a canonical and universal
nature. It is only a legitimate consequence of the use of the
Schr\"odinger's equations
$$\psi(x)\;=\;\frac{1}{\sqrt{2\pi}}\int_{-\infty}^{+\infty}g(k)e^{-ikx}d\,k$$
and
$$g(k)\;=\;\frac{1}{\sqrt{2\pi}}\int_{-\infty}^{+\infty}\psi(x)e^{ikx}d\,x$$
in order to find out the probability of localizing a given
particle, under some constraints, in a given position and
determining its momentum. The uncertainty principle is stated
after using the probability distributions
$$\frac{dP}{dx}\;=\;|\psi(x)|^2\qquad\mbox{and}\qquad\frac{dP}{dk}\;=\;|g(k)|^2$$
and their standard deviations $\sigma_x$ and $\sigma_k$ as well as
the De Broglie relation $p=\frac{h}{\lambda}.$ Therefore, this
principle states simply that this particular approach and the use
of this particular method hold within themselves the uncertainty
so
quantized.\\\\
Moreover, the very definition of $\sigma_x$ and $\sigma_k$ only
means that there is a large probability for the position $x$ to be
within a distance less than $\sigma_x$ to the mean value $\langle
x \rangle$ and for the component $p_x$ of the momentum $p$ to be
within an interval less than $\sigma_k$ about the mean value
$\langle p_x \rangle$. However, it is obvious that there is lesser
probability of finding $x$ within a distance lesser than
$\sigma_x$ to $\langle x \rangle$ and a non negligible probability
for $x$ to be at a distance larger than $\sigma_x$ to $\langle x
\rangle$. Similarly, we can assert the same properties for $p_x$ and
$\sigma_k$. So, $\sigma_x$ and $\sigma_k$ only determine a
probabilistic estimation and they can not institute a sharp
limiting for the uncertainty of both position and momentum and of
their product.\\
The same reasoning can be produced when commenting on the
Heisenberg relation $\Delta A \Delta B \geq \frac{1}{2} | \langle
M \rangle |$ for any two observables $A$ and $B$ where $M$ is
defined by $\widehat{M} = -i [\widehat{A},\widehat{B}]$
([3],11.021) and particularly for $x$, $p_x$ and $-i
[\widehat{x},\widehat{p_x}] = \overline{h}.$\\\\
Nevertheless, this does not mean that the position $x$ of the
particle or its momentum $p_x,$ at a given time, can not be well
defined or can not be determined with more precision when more
physical information are specified by more efficient theoretical
or
experimental processes.\\

\subsection*{ Scaling problem}

More generally, the problem of determining the position, the
trajectory and other characteristics (such as momentum and energy,
for instance) of subatomic particles, which move with very large
speed, was one basic problem in the heart of the fundation of
quantum theory.\\
Indeed, in spite of our fantastic technical progress, we are,
until now, incapable of visualizing or perceiving these minuscule
particles and their movement and even of distinguishing between
them. The time and distance scales that suit our perception
actually are infinitely large regarding their infinitely small
world. Our centimeters and grams and our seconds are really
gigantic and inappropriate for analyzing this microworld (or
rather this nano or femtoworld).\\In spite of using the most
sophisticated means, the electron motion around the nucleus
appears for us as a foggy scene because of the infinitely small
size of the electron orbit and the infinitely large speed of the
electron. Not only we are incapable of determining its trajectory,
but we are still at the stage of contenting ourself with
determining the probability of finding it at such and such region
of the minute space around the nucleus.\\\\As for quarks, the
ultrasophisticated means and the ultraclever methods are necessary
in order to get some scanty information concerning their existence
and their characteristics which are ultra-fluctuating and even
ultra-ephemeral. However, all that does not prevent us from
conceding that the electron, for example, has, at every fixed
time, a precise position and that it has a well defined speed and
trajectory during an infinitesimal fraction of nanosecond in spite
of all evolutions it may undergo. \\\\
In order to convince ourselves that this nanoworld respects
mechanical and physical laws during infinitesimal time interval,
we can imagine that a minicreature (or a nanocreature) that is as
intelligent as us but infinitely more sensitive than us regarding
the infinitesimal distances and time-intervals making them (when
living inside the nanoworld of atoms) capable of discerning
(without using sophisticated technical means that would alter
physical characteristics) between infinitesimal particles and
noting the fractions of nanodistances between them as well as the
fractions of nanoseconds separating two minute events and finally
of perceiving the tiny transformations and fluctuations that occur
within any infinitesimal space and time. Moreover, we have to imagine
that these intelligent creatures possess the means and the good
will of communicating us their observations along infinitesimal
time - intervals after registrating and schemetizing them and
above all after enlarging and rescaling them in order to make us
capable of reading the slightest details concerning positions at
very precise time and trajectories (during infinitesimal time
intervals) of the nanoparticles of this nanoworld. This has to be
done in such a manner that, for instance, the foggy scene of the
electron motion transforms for us into interlacing lines. All that
we need is to
enlarge the distances and to slow down the motions.\\\\

In other respects, we can say, for instance, that the ground state
energy of the hydrogen atom in the Bohr model is determined by the
finiteness of the electron energy and has nothing to do with the
uncertainty principle. Indeed, when an electron moves from an
energy level corresponding to a $V_1$ potential energy to another
level corresponding to a $V_0$ potential energy then it releases a
photon $\gamma$ with $E(\gamma)$ energy. If $m_i(t)$, $m^{'}_i(t)$
and $v_i(t)$ denote respectively the electron's apparent mass, the
electron's reduced mass and the electron's speed that correspond
to the $V_i$ levels, for $i=0,1,$ then we must have
$$m_1(t)c^2+\frac{1}{2}m^{'}_1(t)v_1^2(t)+V_1-E(\gamma)=m_0(t)c^2+\frac{1}{2}m^{'}_0(t)v_0^2(t)+V_0,$$
which yields
$$\Delta
V=V_1-V_0=m_0(t)c^2+\frac{1}{2}m^{'}_0(t)v_0^2(t)-m_1(t)c^2-\frac{1}{2}m^{'}_1(t)v_1^2(t)+E(\gamma).$$
This shows that $\Delta V$ and consequently $V_0$ are finite.\\
Likewise, we can state that, when we use our proper expression for
the kinetic energy $E_k$ of the Bohr atom electron (for instance),
the energy and the kinetic energy can not exceed the absolute
value of the potential energy for arbitrary $r$ because then we
would have
$$\frac{ke^2}{r} \leq \gamma(t)m_0 (c^2+\frac{v^2}{2}) <
\gamma(t)m_0(c^2+\frac{c^2}{2})$$which is impossible for
sufficiently
small $r$.\\
Therefore, there exists a finite minimal potential energy
corresponding to a finite minimal energy level for the electron
inside the hydrogen atom. This level is, as experiments
show, $V_0\simeq -13,6$ ev.\\
Besides, we notice that the inverse process to the above one takes
place after an electromagnetic or a thermal energy absorption
which leads the atom to an excited state and can even lead the
electron to a pure "separation" from its original atom and even
(occasionally) with a large kinetic energy. In that case we have
(using obvious notations)
$$m_0(t)c^2+\frac{1}{2}m^{'}_0(t)v_0^2(t)+\Delta E+V_0=m_e
c^2+\frac{1}{2}m^{'}_ev^2.$$ We mention that, for a non uniform
movement, the mass $m_0(t)$ is variable because of the radiation
phenomenon that comes with such a movement.\\
Theoretical and experimental measurements of the hydrogen atom
ground state energy show that this energy is characterized by the
planck constant $\overline{h}$ and the Bohr radius
$$r\;=\;a_0\;=\;\frac{4\pi\varepsilon_0\overline{h}^2}{me^2}$$
where $m$ is the electron mass corresponding to this energy level.
The ground state energy actually is determined by the minimal
value
$$E_m\;=\;\left(\frac{1}{2}mv^2-\frac{e^2}{4\pi
\varepsilon_0r}\right)_m\;=\;\left(\frac{\overline{h}^2}{2\pi
r^2}-\frac{e^2}{4\pi \varepsilon_0r}\right)_m.$$
\subsection{Classical versus quantum Mechanics}
It is well known that classical Mechanics and Physics are based on
some principles and laws that derive from a theoretical
formulation essentially obtained from idealizing real physical
systems and phenomena. This does not mean that observation and
experiments are less important than theoretical formulations of
classical Physics since these formulations stem from those
observations and then are adopted and improved after many
confrontations, inspections and verifications. Quantum Mechanics
consists of several predictive rules that derive from a huge
number of experiments and ends up by founding the powerful
probabilistic quantum Statistics. Some rules and results become
postulates, principles or laws because none has
observed exceptions that contradict them.\\

For our part, we maintain that the wave quantum theory structure,
which leans upon Schr\"odinger equation, is essentially
established with the (declared or undeclared) aim to be unified
with the Lagrangian and Hamiltonian Mechanics by the intermediate
of the Hamilton-Jacobi equation:
$$\mathcal{H}\left(q_j,\frac{\partial S}{\partial q_j},t\right)+\frac{\partial S}{\partial
t}=0$$ where $S$ is the Hamilton's principal function. This
equation reduces, in the well known particular case, where the
Hamiltonian is written as
$$\mathcal{H}=\frac{1}{2m}p^2+V(r,t),\quad\mbox{with}\quad p=\nabla
S\quad\mbox{and}\quad\mathcal{H}=-\frac{\partial S}{\partial t},$$
to
$$\frac{1}{2m}\left|\nabla S\right|^2+V(r,t)+\frac{\partial S}{\partial
t}=0$$ that is
$$\mathcal{H}=\frac{1}{2m}\left|\nabla S\right|^2+V(r,t).$$
Thus, the wave quantum theory is based, on one hand, upon the
notion of Schr\"odinger's wave functions (having the general form
of $\Psi(r,t)=A_0(r,t)\exp\left(i\sigma(r,t)\right)$), stationary
waves, plane, quasiplane and packet waves and, on the other hand,
upon the following eikonal equation (which is obtained when
putting $S=\overline{h}\sigma$):
$$\frac{\overline{h}^2}{2m}\left|\nabla \sigma\right|^2+V(r,t)+\overline{h}\frac{\partial \sigma}{\partial
t}=0$$ and finally upon the Schr\"{o}dinger's equation:
$$-\frac{\overline{h}^2}{2m}\nabla ^2\Psi+V(r,t)\Psi=i\overline{h}\frac{\partial \Psi}{\partial t}.$$
This latter equation is written, for a time-independent potential
$V$ and for $\Psi(r,t)\;=\;\psi(r)\exp(-i\omega t),$ as the
classical time-independent Schr\"{o}dinger's equation
$$-\frac{\overline{h}^2}{2m}\nabla ^2\psi+V\psi\;=\;\overline{h}\omega\psi\;=\;E\psi.$$
Then starts the mechanism that relates wave quantum Mechanics to
Hermitian operators (associated with Observables) and to
expectation values by means of relations such as
$$\hat{p}=-i\overline{h}\nabla\qquad,\qquad\hat{\mathcal{H}}\left(\hat{q}_j,\hat{p}_j,t\right)\Psi=i\overline{h}\frac{\partial \Psi}{\partial
t},$$
$$\hat{\mathcal{H}}\left(\hat{q}_j,\hat{p}_j\right)\psi=E\psi\qquad\mbox{for}\qquad\mathcal{H}=E=\overline{h}\omega$$
and (as a particular case)
$$\hat{\mathcal{H}}=-\frac{\overline{h}^2}{2m}\nabla ^2\;+\;V(r,t),$$
as well as the relations
$$<r>\;=\;\int \psi^*\hat{r}\psi\,d\tau$$
and
$$<p>\;=\;\int \psi^*\hat{p}\psi\,d\tau.$$
Moreover, when $\psi$ is represented with the Hamiltonian
eigenfunctions (i.e. $\hat{\mathcal{H}}\psi_n\;=\;E_n\psi_n$ for
$\psi\;=\;\sum\alpha_n\psi_n$), we get
$$<\mathcal{H}>\;=\;<E>\;=\;\sum|\alpha_n|^2E_n.$$
All this is accompanied by the uncertainty principle and extended
by
the Heisenberg matrix quantum theory.\\
It is very convenient to write down here the following quotation
of
$[3]$ that illuminates the preceding with a specific example:\\
"Attention is now directed to wave Mechanics and the immediate
objective is to derive the fundamentals of this branch of quantum
theory in a way that takes inspiration from one of Schr\"odinger's
lines of thought. As a specific example, from which broader
conclusions may be readily deduced, consider an electron moving in
a prescribed field characterized by a scalar potential
$\varphi(r,t)$ and at most a negligible vector potential $A(r,t).$
The wave which, according to experimental evidence, is in some way
associated with this electron is called the wave function and is
denoted $\Psi(r,t).$ The program of derivation begin by assuming
properties for the $\Psi-$wave such that, in a classical
situation, a packet of these waves moves according to the laws of
Newtonian mechanics and thereby "explains" the motion of the
electron. This is the spirit of the correspondence principle since
it expects as a first requirement that the new Mechanics should
predict,in a classical context, behavior appropriate to that
context. The hypotheses involved in this program are by no means
gratuitous but are suggested by Hamilton-Jacobi theory and by De
Broglie's results. Once the fundamental properties of the
$\Psi-$wave have been determined in this way, it is an easy matter
to derive the linear wave equation which $\Psi$ must obey. This
equation stands at the apex of wave mechanics; from it an enormous
number of deductions, some within the domain of classical
Mechanics but most going far beyond that domain, can be made. It
is of course, in the agreement between such deductions and the
results of experimentation that the ultimate justification of the
theory
lies".\\

The successful reconciliation between both theories has gone
beyond the status of a justification process and has led to a
hurried and non justified conclusion asserting that there exists,
in fact, a unique Mechanics which is "naturally" the quantum
Mechanics having two branches that are the wave and the matrix
quantum Mechanics; the latter, initiated by Heisenberg, is
considered as more general than the former. Moreover it is
declared that classical Mechanics is a particular case of the
quantum one and it has  to be limited to macroscopic situations.
For our part, we think that there is actually a unique theoretical
Mechanics based upon well approved mechanical and physical laws,
even though there are other ones to be discovered, checked and
improved. Many fundamental laws have been established by Newton,
Lagrange, Hamilton, Maxwell and his predecessors, Einstein, Planck
and Bohr beside of a large number of physicists and mathematicians
such as Gauss, Euler, Riemann, Fourier, Laplace, Hilbert,
Schr\"{o}dinger, Dirac and
many others.\\
We have to admit that this Mechanics is not presently completely
adapted for studying infinitesimal phenomena and therefore it must
be superseded by quantum Mechanics as an efficient means for
studying microscopic phenomena such as the dynamic behavior, the
energy and the structure of particles. These phenomena are
presently beyond the reach of our measurement means and tools and
of our analyzing capacity. Until further decisive technological
and theoretical progress, the analysis of these phenomena needs
the predictive and probabilistic methods of the quantum Statistics
guided by the quantum theory of Schr\"{o}dinger, Heisenberg, Born,
Fermi, Dirac, Pauli and many others. This theory was in fact
inaugurated by Einstein, Planck and Bohr who have definitely
proved the quantum nature of waves and energy levels beside of the
quantization of electrical charges. The efficiency of these
methods are fortunately increased by numerical methods progress
and the presently huge capacity of
empirical data treatment.\\
However, we can state that, although some natural phenomena are
quantized, there are a lot of others that are not. Electrical
charges and energy levels inside the atom, for instance, are
quantized. Electromagnetic waves are constituted with integer
numbers of photons but wavelengths, speed, masses and energies,
for instance, are continuous variables evolving (themselves)
continuously with the variable that essentially gives the
continuity meaning: the
time.\\

\noindent\textbf{Remark $2.$}$\;\;$ In the previous sections, we
have established that the material and immaterial point energy is
given by $E(t)=h(t)f(t)$ where $h(t)$ and $f(t)$ depend on time
and $E(t)$ depends also on time by the intermediate of the
temperature and environment. Frequency, wavelength and energy are
then continuous mathematical objects.
\subsubsection*{Schr\"odinger probability density and classical
probability} The general Schr\"odinger equation, where
$\Psi(r,t)=A_0(r,t)\exp i\sigma(r,t),$ implies the following
equation
$$\nabla.\left(A_0^2\overline{h}\frac{\nabla\sigma}{m}\right)\;+\;\frac{\partial
A_0^2}{\partial t}\;=\;0.$$ Comparison of this equation with the
continuity equation of a substance of density $\rho$ having a
current density $\emph{\textbf{J}}=\rho \emph{\textbf{v}}$ (i.e. $\nabla\cdotp(\rho\emph{\textbf{v}})+\frac{\partial\rho}{\partial t}=0$) has led
Born to identify $\Psi^*\Psi=A_0^2$ to an imaginary substance density
$\rho.$ Then, he interpreted $\Psi^*\Psi$ as being the probability
of localizing the particle having $\Psi$ as its wave function.
Namely, the probability of finding the particle at time $t$ in a
given volume element $d\,\tau$ at position \textbf{r} is
$$dP(\textbf{r},t)\;=\;\Psi^*\Psi(\textbf{r},t)d\,\tau.$$
Since $\Psi^*\Psi$ is interpreted as a probability density, it
must obey the normalization condition
$$\int_{\mathbb{R}^3}\Psi^*\Psi d\,\tau\;=\;1.$$
However, this fundamental notion joined to another fundamental one
in Quantum theory which is the quantum measuring apparatus leads
to a paradox which is clearly explained in the following quotation
of
$[3]:$\\
"Such an apparatus does not detect that a particular system is in
a certain final state, rather it places the system in its final
state and does so with a probability that depends upon the degree
to which the final state was involved in the composition of
the initial state $!$\\
The basic paradox of quantum Mechanics exhibits itself here with
unusual clarity; a distribution of measurement results is
generated obeying a known calculus of probabilities without any
apparent internal mechanism to explain how such a distribution
comes into being. Many physicists accept this at face value,
reasoning that the ultimate theory of the universe will probably
contain elements which are incomprehensible in terms abstracted
from macroscopic experience; hence, if Quantum theory is the
ultimate theory, it is not surprising that a paradox of the type
just described should be incorporated in its makeup. Others, not
satisfied with such a state of affairs, incline toward «hidden
variable» theories. On this view point, the pre-measurement
systems of such apparatus, although quantum mechanically
indistinguishable, are actually
distinguishable in some yet more fundamental ways".\\

\noindent\textbf{Remark $3.$}$\;\;$ In the next section, we will
give a general classification of fundamental particles. Using
Dirac operator, we show that there are originally two types of
electrons that have two opposite "spins". This classification
gives a coherent explanation of the Stern-Gerlach experiment
results
which conversely give an argument that sustains it.\\

\noindent Apart from this paradox and this discussion, let us
consider, as an example, the classical case of a particle in a
box. If $\psi_n$ denotes, for large $n,$ the stationary solution
of the Schr\"{o}dinger equation, then the probability distribution
$\frac{dP}{dx}=\left|\psi_n(x)\right|^2$ can be compared to the
classical probability which is in that case equal to
$\frac{1}{L}.$ The reconciliation between these two notions
increases with increasing $n$ (c.f. $[2]$) and ends up by a sort
of justification of the Bohr correspondence principle.
Nevertheless, stationary solutions are generally considered as
being highly improbable and essentially ephemeral and the utmost
probable solutions are constituted with finite or infinite linear
combination of such solutions. For our part, we think that only
the limit cases (i.e. infinite linear combination of stationary
solutions) reveal the real physical probability of finding the
particle at a given position and
this probability is the classical one.\\
Likewise, we consider that, for the harmonic oscillator, only the
limit cases (taking parity into account) have genuine real value
and they clearly give good approximate results as (using here and
below
the notations of $[3]$):\\
$$<x>=0\;\qquad,\;\qquad<F>=0\;\qquad \mbox {and}\;\qquad<E>=\frac{1}{2}KA_0^2.$$
These results are naturally obtained, within idealized conditions,
from the well established laws of classical Mechanics and Physics.
\subsubsection*{Wave packet and Born statistical interpretation}
It is generally admitted that a wave mechanical packet represents
the center of mass of a system of particles rather than a single
particle since, in that case, there may be no particle present at
the site of the packet. This point of view which excludes the
identification of a packet wave with a particle is called the Born
statistical interpretation. However, when we attribute to a wave
packet a definite centroid to which we associate the expectation
values $<r>$ for the position and $<p>$ for the average momentum
of all individual momenta of the packet wave components, we
obtain, according to Ehrenfest's theorem that $<p>$ is equal to
the particle mass times the velocity of the centroid, and both
$<r>$ and $<p>$ obey the laws of classical Mechanics. Contrary to
the discussion about centroid of probability, hidden variables,
multiple worlds or the real existence of particle entities, we
maintain that what precedes gives only a new justification to the
legitimacy of using wave Quantum approach when studying dynamical
phenomena where classical Mechanics formulations are unreachable.
For us Ehrenfest's theorem states that statistical wave Quantum
approach is, as well as the idealizing classical Physics one, just
an approximate description of
the real physical phenomenon.\\
\subsubsection*{Relationship between wave functions and
trajectories} It is clear that a wave function $\Psi (r,t) =
A(r,t) e^{i \sigma (r,t)}$ associated with a particle (such as an
electron moving around a nucleus) that satisfies a
Schr\"{o}dinger's equation is specified by its eikonal function
$\sigma$ and its normalized amplitude $A$. The eikonal $\sigma$
which satisfies the eikonal equation determines the Hamilton's
principal function $S$ which (theoretically) determines the exact
trajectory of (the center of mass of) the particle $(q_j(t))_j$.
The particle trajectory can not be clearly perceived or specified
with our present means. All we can perceive is its gross location
at some fraction of time without discerning the particular line
that is described by it because of the too many loops that are
carried out by (the center of mass of) the particle during any
fraction of time. So the role of the amplitude of the wave
function $\Psi$ is to indicate the probability of finding the
particle in a given region within the clouded region formed by the
very swift particle into movement. Therefore $S$ is associated
with the classical Newton-Lagrange-Hamilton Mechanics whereas
$\Psi$ is associated with the quantum wave Mechanics and $\sigma$
is the connection between them.\\
Now, when we are dealing with two particles into movement, for
instance, there are two wave functions $\Psi_1$ and $\Psi_2$, two
Hamilton's principal functions $S_1$ and $S_2$ and possibly a wave
function $\Psi$ associated with the system formed with both
particles and the Hamilton's principal function $S$ associated
with the center of mass of the system. If the two particles are
distinguishable there are two trajectories and two probabilities
and as usual the probability of finding each of them inside two
pre-indicated regions is the product of the two probabilities. If
the two particles are indistinguishable bosons, then $\Psi$ is
symmetrical and the two trajectories can be arbitrarily close to
each other and they form a dense cloud which is more dense than
the cloud formed by two indistinguishable fermions in virtue of
the Pauli exclusion principle. This fact may explain the smaller
probability of finding the (indifferently located) two fermions in
a given region than that of finding two (indifferently located)
bosons in a comparable region. Each of these probabilities is
obtained by adding the amplitudes before squaring the resulting
amplitude. It is normal that the new probabilities are related to
the trajectory of each pair of particles as well as to the
properties
of each of them.\\\\
Finally, we can state that there is no antagonism between the
results of quantum and classical Mechanics. The former deals only
with microscopic physical situations (that roughly involve
moderately small "wavelengths") where classical Physics is
presently not efficient enough. Both quantum and classical
Mechanics are applicable in macroscopic physical situations,
roughly characterized by very small "wavelengths". In that case,
if classical Physics which involves the Newtonian, Lagrangian and
Hamiltonian confirmed idealizing laws is not easy to use, we can
use quantum Mechanics which involves the wave packet approximate
notion (with the imprecise De Broglie wavelength notion and the
quantum Statistic approach).\\

Let us consider, for instance, the case of two rectangular
barriers, one with relatively abrupt inclines and the other with a
relatively gradual inclines for the potential levels (c.f. [3],
p.170). The incident microscopic particle (or the packet wave) has
a relatively large wavelength for the former barrier and a
relatively small one for the second barrier. Classical Mechanics
and quantum Mechanics both give a very little probability for the
occurrence of tunneling phenomenon for the second barrier. For the
first barrier type (if a sharp incline could really exist),
classical Mechanics gives a null probability for both macroscopic
and microscopic particles. When such a tunneling does exist, we
can explain it, and other similar phenomena, by noticing that
kinetic and mass energies transform easier into potential energy
for abrupt potential energy inclines than for gradual ones,
provided high energies are involved. This also means that such
transformations are easier in smaller fraction of time. This
proves, by the way, that if we have better knowledge of the
particle physical situations, we can refine our probabilistic
expectations. The uncertainty principle corresponds to the case
where minimal physical conditions are known
about particles and systems.\\

Conversely, determinism in Mechanics is achieved when all physical
conditions are exactly known and practically realized. Initial
conditions then imply, as Laplace asserted, a unique solution that
extends wherever and whenever all conditions are known and
satisfied. If, for example, we consider a ball that is in an
actual stable equilibrium on a punctual vertex of a cone, then it
must (in ideal conditions) stay indefinitely in that state. The
solution is unique. If it is not really in a stable equilibrium
(as it is probably the case), then it rolls downward along a cone
ray in a given direction. In the idealized former case, only a
given (yet infinitesimal) applied force can make it roll down in a
given direction. This force can be determined, a posteriori,
according to Newton's laws. Then, we can not state in any case
that a well determined initial conditions for a physical system
can result, unexpectadly, on several solutions. The real problems
is the possibility of defining entirely and exactly the initial
conditions in order to predict the solution yet
in an ideal surrounding situation.\\

\subsection*{Conclusion}
We can now summarize the preceeding study by stating that:
\begin{description}
    \item[$\bullet$] A particle is never reduced to a single point.
    \item[$\bullet$] Any particle has at any time $t$ a centroid.
    \item[$\bullet$] Even for a pointlike particle the centroid can not have a
    definite geometrical position inside the particle during any
    small time interval since a particle is permanently evolving due
    to internal and external interactions and energy
    transformations.
    \item[$\bullet$] If we use the Hamilton-Jacobi equation and a specific
    Schr\"odinger function $\Psi$ that determines an eikonal
    function $\sigma$ which is proportional to a Hamilton's principal function $S,$ we can determine $\sigma$ as the solution of the equation
    $$\frac{\overline{h}^2}{2m}\left|\nabla \sigma\right|^2+V(r,t)+\overline{h}\frac{\partial \sigma}{\partial
t}=0.$$ $S$ can then be theoretically determined. Therefore, if we
assume that the particle centroid is fixed relatively to the
particle and if the initial conditions are well defined as well as
all surrounded conditions, then $S$ can be used to exactly
determine the trajectory $q=(q_j)_j$ and the momentum $p=(p_j)_j.$
    \item[$\bullet$] Since such specifications are quasi impossible, we have to
    content ourselves with using the probability density
    $$dP=\Psi^*\Psi d\,\tau$$and then our knowledge of both position
    and momentum are limited by the uncertainty principle.
    \item[$\bullet$] Nevertheless, if we could have some specific information
    about the particle and some surrounding physical conditions, we
    can hope to set down some constraints on the centroid and the momentum. Then,
    when we take a given point as approximate centroid, we can
    determine a fictitious trajectory for this point and deduce,
    using some estimates, that the trajectory of the real centroid is
    within a space tube about the fictitious trajectory during a
    reasonable time interval. If this theoretically possible
    situation is realized in practice, then (using in a similar way some estimates for the momentum) we can obtain a smaller
    uncertainty than the limit given by the uncertainty principle.
    \item[$\bullet$] Finally, when we use a wave packet for a particle (in
    macroscopic cases or in the short wave limit and the Bohr's
    correspondence principle case), we may have approximate values $<r>$
    for the position and $<p>$ for the momentum of the particle but
    then, it is sometimes possible to use the idealizing classical Physics for
    getting better approximate values.
    \item[$\bullet$] {\bf About Schr\"odinger equation:} We have already noticed that the trajectory $X(t)$ of an electron inside atoms is a geodesic with respect to the
ambiant metric $g_t$ (i.e. $\nabla^{g_t}_{X'(t)}X'(t)=0$) as long as it does not change its orbit where it has a well determined energy and approximatly a
constant speed $v$. In other respects, if $V(r,t)$ is the potential energy of the electron, then the Hamiltonian is given by $\mathcal{H}=\frac1{2m}p^2+V(r,t)=T+V$.
Letting $S(r,t)$ be the Hamilton's principal function then we have approximatly:
$$\overrightarrow{p}=\nabla S\qquad\mbox{and}\qquad \mathcal{H}=-\frac{\partial S}{\partial t}$$
and so the Hamilton-Jacobi equation can be written as
$$\frac1{2m}|\nabla S|^2+V(r,t)+\frac{\partial S}{\partial t}=0.$$
Let $\displaystyle \Psi(r,t)=A_0(r,t)e^{i\sigma}(r,t)$ be the Schr\"odinger's wave function, i.e. the average function of the wave packet associated to the electron. As
explained in Section 6.03 of [3] it is possible, under classical conditions and the non relativistic cases, to characterize the wave function $\Psi$ by its eikonal function $\sigma$ and, as it is proven in [3], the Schr\"odinger equation (6.021), i. e.
$$-\frac{h}{2m}\nabla^2\Psi+V\Psi=i\overline{h}\frac{\partial\Psi}{\partial T}$$\\
for $\Psi$ is equivalent to the eikonal equation (6.015), i. e.
$$\frac{\overline{h}^2}{2m}|\nabla \sigma|^2+V(r,t)+\overline{h}\frac{\partial \sigma}{\partial t}=0.$$
which is nothing but the Hamilton- Jacobi equation when using $S=\overline{h}\sigma$ where $\overline{h}$ is the Planck constant,
which can be considered as being constant along the orbit of the electron before changing its energy level. Therefore, we have just proved that,
under classical conditions (i.e. in the short wave length limit), the Shr\"{o}dinger equation is nothing but the above Hamilton-Jacobi
equation which can be written as $$\frac1{2m}|\nabla S|^2+V(r,t)=-\frac{\partial S}{\partial t}=\mathcal{H}=T+V(r,t)$$
which gives
$$\frac1{2m}|\nabla S|^2=T=\frac12mv^2,$$
that is, the trivial identity $$\frac1{2m}|\nabla S|^2=m^2v^2=p^2.$$
So the Schr\"odinger equation is an approximate result of the Lagrange-Hamilton-Jacobi Mechanichs and since the latter gives the Newtonian Mechanics
and the Heisenberg matricial Mechanics is equivalent to the wave quantum one, the quantum Mechanics is only a practical approximate result of the
unique theoretical Newton-Lagrange-Hamilton Mechanics.
\end{description}}

\subsection{Remarks on the quantum Statistics foundation} The aim
of this section is to show that only physical characteristics of
an interference problem (particle types, momenta, distances,
symmetries) determine the general quantum Statistics schemes.

\subsubsection*{The two slits problem scheme}
Assume that a large number $n$ of identical particles having a given momentum $p$ are directed perpendicularly toward two parallel slits $S_1$ and $S_2$ extremely close to each other and having both the same infinitely small width. We further assume that we get, on a screen located behind the slits, only two possible outcomes $E_1$ and $E_2$ having respectively $n_1$ and $n_2$ events such as $n=n_1+n_2.$ Finally, we assume that, between the $n_1$ particles reaching $E_1,$ $r_1$ particles originate from the slit $S_1$ and $s_1$ particles originate from the slit $S_2$ and that between the $n_2$ particles reaching $E_2,$ $r_2$ particles originate from $S_1$ and $s_2$ particles originate from $S_2.$\\
We then have
$$n_1=r_1+s_1\quad\hbox{and}\quad n_2=r_2+s_2.$$
Let $A_1$ and $A_2$ be two complex numbers such as
$$|A_1|^2=\frac{n_1}{n_1+n_2}\quad\hbox{and}\quad |A_2|^2=\frac{n_2}{n_1+n_2}$$
and $\alpha\in[0,2\pi]$ such that
$$|A_1|=\cos\alpha=\sqrt{\frac{n_1}{n_1+n_2}}=\sqrt{\frac{n_1}{n}}$$
and
$$|A_2|=\sin\alpha=\sqrt{\frac{n_2}{n_1+n_2}}=\sqrt{\frac{n_2}{n}}.$$
We then have
$$A_1=|A_1|e^{i\theta_1}=\cos\alpha e^{i\theta_1}$$
and
$$A_2=|A_2|e^{i\theta_2}=\sin\alpha e^{i\theta_2}.$$
The probability that an $E_1$ event (resp. $E_2$ event) originates
from $S_1$ is
$$|B_1|^2=\frac{r_1}{r_1+s_1}=\frac{r_1}{n_1}\quad(\hbox{resp.}\;|B_2|^2=\frac{r_2}{r_2+s_2}=\frac{r_2}{n_2})$$
and the probability that an $E_1$ event (resp. $E_2$ event)
originates from $S_2$ is
$$|C_1|^2=\frac{s_1}{r_1+s_1}=\frac{s_1}{n_1}\quad(\hbox{resp.}\;|C_2|^2=\frac{s_2}{r_2+s_2}=\frac{s_2}{n_2})$$
for $B_1, B_2, C_1, C_2 \in \mathbb{C}$.\\
We then have
\begin{eqnarray*}
 |B_1|&=&\sqrt{\frac{r_1}{n_1}}=\cos\beta\quad\quad |B_2|=\sqrt{\frac{r_2}{n_2}}=\cos\gamma\\
  |C_1|&=&\sqrt{\frac{s_1}{n_1}}=\sin\beta\quad\quad |C_2|=\sqrt{\frac{s_2}{n_2}}=\sin\gamma
\end{eqnarray*}
and
\begin{eqnarray*}
B_1&=&|B_1|e^{i\beta_1}=\cos\beta e^{i\beta_1}\quad\quad B_2=|B_2|e^{i\beta_2}=\cos\gamma e^{i\beta_2}\\
C_1&=&|C_1|e^{i\gamma_1}=\sin\beta e^{i\gamma_1}\quad\quad
C_2=|C_2|e^{i\gamma_2}=\sin\gamma e^{i\gamma_2}.
\end{eqnarray*}
Under these conditions, the relations
$$\left\{\begin{array}{c}
|A_1|^2=|B_1+C_1|^2\\
|A_2|^2=|B_2+C_2|^2
\end{array}
\right.
$$
are equivalent to the system
$$\left\{\begin{array}{c}
\cos^2\alpha=\cos^2\beta+\sin^2\beta+2\cos\beta\sin\beta\cos(\beta_1-\gamma_1)\\
\sin^2\alpha=\cos^2\gamma+\sin^2\gamma+2\cos\gamma\sin\gamma\cos(\beta_2-\gamma_2)
\end{array}
\right.
$$
or also to the system
$$\left\{\begin{array}{c}
\displaystyle\frac{n_1}{n}=1+2\sqrt{\frac{r_1s_1}{n_1^2}}\cos(\beta_1-\gamma_1)\\
\displaystyle\frac{n_2}{n}=1+2\sqrt{\frac{r_2s_2}{n_2^2}}\cos(\beta_2-\gamma_2).
\end{array}
\right.
$$
If we assume a perfect symmetry of the physical system, we can
state that equal number of particles passes through $S_1$ and $S_2$ which gives
$$r_1+r_2=s_1+s_2=\frac{n}{2}=:m$$
and that $r_1=s_2$ and $r_2=s_1$ which yield
$$r_1+s_1=r_2+s_2=m=n_1=n_2$$
and the above system is reduced to
$$\left\{\begin{array}{c}
\displaystyle\frac{m}{2m}=1+\frac{2}{m}\sqrt{r_1s_1}\cos(\beta_1-\gamma_1)\\
{}\\
\displaystyle\frac{m}{2m}=1+\frac{2}{m}\sqrt{r_2s_2}\cos(\beta_2-\gamma_2)
\end{array}
\right.
$$
and, putting $a=\cos(\beta_1-\gamma_1)$ and
$b=\cos(\beta_2-\gamma_2),$ to
$$\left\{\begin{array}{c}
\displaystyle\frac{m}{2}=m+2a\sqrt{r_1s_1}\\
{}\\
\displaystyle\frac{m}{2}=m+2b\sqrt{r_2s_2}.
\end{array}
\right.
$$
Adding these two equations, we get
$$2a\sqrt{r_1s_1}+2b\sqrt{r_2s_2}=-m$$
or
$$2b\sqrt{(m-r_1)(m-s_1)}=-2a\sqrt{r_1s_1}-m,$$
which gives
$$4b^2(m-r_1)(m-s_1)=4a^2r_1s_1+m^2+4a\sqrt{r_1s_1}m$$
that is
$$(4b^2-1)m^2-4[(r_1+s_1)b^2+a\sqrt{r_1s_1}]m+4(b^2-a^2)r_1s_1=0.$$
As $m$ is assumed to be an arbitrary large number, we obtain
$$\left\{\begin{array}{c}
\displaystyle 4b^2-1=0\\
\displaystyle b^2(r_1+s_1)+a\sqrt{r_1s_1}=0\\
\displaystyle b^2=a^2
\end{array}
\right.
$$
and then
$$\left\{\begin{array}{c}
\displaystyle b=\pm\frac{1}{2}\\
\displaystyle a=\pm b\\
\displaystyle b^2(r_1+s_1)+a\sqrt{r_1s_1}=0.
\end{array}
\right.
$$
These relations imply successively
\begin{eqnarray*}
  a &=& -\frac{r_1+s_1}{\sqrt{r_1s_1}}b^2= -\frac{r_1+s_1}{\sqrt{r_1s_1}}a^2,\\
  1 &=& -\frac{r_1+s_1}{\sqrt{r_1s_1}}a,
\end{eqnarray*}
$$a=-\frac{1}{2}\quad \hbox{and}\quad \frac{r_1+s_1}{\sqrt{r_1s_1}}=2$$
and finally $r_1=s_1$ which implies
$$r_1 = s_1 = r_2=s_2=\frac{m}{2}\quad n_1=n_2=m.$$
Reciprocally, $n_1=n_2=m,$ with $r_1+s_1=r_2+s_2=m,$ is the only solution to the considered system, what is perfectly legitimate as we have considered a perfect symmetrical system.\\
We notice that, the solution of this problem can not be given by
$$\left\{\begin{array}{c}
\displaystyle|A_1|^2=|B_1|^2+|C_1|^2\\
\displaystyle|A_2|^2=|B_2|^2+|C_2|^2
\end{array}
\right.
$$
or equivalently by
$$\left\{\begin{array}{c}
\displaystyle\cos^2\alpha=\cos^2\beta+\cos^2\gamma\\
\displaystyle\sin^2\alpha=\sin^2\beta+\sin^2\gamma
\end{array}
\right.
$$
since this latter leads to
$$1=1+1.$$
We can sum up the preceding by noting that the above physical
problem reduces to the determination of $6$ unknowns:
$$|A_1|^2,\;|A_2|^2,\;|B_1|^2,\;|B_2|^2,\;|C_1|^2,\;|C_2|^2,$$
knowing $5$ equations:
\begin{eqnarray*}
   &{}&|A_1|^2+|A_2|^2=|B_1|^2+|B_2|^2=|C_1|^2+|C_2|^2=1, \\
   &{}& |A_1|^2=|B_1+C_1|^2,\quad |A_2|^2=|B_2+C_2|^2. 
\end{eqnarray*}
This system is equivalent to a system of six unknowns, $n_1$, $n_2$, $r_1$, $r_2$, $s_1$ and $s_2$, that satisfy the three relations:
$$n_1+n_2=n\qquad r_1+s_1=n_1\qquad r_2+s_2=n_2.$$
The symmetry of the physical problem reduces the unknowns by $2$ and the arbitrariness of $n$ permits to uniquely resolve the system.\\

\noindent We assume now that there are, as previously, two slits $S_1$ and
$S_2$ and that only three possible outcomes $E_1,$ $E_2$ and
$E_3.$ By processing as before, we notice that $9$ variables are
associated to this problem, which are (using obvious notations)
the following:
\begin{eqnarray*}
&{}&|A_1|^2,\;|A_2|^2,\;|A_3|^2,\;|B_1|^2,\;|B_2|^2,\;|B_3|^2,\;|C_1|^2,\;|C_2|^2,\;|C_3|^2.
\end{eqnarray*}
The equations relating them are only $6$:\\
\noindent$\bullet$\;$|A_1|^2+|A_2|^2+|A_3|^2=|B_1|^2+|B_2|^2+|B_3|^2=|C_1|^2+|C_2|^2+|C_3|^2=1$.\\
\noindent$\bullet$\;$|A_1|^2=|B_1+C_1|^2,$ $|A_2|^2=|B_2+C_2|^2,$ $|A_3|^2=|B_3+C_3|^2$.\\
This system is equivalent to a system of $9$ unknowns, $n_1$, $n_2$, $n_3$, $r_1$, $r_2$, $r_3$, $s_1$, $s_2$ and $s_3$, that satisfy the $4$ relations:
$$n_1+n_2+n_3=n\qquad r_1+s_1=n_1\qquad r_2+s_2=n_2\qquad r_3+s_3=n_3.$$
The symmetry of the physical problem reduces the unknowns by $3$ ($n_1=n_3$, $r_1=r_3$, $s_1=s_3$) and the arbitrariness of $n$ permits to uniquely resolve the system.\\
Thus, pushing this reasoning to the very end, we can show that the basis of the Probability and Statistics associated with the above problem is determined by its physical characteristics. It is also shown that if the occurrence of an event $E_i$, to which is associated the complex number $A_i$, can be realized by two different ways then the probability of this occurrence, which is the amplitude of $A_i$ squared, is obtained by first adding the complex numbers that are associated to both probabilities and then squaring the amplitude of the sum.\\ 

\subsubsection*{Quantum Statistics versus classical Statistics}
Let us  consider two particles $P_1$ and $P_2$ that can only occupy two independent physical states $a_1$ and $a_2.$\\
Several questions can be formulated concerning the occupation distribution and the answer to these questions fundamentally depends on the physical characteristics of these particles.\\

\noindent$1^\circ)\;$ If the two particles are distinguishable and each of them can occupy with the same probability each of both states, we can assert that the probability that $P_1$ be in $a_1$ and $P_2$ be in $a_2$ is equal to the probability that $P_1$ be in $a_2$ and $P_2$ be in $a_1$ which is equal to the probability that $P_1$ and $P_2$ be both in $a_1$ or in $a_2.$ All these probabilities are then equal to $\frac{1}{2}\times\frac{1}{2}=\frac{1}{4}.$\\

\noindent$2^\circ)\;$ Under the same conditions as previously, we can assert that if we know that particle $P_1$ is in state $a_1,$ for example, then the probability that $P_2$ be in $a_1$ is equal to that of $P_2$ be in $a_2$ and both probabilities are equal to $\frac{1}{2}.$\\

\noindent$3^\circ)\;$  Always under the same conditions, we can also assert that the probability that both particles be in the same state $a_1$ (or $a_2$) is $\frac{1}{4}$ and the probability that $P_1$ be in $a_1$ and $P_2$ in $a_2$ (or $P_1$ be in $a_2$ and $P_2$ in $a_1$) is $\frac{1}{4}.$ Finally we can assert that the probabilities that both particles be in the same state ($a_1$ or $a_2$) and that these two particles be in different states ($a_1$ and $a_2$ or $a_2$ and $a_1$) are both equal to $\frac{1}{2}.$\\

We now suppose that particles $P_1$ and $P_2$ are indistinguishable and that the question is:\\
\noindent What is the probability that both particles be in the same state (without specifying which of the two states)? The answer then depends on the physical nature of the particles.\\

\noindent$4^\circ)\;$ If both particles are bosons (i.e. of integer spin or also having symmetrical wave function), then experimentation shows that the probability that both particles be in the same state ($a_1$ or $a_2$) is twice the probability of being in two different states $(P_1\;\hbox{in}\;a_1\;\hbox{and}\;P_2\;\hbox{in}\;a_2\;\hbox{or}\;P_1\;\hbox{in}\;a_2\;\hbox{and}\;P_2\;\hbox{in}\;a_1).$ Thus the probability of each of the two first cases is $\frac{1}{3}$ and that of each of the two last cases is $\frac{1}{6}.$ We can then state that, in accordance with Bose-Einstein Statistics, the probability that both particles be in the same state is $\frac{2}{3}$ and the probability of being in two different states is $\frac{1}{3}.$\\

\noindent$5^\circ)\;$ If both particles are fermions (i.e. of fractional spin or also having an antisymmetrical wave function) then experimentation shows that (in accordance with the Pauli exclusion principle) the probability that both particles be in the same state is null and the probability that $P_1$ be in $a_1$ and $P_2$ in $a_2$ is the same as the probability that $P_1$ be in $a_2$ and $P_2$ in $a_1$ which is $\frac{1}{2}.$ So, the probability that these two particles  be in two different states is $1.$\\

We consider now the triple experiment of collisions between particles $^4$He and $^3$He ([2],p.$340$) where we study the probability of the right angle scattering of these types of particles; the first is a boson and the second is a fermion.\\

\noindent$1^\circ)\;$ For the $^4$He$\;$-$\;$$^3$He scattering, we
have two distinguishable particles and we naturally ask for
determining the probability that the particle $^4$He be scattered
upward and the particle $^3$He be scattered downward and vice
versa. These two probabilities are equal and then we can as well
ask for determining the global probability $P_{34}$ of the right
angle scattering. This problem must be resolved with classical
Statistics. If $P_{34}$ is the probability of right angle
scattering that is observed after a large number of scattering
experiments, we can conclude that the number of events that the
particle $^4$He is scattered upward or downward is the same (this
is due to the physical symmetry of the collision:$\;$An equal
number of $^4$He particles comes slightly above or slightly under
the collision axis) and we have
$$P_{34}=a^2+a^2.$$
If we associate the amplitude $A$ to each of these probabilities
and the amplitude $B$ to the global probability of the right angle
scattering, we obtain
$$P_{34}=|B|^2=|A|^2+|A|^2=2|A|^2.$$
In other respects, for identical (indistinguishable) particles, the natural question is:\\
\noindent What is the probability of a right angle scattering of these particles independently of knowing the origin (from the right or the left) of those that are scattered upward and those that are scattered downward?\\

\noindent$2^\circ)\;$ When we consider the $^3$He - $^3$He
collision, experiments show that the probability of a right angle
scattering is null and if the amplitude $A$ is associated with the
probability of an hypothetical upward right angle scattering of
$^3$He particles coming from the right or from the left and the
amplitude $B$ with the global probability of a right angle
scattering, we have
$$P_{33}=|B|^2=|A-A|^2=0.$$
\noindent$3^\circ)\;$ Conversely, when we consider the $^4$He -
$^4$He collision, we can attribute the positive number $a^2$ to
the probability of the upward right angle scattering for $^4$He
particles coming from the right (or from the left) and we can
conclude that the global probability of a right angle scattering
is
$$P_{44}=a^2+a^2+a^2+a^2=4a^2.$$
If we associate the amplitude $A$ to the probability of an upward
right angle scattering of particles coming from the right and from
the left, and the amplitude $B$ to the probability of the global
right angle scattering, we obtain
$$P_{44}=|B|^2=|A+A|^2=4|A|^2.$$
Consequently, the physical reality of the experiment has determined that the probability of a right angle scattering of the $^4$He-$^4$He collision is twice the $^4$He-$^3$He collision and this is independent of the fact of knowing or no the number of the particles that have followed any one of the possible trajectories. The fact of knowing such details can not obviously alter the answer to any question of the type:\\
\noindent What is the probability of a right angle scattering for two beams of particles having well defined physical characteristics (momentum, spin, charge, mass)?  Only these characteristics hold the answer.\\

Concerning the above collisions, we notice that the charges, the
masses and the spin have made the difference. The Pauli exclusion
principle prevents the two particles $^3$He to get sufficiently
closer to each other in order to cause a right angle scattering.
The masses' inequality of the two particles $^3$He and $^4$He
disadvantages them to get sufficiently closer to each other in order to
cause such a scattering as it would be the case for two identical particles ${}^4$He.
\subsubsection*{Interference Bragg condition}
Let us consider the light diffraction experiment through a slit of
width $d$ $([2],\;\hbox{p.}5)$. The destructive interference is
achieved for
$$n\lambda=d \sin \theta_n$$
where $\lambda$ is the wavelength of the used light and $\theta_n$
is given by
$$\sin \theta_n=\frac{\Delta L}{d/2}=\frac{2\Delta L}{d}=n\frac{\lambda}{d}.$$
If we use the De Broglie relation $p=\frac{h}{\lambda},$ we can
write it as
$$n\frac{h}{p}=d \sin\theta_n$$
or
$$|\overrightarrow{p}.\overrightarrow{d}|=nh.$$
The destructive (or constructive) interference condition is then
traduced by a momentum quantization condition on the light's
photon. The momentum $p$ of the photon is inversely proportional
to the real wavelength of the light's ray. When we consider the
Bragg scattering of $X$ rays through a given crystal
$([2],\;\hbox{p.}142),$ we recover the same condition for a
constructive interference, that is
$$n\lambda=2d\sin\theta_n$$
where $\lambda$ is the wavelength of the used $X$ ray, $d$ is the distance between two adjacent layers of the crystal and $\theta_n$ is the angle that makes the ray with the plane of the crystal.\\
Again this condition can be written as
$$|\overrightarrow{p}.\overrightarrow{d}|=n\frac{h}{2}$$
which is a sort of a quantization on the momentum $p=\frac{h}{\lambda}$ of the used photon.\\\\
We consider now the Davisson-Germer experiment. The obtained
condition on the scattered electrons' maxima is exactly the same
as the previous one, namely
$$n\lambda=2d\sin\theta_n$$
where $\lambda$ denotes here the De Broglie «wavelength» attributed to the electron:$\;\lambda:=\frac{h}{p}.$\\
We have already seen that this practical and useful notion is derived from the fundamental notion that characterizes the used electrons (as well as all other particles) namely the momentum $p.$\\
The «constructive interference» intrinsic condition, which is
traduced here by the reflected electrons' maxima in some directions
is in fact the quantization relation
$$|\overrightarrow{p}.\overrightarrow{d}|=n\frac{h}{2}.$$
A similar interpretation can be furnished concerning the Thomson-Reid experiment. This phenomenon contributes to consider that electrons possess a wavy nature similar to electromagnetic waves. This is true in a certain sense but we do not have to deduce that the interference phenomenon here is identical to that of the electromagnetic waves and that every particle possesses a real wavelength identical to that of the electromagnetic wave. The only two fundamental common points between particles and waves (or more exactly photons) is the momentum and its quantization which is associated with the experiment physical characteristics. Recall that material particles possess other characteristics (mass, charge, spin) that photon does not possess.\\

The interference problem during a scattering from crystal is
related, beside of the physical nature of the electron, to the
atomic structure of the crystal and to the layout of the energy
bands within the crystal and to the Fermi gap of the material as
it is shown by the fact of recovering the Bragg condition when
analyzing the wave numbers
$$k=\pm\frac{2\pi}{\lambda}=\pm\frac{n\pi}{a}$$
where $a=n\frac{\lambda}{2}$ characterizes the gaps between the crystal energy bands ([2], p.373). This clearly shows that the electrons scattering (or their reflection similar to the electromagnetic wave reflection) is advantaged for some angles that are determined by a given momentum of the electrons, the energy bands of the crystal (the relation $a=n\frac{\lambda}{2}$ is, in fact, $p=n\frac{h}{2a}$) and the Fermi gap that characterizes the material taking into account that all three factors are readily quantized.\\

Finally, we notice one of the numerous contradictions to which
leads the formula $\lambda=\frac{h}{p}$ when stated for material
particles. Indeed, when we attribute to the electron inside the
hydrogen atom (in accordance with the Bohr model) the wavelength
$\lambda$ that satisfies
$$\lambda\simeq 2\pi<r>,$$
we get ([2],p.139)
$$<\frac{1}{r}>\simeq\frac{2\pi}{\lambda}=\frac{p}{\overline{h}}.$$
Thus, using the formula
$$E_k=\frac{1}{2}mv^2$$
we obtain
\begin{eqnarray*}
  -\frac{1}{2} &=& \frac{<E_k>}{<V>}=-\frac{\frac{1}{2}m<v^2>}{ke^2<\frac{1}{r}>}\simeq- \frac{\frac{1}{2}m<v^2>}{ke^2<\frac{m<v>}{\overline{h}}>} \\
   &=& -\frac{<v>\overline{h}}{2ke^2}=-\frac{<v>\overline{h}}{2\alpha\overline{h}c}=-\frac{1}{2\alpha c}<v>
\end{eqnarray*}
which gives
$$\frac{1}{\alpha c}<v>\simeq 1$$
or
$$<v> \simeq \alpha c.$$
This approximate relation is obtained independently of the electron mass (in both cases: constant or depending on the speed) and independently of the energy levels, the momentum and the mean radius $<r>.$\\
But, we know that $\alpha\simeq\frac{1}{137}$ is quasi-constant for the significative energy scale of the hydrogen atom. Nevertheless, the relation $<v>=\alpha c$ is correct only for the ground state of the hydrogen atom.\\\\

\section{Matter, antimatter and fundamental forces}
$\hskip 0.5cm$Let us consider the dynamical universe $U(t)$ as
being the Riemannian space $(B_e(O,t),g_t),$ where $g_t$ is the
physical metric at time $t > 0$, and the Laplace operator
$-\Delta$ on ($B_e(O,t),g_e$). If $E(t,X)$ is the universe
matter-energy distribution at time $t,$ then $E$ satisfies the
matter-energy equation:
$$\Box E(t,X) = \frac{\partial^2}{\partial t^2}E(t,X) -
\Delta E(t,X) = 0\hskip 0.6cm \mbox{for} \hskip 0.3cm X \in B(O,t)
\hskip1.5cm (E^*)$$
 $$\mbox{with}\qquad E(t,X)_{|S(O,t)} = 0\qquad\mbox{ for every}\;\; t>0.$$\\
Let $E_\mu(t,X)$ be a solution of $(E^\ast)$ written as
$$E_\mu(t,X) = g_\mu(t)\psi \left(\frac{X}{t}\right) \quad\mbox { for}\quad X\in B_e(O,t),$$where $\mu$ is an eigenvalue associated to
the Dirichlet problem on $B_e(O,1)$ with respect to $-\Delta$ and
$\psi$ is the associated eigenfunction and let $D$ be the Dirac
operator defined by the spinorial structure of the space
$(B_e(O,t),g_e).$ The spinor fields are, in this case, the
sections
$$\Phi:B_e(O,t) \longrightarrow B_e(O,t)\times \Sigma_3$$
where $\Sigma_3$ $\simeq$ $\mathbb{C}^{2^{[\frac{3}{2}]}}$ =
$\mathbb{C}^2$. These spinor fields are then identified with the functions
\begin{eqnarray*}
\Phi:&B_e(O,t)& \longrightarrow \mathbb{C}^2\simeq \mathbb{R}^4\\
& X& \longrightarrow (\Phi_1(X),\Phi_2(X))\begin{array}{l}
{}\\
 =(\varphi_1(X)+i\varphi_2(X),\varphi_3(X)+i\varphi_4(X))\\
\simeq(\varphi_1(X),\varphi_2(X),\varphi_3(X),\varphi_4(X))\\
\end{array}
\end{eqnarray*}
and, in that case, we have:\\
\begin{description}
    \item[i)] $\;D^2 := D\circ D = -\left(%
\begin{array}{cc}
  \Delta & 0 \\
  0 & \Delta \\
\end{array}%
\right)\simeq -\left(%
\begin{array}{cccc}
  \Delta & 0 & 0 & 0 \\
  0 & \Delta & 0 & 0 \\
  0 & 0 & \Delta & 0 \\
  0 & 0 & 0 & \Delta \\
\end{array}%
\right)$
    \item[ii)] $\;D$ is an elliptic
operator formally selfadjoint of order 1.\\
\end{description}
Consequently the set of solutions to the equation
$$
\frac{\partial^2}{\partial t^2}\overrightarrow{E}(t,X) - D
\overrightarrow{E}(t,X) = 0\;\; \mbox { with }\;\;
\overrightarrow{E}_t(X)|_{S_e(O,1)} = 0 \hskip 2cm (D)
$$
determines a hilbertian space having a
hilbertian basis $\Phi^p=(\varphi_1^p,\varphi_2^p,\varphi_3^p,\varphi_4^p),$ for $p\in\mathbb{Z},$ of eigenvectors associated to the Dirichlet problem defined by using the Dirac operator instead of the Laplace-Beltrami operator on the unit ball $B_e(O,1).$\\
Moreover, for a simple eigenvalue $\mu_n>0$ for $-\Delta$, $\lambda_n$ = $\sqrt{\mu_n}$ is an eigenvalue for $D$ to which is associated the eigenvector
$$\Phi^n=(\varphi_1^n,\varphi_2^n,\varphi_3^n,\varphi^n_4)$$
and we have
$$D\;\Phi^n=\lambda_n\Phi^n$$
and
$$D\circ D\;\Phi^n=-\Delta\;\Phi^n=\mu_n\Phi^n.$$
Therefore, we have
\begin{eqnarray*}
D\circ D (\varphi_1^n, \varphi_2^n\varphi_3^n,\varphi_4^n)&=& -(\Delta\varphi_1^n, \Delta\varphi_2^n,\Delta\varphi_3^n,\Delta\varphi_4^n)\\
&=&\mu_n(\varphi_1^n, \varphi_2^n,\varphi_3^n,\varphi_4^n);
\end{eqnarray*}
which implies that $\varphi_i^n,$ $i=1,2,3,4,$ is an eigenfunction associated with the eigenvalue $\mu_n$ of the classical Dirichlet problem on $B_e(O,1)$ and we have:
$$\varphi_2^n=a_n\varphi_1^n\quad\varphi_3^n=b_n\varphi_1^n\quad\varphi_4^n=c_n\varphi_1^n.$$
We notice that, when we deal with the Laplace-Beltrami operator
$-\Delta_{g_{t}}$ on $B_e(O,t),$ we can not have a simple
relationship between the eigenvalues of $-\Delta_{g_{t}}$ and
those of the Dirac operator $D_{g_{t}}.$ This is only possible
when the scalar curvature associated with $g_t$ is constant, which
could be the case of the
very early universe.\\\\
Now, we fix three simple eigenvalues $\mu_1,\mu_2$ and $\mu_3$ of the Dirichlet problem on
the unit ball $B_e(O,1)$ with respect to the Laplace-Beltrami operator
$-\Delta$ and we put $\lambda_i=\sqrt{\mu_i}$ for $i=1,2,3.$ We notice that we think that the
following mathematical modeling aiming to classify the different types of matter and antimatter does
not depend of the choice of these three eigenvalues. We think that
this choice corresponds to the instoration of a
measure scale concerning all fundamental notions related to the
matter-energy, the time and the distances. We assume then that our choice
corresponds to the international system (I.S.) which leads
to the classical universal constants of Physics.\\

According to this choice, we can reconsider the following
equalities (previously and successively stated) concerning the
energy $E(t,X(t))$:
$$E_\mu(t,X(t)) = h_\mu(t)f_\mu(t) = c(\mu) \sqrt{\mu} = e(\mu),$$
$$E_\mu(t,X(t),T(t)) = h_\mu(t,T(t))f_\mu(t,T(t)) = c(\mu,t) \sqrt{\mu} = e(\mu,t),$$
and $$h_P = \frac{e(\mu,t)}{f_D(\mu,t)} =
\frac{E_\mu(t,X(t))}{f_D(\mu,t)}$$where $h_P$ is the Planck
constant and $f_D$ is the De Broglie frequency.\\
Then, with our fixed choice of $\mu$, we can write the second
relation as
$$E(t) := E(t,X(t)) = h(t)f(t)$$where \emph{h}, \emph{f} and \emph{E}
depend naturally on time. The last equality can be written as
$$h_P(t) = \frac{E(t,X(t))}{f_D(t)}$$
or $\hskip 4cm$  $E(t,X(t)) = h_P(t) f_D (t)$\\\\which is the
classical relation
$$E = h_Pf_D$$
The only difference is that $E$, $h_P$ and $f_{D}$ depend on time through
the temperature $T(t)$ dependence. The variations of this dependence is
really negligible at our temporal and cosmic scale.\\
We notice also that, according to our model, $h(t)$ and $f(t)$
change, at fixed temperature, with distance (or time) but their
product keeps constant. However $h(t),$ $f(t)$ and $h(t)f(t) =
E(t)$ change with temperature. Moreover $f_D$ changes with energy
and therefore with temperature. Otherwise, we think that what we
measure in most experiences are in fact the quantities $f(t)$ and
$\lambda(t)$ and not the
classical De Broglie's quantities $f_{D}$ and $\lambda_{D}.$\\\\

\subsection*{Classification of matter, antimatter and energy}
We consider now the eigenvector subspace $E_{\lambda_1},$ which is
generated by
$$\Phi_1=(\varphi_1,a_1\varphi_1,b_1\varphi_1,c_1\varphi_1)=:(\psi_1,\psi_2,\psi_3,\psi_4)$$
and we put
\begin{eqnarray*}
\overline{\Phi}_1=(\overline{\varphi}_1,a_1\overline{\varphi}_1,b_1\overline{\varphi}_1,c_1\overline{\varphi}_1)&:=&(-\varphi_1,-a_1\varphi_1,-b_1\varphi_1,-c_1\varphi_1)\\
&=:&(\overline{\psi}_1,\overline{\psi}_2,\overline{\psi}_3,\overline{\psi}_4)
\end{eqnarray*}
We then take the vectors of $\mathbb{R}^8$
$$\Gamma_1= (\psi_1, \psi_2,
\psi_3, \psi_4, \overline{\psi}_1, \overline{\psi}_2,
\overline{\psi}_3, \overline{\psi}_4)$$
 and
$$\Gamma_2 = (\overline{\psi}_1, \overline{\psi}_2, \overline{\psi}_3,\overline{ \psi}_4,
\psi_1, \psi_2, \psi_3, \psi_4).$$
Thus, we obtain
\begin{eqnarray*}
D\times D\;\Gamma_1&=&\lambda_1\Gamma_1,\\
D\times D\;\Gamma_2&=&\lambda_1\Gamma_2
\end{eqnarray*}
and
$$\Delta \psi_i=\mu_1\psi_i,\qquad\Delta \overline{\psi}_i=\mu_1\overline{\psi}_i.$$\\
\noindent Now, we replace in $\Gamma_1,$ $\psi_1$ by $e^{-}_{1/2},$ $\psi_2$ by $\nu_e,$ $\psi_3$ by $u_{1/2},$ $\psi_4$ by $d_{1/2},$ $\overline{\psi}_1$ by $e^+_{-1/2},$ $\overline{\psi}_2$ by $\overline{\nu}_e,$ $\overline{\psi}_3$ by $\overline{u}_{-1/2}$ and $\overline{\psi}_4$ by $\overline{d}_{-1/2}.$ Likewise, in $\Gamma_2$ we replace $\psi_1$ by $e^{-}_{-1/2},$ $\psi_2$ by $\nu_e,$ $\psi_3$ by $u_{-1/2},$ $\psi_4$ by $d_{-1/2},$ $\overline{\psi}_1$ by $e^+_{1/2},$ $\overline{\psi}_2$ by $\overline{\nu}_e,$ $\overline{\psi}_3$ by $\overline{u}_{1/2}$ and $\overline{\psi}_4$ by $\overline{d}_{1/2}.$ By rearranging the components of the vectors $\Gamma_1$ and $\Gamma_2,$ we obtain the two energy vectors
$$\Gamma_1=\left(  e_{1/2}^-, e^+_{-1/2}, \nu_e, \overline{\nu}_e, u_{1/2}, \overline{u} _{-1/2},  d_{1/2}, \overline{d}_{-1/2}\right)$$
and
$$\Gamma_2=\left(  e_{1/2}^+, e^-_{-1/2}, \overline{\nu}_e, \nu_e, \overline{u}_{1/2}, u _{-1/2}, \overline{d}_{1/2}, d_{-1/2}\right)$$
each of which represents one of the two polarizations of the same electromagnetic wave or the same photon. \\
In that way, we have associated to the Dirichlet problem solution $\psi_1$ the electron $e^-_{1/2}$ in $\Gamma_1$ and $e^-_{-1/2}$ in $\Gamma_2,$ to the solution $\psi_2$ we have associated the neutrinos $\nu_e,$ to the solution $\psi_3$ (resp. $\psi_4$) we have associated the quark $u_{1/2}$ in $\Gamma_1$ and the quark $u_{-1/2}$ in $\Gamma_2$ (resp. the quark $d_{1/2}$ in $\Gamma_1$ and  $d_{-1/2}$ in $\Gamma_2$) and finally to each solution $\overline{\psi}_i$ we have associated the antiparticle of the particle associated with $\psi_i$ with an opposite spin to the particle spin.\\
Moreover, we think that if we fix, in $\Gamma_1$ (resp. $\Gamma_2$), the couple $(e^-_{1/2},e^+_{-1/2})$ (resp. $(e^+_{1/2},e^-_{-1/2})$) in the first box, then all the couples involving the neutrinos-antineutrinos and the quarks-antiquarks can be located in any one of the other boxes of $\mathbb{R}^8\simeq\mathbb{R}^2\times\mathbb{R}^2\times  \mathbb{R}^2\times\mathbb{R}^2.$\\
This possibility evokes the colors symmetry of the standard model and could explain the existence of each quark and antiquark under three distinct variants. The existence of three colors attributed to each flavor of quarks is well confirmed by the rate of the hadrons' formation during the electron-positron annihilation experiments.
This also could explain the existence of several kinds of mixed colored gluons.\\
By processing in the same way with the eigenvalues $\mu_2$ and $\lambda_2$ as well as with $\mu_3$ and $\lambda_3,$ we obtain the pure energy vectors
\begin{eqnarray*}
\Gamma'_1&=&\left(\mu^-_{1/2},\mu^+_{-1/2}, \nu_\mu,\overline{\nu}_\mu,s_{1/2},\overline{s}_{-1/2}, c_{1/2},\overline{c}_{-1/2}\right)\\
  \Gamma'_2&=&\left(\mu^+_{1/2},\mu^-_{-1/2},\overline{\nu}_\mu,\nu_\mu,\overline{s}_{1/2}, s_{-1/2}, \overline{c}_{1/2},c_{-1/2}\right)\\
  \Gamma''_1&=&\left(\tau^-_{1/2},\tau^+_{-1/2}, \nu_\tau, \overline{\nu}_\tau,b_{1/2},\overline{b}_{-1/2}, t_{1/2},\overline{t}_{-1/2}\right)\\
  \Gamma''_2&=&\left(\tau^+_{1/2},\tau^-_{-1/2},\overline{\nu}_\tau,\nu_\tau,\overline{b}_{1/2}, b_{-1/2}, \overline{t}_{1/2},t_{-1/2}\right).
    \end{eqnarray*}
Thus, we obtain (within the framework of our model) that the fundamental particles are precisely $24,$ twelve particles:
$$\;e^-,\;\mu^-,\;\tau^-,\;\nu_e,\;\nu_\mu,\;\nu_\tau,\;u,\;d,\;s,\;c,\;b\;\mbox{and}\; t$$
to which are associated their antiparticles. Each of the three former particles (and antiparticles) exists under two variants that correspond to two opposite spins. Each of the three following particles (resp. antiparticles) exists only with a negative spin $-1/2$ (left particles) (resp. positive spin $+\frac{1}{2}$). Each of the six flavors of quarks exists with two opposite spins and under three variants which correspond to three colors.\\
Anyone of these particles and antiparticles (except probably the neutrinos and antineutrinos) is formed with a distribution $E(t,X)$ on a domain $D_t$ giving the particle (or the antiparticle) a material (or antimaterial) mass $m(t)$ with a given density $\rho(t).$\\
Thus, the fundamental particles are (according to our model) originated by the solutions $\Phi^n$ of the Dirichlet problem (\emph{D}) associated to the Dirac operator on the unit ball $B_e(O,1),$ each of them determines $4$ solutions of the Dirichlet problem associated with the Laplace-Beltrami operator on $B_e(O,1).$ These solutions determine, on one hand, the $\Gamma$ energy vectors which are associated to the three generation of leptons and quarks and, on the other hand, determine the solutions of the matter-energy equation $(E^*)$ giving birth to particular distributions $E(t,X)$ on domains $D_t$ constituting in that way all fundamental particles (and antiparticles) having given mass-energies that evolves with time, temperature and all sorts of interaction.\\
All of these particles (except the neutrinos) undergo interactions between themselves and with photons. All of them can annihilate with their antiparticles in order to create photons and gluons.
Conversely, photons and gluons can give birth to pairs of particles.\\
Only the electron, the neutrinos and the $u$ quark are absolutely
stable against all disintegrations. The five more massive quarks
$(t,b,c,s\;\mbox{and}\;d)$ can transform by means of natural
disintegration or weak or electromagnetic interactions in order to
give birth to less massive quarks. The less massive of all quarks,
namely the $u$ quark, need to be supplied with energy to transform
into another quark. This supply can be achieved during
interactions with electrons or antineutrinos (for instance). The
\emph{u} quark transformation can also be achieved during a
nuclear transformation resulting on a binding energy increase. The
leptonic families  $\mu$ and $\tau$ always give birth to electrons
or positrons together with neutrinos or antineutrinos. Neutrinos
are created during annihilations, disintegrations and weak
interactions between all types of particles. They are absolutely
stable and electrically neutral and we do not know if they are
material (or antimaterial) particles or a type of a pure energy
particles which, for us, is more probable. The infinitesimal mass
of these particles (in case where it really exists) is not yet
precisely determined. We notice that the only absolutely stable
fundamental particles (i.e. having an illimited lifetime) are the
electron, the quark $u$ and the neutrinos. To these stable
elementary particles we must add the only composite material
particle that is considered until now as being stable against all
sorts of disintegration (or more precisely having a lifetime that
is more than $10^{32}$ years): the proton. The immaterial
particles (i.e. having a null mass) of pure energy, namely the
photons (and gluons) are equally stable. They hold by themselves
the pre-existence of matter and the capacity of creating all sorts
of matter and antimatter and interacting with all sorts of matter
and the capacity to transform into all sorts of energy.\\

\noindent{\bf Remark:} The above study could be achieved for $\mu_1=\mu_2=\mu_3=:\mu$ where $\mu$ is a triple eigenvalue of the Dirichlet problem on the unit ball to which correspond three one dimensional eigenspaces spanned successively by the eigenfunctions $\varphi_1,\varphi_2$ and $\varphi_3$.

\subsection*{Natural forces}

Concerning natural forces, we deduce from our global study that
there essentially exist two fundamental forces that are inherent
into the creation, the formation and the evolution of the
universe. The first one is the electromagnetic force which is
originated in the formation of two sorts of matter (and
antimatter) since they were differently charged. The arithmetical
addition of the electron and the six quarks charges leads to the
equality
$$1 + \frac{1}{3} + \frac{1}{3} +\frac{1}{3} = \frac{2}{3} +
\frac{2}{3} + \frac{2}{3}$$The two differently charged particles
attraction recalls the original unity of the matter-energy. The
two likewise charged particles repulsion recalls the expansion,
the original movement and the original dispersion of matter. This
force has an unlimited range and is proportional to the product of
charges and to the inverse of the distance squared. It takes place
between all types of charged matter (and antimatter): quark-quark,
quark-lepton, lepton-lepton and particularly electron-electron,
electron-proton, proton-proton inside atoms and between atoms and,
more generally, between all charged material agglomerations. The
electromagnetic force results from electrical and electromagnetic
fields which naturally exist around charged static matter and
charged matter into movement. It did not need to be conveyed by
any type of intermediaries. The electromagnetic waves are simply
radiations that are emitted by accelerated charges or by electron
changes of energy levels (to speak only about atomic radiations
without speaking about nuclear ones) regardless of the possible
receivers of these radiations. The energy exchange, conveyed by
electromagnetic waves (exchange of a large number of photons),
leads only to the modification of the surrounding electromagnetic
field which modifies the surrounding force field that is the
gradient field of the generalized energy distribution $E_t(X)$.
This distribution, as we have seen before, is reflected by the
real physical metric $g_t$ which characterizes the physical
universe,
permanently evolving.\\

The second fundamental force is the generalized gravity which is
based on the tendency of material particles to attract each other
and matter to contract in order to recover its quasi-original
ultracondensed state as the neutron stars or to transform into
black holes. This tendency is counterbalanced by the
electromagnetic repulsion and the fermionic exclusion, on one
hand, and the kinetic energy related to the movement that has been
initiated by the Big Bang on the other hand. The gravitational
force takes place naturally for all material particles or bodies:
quark-quark, nucleon-nucleon, nucleus-nucleus, nucleus-electron,
atom-atom, star-meteorite, star-star, galaxy-galaxy, etc. It is
proportional to the product of masses and to the inverse of the
distance squared. It has an unlimited range and did not need to be
conveyed neither by gravitons nor by gluons. A given mass is
surrounded by a gravitational field which attracts proportionally
any other mass
located around it. It is the same for the invisible black holes which probably constitute a large part of the mass equivalent of the universe.\\

Although these two forces take place simultaneously and do not
exclude each other (both fields exist around charged matter), the
study of laws that govern them shows that the electromagnetic
force is predominant over the gravitational one except on the
cosmic level and in the case of two large masses weakly charged
(such as an apple a few meters above the earth). However, we
notice that both of static formulas that characterize these two
forces, namely $F = k\frac{q_1q_2}{r^2}$ and $F =
G\frac{m_1m_2}{r^2}$, are inadequate at subatomic scale. Indeed,
concerning the gravitational force, we have to take into account
the important (vibrational and rotational) movement of all very
energetic subparticles that are extremely close each to other and
introduce a correction (generally qualified as relativistic) to
the above formula for gravity involving speeds, accelerations,
frequencies, momenta and mass evolutions. For the electromagnetic
force, the corrections that have to be operated to the Coulomb
formula are decisive. Inside atoms and nucleons, we have to use
Lorentz formula, namely $F$ = $q$($E$ + $v \wedge B$). The global
(orbital and intrinsic) magnetic field \emph{B} produces the
indispensable correction to the attractive and repulsive forces
inside atoms, nucleus and nucleons taking into account the
important fact that moving charges, even if they are both positive
or negative, give birth to force fields that can be attractive or
repulsive according to the movement directions; both effects may
be extremely strong when the speed $v$ is very large or
"relativistic". The association of these force fields (including
the screen effect) with gravitational force (corrected by dynamic
effects and their multi-energetic consequences) leads to the
internal force field at all levels of material particles to which
we have to add external
forces and the internal and external interactions.\\

Concerning the classical factors $\alpha$, $\alpha_g$, $\alpha_s$,
and $\alpha_w$, only the electromagnetic factor $\alpha$ is
naturally defined by (c.f.[2])
$$ \alpha = \frac{2\pi k e^2}{E_{photon}\lambda_{photon}}=\frac{k e^2}{\overline{h}c}.$$
It is obvious that the definition of $\alpha_g$ by
$$\alpha_g = \frac{G m_p m_p}{1240}$$
and
$$\alpha_g(E)=\frac{G(\frac{E}{c^2})^2}{\overline{h}c} $$
is less natural.

\noindent It is the same for the definitions
$$\alpha_w(E)\simeq
(\frac{kg^2}{\overline{h}c})\left(\frac{E}{m_wc^2}\right)^2\simeq\alpha\left(\frac{E}{m_w
c^2}\right)^2$$ and $$\alpha_s(E)\simeq
\frac{12\pi}{(33-2m_f)\ln(\frac{E^2}{\Lambda^2})}$$ for $g\simeq e$
and $\Lambda = 0,2$ GeV.

\noindent If we admit the legitimacy of these definitions, taking
into account the dependence of these last factors on the energy and
the distance, we obtain, in place of the figure $(15)$ (c.f. [6]), the
scheme of the figure $(15')$ for the unified factor $\alpha$ which
strongly depends on the considered distance.

\subsection*{Strong and weak interactions}

Let us, for instance, consider a neutron constituted with a bound
state of two negatively charged quarks and a positively charged
one symbolized by \emph{udd}. Two fundamental reasons push these
quarks to form such a bound state owning some cohesion inside this
particle: the original unity of matter and the fractional charges
of these quarks. An individual quark does not exist in Nature a
little time after the Big Bang which induces the confinement
effect. It is the same for a particle with fractional charge. The
internal force field, that prevents the particle against the
dispersion of these three quarks or against the separation of one
of them, comes from the corrected tripolar gravity field and the
total electromagnetic one. The global field is responsible of the
neutron inner cohesion . The binding energy of this bound state is
equal to
$$m_u c^2 + 2 m_d c^2 - m_n c^2$$This bound state is not static (as
well as for all hadrons); it is not stable too (as well as for all
hadrons except the proton). Indeed, there is, inside the neutron,
a permanent exchange of pure energy particles (i.e. massless
particles), called gluons, in the same manner as the permanent
exchange of photons inside atoms. Furthermore, there is an
electromagnetic and gravitational interaction between each of the
quarks inside the neutron and the neighboring quarks inside
nucleons of neighboring nucleus in the same manner as there is an
electromagnetic interaction between protons and electrons of two
neighboring atoms inside molecules. To all these forces is added
an electromagnetic attraction between nucleons inside the nucleus,
doubled with the electromagnetic interactions between the nucleus
and electrons inside the atom and between all particles inside and
outside the atom. Electromagnetic interactions can lead to a
neutron decay (qualified as weak decay) that transforms neutrons
into protons by the $\beta^- -$ decay process symbolized by
$$n \longrightarrow p + e^- + \overline{\nu_e}$$
This decay actually is realized by the emission, from the quark
\emph{d}, of a virtual (for us a real) particle denoted by
$W^{\ast^-}$ which gives birth to an electron and an antineutrinos
symbolized by
$$W^{\ast^-} \longrightarrow e^- + \overline{\nu_e}.$$However, we
notice that in a large number of interaction and decay processes
(called weak) three particles denoted by $W^+, W^-$ and $Z^0$ are
emitted; they are material bosons having well determined mass
energies (although the values of which are approximately
determined essentially because of their dynamic and evolutive
nature) and very short lifetimes. We produce below (fig.7), as
representative examples, some schemes (called Feynman diagrams) of
interactions and decays in which are involved quarks, leptons and
neutrinos beside of these three particles acting as weak interaction intermediates (c.f.[2]).\\

The previous description of neutron bound state is also valid for
a proton bound state \emph{uud} inside a nucleus of a given atom.
The electrical repulsion effect between two positively charged
quarks inside a proton or between two protons inside an atom
nucleus is counterbalanced by the attractive forces generated by
the movement and the electromagnetic (especially the magnetic) and
gravitational fields resulting in an energy equilibrium at every
time (even though evolving permanently) between mass, potential
and kinetic energies. The binding energy of the proton bound state
is
$$2 m_u c^2 + m_d c^2 - m_p c^2$$and inside the nucleus is
$$\Sigma m_p c^2 + \Sigma m_n c^2 - m_N c^2$$
We notice that, for light nucleus, the nucleus stability against
strong decay, such as, for instance, natural fission and $\alpha-$
decay requires an approximate equality of the atomic numbers $Z$
and $A-Z$ whereas, for heavy nucleus, the stability requires a
significantly larger number of neutrons than protons. The
repulsion effect is then more important when $Z$ is close to $A-Z$
and it can not be durably counterbalanced by the electromagnetic
attractive component together with gravity. The $\beta$ (called
weak) decay and the \emph{X} - rays emission are always possible
in a natural or artificial ways (by means of weak or
electromagnetic interactions) except for protons as it is
generally admitted. We notice however that the so called strong
interactions are actually conveyed by the gluons' intermediary
which are massless bosons, exchanged (emitted and received) at
short range inside nucleons and nucleus contrary to
the $\alpha$, $\beta$ and $\gamma$ decays which are of a different nature.\\
Do we have to consider gluons as being fundamentally different
from electromagnetic photons?\\
Our answer is No: virtual or real, involved into strong
interactions such as those represented in fig.8 or into
interactions such as $g \longleftrightarrow q + \overline{q}$,
they are not essentially different from photons. Indeed, photons
are involved, in a similar way, into many interaction, exchange
and annihilation processes. We can invoke , for instance,
annihilations such as $\gamma \longleftrightarrow e^+ + e^-$,
annihilations with pair production such as those represented in
fig.9, or jet phenomena produced by electron-positron's collision
represented by
$$e^+ + e^- \longrightarrow q +
\overline{q} + g$$ The difference between these two pure energy
particles consists on the fact that gluons are only exchanged by
particles with extremely short range and lifetime while photons
are exchanged within any range but also can be emitted in many
circumstances independently of any potential receiver.\\\\
Consequently, we do not think that there exists a third
fundamental force that would be independent of the two previously
depicted fundamental forces, based on strong charges (or colors)
inherent in quarks and conveyed by intermediary gluons (which are
for us, as the photons, intermediary particles for strong
interaction that do not transmit charges). Actually, each quark
does exist with three different colors in much the same manner as
some particles exist with two opposite spins. These colors only
constitute three different kinds of the same particle without
(until now) any detectable physical difference between them
contrary to the clear physical difference between the two opposite
spins of the same particle. Our judgment is however sustained by
the fact that there is no significantly stable elementary particle
(such as a nucleon, for instance) which is constituted with
similarly charged quarks nor a nucleon, for example, that is
constituted with uniquely two protons or two neutrons. Gravity and
electromagnetic forces having not permitted such a bound state, we
can wonder why the ''strong charges'' or the ''strong forces''
would not have played a favorable role?\\\\
Thus, we can maintain that, although strong forces and charges do
not really exist, strong interactions and bounds do exist. Strong
bounds, organically related to the two fundamental forces,
contribute together (within all energy rules' agreement) to form
all bound states between quarks inside hadrons and between hadrons
inside nucleus. The stability of these bound states is directly
related to the stability of the energy equilibrium (which is more or less temporary) between the (electromagnetic and
gravitational) potential energy, the vibrational and rotational
(thermodynamic) kinetic energy and the masses' energy of all
components. The strong bounds (and interactions) essentially rely
on the transformation matter-energy and reciprocally. These
transformations are conveyed by intermediary gluons that are
exchanged between quarks favorizing many energy transformations
which change their nature and then the nature of hadrons
containing them. We can have changes of masses, of binding energy
and of electrical charges (which are accompanied with emission of
electrons or positrons, for example) leading to all sorts of
nuclear transformations. That is what explain the extremely short
range and lifetime of gluons contrary to the photons which are
emitted without any prior exchange
condition.\\\\

Likewise, all weak interactions involving the $W^+$, $W^-$ and
$Z^0$ bosons (as those we have showed in fig.7) and weak decays
(when studied parallel to interactions such as
$$e^+ + e^- \longrightarrow \psi' \longrightarrow \pi^+ + \pi^- +
\psi \hskip 0.3cm \mbox { with } \hskip 0.3cm \psi \longrightarrow
e^+ + e^- ,$$
$$e^+ + n \longrightarrow \overline{\nu_e} + p  \hskip 0.5cm , \hskip 0.5cm e^- + p
\longrightarrow n + \nu_e ,$$
$$p + p \longrightarrow d + e^+ + \nu_e \hskip 0.3cm \mbox { and } \hskip 0.3cm \gamma + d
\longrightarrow n + p)$$do not indicate the existence of another
fundamental force (called weak force) that would be based on weak
charges carried by hadrons and leptons and would be conveyed by
the \emph{W} and \emph{Z} bosons. These particles have a very short lifetime, an extremely short range and are, essentially, the weak interactions' intermediates between quarks and leptons at extremely short distances.\\
 Electromagnetic interactions,
conservation laws and  the Pauli exclusion law are sufficient for
explaining all these phenomena as well as some others such as the
formation of deuteron in the early universe from neutron and
proton and the nonexistence of a bound state of two protons or two
neutrons, for instance. They are also sufficient for explaining
the (solar) cycle:
$$p + p \longrightarrow d + e^+ + \nu_e$$
$$p + d \longrightarrow  {}^3He + \gamma$$
$${}^3He + {}^3He \longrightarrow p + p + \alpha$$
and the interaction $$n + {}^3He \longrightarrow \alpha + \gamma,$$
for instance.\\
\textbf{Remark:} Other discussions on the bound states and
fundamental forces will be given in sections 11 and 13.\\

\subsection*{Binding energy and Matter-Energy}

The global energy of the universe $E_0$ was concentrated, before
the Big Bang, at a singular point of the space-time (i.e. at the
vertex of the space-time semi cone) where we can consider that all
of the energy was a sort of a binding energy. The appearance,
after the Big Bang, of the matter under the form of hadrons with
their internal binding energy and particularly of protons and
neutrons which (later on) formed the nuclei with their proper
binding energy was not the only transformation of the initial
energy $E_0$. Indeed, we have to add to the mass energy of these
particles and agglomerations of particles and to their binding
energy that is inherent in their formation, the kinetic energy of
the matter into movement and their interaction potential energy
which essentially consists on the gravitational and
electromagnetic potential energies. These two sorts of potential
energy are in fact two forms of binding energy. For an atom (for
instance), the quantity
$$ (m_N + \Sigma {m_i} - m_a) c^2,$$where $m_N, \Sigma {m_i}$ and
$m_a$ are respectively the nucleus, the electrons and the atom's
masses, is the binding energy of the atom (i.e. the mass
difference ($m_N + \Sigma {m_i}$) - $m_a$ that is transformed into
energy). This energy is tightly related to the electromagnetic
potential energy inside the atom (the gravitational potential
energy inside the atom is negligible). The passage of an electron
from a potential level to another necessarily is accompanied with
a change of the binding energy inside the atom. The passage of the
electron's hydrogen atom from the ground state to an excited state
is expressed by a weaker attractive electromagnetic potential
energy and a weaker binding energy in such a way that
$$ e^{'}_l := (m^{'}_p + m^{'}_e - m^{'}_H)c^2 < (m_p + m_e - m_H)c^2 =:
e_l$$ We notice that the mass $m^{'}_H$ of the atom and the mass
$m^{'}_e$ of the electron inside the atom have increased; the
electron's energy increase is due to the potential energy increase which is larger than the kinetic
energy decrease.\\\\
The absorption of a photon with energy $e_p$ by the electron
increases its (negative) potential energy and slightly increases
its mass by decreasing its speed and its kinetic energy. However,
the total electron's energy (potential + kinetic) has increased
and obviously the global mass of the atom has increased by
$\frac{e_p}{c^2}$. As the electron has absorbed the photon energy
$e_p$, the increase of the atom mass originates from the
transformation of the binding energy difference $e_l - e^{'}_l$
into mass:
$$ \frac{e_l - e^{'}_l}{c^2} = m^{'}_H - m_H = \frac{e_p}{c^2}$$
The gravitational potential energy can be interpreted in the same
manner in terms of binding energy that is related to the
transformation of mass into energy and vice versa.\\\\
For a planet that has a stable orbit around  gravitational pole
(such as a star), there is a stable equilibrium between the mass
energies, the kinetic energies, the gravitational potential energy
and the binding energy when neglecting the thermic and
gravitational radiations. When the planet orbit becomes closer to
the pole, the gravitational potential energy becomes more and more
negative, the orbital speed and the kinetic energy of the planet
continually increase; the binding energy also increases whereas
the planet mass decreases as well as the mass of the system
constituted of the gravitational source and the planet. Part of
the mass difference transforms into additional binding energy of
the system that is caused by the increasing intensity of the
gravitational field; the other part transforms into kinetic
energy. In the extreme case where a planet is swallowed by a black
hole, the mass energy is entirely transformed into a binding
energy traduced by the black hole energy increase.\\
A similar phenomenon happens when the orbits of a binary system
are slightly and continuously reduced; the absolute value of the
negative gravitational potential energy increases, the kinetic
energy also increases and the total mass of
the system decreases whereas the binding energy increases.\\\\
We also notice that the passage of the hydrogen atom's electron
from an orbit to another that is closer to the nucleus displays a
similar scheme where the electromagnetic potential energy plays
the role of the gravitational potential one and the (temporary)
energetic equilibrium is again insured by the intermediate of the
new orbital motion after the emission of a photon with a well
determined energy: the electron's total energy decreases, its
electromagnetic potential energy decreases, its speed and kinetic
energy increase, its mass decreases, the energy and the mass of the
atom decrease whereas the binding energy increases. A similar
process happens inside nuclei where the binding energy between
nucleons is created at the expense of the decrease of all
individual masses. In that case, the electromagnetic and
gravitational potential energies are caused by the charged
constituents of the nuclei, i.e. the quarks, although each nucleon
forms a bound state between three quarks having different flavor, spin or color.\\\\
To sum up, the universe initial energy $E_0$ has, after the Big
Bang, the following forms: Mass energy, kinetic energy, binding or
interaction energy having gravitational or electromagnetic
potential energy character, the pure energy of photons and neutrinos and the pure energy (that can be
considered as a sort of binding energy) of the black holes.
However, the mass energy of the visible matter is only a small
part of the total universe's mass energy and we think that what is
generally called dark matter or dark energy is constituted with
black holes (whose energy has a gravitational mass energy
character), neutron stars, brown dwarves and other invisible stars
associated with a large number of binary systems and finally with
invisible ordinary matter such as planets. Nevertheless, the
classification of the background energy associated with the
neutrinos still is enigmatic. Neutrinos have probably no mass;
although they have no electromagnetic interactions they
significantly contribute to the
radiational energy of the universe.\\\\

\subsection*{Brief description of the universe}

Finally, we notice that our global model is compatible with the
classical description of the different stages of the universe
evolution and the evolution of matter, antimatter and
energies.\\\\
1.At the beginning of the expansion (for \emph{t} $\ll$ 1), the
infinitely small sized universe was dominated by ultra-energetic
radiations, in perfect thermal equilibrium state, having
infinitely large frequencies (i.e. infinitely small wavelengths)
with infinitely large radiations' density under infinitely large
pressure and at an infinitely large
temperature; all of them decreasing very quickly.\\\\
2.Then began the stage qualified as a quark and lepton soup
(without any doubt, with their antiparticles) having two opposite
sorts of electrical charges, followed by the formation of protons,
neutrons and (without doubt) neutrinos with their antiparticles.
This formation has become possible thanks to the relative
attenuation of the gigantic original radiations energy and
original
pressure and temperature.\\\\
3.The conservation and exclusion rules permitted then for some
electromagnetic interactions to take place more than others
resulting in a fall of neutrinos' formation (neutrinos' freeze)
and favored the progressive disappearance of antimatter (such as
positrons) for the benefit of more protons than neutrons'
production. All of this is governed by energetic equilibriums
involving a stability and lifetime disparity.\\\\
4.After that occurred the formation of light stable nucleus and
other, more or less stable, material particles coming with the
decrease of radiations' density for the benefit of matter density,
so permitting the formation of all material agglomerations from
atoms and molecules to galaxies. From this stage qualified as
photons' freeze, the tendency toward matter density predominance
over the radiation one although the general density,
the general pressure and the cosmic temperature are decreasing
because of the continual expansion which occurs at a speed very
close to its limit 1.\\\\

Finally, according to our previous global study, we can suppose
that we have, for $t>0$ belonging to a small interval of time, the formula
$$v(t) =
\lambda(t) f(t) = \frac{1}{2 \pi} \frac{\sqrt{\mu}}{t}
\lambda(t)$$where \emph{v}(\emph{t}), $\lambda$(\emph{t}) and
\emph{f}(\emph{t}) = $\frac{1}{2 \pi} \frac{\sqrt{\mu}}{t}$ are
respectively the expansion speed, the wavelength and the frequency
of the expansion waves creating the universe geometrical space. We
can then assume that
$$\lambda(t) = \frac{2 \pi}{\sqrt{\mu}} \hskip 0.1cm t \hskip
0.1cm \frac{a(t)}{b(t)} \mbox { and } v(t) =
\frac{a(t)}{b(t)}$$with\\
(i) \emph{a}(\emph{t}) is an increasing function satisfying the
following properties:
$$\lim_{t\rightarrow 0^+} a(t) = 0 \mbox { and }\lim_{t\rightarrow
+\infty} a(t) = +
\infty.$$\\
(ii) \emph{b}(\emph{t}) is also increasing and satisfies the
properties:
$$\lim_{t\rightarrow 0^+} b(t) = b > 0 \mbox { and
}\lim_{t\rightarrow +\infty} b(t) = +\infty.$$\\
(iii) \emph{v}(\emph{t}) = $\frac{a(t)}{b(t)}$ is equally
increasing and satisfies the properties:$$\lim_{t\rightarrow 0^+}
\frac{a(t)}{b(t)} = 0 \mbox { and } \lim_{t\rightarrow +\infty}
\frac{a(t)}{b(t)} = 1.$$(Such a function exists; indeed we can
obviously take $$v(t) = \frac{\ln(1+t)}{\ln(1+\alpha+t)} \mbox {
with
} \alpha > 0 \mbox { and } \ln(1+\alpha) = b).$$\\
In order to determine approximately \emph{a}(\emph{t}),
\emph{b}(\emph{t}) and \emph{b}, we have to collect a large number
of extremely precise measures using many sophisticated technical
means such as ultra-powerful telescopes and ultra-high-energy
nuclear accelerators in order to go forward in the knowledge of
the most earlier universe and the understanding of the original
matter-energy structure. However, we notice that the value of the
real number \emph{b} of the above example, for instance, is
decisive: For small or infinitely small \emph{b}, the universe age
is close to what is now generally admitted; whereas, if \emph{b}
is large or infinitely large, then the universe age is larger then
what we are presuming (this age will be approximately determined
at section $11$) and its evolution until the earliest stage that
we can scrutinize at present has taken a long time. In this case,
the time $T_0$ (expressed in seconds) needed for the universe to
attain the size $B_e$(O,1) (where 1 represents here 3 $\times$
10$^8$m), which would correspond to a significant expansion speed
(becoming later on close to 1), is rather large and it could even
be very large ($T_0 \gg$ 1). Therefore, if we assume that the
present universe size is approximately $t \times(3 \times 10^8m),$
then the effective universe age, starting from $t = 0,$ would be
nearly $t + T_0.$ Conversely, if we assume that the time interval
from the Big Bang until now is \emph{t}, then the
universe size would be close to $t - T_0.$\\\\

This eventuality is supported by the validity of general
relativity theory concerning the electromagnetic waves -
gravitational field interaction. The influence of gravitational
force on waves is confirmed by many natural phenomena's
observation and many experiments such as the Pound and Rebka one.
This influence is moreover reconfirmed by our model as the
gravitational field curves geodesics and contracts distances. If
$X(t)$ describes an electromagnetic wave trajectory (with respect
to a virtual fixed frame), we have $\nabla_{X^{'}(t)}^{g_t}
X^{'}(t)$ = 0 and generally $\|X^{'}(t)\|_{g_e}$ = \emph{c} = 1
according to special relativity first postulate of Einstein
whereas we have, inside a significative gravitational field,
$\|X^{'}(t)\|_{g_t} < \|X^{'}(t)\|_{g_e}.$ Thus, although the
Pound-Rebka experiment interpretation is, according to our model,
different from the generally accepted one, it proves the existence
of the gravitational force action on photons. Indeed, according to
Pound and Rebka, if the gravity does not cause a blueshift
$\Delta_{1} E = gL$ when the photon $\gamma$ is propagating toward
the earth and if the temperature is actually constant and the
vacuum is actually absolute, then we would obtain an optimal
resonance for a fixed emitter. The fact that such a resonance is
obtained when moving the emitter implies that, according to them,
it is necessary to cause a redshift $\Delta_{0}E=-\beta E$ with
$\beta = \frac{v}{c}$ and the movement has to be upward. When the
photon is propagating upward, we have then to move the emitter
upward too in order to cause a blueshift $\Delta_{0}E=\beta E$
which would balance the redshift $\Delta_{1} E= -gL.$\\
Our interpretation coincides partially with the Pound and Rebka
one. If there is no a blueshift (or a redshift) $\Delta_{1} E=gL$
and if the temperature is absolutely constant and the absolute
vacuum is perfectly respected, then the photon energy does not
change and the resonance phenomenon would happen without moving
the emitter. The photon energy $E$ would be, according to our
model, constant and it is given, for all distances (under the same
conditions), by $E = h(t)f(t),$ even though $h$ and $f$ are
depending on time. This quantity is moreover equal to $h_{P}f_{D}$
as it has been noticed previously. So, if $E$ and $f_{D}$ were
constant, then Pound and Rebka would not need to move the source.
Conversely, if the previous conditions are not respected, then
even if the gravitational blueshift (or redshift) does not exist,
we would need to cause a blueshift in both cases. Finally, since
there is an energy change $\Delta_{1}E= \pm gL,$ we need to cause
an appropriate $\Delta_{0} E.$ For a fixed position of the photon
source, this could be supplied (in case of propagation toward the
earth) by temperature fluctuations and the lake of absolute
vacuum, but, in the other case, this can not be achieved since
both $\Delta_{i} E (i = 0,1)$ would have the same sign. We will
need then to operate a correction $\Delta_{0} E$ (which is
necessarily a blueshift in the case of upward propagation) which,
according to our model, can be produced only by speed fluctuations
(accelerations) or distance variations. The distance variation
implies a variation of the temperature effect and of the lake of
absolute vacuum effect.
\subsection*{General modeling of the universe}
We notice also that our study could have been entirely
reconstructed by taking as a starting point the universe
$U(t_0)=(B_e(O,R(t_0)),g_{t_0}),$ for any fixed $t_0,$ provided
that we can recover back this privileged instant and that we have
a sufficient knowledge on how the universe goes back and
forward in time.\\

\noindent Let us consider, for instance, the universe at time $t
>0$ as being the ball $B_{e}(O,R(t))$ equipped with the physical
metric $g_t(X)$ for $|X| \leq R(t).$ This metric is determined by
the generalized energy distribution $E_{t}(X)= E(t,X).$ The
matter-energy equation $(E^{*})$ on the ball $B_{e}(O,R(t))$
becomes
$$\square E(t,X) = \frac{\partial^2 E}{\partial t^2}(t,X) - \Delta
E(t,X) = 0$$with$$E(t,X)_{|S_e(O,R(t))} = 0.$$When we bring back
the resolution of this problem to that of the Dirichlet problem on
the unit ball $B_{e}(O,1)$ and when we choose a particular
eigenvalue $\mu$, we obtain the pseudo-periodic solution
$$E_\mu(t,X) = \left(f_\mu(0)\cos \sqrt{\mu} \frac{t}{R(t)} +
\frac{R(t)}{\sqrt{\mu}}f^{'}_\mu(0)\sin \sqrt{\mu}
\frac{t}{R(t)}\right)\psi_\mu \left(\frac{X}{R(t)}\right)$$having
$T_\mu(t) = 2 \pi \frac{R(t)}{\sqrt{\mu}}$ and $f_\mu(t)
=\frac{1}{2 \pi} \frac{\sqrt{\mu}}{R(t)}$ as a pseudo-period and a
pseudo-frequency respectively. This function satisfies, for any
geodesic (relatively to $g_{t}$) trajectory $X(t),$ the relation
$$E(t) := E_\mu(t,X(t)) = h_\mu(t) f_\mu(t) =: h(t)f(t) = h_P
f_D$$where $h_{P}$ is the Planck constant and $f_{D}$ is the De
broglie frequency which depends on time naturally. All our above
formulas which have been established in simplified contexts can be
adapted to the definitive setting when we consider the dynamical
universe $U(t)$ as being, for every time $t,$ modelized by the
Riemannian space $(B_{e}(O,R(t)),g_{t})$ where $R(t)$ is the
actual radius of the universe and $g_{t}$ is the physical real
metric, at time $t$, which is determined by the matter-energy
distribution, the time and the temperature (or the
pressure).\\
Finally, we can state that, for every time $ t > $ 0,
we have, for any fixed $t_{0} > 0$ :
$$R(t) = R(t_0) + k(t)(t - t_0)$$ with $k(t) \sim 1 $ for $t \gg $
1, $k(t)$ is increasing and $$\lim_{t \rightarrow +\infty}k(t) =
1.$$

According to our study, we can finally notice that the space and
time semi-cone is rather of the form sketched in fig.10 unless the
speed of  electromagnetic waves'propagation and that of the space
expansion were always equal to 1 and then the semi-cone of space
and time would have really the
form sketched in fig.3. Nevertheless, we think that the first case is logically much more probable. Then, the matter-energy equation has to be of the form
$$\frac{1}{v^2(t)}\frac{\partial^2 E}{\partial t^2}(t,X)-\Delta E(t,X)=0$$
where $v(t)$ is originally very small $(v(t)\ll 1)$ for an unknown interval of time. Moreover, $v(t)$ is increasing and becomes afterword near to $1$ and $\lim\limits_{t\rightarrow+\infty}v(t)=1$. This is due to the gigantic gravitational field in the very early universe when the global energy was concentrated in a singular point much like a gigantic black hole.\\\\

Our progressive approach has been (subjectively) imposed for us in
order to avoid some complications (that could result from a large
number of factors that are involved) along the construction of our
model and also in order to achieve it with most possible
simplicity and clarity. However, we are perfectly aware that there
are inside this study many points to be detailed, specified,
clarified, added and discovered.\\\\

\section{The reviewed Einstein's General Relativity Theory}

In this section, we propose to prove that our dynamical universe
is overall described by readjusting general relativity theory to
our model. Once this adaptation is achieved, a large number of
problems related to theoretical Physics and Cosmology, for
instance, can be unequivocally well posed and
clearly resolved in a most simpler way.\\

We start by noticing that our mathematical and physical background
along this section will essentially be the results established in
the previous sections on one hand and on the great classical work
on general relativity theory of R. Wald ([4], General relativity,
1984).

\subsection*{Preliminaries}

Our model briefly consists of considering the physical universe as
being, at every time $ t >$ 0, the Riemannian space
(\emph{B}(\emph{O},\emph{R}(\emph{t})),$g_t$), where
\emph{R}(\emph{t}) $\sim$ \emph{t} for $ t \gg$ 0,
\emph{B}(\emph{O},\emph{R}(\emph{t})) is the ball of Euclidean
radius \emph{R}(\emph{t}) and $g_t$ is the regularized Riemannian
metric that reflects, at time \emph{t}, the whole physical
consistence of the universe. This consistence is entirely
specified by the generalized matter-energy distribution
$E_t$(\emph{X}) on \emph{B}(\emph{O},\emph{R}(\emph{t})), which
includes the matter distribution $m_t$(\emph{X}) as well as all
manifestations of matter-energy. This distribution is reflected by
the curvature of the position and time-dependent metric $g_t$. We
then define on \emph{B}(\emph{O},\emph{R}(\emph{t})) the measure
$$\nu_t(X) = E_t(X)dX$$
i.e. the measure of density $E_t$(\emph{X}) with respect to the
Lebesgue measure \emph{dX} on
\emph{B}(\emph{O},\emph{R}(\emph{t})). We then have
$$dv_{g_t}(X) =: \mu_t(X) = dX - \nu_t(X) = (1 - E_t(X))dX$$
This property expresses that $\mu_t$ measures the physical volume
of a domain in $\mathbb{R}^3$ that contains a distribution of
matter-energy and $\nu_t$ measures the failure of the physical
volume to be equal to the volume of the same domain when supposed
being empty. This latter is conventionally measured by the
(Euclidean) Lebesgue measure. Thus the matter-energy filling the
space contracts distances and volumes. The dynamical feature of
the permanently expanding universe whose expansion occurs (for $t\gg1$) nearly
at the speed of light in the vacuum \emph{c} (=1) is described by
the semi-cone of space-time:
$$C = \{ (x, y, z, t); x^2 + y^2 + z^2 \leq R^2(t), t > 0 \} =
\bigcup_{t > 0} B(O,R(t))\times \{t\}$$which will be considered
(for sake of simplicity and clarity) as being
$$C = \{ (x, y, z, t); x^2 + y^2 + z^2 \leq t^2, t > 0 \} =
\bigcup_{t > 0} B(O,t)\times \{t\}$$equiped with the Riemannian
metric
$$h = dt^2 - g_t$$
The universe will then be represented, at any time $t_0$, as the
hypersurface-intersection of \emph{C} with the hyperplane of
equation $t = t_0$ in $\mathbb{R}^4$. This hypersurface will be
denoted by $\Sigma_{t_0}$ and \emph{C} by \emph{M}. Then, within
the framework of our modeling, $\Sigma_{t}$ is a compact Cauchy
surface (with boundary) and our space-time manifold ($M$,
$h_{ab}$) is stably causal and asymptotically flat. The vector
field $(\frac{\partial}{\partial t})^a$ on \emph{M} is
$\Sigma_{t}$-hypersurface orthogonal. These properties shall
considerably simplify the foundation of the general relativity
theory as well as its use for explaining the
universe dynamics.\\

\subsection*{Einstein's tensor equation - Lagrangian
formulation}

Our starting point is the equation
$$\hskip 2cm ^{(3)}G_{ab}(t) := \hskip 0.1cm ^{(3)}R_{ab}(t) - \frac{1}{2} \hskip 0.2cm ^{(3)}R \hskip 0.2cm ^{(3)}g_{ab}(t) =: \hskip 0.1cm^{(3)}T_{ab}^{*}(t) \hskip 1.7cm (23) $$
defined on (\emph{B}(\emph{O},\emph{t}),$g_t$), where $^{(3)}
R_{ab}$ and $^{(3)}R$ respectively designate the Ricci curvature
and the scalar curvature associated with the physical metric
$^{(3)} g_{ab}$(\emph{t}) := $g_t$ and
$^{(3)}T_{ab}^{*}$(\emph{t}) is the matter-energy tensor that
specifies, at every time $t > $0, the physical consistency of the
universe. This consistency is fundamentally related to the very
existence of the permanently evolving matter-energy which is
filling the virtual empty space \emph{B}(\emph{O},\emph{t}).
According to our definitions, $^{(3)}T_{ab}^{*}$(\emph{t})
$\equiv$ 0 on a domain $D \subset B$(\emph{O},\emph{t}) means that
\emph{D} is absolutely empty (i.e. \emph{D} is quasi-free of all
matter-energy manifestations and effects : gravity,
electromagnetism, thermodynamic effects...) and then we have
$$g_{ab}(t) \equiv g_e \;\;\;\;\;\; \mbox { on  \emph{D}.} $$
Next, we consider on (\emph{M}, $h_{ab}$), the (slightly modified)
Einstein's equation:
$$\hskip 2cm  ^{(4)} G_{ab} := \hskip 0.1cm ^{(4)} R_{ab} - \frac{1}{2} \hskip 0.2cm
^{(4)}R\hskip 0.1cm h_{ab} = \hskip 0.1cm ^{(4)} T_{ab}^{*} \hskip
4.3cm (24)$$ where $^{(4)}R_{ab}$ and $^{(4)}R$ are respectively
the Ricci and scalar curvatures associated with $h_{ab}$ on
\emph{M}. Here, $^{(4)}T_{ab}^{*}$ represents ( up to a constant)
the generalized Einstein's stress-energy tensor. Thus, with the
previous notations, we have
$$^{(3)}T_{ab}^{*} = 0 \Rightarrow g_t = g_e \Rightarrow h = dt^2
- g_e = \eta,$$ where $\eta$ is, up to sign, the flat Minkowsky
metric on the semi-cone \emph{C} and then we have
$^{(4)}T_{ab}^{*}$ = 0. We notice that our manifold \emph{M} is
evolving with time since we have, for every time $t_0$:
$$M = C(t_0) = \{ (x, y, z, t); x^2 + y^2 + z^2 \leq t^2, 0 < t
\leq t_0 \}. $$In the following, we shall intensively use the
results presented in [4] for sake of greatly valuable economy.
Thus, for further facility and clarity we shall slightly modify
our notations in order to conform to those used in this capital
reference. Consequently, our metric on the dynamical space-time
\emph{M} = \emph{C}(\emph{t}) will be denoted by $g_{ab}$. It is
chosen to be of a lorentzian signature and is defined by
$$^{(4)}g_{ab} = - dt^2 + \hskip 0.1cm ^{(4)}h_{ab}$$
where $^{(4)}h_{ab}$ denotes here the induced metric on $\Sigma_t$
by $^{(4)} g_{ab}$ so that our physical metric on the universe
\emph{B}(\emph{O},\emph{t}), previously denoted as $g_t$,
identifies with the metric $^{(3)}h_{ab}$ obtained by restricting
$^{(4)}h_{ab}$
on $\Sigma_t$.\\

Now, we are ready to obtain the adequate Lagrangian and
Hamiltonian formulations which will be well adapted to the new
context within which we shall present the general relativity
theory that globally describes the general laws of our dynamical,
permanently evolving universe. Naturally, these laws correspond to
an idealization of the universe by the intermediate of the metric
$g_t$ regularization on \emph{B}(\emph{O},\emph{t}). Indeed, the
real physical metric is far from being of class $C^2$ because of
singularities that are essentially reduced to black holes.\\
In the light of these adaptations, our two equations (24) and (23)
become
$$\hskip 3cm ^{(4)} R_{ab} - \frac{1}{2} \hskip 0.2cm ^{(4)}R \hskip 0.1cm
^{(4)}g_{ab} = \hskip 0.1cm ^{(4)} T_{ab}^{*} \hskip 5cm (E)$$and
$$\hskip 3cm ^{(3)} R_{ab} - \frac{1}{2} \hskip 0.2cm ^{(3)}R
\hskip 0.1cm ^{(3)}h_{ab} = \hskip 0.1cm ^{(3)} T_{ab}^{*} \hskip 5cm
(E^*)$$with
$$^{(3)} T_{ab}^{*} = 0 \Leftrightarrow \hskip 0.1cm ^{(3)} h_{ab}
= \hskip 0.1cm ^{(3)} g_e \Leftrightarrow \hskip 0.1cm ^{(4)}
g_{ab} =
\eta_{ab}$$where $\eta_{ab}$ is the Minkowsky metric.\\
Thus \emph{B}(\emph{O},\emph{t}) is the virtual vacuum space
within which leaves our real physical universe and
\emph{C}(\emph{t}) is the virtual space within which evolves the
dynamical universe. We notice that equation (\emph{E}) (using an
orthonormal basis with respect to $h_{ab}$ together with the
vector field ($\frac{\partial}{\partial t}$)$^a$ = (1,0,0,0))
implies
$$ ^{(4)} R = - \hskip 0.2cm ^{(4)} T^{*} := - \hskip 0.2cm ^{(4)}
T^{* \hskip 0.1cm a}_{\hskip 0.2cm a}$$and equation ($E^*$)
implies
$$ ^{(3)} R = -2 \hskip 0.2cm ^{(3)} T^{*} := - 2\hskip 0.2cm ^{(3)}
T^{* \hskip 0.1cm a}_{\hskip 0.2cm a}.$$We notice also that the
volume form $^{(3)} \varepsilon_{abc}$ =: $^{(3)} \varepsilon$
associated with $^{(3)} h_{ab}$ is equal to $\sqrt{h} e_{abc}$,
where $e_{abc}$ =: $^{(3)}e$ is the canonical (Euclidean) volume
form of $\mathbb{R}^3$. In the same way, the volume form
$^{(4)}\varepsilon_{abcd}$ =: $^{(4)}\varepsilon$ associated with
$^{(4)} g_{ab}$ is equal to $\sqrt{-g} e_{abcd}$ = $\sqrt{h}
e_{abcd}$, where $e_{abcd}$ =: $^{(4)} e$ is the canonical
(Euclidean) volume form on $\mathbb{R}^4$ (here $g$ and $h$ are
respectively the determinants of the matrices associated with
$g_{ab}$ and $h_{ab}$ when expressed by using the canonical bases
of $\mathbb{R}^4$ and $\mathbb{R}^3$). We further have
$$\hskip 4cm^{(3)} \varepsilon = i_{\frac{\partial}{\partial t}} \hskip 0.2cm ^{(4)}
\varepsilon \hskip 3cm \mbox { (interior product) }$$as, in the
framework of our model, the unitary orthogonal vector field to the
hypersurfaces $\Sigma_t$ is $\overrightarrow{n}$ =
$(\frac{\partial}{\partial t}) ^ a$. Finally, we notice that the
flow of time function is nothing but the fourth coordinate \emph{t}
since, within our model, there is no a relativistic proper time
notion.\\
Now, following R.Wald, the Hilbert action associated with the
Einstein's vacuum equation (where here $g_{ab}(t)$ is the
Einstein's vacuum metric on the spacetime $M=C(t)$):
$$\hskip 2cm ^{(4)} R_{ab} - \frac{1}{2} \hskip 0.2cm ^{(4)}R\hskip 0.1cm
g_{ab} = 0 \hskip 5cm (E_0)$$is given by
$$S_G\hskip 0.1cm[ g^{ab} ] = \int_{M} {\cal L}_G  \hskip 0.2cm ^{(4)}
e$$where
$$ {\cal L}_G = \sqrt{-g} \hskip 0.1cm ^{(4)} R =
\sqrt{h}
\hskip 0.1cm ^{(4)} R$$.\\
We then have (for a one parameter family $(g_{ab})_{\lambda}$
([4], E.1.18)):
$$\frac{dS_G}{d\lambda} = \int \frac{d{\cal L}_G}{d\lambda} \hskip 0.1cm ^{(4)} e =
\int \nabla^a \upsilon _a \sqrt{-g}\hskip 0.2cm ^{(4)}e + \int
(^{(4)} R _{ab} - \frac {1}{2} \hskip 0.1cm ^{(4)} R \hskip 0.1cm
g_{ab}) \delta g^{ab} \sqrt{-g} \hskip 0.2cm ^{(4)}e $$and,
neglecting the first term of the right hand side of this equation
as being the integral, with respect to $^{(4)} \varepsilon$, of a
divergence term, we obtain (E.1.19):
$$ \frac{\delta S_G}{\delta g^{ab}} = \sqrt{-g}(\hskip 0.1cm
^{(4)} R_{ab} - \frac{1}{2} \hskip 0.1cm ^{(4)} R \hskip 0.1cm
g_{ab}),$$ where $\delta
g_{ab}=\frac{d(g_{ab})_\lambda}{d\lambda}|_{\lambda=0}$ and
$\upsilon_a=\nabla^b(\delta g_{ab})-g^{cd}\nabla_a(\delta
g_{cd}).$ We then have
$$ \frac{\delta S_G}{\delta g^{ab}} =0\quad\Longleftrightarrow\quad (E_0).$$
 But, when we take into account the contribution of the
boundary term of this equation, the real action must be modified
in order to become (E.1.24):
$$ S^{'}_G \hskip 0.1cm = \hskip 0.1cm S_G + 2 \int_{\dot{U}}
K.$$Here, $\dot{U}$ is the boundary of the part of the space-time
semi-cone \emph{C} located between two Cauchy hypersurfaces
$\Sigma_{t_1}$, $\Sigma_{t_2}$ and the lateral boundary. On the
last part, the curvature vanishes and the action $S^{'}_G$ becomes
$$S^{'}_G = S_G + 2 \int_{\Sigma_{t_2}} K \hskip 0.1cm ^{(3)}
\varepsilon - 2 \int _{\Sigma_{t_1}} K \hskip 0.1cm ^{(3)}
\varepsilon.$$We recall that (E.1.39) here \emph{K} is
$$K = K^a_{\hskip 0.1cm a} = h^a_{\hskip 0.1cm b} \nabla_a
(\frac{\partial}{\partial t})^b$$where
$$K_{ab} = \nabla_a (\frac{\partial}{\partial t})_b = B_{ba}$$
$$\hskip 3cm = \frac{1}{2} {\cal L}_{\frac{\partial}{\partial t}} \hskip 0.1cm g_{ab}
= \frac{1}{2} {\cal L}_{\frac{\partial}{\partial t}} \hskip 0.1cm
h_{ab} = \frac{1}{2} {\dot{h}}_{ab}$$is the extrinsic curvature
tensor of the hypersurface $\Sigma_t$ (c.f. E.2.30). This implies
that the scalar extrinsic curvature of $\Sigma_t$ is
$$ \hskip 3cm K = \frac{1}{2} \dot{h} = \frac{1}{2} \dot{g} \hskip 1.5cm (K(t)
= \frac{1}{2}\dot{h}(t) = \frac{1}{2}\dot{g}(t))$$where $\dot{h}$
and $\dot{g}$ are the common trace of $^{(4)} \dot{g}_{ab}$ =
$^{(4)} \dot{h}_{ab}$ or $^{(3)} \dot{h}_{ab}$.\\

\subsection*{Hamiltonian formulation}

We now consider the Hamiltonian formulation associated with the
equation ($E_0$). Using the notations of [4], we notice that, in
the context of our model, we have
$$ \hskip 3cm N = 1 \hskip 3cm \mbox{ (the lapse function
is 1) } $$and
$$ \hskip 3cm N^a = 0 \hskip 3cm \mbox{ (there is no shift
vector).}$$We also have ((E.2.26), (E.2.27), (E.2.28) and
(E.2.29))
$$ R = 2 \left(G_{ab} \left(\frac{\partial}{\partial t}\right)^a \left(\frac{\partial}{\partial
t}\right)^b - R_{ab} \left(\frac{\partial}{\partial t}\right)^a
\left(\frac{\partial}{\partial t}\right)^b\right),$$
$$ G_{ab} \left(\frac{\partial}{\partial t}\right)^a \left(\frac{\partial}{\partial
t}\right)^b = \frac{1}{2} \left({}^{(3)} R - K_{ab} K^{ab} + K^2\right),$$
$$ R _{ab} \left(\frac{\partial}{\partial t}\right)^a
\hskip 0.1cm \left(\frac{\partial}{\partial t}\right)^b = K^2 - K_{ac}\hskip
0.1cm K^{ac}-\nabla_a\left(\left(\frac{\partial}{\partial t}\right)^a\nabla_c\left(\frac{\partial}{\partial t}\right)^c\right)+\nabla_c\left(\left(\frac{\partial}{\partial t}\right)^a\nabla_a\left(\frac{\partial}{\partial t}\right)^c\right)$$
and
$${\cal L}_G = \sqrt{h} (\hskip 0.1cm ^{(3)} R + K_{ab}\hskip 0.1cm
K^{ab} - K^2)$$and always
$$K_{ab} = \frac{1}{2} \dot{h}_{ab}$$
By defining the canonically conjugate momentum to $h_{ab}$ by
((E.2.31))
$$ \Pi^{ab} = \frac{\partial {\cal L}_G}{\partial \dot{h}_{ab}} =
\sqrt{h} ( K^{ab} - K \hskip 0.1cm h^{ab})$$and the configuration
space as the set of all asymptotically flat Riemannian metrics on
$\Sigma_t$, we define the Hamiltonian density, that is associated
to the gravitational action $S_G$ (the difference of the two other
terms in the $S'_G$ expression may be neglected for $t_2$ near to
$t_1$), by ((E.2.32))
$${\cal H}_G = \Pi^{ab} \hskip 0.1cm \dot{h}_{ab} - {\cal L}_G = \sqrt{h} ( - \hskip 0.1cm ^{(3)} R + \frac{1}{h}
\Pi^{ab} \hskip 0.1cm \Pi_{ab} - \frac{1}{2h} \Pi^2)$$where
$$\Pi = \Pi^a_a.$$
The Hamiltonian \emph{H} is then the function defined, for each
$\Sigma_t$, by
$$ H (g_{ab}, \Pi^{ab}) = \int_{\Sigma_t} {\cal H}_G \hskip 0.2cm
^{(3)} e.$$The Hamiltonian formulation resulting from variations
of $h_{ab}$ that satisfy $\delta h_{ab}$ = 0 on the $\Sigma_t$
hypersurfaces is equivalent to ($E_0$). The metrics $h_{ab}$ being
asymptotically flat ($h_{ab} \cong g_e$ on a neighborhood of
$S(O,t)$ for $t \gg$ 1), the solutions of ($E_0$) are the
solutions of the following constraint free Hamiltonian system:
$$\dot{h}_{ab} = \frac{\delta H_G}{\delta \Pi^{ab}} =
\frac{2}{\sqrt{h}} (\Pi_{ab} - \frac{1}{2} \Pi \hskip 0.1cm
h_{ab})$$
$$\dot{\Pi}^{ab} \hskip 0.2cm = \hskip 0.2cm -\frac{\delta H_G}{\delta h_{ab}} = - \sqrt{h}
(\hskip 0.1cm ^{(3)} R^{ab} - \frac{1}{2} \hskip 0.1cm ^{(3)} R
\hskip 0.1cm h^{ab})$$
$$\hskip 3cm + \frac{1}{2} \frac{1}{\sqrt{h}} h^{ab} (\Pi_{cd} \hskip 0.1cm \Pi^{cd} - \frac{1}{2}\Pi^2)$$
$$\hskip 3cm - \frac{2}{\sqrt{h}} (\Pi^{ac} \hskip 0.1cm \Pi_c^b -
\frac{1}{2} \Pi \hskip 0.1cm \Pi^{ab}).$$This system reduces,
within our framework, to twelve equations of twelve independent
unknowns. A solution $h_{ab}$ of this system is nothing but our
original metric $g_t$ on \emph{B}(\emph{O},\emph{t}) in the
(virtual) particular case when the tensor $^{(3)} G^{*}_{ab} =
\hskip 0.1cm
^{(3)} T_{ab}^{*}$ describes only the gravity effect.\\

\subsection*{Remarks:}
1$^{\circ}$) Both constraints (E.2.33) and (E.2.34), i.e. 
$${}^{(3)}R+h^{-1}\Pi^{ab}\Pi_{ab}-\frac{1}{2}h^{-1}\Pi^2=0$$ 
and 
$$D_a(h^{-\frac{1}{2}}\Pi^{ab})=0,$$
originate from the variations of $H_G$ with respect to $N$ and $N_a$. However, we
can get rid of constraint (E.2.34) by using the notion of
wheeler's superspace. Within the framework of our model, both
constraints have no existence, as in our model there exists only one canonical time's ``slicing'' of the space-time. Actually, they constitute the
heritage of special relativity into the standard general
relativity theory. Besides, even for this theory, $N$ and $N_a$ do
not constitute genuine dynamical variables.\\
Actually, constraints (E.2.33) and (E.2.34) are equivalent to
equations (10.2.28) and (10.2.30) of [4], i.e.
$$ D_b K^b_a - D_a K ^b_b = 0$$
and
$$ ^{(3)} R + (K_a^a)^2 - K_{ab} K ^{ab} = 0$$
called general relativity's initial values constraints.\\
These same constraints can also be expressed by means of the
second fundamental form (equally denoted in [5] by $K_{ij}$) of
the spacelike hypersurface $S$ of the Lorentzian manifold $M$
which has a vanishing Ricci curvature, in the following manner
(c.f. (9.7) and (9.8) of [5])
$$R_S - K_{jk} K^{jk} + K_j^j K^k_k = 0$$
and
$$ K^k_{j;k} - 3H_{;j} = 0.$$
We notice that the relation (10.3) of [5], i.e. $\dot{g}_{jk}$ =
-2$\lambda K_{jk}$, where $\lambda$ is the coefficient of $-dt^2$
in the Lorentzian metric (10.1) of [5], which reduces, in the
framework of our model to 1, shows that (within our model) the
extrinsic curvature of $S$ inside $M$ is the same as the second
fundamental form of $S$.\\\\
Let now $\displaystyle M_\infty=\lim_{t\rightarrow+\infty}C(t)=C\subset\mathbb{R}^4$. Since $\displaystyle\lim_{t\rightarrow+\infty}\rho(t)=0$, the non negative ``Lorentzian'' metric of $M_\infty$ becomes $h=dt^2-g_e=-\eta$ which actually is a Riemannian metric on the interior of $M_\infty$. If $\Sigma_t\simeq B(O,t)$ is provided with the Riemannian metric $g_t$, the extrinsic curvature tensor of $\Sigma_t$ into $M_\infty$ is given (c.f. [8]) by the Gauss theorem
$$R(x,y,u,v)=\ell(x,u) \ell(y,v)-\ell(x,v)\ell(y,u)$$ 
since here $\tilde{R}(x,y,u,v)=0$, and the scalar curvature is given by
$${}^{(3)}R_{\Sigma_t}=\sum_{i,j=1}^{3}R(e_i,e_j,e_i,e_j)=\sum_{i,j=1}^{3}(\ell(e_i,e_i)\ell(e_j,e_j)-\ell(e_i,e_j)^2)$$
where $e_i$, $i=1,2,3$ is an orthonormal basis of $T_m\Sigma_t$ and $\ell$ is the second fundamental form of $\Sigma_t$. This relation is nothing but the above first presumed constraint 
$$R_s=K_{jk}K^{jk}-K_j^j K^k_k;$$
the sign difference being the result of the signature difference of our metric $h_t=dt^2-g_t$.\\
Similarly, the Gauss-Codazzi equation ([8], 5.8.e)) becomes within the framework of our model 
$$\sum_{i,j=1}^{3}\tilde{R}(e_i,e_j,e_i,\frac{\partial}{\partial t})=\sum_{i,j=1}^{3}(D\ell(e_j,e_i,e_i)-D\ell(e_i,e_j,e_i))$$
that is
\begin{eqnarray*}
0&=&\sum_{i,j=1}^{3}(D_{e_j}\ell(e_i,e_i)-D_{e_i}\ell(e_j,e_i))\\
&=&\sum_{i,j=1}^{3}(D_{e_j}K_{ii}-D_{e_i}K_{ji})\\
&=&\frac{1}{2}\sum_{i,j=1}^{3}(D_{e_j}\dot{g}_{ii}-D_{e_i}\dot{g}_{ji})
\end{eqnarray*}
which is always satisfied.\\
If now the metric is represented by the matrix 
$$\left(\begin{array}{lll}
g_{11}&g_{12}&g_{13}\\
g_{21}&g_{22}&g_{23}\\
g_{31}&g_{32}&g_{33}\\
\end{array}
\right)
$$
then we have
$$K_{ij}=\frac{1}{2}\left(\begin{array}{lll}
\dot{g}_{11}&\dot{g}_{12}&\dot{g}_{13}\\
\dot{g}_{21}&\dot{g}_{22}&\dot{g}_{23}\\
\dot{g}_{31}&\dot{g}_{32}&\dot{g}_{33}\\
\end{array}
\right)
$$
and
\begin{eqnarray*}
{}^{(3)}R&=&K_{jk}K^{jk}-K_j^jK_k^k\\
&=&\frac{1}{4}(\dot{g}^2_{11}+\dot{g}^2_{22}+\dot{g}^2_{33})+\frac{1}{2}(\dot{g}^2_{12}+\dot{g}^2_{13}+\dot{g}^2_{23})-\frac{1}{4}(\dot{g}_{11}+\dot{g}_{22}+\dot{g}_{33})^2\\
&=&\frac{1}{2}(\dot{g}^2_{12}+\dot{g}^2_{13}+\dot{g}^2_{23}-\dot{g}_{11}\dot{g}_{22}-\dot{g}_{11}\dot{g}_{33}-\dot{g}_{22}\dot{g}_{33}).
\end{eqnarray*}

2$^{\circ}$) In order to include all other effects of the
matter-energy distribution, filling the universe, within a
realistic Lagrangian formulation, we have to consider a Lagrangian
density ${\cal L}$ given by
$$ {\cal L} = {\cal L}_G + {\cal L}_M.$$
This density determines an action
$$S = S_G + S_M$$
whose extremization is equivalent to the resolution of (\emph{E}).
In the particular case of a coupled gravitational field and
Klein-Gordon scalar field, then ${\cal L}$, ${\cal L}_G$, ${\cal
L}_{KG}$, $T_{ab}^{KG}$ and $S_{KG}$ are explicitly given and
related by the relations (E.1.22) and (E.1.24)-(E.1.26) of [4].
For the coupled Einstein-Maxwell equation we can refer to the
relations (E.1.23) - (E.1.26) of [4].\\
In a more general context, when we consider the equations that
govern the interaction of an electromagnetic field with a charged
dust matter modelized by a continuous charged matter within a
gravitational field, we are led to consider the Lagrangian
$$ L = L_1+L_2+L_3$$
with ($S$ being the scalar curvature)
\begin{eqnarray*}
 &L_1& = - \frac{1}{8 \pi} < \cal{F}, \cal{F} > + < A, J >\\
 &L_2& = \frac{1}{2} \mu < u,u >\\
 &L_3& = \alpha S
\end{eqnarray*}
where we have used the notations of [5] together with the
conditions (1.6), (1.7) and (1.10) of [5]. Variations with respect
to the metric $g_{jk}$ give
$$\hskip 3cm \delta \int (L_1+L_2) dV = \frac{1}{2} \int T^{jk} (\delta
g_{jk}) dV \hskip 3cm (1.15)$$ with
$$ \hskip 2cm T^{jk} = \mu u^j u^k + \frac{1}{4 \pi} ( {\cal{F}} ^j_l
{\cal{F}} ^{kl} - \frac{1}{4} g^{jk} {\cal{F}} ^{il} {\cal{F}}
_{il}) \hskip 3cm (1.24)$$ and
$$ \hskip 2cm \delta \int S dV = - \int G^{jk} (\delta g_{jk})dV \hskip 5cm
(1.26)$$ which gives the Standard Einstein's tensor equation in
the considered case.\\\\
In other respects, our physical metric $g_t$ on the universe
$B(O,t)$ and our dynamical metric $h = dt^2 - g_t$ on $C^* = C
\setminus O$ are respectively the solutions of $(E^*)$ and $(E)$
where the tensors $^{(3)} T_{ab}^{*}$ and $^{(4)} T_{ab}^{*}$
reflect all forms of matter-energy (local mass and gravitational
field, cosmic mass, black holes and gravitational field, local
electromagnetic field, cosmic radiations, neutrinos and pressure).
All free movements into $B(O,t)$ and $C$ are geodesic for $g_t$
and $h$ respectively.\\
Lagrangians and tensors that are considered in classical general
relativity constitute (in particular cases) good approximations
for
the Lagrangians and tensors $^{(4)} T^{*}_{ab}$ of our model.\\\\
To sum up, we can state that the idealized physical universe
is equivalently characterized by:\\\\
1. The matter-energy distribution $E_t$(\emph{X}) for all couples
(\emph{t}, \emph{X}) such that \emph{t} $>$ 0 and \emph{X} $\in$
\emph{B}(\emph{O},\emph{t}).\\\\
2. The tensor $^{(3)} T_{ab}^{*}$ defined on
\emph{B}(\emph{O},\emph{t})
for \emph{t} $>$ 0.\\\\
3. The Riemannian metric $g_t$ defined, for every \emph{t} $>$ 0,
on \emph{B}(\emph{O}, \emph{t}).\\\\
4. The distribution \emph{E}(\emph{t}, \emph{X}) = $E_t$(\emph{X})
for \emph{t} $>$ 0 and \emph{X} $\in$ $\Sigma_t$.\\\\
5. The metric $^{(4)} g_{ab}$ defined on \emph{C} by
$$ ^{(4)} g_{ab} = - dt^2 + \hskip 0.1cm ^{(4)} h_{ab}$$where
$$^{(4)}h_{ab} ( \frac{\partial}{\partial t},
\frac{\partial}{\partial t}) = 0 \hskip 1cm , \hskip 1cm  ^{(4)}
h_{ab} (\frac{\partial}{\partial t}, \frac{\partial}{\partial
x_i}) = 0 \mbox{ for } i = 1,2,3 $$and$$^{(4)} h_{ab} = g_t
\mbox{ on } \Sigma_t = \emph{B}(O,\emph{t}) \times \{ t \}$$.\\\\
6. The generalized Einstein's mass-energy tensor $^{(4)}
T^{*}_{ab}$ satisfying the (modified) Einstein's equation:
$$^{(4)} R_{ab} - \frac{1}{2} \hskip 0.1cm ^{(4)} R \hskip 0.1cm
^{(4)}g_{ab} =  ^{(4)} T^{*}_{ab}.$$\\\\
7. The Lagrangian density
$${\cal L} = {\cal L}_G + {\cal L}_M$$where$${\cal L}_G = \sqrt{-g} \hskip 0.1cm ^{(4)}R$$and
${\cal L}_M$ is the Lagrangian density corresponding to all fields
apart from the gravitational one, which in that way guarantees
that the extremization of the action
$$S^{'}_G = S_G + 2 \int_{\Sigma_t} K - 2
\int_{\Sigma_{t_0}}K$$with respect to variations satisfying
$\delta g_{ab}$ = 0 or $\delta h_{ab}$ = 0 on both $\Sigma_t$ and
$\Sigma_{t_0}$ gives the solutions of (\emph{E}) and ($E^*$).\\\\
8. The Hamiltonian density ${\cal H}$ and the Hamiltonian \emph{H}
= $\int_{\Sigma_t} {\cal H} $ defined by means of the Lagrangian
${\cal L}$ by
$${\cal H} = \Pi^{ab} \hskip 0.1cm h_{ab} - {\cal L}$$where$$\Pi^{ab} =
\frac{\partial {\cal L}}{\partial \dot{h}_{ab}}$$
\bigskip
$h_{ab}$ is then the solution of the constraint free Hamiltonian
system
$$\dot{h}_{ab} = \frac{\delta H}{\delta \Pi^{ab}}$$
$$\dot{\Pi}^{ab} = - \frac{\delta H}{\delta h_{ab}}.$$

\subsection*{General features of the solution}

Consequently, our modeling leads to a determinist solution of the
space-time Einstein's equation. Our space-time will be, for every
time \emph{t}, the Riemannian manifold (\emph{C}(\emph{t}),
$^{(4)} g_{ab}$) realized as the Cauchy maximal development
associated to the initial conditions
$$^{(3)} h_{ab} (t_0) \hskip 1cm \mbox{ and } \hskip 1cm
\frac{1}{2} \hskip0.1cm ^{(3)} \dot{h}_{ab}(t_0)$$defined on an
arbitrary Cauchy surface $\Sigma_{t_0}$ of \emph{C}(\emph{t}) for
$t_0 < t$ in such a manner that the Riemannian metric
$^{(3)}h_{ab}(t_0)$ on $\Sigma_{t_0}$ identifies to the physical
metric $g_{t_0}$ on \emph{B}(\emph{O},$t_0$) of the previous
sections and $\frac{1}{2} \hskip 0.1cm ^{(3)} \dot{h}_{ab} (t_0)$
= $\frac{1}{2}$ ($\dot{g_{t_0}}$)$_{ab}$ constitutes the extrinsic
curvature tensor of $\Sigma_{t_0}$ into the space-time
(\emph{C}(\emph{t}), $^{(4)} g_{ab}$). This solution could be a
good approximation of the real metric on a time interval as large
as our approximation and regularization of the initial conditions
$^{(3)} h_{ab}$($t_0$) and $\frac{1}{2} \hskip 0.1cm ^{(3)}
\dot{h}_{ab}$($t_0$) on $\Sigma_{t_0}$ are close to the physical
reality of our universe at time $t_0$. A permanent readjustement
of initial conditions based on the enlargement and refinement of
available data is necessary.\\
Our model contains the right degree of liberty number and does not
undergo the gauge freedom notion. Indeed, the physical nature of
the universe forces it, in virtue of the constancy of the
propagation speed of electromagnetic waves and its isotropic
character, to exist, at every time \emph{t}, under the form of a
ball having a Euclidean radius \emph{t} and to evolve within the
space-time semi-cone \emph{C}. This forces any gauge
diffeomorphism $\psi$ to transform \emph{C}(\emph{t}) into a
semi-cone \emph{C}($t^{'}$) and to be of the form $\psi$ =
(\emph{t}, $\varphi_{t}$) where $\varphi_{t}$ is a diffeomorphism
of \emph{B}(\emph{O},\emph{t}) on \emph{B}(\emph{O},$t^{'}$)
satisfying $g_t$ = $\varphi^{*}_{t} g_{t^{'}}$. This simply
reduces to a purely conventional rescaling unless $\psi$ being an
isometric transformation of \emph{C} in $\mathbb{R}^4$ and
$\varphi$ being an isometric transformation of
\emph{B}(\emph{O},\emph{t}) in $\mathbb{R}^3$. In that case $\psi$
constitutes a trivial gauge diffeomorphism:
$$\psi^{*} g_{ab} = g_{ab} \hskip 1cm \mbox{ and } \hskip 1cm
\varphi^{*}_{t} g_t = g_t$$
\subsection*{Remarks}
\noindent According to our model, we can state the following properties:\\\\
1. The real physical metric of the universe can not be globally
determined by a linearization process: the real perturbations of
the Minkowsky metric on \emph{C} and the Euclidean metric on
\emph{B}(\emph{O},\emph{t}), due to the matter-energy effects, are
far from being "small", especially around
black holes.\\\\
2. Our universe is evidently not homogeneous nor isotropic. There
indeed exists a foliation family ($\Sigma_t$)$_{t>0}$ of the
space-time, but for arbitrary \emph{p},\emph{q} $\in \Sigma_t$,
there can not exist an isometry of $\Sigma_t$ which transforms
\emph{p} into \emph{q}. Furthermore, there does not necessarily
exist an isometry of \emph{C} that leaves $p \in \Sigma_t$ fixed
and transforms a unitary spatial vector at \emph{p} into another
vector having the same properties. Consequently, our model is
totally different from the \emph{K} = $\mp$ 1 cases of the
Robertson-Walker's model even though it is, in case of extreme
idealization, very similar to the case \emph{K} = 0 of this
model.\\Indeed, the corresponding metric to this latter case
reduces, within the framework of our model, to
$$^{(4)}g = -dt^2 + a^2(t)(dx^2 + dy^2 + dz^2)$$where $dx^2 + dy^2 +
dz^2$ is the restriction to the ball
\emph{B}(\emph{O},\emph{R}(\emph{t})) of the euclidean metric on
$\mathbb{R}^3$.\\
If $g_t=a^2(t)g_e$ and $h_t=dt^2-g_t$ then 
$$\dot{g}_t=\left(\begin{array}{lll}
2a\dot{a}&\;\;0&\;\;0\\
\;\;0&2a\dot{a}&\;\;0\\
\;\;0&\;\;0&2a\dot{a}\\
\end{array}
\right)
$$
and
$${}^{(3)}R=\frac{1}{2}(4a^2\dot{a}^2+4a^2\dot{a}^2+4a^2\dot{a}^2)=6a^2\dot{a}^2.$$
Moreover, the evolution equations for the homogeneous isotropic
Cosmology are written down (c.f. (5.2.14) and (5.2.15) of [4]) as
$$ 3 \frac{\dot{a}^2}{a^2} = 8 \pi \rho - \frac{3K}{a^2}$$and
$$ 3 \frac{\ddot{a}}{a} = -4 \pi (\rho + 3P),$$
where $K$ is the curvature parameter (which is, contrary to our
model, a constant), $\rho$ is the average matter density of the
universe and $P$ is the average pressure associated with the
massless thermic radiations filling the universe, which together
constitute the stress-energy tensor of Einstein. This implies
(5.2.18)
$$\dot{\rho} + 3 (\rho + P) \frac{\dot{a}}{a} = 0$$which,for standard dust
model (\emph{P} = 0) leads to
$$\rho_m a^3 = cte$$
(where $\rho_m$ is the matter mean density for this model) and,
for standard radiations model (\emph{P} = $\frac{\rho}{3}$),leads
to
$$\rho_r a^4 = cte,$$(where $\rho_r$ is the radiation mean density
for this last model).\\
Thus, we have, for the standard dust model (5.2.21):
$${\dot{a}}^2 - \frac{C}{a} = 0 \;\;\;\;\;\; \mbox{ with } \;\;\;\;C =
\frac{8}{3} \pi \rho a^3$$and, for the standard radiation model,
we have (5.2.22):
$${\dot{a}}^2 - \frac{C^{'}}{a} = 0 \;\;\;\;\; \mbox{ with } \;\;\;\;
C^{'} = \frac{8}{3} \pi \rho a^4.$$In the first case, which
corresponds to our present universe, we have (Table 5.1)
$$a(t) = (\frac{9C}{4})^{\frac{1}{3}} \hskip 0.2cm t^{\frac{2}{3}}$$
and in the second one, which corresponds to the early universe
just after the (hot) Big Bang, we have ((5.4.1) and (5.4.2)):
$$a(t) = (4C^{'})^{\frac{1}{4}} \hskip 0.2cm t^{\frac{1}{2}}$$
and
$$\rho_r(t) = \frac{3}{32 \pi G t^2}.$$
Then, using quantum statistics, we find that, for infinitely small
\emph{t}, the temperature \emph{T} of the universe is proportional
to $\rho_r^{\frac{1}{4}}$ and to $\frac{1}{a}.$\\\\
Besides, we have, according to our model (for the very early
universe), after a crude simplification within the context of the
homogeneous isotropic Cosmology:
$$g_t = a^2(t) (dx^2 + dy^2 + dz^2)$$on
\emph{B}(\emph{O},\emph{R}(\emph{t})), which implies
$$\int _{B(O,R(t))} dv_{g_t} = \int_{B(O,R(t))} a^3(t)dv_{g_e} =
\int_{B(O,R(t))} dX - E = \frac{4 \pi R^3(t)}{3} - E$$where
\emph{E} is the total energy of the universe. This allows us to
calculate, for instance, $C^{'}$ as a function of \emph{E} and
\emph{R}(\emph{t}) for \emph{t} $\ll$ 1 and $C$ as a function of
\emph{E} and
\emph{t} when assuming $R(t) \simeq t$ for $t\gg1$.\\
Moreover, we also could have some information on the Hubble's
factor (i.e. the time-dependent Hubble's constant) $H(t) =
\frac{\dot{a}}{a}$.\\\\
3. The universe is not stationary. The translation of time vector
field ($\frac{\partial}{\partial t}$)$^{a}$ is not a killing
vector field, although it can be considered approximately, as
being a killing field in all regions of the space-time that
correspond to the regions of $\Sigma_t \simeq B(O,t)$ that can be
considered as durably deprived of matter and its effects (i.e.
where $h_{ab}=g_t \simeq g_e$). The universe is not static too, even
though the family $\Sigma_t$ is orthogonal to
($\frac{\partial}{\partial t}$)$^{a}$ and we have
$$g_{ab} = -dt^2 + h_{ab}.$$
\bigskip
4. Our universe is not spherically symmetric nor axisymmetric.
Hence the Schwarzschild's solution can not be other than an
idealization of the universe reducing it to a gravitational field
resulting from a material spherical and static core.\\\\
5. \textbf{(Cosmological constant)} When we write down the Einstein's tensor equation with a
non vanishing cosmological constant, we get:
$$ R_{ab} - \frac{1}{2} R g_{ab} = 8 \pi T_{ab} - \Lambda g_{ab}.$$
The Einstein's vacuum equation becomes
$$ R_{ab} -\frac{1}{2} R g_{ab} = - \Lambda g_{ab}$$which (by
contracting) gives
$$ R = 4 \Lambda.$$
Then the curvature of the Einstein's vacuum, characterized by
$\rho=0$ and $T_{ab} =0$, is non vanishing. In our setting, the identity $T^*_{ab}=0$ implies $T_{ab}=0$ but $T_{ab}=0$ does not imply
$T^*_{ab}=0$. This comparison shows that our condition $T^*_{ab}=0$ (and then $^{(3)}T^*_{ab}=0$ and $g_t=g_e$) is very restrictive and, in fact, idealistic. It corresponds to an absolute
vacuum region whose existence is highly improbable.\\
The $\Lambda$ "constant" appears into our cosmological setting as
being the result of the influence of the cosmic matter and cosmic
radiations and gravity on regions that are deprived of matter and
are not lying within a direct gravitational and electromagnetic
field i.e. the regions of $B(O,t)$ characterized by
$^{(3)}T^*_{ab}=\Lambda g_e$. Our field tensorial equation then is
$$ ^{(3)} R_{ab} - \frac{1}{2} \; ^{(3)}R g_t = \Lambda g_e$$ where
$\Lambda$ probably depends on the considered region and on
time.\\
Consequently, for our metric $g_t$ (as well as for the metric $h =
dt^2 - g_t$), we hardly can have $R_{ab} - \frac{1}{2} R g_t = 0$
in any (intergalactic) region of the universe. Therefore, the
scalar curvature $R$ and the Ricci tensor $R_{ab}$ can not
rigorously vanish in any such a region although $\Lambda$ is necessarily extremely small there.\\\\

Nevertheless, a large number of results obtained from such
 hypothesis, reductions and idealizations still continue to be
 qualitatively and quantitatively (more or less) valid. This is
 particularly true for the results obtained by means of what is qualified as
 Newtonian limit, homogeneous isotropic cosmology or, simply,
 spatially homogeneous cosmology and their consequences on some
 features related to the universe evolution and to its causal
 structure. It is the same for some consequences of the Kruskal
 extension of the Schwarzschild solution concerning stationary
 black holes in the vacuum and charged Kerr black holes associated
 to Einstein-Maxwell equation as well as for thermodynamics-like
 properties of black holes.\\

Naturally, there is always a gigantic work to be done about these
important problems. We can talk also about the necessary
modifications to be made for adapting the Einstein-de
Sitter-Friedmann cosmological model to our framework. All this
demands avoiding some erroneous and unjustified postulates and
principles and adapting everything to the dynamical expansion
reality of our universe which is, even though gigantic, perpetually finite.\\
We can add that our dynamical model makes possible the
construction of a quantum theory of the general relativity (using the canonical quantization method) after
getting rid of useless constraints $((E.2.33)$ and $(E.2.34)$ of [4]) previously imposed on the
Hamiltonian system specifying the dynamical evolution of the
universe as well as of the constraints imposed to the initial
values leading, in this way, to a well posed initial values
formulation for
general relativity theory.\\\\

\section{Introduction to a reviewed cosmology}
\subsection*{ The Friedmann equation}

In this section, we will adapt the (Einstein - Friedmann - Hubble
- de Sitter) homogeneous isotropic Cosmology to our model in order
to specify some approximate results, that have been established
half theoretically and half experimentally but not rigorously.\\
We start by recalling that the universe is assimilated, at any
time \emph{t} $\gg$ 0, to the Riemannian space \emph{U}(\emph{t})
= (\emph{B}(\emph{O},\emph{ct}),$g_t$) where $g_t$ is a position
and time-dependent metric. We then adopt the global macroscopic
cosmological model within which the present universe reduces to a
dust of galaxies that is distributed in a homogeneous and
isotropic manner into the ball \emph{B}(\emph{O},\emph{ct})
(although this is not rigorously exact). So, we will follow the
Hubble - Friedmann's work (c.f. [2]) by putting
$$ r = r_0 R(t)$$where \emph{R}(\emph{t}) denotes here the
expansion parameter.\\\\This equation becomes, within our
framework,
$$c t = c t_0 R(t)$$where \emph{c} is the speed of light in the
vacuum. We then have
$$R(t) = \frac{t}{t_0}$$
$$\frac{dR}{dt} = \frac{1}{t_0}$$and
$$H := \frac{\frac{dR}{dt}}{R} = \frac{1}{t}.$$
This result is perfectly conforms with our readjustment of the macroscopic homogeneous and isotropic cosmology to our model which gives, on one hand, that the radius of the universe (for $t\gg0$) is equivalent to $ct$ and, on the other hand, that the metric of the universe at time $t$ is given by
$$g_t=a^2(t)(dx^2+dy^2+dz^2).$$
Actually, the computation of the Euclidean volume of the universe $B(O,ct)$ implies $a=ct$ an $H=\dfrac{\dot{a}}{a}=\dfrac{1}{t}$.\\
This leads to the Friedmann's equation
$$\hskip 4.5cm {(\frac{dR}{dt})}^2 = \frac{ 8 \pi G \rho R^2}{3 c^2} - K(t)
\;\;\;\;\;\;\;\;\;\;\;\;\;\;\;\; \hskip 1.8cm (25)$$where here
\emph{K}(\emph{t}) denotes the time-dependent curvature parameter
associated with the space curvature which is originated in the
matter distribution throughout the universe and reflected by the
Riemannian metric $g_t$ and $\rho=\rho(t)$ is the mean density of the matter-energy associated with the matter-energy distribution $E_t(X)$ at time $t\gg0$ in the universe i.e.
$$\rho(t)=\int_{B(I,r_0)}E_t(X)\, dX$$
where $B(I,r_0)$ is a mean ball of $B(O,t)$ with Euclidean volume equal to the unit of volume. This process is made possible thanks to
our homogeneity and isotropy hypothesis which is valid when we are
dealing with general macroscopic results.\\\\
\textbf{Remark:} In our model, the curvature parameter that
appears in the Friedmann equation ([2], 19.58)
$$ (\frac{dR}{dt})^2 = \frac{8\pi G \rho R^2}{3 c^2} - K$$depends
on time (as well as does the gravitational constant $G$). The fact
of considering $K$ as being an absolute constant (i.e. independent
of time) as it is generally accepted, leads to flagrant
contradictions. Actually, in order to establish this equation,
Friedmann has taken a ball of the universe of radius $r=:r_0 R(t)$
having the mass $M$ and a galaxy that is located on the
corresponding sphere having the mass $m$ and established the
relation ([2],19.56) that specifies the total energy of the galaxy:
$$ E = -\frac{K m r_0^2}{2}.$$
Now, the constant $K$ as introduced here necessarily depends on
the arbitrary chosen instant $t_0$ (i.e. $K = K(t_0)$). Otherwise,
if we choose another privileged instant $t_1 \neq t_0$ to which
corresponds another radius $r_1 \neq r_0$, we obtain by proceeding
similarly the relation
$$ E = -\frac{K m r_1^2}{2}$$which, in case of taking $K$ as an
absolute constant, contradicts the energy conservation law.\\
In other respects, the fact of considering $K$ and $G$ as being
absolute constants leads to a contradiction between Friedmann
equation and the second cosmological principle which stipulates
that the relative speed of galaxies is proportional to their
relative distance. Indeed, as $\frac{dR}{dt}$ is proportional to
$\frac{dr}{dt} = v_r$ and as $\rho$ is proportional to
$\frac{1}{r^3}$ which is proportional to $\frac{1}{R^3}$, then
this equation implies that $\frac{dR}{dt}$ decreases with
increasing $R$ as $\frac{\alpha}{\sqrt{R}}$ and consequently $v_r$
is decreasing as $\frac{\beta}{\sqrt{r}}$.\\ 
Similarly, the
relations (19.66) and (19.67) of [2], namely
$$ \frac{dR}{dt} = \sqrt{\frac{8\pi G\rho_c}{3c^2}}
R^{-\frac{1}{2}}$$where $\rho_c$ is the critical density (19.65)
of [2] and
$$ R = (\frac{3}{2})^ {\frac{2}{3}} (\frac{8\pi G
\rho_c}{3c^2})^{\frac{1}{3}} t^{\frac{2}{3}}$$show clearly the
deficiency of the Einstein - de Sitter - Friedmann model. Indeed,
these relations imply
$$ \frac{dr}{dt} \propto \frac{dR}{dt} \propto
\frac{1}{R^{\frac{1}{2}}} \propto \frac{1}{t^{\frac{1}{3}}}$$which
shows that $r$ is increasing with time whereas $v_r =
\frac{dr}{dt}$ is decreasing.\\\\
The equation (25) can be written as
$$\frac{1}{t_0^2} = \frac{ 8 \pi G}{3} \frac{E}{\frac{4\pi c^3
t^3}{3}} \times \frac{R^2}{c^2} - K$$
$$\hskip 1.5cm = \frac{2 G E}{c^5 t^3}R^2 - K = \frac{2 G E}{c^5
t\hskip 0.1cm t_0^2} - K.$$So, we get
$$\frac{2 G E}{c^5 t} = 1 + K t_0^2$$
and then
$$\hskip 2cm E = \frac{c^5 t (1 + Kt_0^2)}{2 G} \equiv \frac{c^5 (t_0 + K_0
t_0^3)}{2 G_0} = \frac{c^5 C_0}{2 G_0} \hskip 3cm (26)$$where
$G_0$ is the gravitational constant calculated at the time $t_0$
and
$$ C_0 = t_0 + K_0 t_0^3 = \frac{2G_0
E}{c^5}.$$ Furthermore, we can write, within a significant time
interval about $t_0$ :
$$t (1 + K t_0^2) \cong C_0.$$
By differentiation, we obtain
$$1 + K t_0^2 + t_0^2 t K^{'} = 0$$which yields
$$(K + K^{'} t)t_0^2 = -1$$
i.e.
$$K + K^{'} t = -\frac{1}{t_0^2}$$
and then
$$(K t)^{'} = -\frac{1}{t_0^2}.$$
Therefore, we have
$$K t = -\frac{t}{t_0^2} + b$$
with
$$b = K_0 t_0 + \frac{1}{t_0}.$$
So, we get
$$K t = -\frac{t}{t_0^2} + K_0 t_0 + \frac{1}{t_0}$$
and then
$$\hskip 4cm K = -\frac{1}{t_0^2} + \frac{1+K_0 t_0^2}{t_0 t}. \hskip
5cm (27)$$
\subsection*{The density equation}

 Now, we write down the equality
$$\rho(t) = \rho_m(t) + \rho_r(t)$$
where $\rho_m$(\emph{t}) and $\rho_r$(\emph{t}) respectively
denote the mean mass energy and radiation energy
densities.\\

Besides, our model implies
$$\rho(t) = \frac{E}{vol(B(O,ct)} = \frac{E}{\frac{4}{3}\pi (ct)^3} = \frac{3 E}{4 \pi c^3 t^3}$$
and the homogeneous cosmology leads, for \emph{t} $\gg$ 1, to
$$\rho_m(t) = \frac{c^2}{6 \pi G t^2}$$
(c.f. [4], Table 5.1 with $C=\frac{8\pi\rho_ma^3}{3c^2}$ (p.101)).\\
Actually, this value can be considered as an approximate value for
$\rho_m(t)$ within the framework of our model.\\

\noindent Therefore, the above density equation gives, for a given constant $a_0$,
$$\frac{3 E}{4 \pi c^3 t^3} = \frac{c^2}{6 \pi G t^2} +
\frac{a_0}{t^3}$$ that is
$$ \frac {3 E}{4 \pi c^3} = \frac{c^2}{6 \pi G} t + a_0$$
with
$$ a_0 = -\frac{c^2}{6 \pi G} t + \frac{3 E}{4 \pi c^3}.$$
Then, for $t\gg 1,$ we have
$$
 \hspace{2cm}\rho_r (t) = \frac{a_0}{t^3}=-\frac{c^2}{6 \pi G} \frac{1}{t^2}+\frac{3E}{4\pi c^3}\frac{1}{t^3} = -
\rho_m(t)+\frac{3E}{4\pi c^3}\frac{1}{t^3} .\hspace{2.5cm}(\rho)
$$
Writing the density equation $(\rho)$ for \emph{t} = $t_0$ $\gg$
1, we obtain
$$ \frac{3 E}{4 \pi c^3 t_0^3} = \frac{c^2}{6 \pi G_0 t_0^2} +
\frac{a_0}{t_0^3}$$ and then
$$ \frac{3 c^5 (t_0 + K_0 t_0^3)}{4 \pi c^3 t_0^3 \times 2 G_0} =
\frac{c^2}{6 \pi G_0 t_0^2} + \frac{a_0}{t_0^3}$$ that is
$$\frac{3 c^2}{8 \pi G_0} (K_0 + \frac{1}{t_0^2}) = \frac{c^2}{6
\pi G_0 t_0^2} + \frac{a_0}{t_0^3}.$$ By putting
$$ K_0 t_0^3 + t_0 = \frac{4}{9} t_0 + b_0 ,$$
we obtain
$$\frac{3 c^2}{8 \pi G_0} (\frac{4}{9 t_0^2} + \frac{b_0}{t_0^3})
= \frac{c^2}{6 \pi G_0 t_0^2} + \frac{a_0}{t_0^3}$$ which yields
$$\frac{3 c^2}{8 \pi G_0} \frac{b_0}{t_0^3} = \frac{a_0}{t_0^3}$$
and
$$ b_0 = \frac{ 8 \pi G_0 a_0}{3 c^2}.$$
We then have
$$ C_0 = K_0 t_0^3 + t_0 = \frac{4}{9} t_0 + \frac{8 \pi G_0
a_0}{3 c^2}$$ and
$$ \hskip 4cm K_0 = -\frac{5}{9 t_0^2} + \frac{8 \pi G_0 a_0}{3 c^2 t_0^3} \hskip 5.4cm (28) $$
Likewise, we have (using $(26)$)
$$ \hskip 4cm \frac{2 G_0 E}{c^5} = C_0 = \frac{4}{9} t_0 + \frac{8 \pi G_0
a_0}{3 c^2} \hskip 4cm (29) $$ and then we recover the (previously
obtained) relation
$$ a_0 = (\frac{2 G_0 E}{c^5} - \frac{4}{9} t_0) \times \frac{3
c^2}{8 \pi G_0}$$
$$ \hskip 4.7cm = \frac{3 E}{4 \pi c^3} - \frac{c^2 t_0}{6 \pi G_0}
\hskip 5.5cm (30)$$

\subsection*{Energy, age and size of the universe}

Now, we take for $t_0$ the present time and we use the Stefan -
Boltzmann law and the generally accepted estimation of the
neutrinos' contribution to the present universe density for
setting
$$\rho_r (t_0) = \frac{a_0}{t_0^3} = \frac{3 E}{4 \pi c^3 t_0^3} -
\frac{c^2}{6 \pi G_0 t_0^2}$$
$$ \hskip 4cm = -0.4 \times {10}^6 eV/m^3 = - 6.4 \times
{10}^{-14} J/m^3.$$ By writing down the second Einstein -
Friedmann's equation
$$ \frac{\frac{d^2R}{dt^2}}{R} = - \frac{4 \pi G}{3 c^2} (\rho +
3P),$$ where \emph{P} is the mean pressure, we obtain (within the
framework of our model) :
$$ \rho + 3 P = 0.$$
But
$$ \rho = \rho_m + \rho_r = \rho_m + 3 P $$
and then
$$ \rho_m + \rho_r + \rho_r = 0$$
which yields
$$\rho_m = -2 \rho_r \hskip 1cm \mbox{ and } \hskip 1cm \rho =
-\rho_r,$$which conforms with the relation ($\rho$) that reduces, after substitution, to $\rho = \frac{3E}{4 \pi c^3 t^3}=-\rho_r$.\\
This clearly shows that when we study the general relativity we
have to consider the mean radiational density and the mean
pressure of the universe as being negative, which is very
normal as their effect is antigravitational.\\

\noindent Thus, we have
$$ \rho_0 = -\frac{a_0}{t_0^3} = 6.4 \times {10}^{-14} J/m^3.$$
Consequently equation (29) gives
$$ E = \frac{c^5}{2G_0} (\frac{4}{9} t_0 - \frac{8 \pi G_0
t_0^3}{3 c^2} \times 6.4 \times {10}^{-14})$$ and
$$ \frac {E}{\rho_0} = \frac{4 \pi c^3 t_0^3}{3} = \frac{c^5}{2
G_0 \times 6.4 \times {10}^{-14}} (\frac{4}{9} t_0 - \frac{8 \pi G_0
t_0^3}{3 c^2} \times 6.4 \times{10}^{-14})$$ Therefore, we obtain
$$ \frac{4 \pi c^3 t_0^3}{3} = \frac{2 c^5 t_0}{9 G_0 \times 6.4
\times {10}^{-14}} - \frac{8 \pi G_0 c^3 t_0^3}{6 G_0}$$ which
yields
$$ \frac {8 \pi}{3} t_0^2 = \frac{2 c^2 \times {10}^{14}}{9 G_0
\times 6.4}$$ and
$$t_0^2 = \frac{3}{8 \pi} \times \frac{2 c^2 \times
{10}^{14}}{9 G_0 \times 6.4}\;\;\;\;\;\;\;\;\;\; $$
$$ \hskip 4cm = \frac{3 \times 2 \times 9 \times {10}^{16} \times
{10}^{14}}{8 \pi \times 9 \times 6.67 \times {10}^{-11} \times
6.4} \simeq 5.595 \times {10}^{38}.$$ Finally, we have
$$ t_0 \simeq 2.365 \times {10}^{19} s.$$
This same result could have been obtained directly by using only
the second Friedmann's equation. Indeed this equation gives
$$ \rho_m + \rho_r + \rho_r = 0$$
which leads to
$$ \rho_m = -2 \rho_r$$
that is
$$ \rho_m = \frac{c^2}{6 \pi G_0 t_0^2} = 12.8 \times {10}^{-14}
J/m^3$$ and then
$$ t_0^2 = \frac {c^2}{6 \pi G_0 \times 12.8 \times {10}^{-14}}\;\;\;\;\;\;$$
$$ \hskip 4cm = \frac{9 \times {10}^{16} \times {10}^{14}}{6
\times 3.14 \times 6.67 \times {10}^{-11} \times 12.8} \simeq
5.595 \times {10}^{38}$$ which gives again
$$t_0 = 2.365 \times {10}^{19} s.$$
Thus, the universe radius is
$$ r_0 = c t_0 \simeq 7.1 \times {10}^{27} m.$$
The total universe energy is
$$ E = \frac{4 \pi c^3 t_0^3 \rho_0}{3} = \frac{2 c^5 t_0}{9 G_0}
- \frac{4 \pi t_0^3 c^3}{3} \times 6.4 \times {10}^{-14}$$
$$ \hskip 1.5cm = 9.57 \times {10}^{70} J
( = 19.147 \times {10}^{70} - 9.57 \times {10}^{70} )J.$$
The above value of $\rho_m$ conforms with its value deduced from the density equation $(\rho)$ which gives (by replacing $\rho_r(t_0)$ by $-\frac{\rho_m(t_0)}{2}$)
$$\rho_m(t_0)=\frac{3E}{2\pi c^3}\frac{1}{t_0^3}.$$
The Hubble's parameter is
$$ H_0 = \frac{1}{t_0} \simeq 4.228 \times {10}^{-20}.$$
The radiation energy density is presently
$$ \rho_r = - 0.4 \times {10}^6 eV/m^3.$$
The mass energy density is
$$ \rho_m = 0.8 \times {10}^6 eV/m^3.$$
The present matter mass (including the black holes) is
$$ M = \frac{\rho_m}{c^2} \times \frac{4 \pi c^3 t_0^3}{3} =
\rho_m \times \frac{4 \pi c t_0^3}{3}$$
$$  = 2.126 \times {10}^{54} Kg.\;\;\;\;\;\;\;\;\;\;$$
The equivalent mass of the total energy is
$$ M_e := \frac{E}{c^2} = 1.063 \times {10}^{54} Kg.$$


\subsection*{Observer vision field}
Let $I$ be an observer (at time $t_0$) of the universe $B(O,t_0)$ located at a distance $d$ from $O$ (fig. $14$). This observer receive at time $t_0$ all signals that are emitted at times $t_0-s$ from objects which are located on the intersection of the sphere $S(I,s)$ and the universe at time $t_0-s$, i.e. on $S(I,s)\cap B(O,t_0-s)$. Therefore, the vision field of the observer $I$ is
$$V=\bigcup_{0<s\leq\frac{d+t_0}{2}}\left(S(I,s)\cap B(O,t_0-s)\right)$$
which is
$$B\left(I,\frac{t_0-d}{2}\right)\bigcup\left(\bigcup_{\frac{t_0-d}{2}\leq s\leq\frac{t_0+d}{2}}\left(S(I,s)\cap B(O,t_0-s)\right)\right)$$
as it is indicated on the figure $14$.

\subsection*{Comparison with the Einstein - de Sitter model}

According to the standard homogeneous isotropic model (with $k$ =
0), we have (c.f.[2], p.555-557)
$$ R(t) \propto t^{\frac{2}{3}} \hskip 2cm H(t) \propto
\frac{2}{3t}$$
$$ \lambda \propto \frac{1}{T} \propto R$$and then (using the well
confirmed Stefan - Boltzmann's result)
$$ \rho_r(t) \propto T^4 \propto \frac{1}{R^4} \propto
\frac{1}{t^{\frac{8}{3}}}$$
$$ \rho_m(t) \propto T^3 \propto \frac{1}{R^3} \propto
\frac{1}{t^2}$$ 
which is contradictory to the fact that $\rho_r+\rho_m=\rho$ is proportional to $\frac{1}{t^3}$.\\
On the other hand, our model clearly shows the fundamental
property
$$ \rho \propto \frac{1}{R^3} \propto \frac{1}{t^3}$$which
conforms with our result
$$ \rho_r \propto \frac{1}{t^3}$$that gives (joined to the
Stefan - Boltzmann result)
$$T^4 \propto \frac{1}{t^3} \hskip 1cm \mbox{ or } \hskip 1cm T
\propto \frac{1}{t^{\frac{3}{4}}},$$
and, within a small interval of time,
$$ \lambda \propto t \propto \frac{1}{T^{\frac{4}{3}}}$$and finally
$$ \rho \propto \rho_r \propto \rho_m \propto T^4 \propto
\frac{1}{t^3}$$which is obviously more consistent.\\

In other respects, the Einstein's tensor $T$ is written, in the
framework of the homogeneous isotropic cosmology, as ([4],5.2.1)
$$ T_{ab} = \rho_m u_a u_b$$where $\rho_m$ is the mean density of
the mass energy. Moreover, the vector field $u^a$ becomes, in the
framework of our model, the coordinate vector field
$(\frac{\partial}{\partial t})^a$ and then we have
$$ T_{ab} = \rho_m dt^2.$$On the other hand, the expression of the
total mass of the universe is established as being, within the
framework of this cosmology, ([4],11.2.10)
$$ M = \frac{1}{4\pi} \int_\Sigma R_{ab} n^a \xi^b dV = 2
\int_\Sigma(T_{ab}-\frac{1}{2}T g_{ab})n^a \xi^b dV $$So, by
adapting this expression to our model, $\Sigma$ becomes $B(O,t)$,
$g_{ab}$ becomes $h_t = dt^2 -g_t$ and $n^a$ identifies with
$(\frac{\partial}{\partial t})^a$ which also constitutes a
reasonable approximation of $\xi^a$. Consequently, we have $T_{ab}
n^a \xi^b = \rho_m, T = \rho_m$ and $g_{ab} n^a \xi^b = 1$ and
then we obtain
$$ M = 2 \int_{B(O,t)} (\rho_m - \frac{1}{2}\rho_m) dV_t =
\int_{B(O,t)} \rho_m dV_t$$which is, for us, the global mass of
the universe including the black holes and the invisible matter
mass.

\subsection*{Comparison with Newtonian gravity}
Comparing our matter-energy equation
$$ \frac{\partial^2}{\partial t^2} E(t,X(t)) - \Delta E(t,X(t)) =
0$$and our identities
$$ X^{''}(t) = \Gamma (t) = -\nabla^{g_e} E (t,X(t))$$with the two
equations that characterize Newtonian gravity (c.f.[4],(4.4.17)
and (4.4.21))
$$ \hskip 3cm \Delta \varphi = 4 \pi \rho_m \hskip 0.3cm \mbox{ (Poisson equation)}$$
$$ X^{''} = \Gamma = -\nabla^{g_e} \varphi ,$$
we obtain (by identification)
$$ \varphi(X(t)) = E(t,X(t))$$
$$ \Delta \varphi (X(t)) = \Delta E(t,X(t)) = 4\pi
\rho_m(X(t)).$$
Consequently, we have
$$\frac{\partial^2E}{\partial t^2}(t,X(t)) = \Delta E(t,X(t)) = 4
\pi\rho_m(X(t)).$$
Now, the total mass of the idealized universe
within the Newtonian gravity theory (readjusted in order to
conform with general relativity and with our model) is given by
(c.f.[4],(11.2.2))
$$ M_N = \frac{1}{4 \pi} \int_{S(O,R)} \overrightarrow{\nabla
\varphi}.\overrightarrow{n} dS.$$Thus, we get
$$ M_N = \frac{1}{4 \pi} \int_{S(O,R)} \overrightarrow{\nabla E}.\overrightarrow{n}
dS$$
$$ = \frac{1}{4\pi} \int_{B(O,R)} \Delta E dX = \frac{1}{4\pi}
\int_{B(O,R)} 4\pi \rho_m\, dX$$
$$ = \int_{B(O,R)} \rho_m\, dX = M$$where $\rho_m$ denotes here the matter density of the universe at time
$t$ (by taking $R \simeq t$) and $M$ is the total
mass of the matter inside the universe according to our model.\\
Moreover, by integrating on a mean ball $B(I,r_0)$ having a volume equal to the unit of volume of the universe at
time $t$, we obtain
$$\int_{B(I,r_0)} \frac{\partial^2 E}{\partial t^2}(t,X_t)dX_t = \int_{B(I,r_0)} \Delta E(t,X_t) dX_t$$
$$ = 4 \pi \int_{B(I,r_0)} \rho_m (X_t) dX_t =4\pi 
\rho_m(t) = 4\pi E_m(t)$$
where $\rho_m(t) = E_m(t)$ here denote
the matter density of the universe at time $t$. Then we have
$$ E^{''}(t) = 4\pi E_m(t) \hskip 1cm \mbox{or} \hskip 1cm
 \frac{1}{2}\rho_m^{''}(t) = 4\pi\rho_m(t)$$which
implies
$$ E_m(t) = \rho_m(t) = C e^{-2 \sqrt{2\pi}t}$$
where $C$ is a constant that can be determined when using the known values $t_0$
and $\rho_0$. Indeed, the relation
$$E_m(t_0) = C e^{-2 \sqrt{2\pi}t_0} = \rho_0$$implies
$$C = \rho_0 e^{2 \sqrt{2\pi}t_0}\;\; .$$So we have
$$\rho_m(t) = \rho_0 e^{2 \sqrt{2\pi}(t_0 - t)}.$$
{\bf Remark:} The above expression shows that $\rho_m(t)$ as well as $\rho(t)$ go exponentially to $0$ whereas $\rho(t)$ actually goes to $0$ proportionally to $\frac{1}{t^3}$ when $t$ goes to $+\infty$. This shows that the Newtonian gravitational potential $\varphi_t=E_t$, which perfectly explains gravitational laws of an isolated static body, does not exactly explain the cosmic macroscopic gravity. This deviation of the decrease of the density $\rho(t)$ with respect to its exact decrease reveals a real slowing down of the expansion process, due to the gravitational binding forces between stars, galaxies and cluster of galaxies inside the universe, which refers to the cosmological ``constants". Another reason for this deviation is the fact that we are often working, along this study, with the Euclidean Laplacian $\Delta_{g_e}$ in place of the Laplace-Beltrami operator $\Delta_{g_t}$. The results so obtained are good approximations on the cosmological level but can not exactly give the real behavior of $\rho(t)$.\\

\noindent In other respects, it is well known that, although the notion of
gravitational density is well defined by $\rho_G = - \frac{1}{8
\pi} | \nabla^{g_e} \varphi |^2$ within the Newtonian theory of
gravity, we can not define a similar notion within the classical
theory of general relativity. However, in the framework of our
theory, there is a good candidate for playing the role of general
energy density including gravitational one (at time $t$); namely
$$\rho _G = - \frac{1}{8 \pi} | \nabla^{g_t} E_t (X) |^2 = -
\frac{1}{8 \pi} | \nabla^{g_t} \varphi | ^2$$ and we think that we logically have
$$ \rho_G = - \rho_m = 2 \rho_r = - 2\rho = - \frac{\Delta
\varphi}{4 \pi} .$$ 
Recall that $g_t$ here is our physical metric that reflects all of the
physical consistence of the universe characterized by our global
matter-energy tensor $T^*_{ab}$ and $\nabla^{g_t} E_t$ here is the
gradient, with respect to $g_t$, of the matter-energy distribution
$E_t(X)$ on $B(O,t)$ at time $t$.\\\\
\textbf{Remark:} The global gravitational force is, according to
classical general relativity, proportional to $\rho + 3P = \rho +
\rho_r$ which is, according to our model, null; this fact explains
and confirms that the expansion of the universe is, beyond a
certain
time, uniform and permanent.\\\\
This implies
$$ - \frac{1}{8 \pi} | \nabla^{g_t} \varphi|^2 = - \frac{1}{4 \pi}
\Delta \varphi$$ or
$$ \Delta \varphi = \frac{1}{2} | \nabla^{g_t} \varphi|^2.$$
This relation implies that, on any trajectory $X(t)$, we have
$$ \Delta \varphi(X(t)) = \frac{1}{2} | \nabla ^{g_t} \varphi
(X(t))|^2$$ or
$$ \Delta_x E_t(X(t)) = \frac{1}{2} | \nabla ^{g_t} E_t
(X(t))|^2$$ which conforms with our model as, when $X(t)$ is the
trajectory of a free movement (i.e. $X(t)$ is a geodesic for
$g_t$) then we have
$$ E_t(X(t)) = E(t,X(t)) = \mbox{const}$$
and
$$ \nabla ^{g_t} E_t(X(t)) = F_{g_t}(X(t)) = \nabla^{g_t}_{X'(t)}
X'(t) = \widetilde{\Gamma}(t) = 0.$$

\subsection*{Remarks}
\noindent The deviations of the above values (for $t_0$, $r_0$,
\emph{E},...) with respect to the classical approximate values
expected by the Einstein - de Sitter's standard model, for
instance,
can be essentially explained by the two following factors:\\\\
$1^\circ$) The generally accepted value of the present Hubble's
parameter $H_0$ is incorrect for two reasons:\\
The first one is the use of some relativistic notions, the
deficiency of which has been already showed, such as the proper
time and the relativistic
formulas used in order to determine the redshift parameter \emph{z}.\\
The second reason is the fact of being based on measurements
established when considering visible galaxies whose positioning,
mutual distances and relative speeds are far from accurately
representing the expansion parameter. We are not at the center of
the universe and there exists a large number of galaxies that are
lying and moving outside of our horizon. Furthermore, the universe space extends far beyond all galaxies and matter agglomerations.\\
The huge difference between the two estimations of the mean matter
energy densities is due
to the large difference between the two estimations of the universe size.\\\\
$2^\circ$)The curvature parameter generally used in the first
Friedmann's equation is assumed to be constant. But this is absurd
since the fundamental notion of curvature (which reflects and
characterizes the matter - energy distribution which is the real
universe essence) is essentially dynamic and evolutive (locally
and globally). Indeed, according to the well confirmed expansion
theory, the universe does not reduce to $\mathbb{R}^3$ nor to a
fixed domain in $\mathbb{R}^3$, but (according to our model) to a
ball \emph{B}(\emph{O},\emph{t}) always expanding. This is the
reason that pushed us to start by using the first equation of
Friedmann in order to show the necessity of using a time-dependent
(macroscopic and global) curvature parameter \emph{K}(\emph{t})
that is associated with a (both local and global) time-dependent
metric $g_t$, although the determination of $t_0$ can be achieved
by using only the second Friedmann's equation.\\\\

$3^\circ$)We also notice that the sign difference between the
densities $\rho_m$ and $\rho_r$ is fundamentally due to the fact
that the first one is associated to the attractive force of
gravity while the second one is associated with the pressure who
gives rise to a force having an essentially opposed nature. These
are the two fundamental forces of Nature, namely the gravitational
and the radiational pressure (or equivalently the electromagnetic)
forces. Thus, the equation
$$\rho + 3P = \rho_m + \rho_r + 3P = 0$$adds a new dimension to
the cosmological energy conservation problem. Indeed, if we
qualify the quantity $3P$ as (mean density of) negative energy and
the quantity $\rho + 3P$ as (mean density of) generalized global
energy, we
can state\\

The generalized global energy is eternally null (c.f. [1]).\\\\
This principle recalls the momentum conservation principle when
applied to the whole universe:\\

The global momentum of the universe is eternally null.\\\\
Nevertheless, we notice that, according to our model, we can only
speak of pressure and negative energy, and then of null
generalized global energy (when assuming that the universe is
originally reduced to an energy $E_0$ concentrated at a given
point), after the Big Bang. Consequently, our model does not
support the theories that sustain that the universe is coming out
from literally nothing. Also, we notice that the meaning of the
term negative energy used
here is completely different from that to which is generally attributed inside the inflation theory and from the so called "negative gravity".\\\\

\section{Fundamental constants of modern Physics}
\subsection*{The quantum Statistics' constant}

We will here establish some relations involving several
fundamental physical constants showing that a large number of them
are dependent on time (and temperature) and leading to the
unification of the fundamental forces as well as to the
unification of all branches of Physics: General relativity (i.e.
Cosmology) Quantum theory, Electromagnetism, Thermodynamics and
Newton - Lagrange - Hamilton Mechanics.\\
Recall that we have showed in the previous section that
$$ E = \frac{c^5 C_0}{2 G_0}$$
where
$$ C_0 = K_0 t_0^3 + t_0 = \frac{4}{9} t_0 + \frac{8 \pi G_0
a_0}{3 c^2}.$$ This implies
$$ G_0 = \frac{ c^5 C_0}{2E} = \frac{c^5}{2E}(\frac{4}{9} t_0 +
\frac{8 \pi G_0 a_0}{3 c^2})$$
$$ = \frac{2}{9} \frac{c^5 t_0}{E} + \frac{4 \pi}{3}
c^3 \frac{G_0 a_0}{E}\;\;\;\;\;\;\;$$
$$ = \frac{2}{9} \frac{c^5 t_0}{E} + \frac{4 \pi}{3}
c^3 \frac{G_0 t_0^3}{E} \frac{a_0}{t_0^3}\;\;\;\;\;$$
$$ = \frac{2}{9} \frac{c^5 t_0}{E} - \frac{4 \pi c^3
t_0^3}{3} G_0 \frac{\rho_0}{E}\;\;\;\;\;$$
$$ = \frac{2}{9} \frac{c^5 t_0}{E} -G_0.\;\;\;\;\;\;\;\;\;\;\;\;\;\;\;\;\;$$
Consequently, we have
$$ 2G_0 = \frac{2}{9} \frac{c^5 t_0}{E}$$
and
$$ G_0 = \frac{c^5 t_0}{9 E} ( = \frac{243 \times {10}^{40} \times 2.365
\times {10}^{19}}{9 \times 9.57 \times {10}^{70}} \simeq 6.67
\times {10}^{-11}).$$ This equality being valid for an arbitrary
$t_0 \gg$ 1, we obtain, for $t \gg$ 1:
$$\;\;\;\;\;\;\;\;\;\;\;\;\hskip 1cm G = \frac{c^5 t}{9 E} \hskip 1cm \mbox { and } \hskip 1cm E =
\frac{c^5 t}{9 G}
\;\;\;\;\;\;\;\;\;\;\;\;\;\;\;\;\;\;\;\;\;\;\;\;\;\;\;\;\hskip 1cm
(31)$$ Besides, we have
$$ \rho_m = -2 \rho_r = 2 \rho = \frac{6 E}{4 \pi c^3 t^3}.$$which
gives by using quantum Statistics (c.f. [4] p.108)
$$ \frac{6 E}{4 \pi c^3 t^3} = \sum_{i = 1} ^{n} \alpha_i g_i
\frac{\pi^2 (K_B \hskip 0.05cm T)^4}{30 \overline{h}^3 \hskip
0.05cm c^5}$$where here the Boltzmann's constant is denoted by
$K_B$.\\Thus, we have
$$  \frac{3 E}{2 \pi c^3 t^3} = \sum_{i = 1}^{n} \alpha_i g_i
\frac{\pi^2}{30 c^5} \frac{(K_B \hskip 0.05cm
T)^4}{\overline{h}^3}$$which gives
$$ \frac{\frac{3 c^5 t}{9 G}}{2 \pi c^3 t^3} = \frac{c^2}{6 \pi G
t^2} = (\sum_{i = 1}^{n} \alpha_i g_i \frac{\pi^2}{30 c^5})
\frac{(K_B \hskip 0.05cm T)^4}{\overline{h}^3}$$that is
$$ \frac{5 c^7}{\pi G t^2} = (\sum_{i = 1}^{n} \alpha_i g_i \pi^2)
\frac{(K_B \hskip 0.05cmT)^4}{\overline{h}^3}$$or
$$ \frac{1}{G t^2} = (\sum_{i = 1}^{n} \frac{\alpha_i g_i \pi^3}{5
c^7}) \frac{(K_B \hskip 0.05cmT)^4}{\overline{h}^3} =: A
\frac{(K_B\hskip 0.05cm T)^4}{\overline{h}^3}.$$We then have
$$ \hskip 3cm G = \frac{1}{A} \frac{\overline{h}^3}{(K_B \hskip 0.05cmT)^4}
\frac{1}{t^2}
\;\;\;\;\;\;\;\;\;\;\;\;\;\;\;\;\;\;\;\;\;\;\;\;\;\;\;\;\;\;\;\;\;\;\;\;\;\;\;\;\;\;\;\;
(32)$$where
$$ A = \frac{1}{G} \frac{\overline{h}^3}{(K_B \hskip0.05cm T)^4}
\frac{1}{t^2}$$can be calculated by using the presently determined
values of the involved constants corresponding to $t_0$ = 2.365
$\times {10}^{19}$.\\\\
Then, equation (31) gives
$$ \frac{c^5 \hskip0.05cm t}{9 E} = \frac{1}{A} \frac{\overline{h}^3}{(K_B \hskip0.05cm
T)^4} \frac{1}{t^2}$$which yields
$$ \hskip 3cm \frac{\overline{h}^3}{(K_B \hskip 0.05cmT)^4} = \frac{A c^5}{9 E} t^3 = A G
t^2.
\;\;\;\;\;\;\;\;\;\;\;\;\;\;\;\;\;\;\;\;\;\;\;\;\;\;\;\;\;\;\;\;\;\;\;\;
(33)$$

In other respects, if we denote by $\alpha$ the classical
electromagnetic force factor and by $K_E$ the electromagnetic
constant denoted usually by \emph{k}, we obtain
$$ \alpha = \frac{K_E \hskip 0.05cm e^2}{\overline{h}\hskip0.05cm c} = \frac{K_E \hskip 0.05cm e^2}{c}
\times(\frac{9 E}{(K_B \hskip 0.05cm T)^4 A c^5
t^3})^{\frac{1}{3}}$$
$$\hskip 2.7cm = K_E \hskip 0.05cm e^2 \hskip 0.1cm(\frac{9E}{Ac^8})^{\frac{1}{3}}\hskip 0.1cm (K_B \hskip 0.05cm
T)^{-\frac{4}{3}}\hskip 0.1cm t^{-1} \hskip 4cm (34)$$and
$$ \alpha = \frac{K_E \hskip 0.05cm e^2}{c} ((K_B \hskip 0.05cm T)^4\hskip 0.1cm A G t^2)^{-\frac{1}{3}} = \frac{K_E e^2}{cA^{\frac{1}{3}}}(K_BT)^{-\frac{4}{3}}
G^{-\frac{1}{3}} t^{-\frac{2}{3}}\;\;\;\;\;\;\;\;\;\;\;\;\;\;\;
(34^{'})$$
\subsection*{Fundamental relations}

Using the above relations, we can obtain several relations that
specify the dependence of the fundamental constants on each
other in addition to time.\\
Indeed, (32) implies
$$\hskip 2cm  \frac{G (K_B T)^4}{\overline{h}^3} = \frac{1}{A}\hskip 0.1cm \frac{1}{t^2} =
\frac{C_1}{t^2} \hskip 2 cm \mbox{ where } \hskip 0.5cm C_1 =
\frac{1}{A} \hskip 1.5cm (35)$$and (33) implies
$$\hskip 1cm \frac{\overline{h}^3}{(K_B T)^4} = \frac{Ac^5}{9E}
t^3 = C_2 t^3 \hskip 1.5cm \mbox{ with } \hskip 0.5cm C_2 =
\frac{Ac^5}{9E}$$ or
$$ \hskip 4cm \frac{\overline{h}}{(K_B T)^{\frac{4}{3}}} =
C_2^{\frac{1}{3}} t \hskip 6cm (36)$$ Likewise, (34) implies
$$\alpha t = k e^2 (\frac{9E}{Ac^8})^{\frac{1}{3}} (K_B
T)^{-\frac{4}{3}} = \frac{C_3}{(K_BT)^{\frac{4}{3}}}$$
where
$$ C_3 = k e^2 (\frac{9E}{Ac^8})^{\frac{1}{3}}$$ and
$$\hskip 4cm \alpha (K_B T)^{\frac{4}{3}} = \frac{C_3}{t} \hskip 6cm
(37)$$which, added to the relation
$$G = \frac{c^5}{9E} t =: C_0 t \hskip 1cm \mbox{ or } \hskip 1cm
t = \frac{G}{C_0}$$gives
$$ \alpha (K_B T)^{\frac{4}{3}} = \frac{C_3 C_0}{G}$$and then
$$ \hskip 4cm \alpha G (K_B T)^{\frac{4}{3}} = C_3 C_0 =: C_5 \hskip 4cm
(38)$$where
$$ C_5 = k e^2(\frac{9E}{Ac^8})^{\frac{1}{3}} \frac{c^5}{9E} = k e^2
(9E)^{\frac{1}{3}}\times(9E)^{-1}\times(Ac^8)^{-\frac{1}{3}}c^5
$$
$$\hskip 1cm = \frac{k e^2
c^{\frac{7}{3}}}{A^{\frac{1}{3}}(9E)^{\frac{2}{3}}}\hskip 7.5cm
$$which, added to (36), implies
$$ \alpha G \frac{\overline{h}}{C_2^{\frac{1}{3}} t} = C_5$$and
then
$$ \hskip 3.6cm\alpha G \overline{h} = C_5 C_2^{\frac{1}{3}} t = C_6 t \hskip
4cm (39)$$ with
$$C_6 = \frac{k e^2 c^4}{9E}$$
(verification: $\alpha \overline{h}$ = $C_6 \frac{t}{G}$ = $C_6
\frac{9E}{c^5}$ = $\frac{ke^2c^4}{9E}\frac{9E}{c^5}$ =
$\frac{ke^2}{c}$ which is an absolute constant although $\alpha$ and $\overline{h}$ are both dependent on time $t$).\\\\
Finally, (35) and (39) respectively imply
$$\alpha G \overline{h} \hskip0.2cm = \hskip 0.2cm \alpha \frac{\overline{h}^3}{A(K_B
T)^4\hskip 0.05cm t^2} \overline{h} \hskip 0.2cm = \hskip 0.2cm
\alpha (\frac{\overline{h}}{K_BT})^4 \frac{1}{At^2}$$and
$$\alpha G \overline{h} = \frac{k e^2 c^4 t}{9E}$$
Therefore
$$E=\frac{k e^2 c^4 t}{9\alpha\overline{h}G}=\frac{k e^2 c^4 t}{9\frac{ke^2}{c}G}=\frac{c^5 t}{9G}$$
which conforms with the previously established relation by using the radiational density that is equal to the opposite to the matter-energy density 
$$\rho(t)=\int_{B(I,r_0)}E_t(X)\, dX.$$
Likewise, we have
$$ \hskip 2cm\alpha (\frac{\overline{h}}{K_BT})^4 = \frac{A}{9E} k e^2 c^4
t^3 \hskip 4.2cm (40)$$
\\
(verification:  $\alpha$ ($\frac{\overline{h}}{K_BT})^4$ =
$\frac{1}{G} \frac{\overline{h}^3}{(K_B T)^4} \frac{1}{t^2} \times
\frac{1}{9E} k e^2 c^4 t^3$ = $\frac{1}{9E} k e^2 c^4
\frac{\overline{h}^3}{G(K_BT)^4} t$\\\\which gives
$$\alpha
\overline{h} = \frac{1}{9E} k e^2 \frac{c^4 \hskip0.05cm t}{G} = k
e^2 \frac{c^4 \hskip 0.05cm t}{9E}
\frac{1}{\frac{c^5\hskip0.05cmt}{9E}}  = \frac{k e^2}{c}).$$
So, we have
$$\alpha\overline{h}=\frac{ke^2}{c}=\frac{2.304\times 10^{-19}}{3\times 10^8}=7.68\times 10^{-28}.$$
Notice that the standard values of $\alpha$ and $\overline{h}$ give 
$$\alpha\overline{h}=\frac{1.06\times 10^{-34}}{137}=7.737\times 10^{-37}$$
which is obviously contradictory.\\
Notice also that the relation $E_t(X)=h(t)f(t)$ (where $f(t)$ is the mean frequency of the matter-energy) implies
$$\int_{B(I,r_0)}h(t)f(t)\, dX=\int_{B(I,r_0)}E_t(X)\, dX$$
which gives
$$h(t)f(t)=\rho(t)$$
and
$$E=\int_{B(O,t)}h(t)f(t)\, dX=\int_{B(O,t)}E_t(X)\, dX=\int_{B(O,t)}\rho(t)\, dX.$$
The fact that $\rho(t)\propto\frac{1}{t^3}$ and $h(t)\propto t$ implies that $f(t)\propto \frac{1}{t^4}\propto T^{\frac{16}{3}}$.\\
If now $\mu$ is the mean eigenvalue on the unit ball of the classical Dirichlet problem (i.e. on $B(O,1),g_e))$ associated with the matter-energy equation, then the relation 
$$f(t)\propto\frac{\rho(t)}{h(t)}\propto\frac{1}{t^4}\quad\mbox{implies}\quad f(t)^2\propto\frac{1}{t^8}$$ 
and 
$$\mu=Cf(t)^2t^8\quad\mbox{which is}\quad f(t)=\frac{1}{C}\frac{\sqrt{\mu}}{t^4}.$$ 
So the relationship that relates $\mu$ to $f_\mu(t)$ obtained when assuming that
$$e(\mu)=E_\mu(t,X(t))=h_\mu(t)f_\mu(t)=\rho(t)$$
is constant (i.e. $f(t)=\frac{1}{2\pi}\frac{\sqrt{\mu}}{t}$) is inexact on the cosmic scale. This is due to the mean permanently cooling of the universe.\\
The relation 
$$\alpha\overline{h}=7.68\times 10^{-28}$$ 
implies
 $$\alpha h=2\pi\times 7.68\times 10^{-28}$$ 
 which is the same order of $10^{-27}$ and then $h$ is, at present, the same order as $10^{-25}$. Then we have
\begin{eqnarray*}
\mu&\simeq& C\frac{(6.4\times 10^{-14})^2}{(10^{-25})^2}\times(2.365\times 10^{19})^8\\
&\simeq& C\times 10^{180}.
\end{eqnarray*}
Moreover
$$h(t_0)\simeq bt_0\simeq 10^{-25}$$
implies $b$ is of order $10^{-44}$.\\

\noindent To sum up, we write down the following fundamental relations:
$$ T \propto \frac{1}{t^{\frac{3}{4}}} \hskip 0.5cm , \hskip 1cm \rho
\propto T^4 \propto \frac{1}{t^3} \propto \frac{1}{V} \hskip 0.5cm
 \mbox{and} \hskip 0.5cm T \propto
 \frac{1}{V^{\frac{1}{4}}}$$(where $V$ is the volume of the
 universe at time $t$)
$$ \alpha \hskip0.1cm \overline{h} = \frac{K_E\hskip0.05cm
e^2}{c}$$
$$ \frac{\overline{h}^3}{(K_B\hskip0.05cm T)^4} \propto t^3 \hskip 1cm
\mbox{ and } \hskip 1cm \overline{h} \propto(K_B \hskip0.05cm
T)^{\frac{4}{3}}\hskip0.05cm t \propto
K_B\hskip0.05cm^{\frac{4}{3}}$$
$$ E = \frac{c^5 \hskip0.05cm t}{9G} \hskip1cm \mbox{ and }
\hskip1cm G = \frac{c^5}{9E} t \propto \frac{c^5}{9E}
\frac{1}{T^{\frac{4}{3}}}.$$ 
The real dependence of the wavelength $\lambda$ on the cosmic temperature $T$ can be expressed as 
$$\lambda\propto\frac{1}{f}\propto t^4\propto\left(\frac{1}{T^{\frac{4}{3}}}\right)^4=\frac{1}{T^{\frac{16}{3}}}\propto V^{\frac{4}{3}}.$$ 
Moreover, concerning the curvature
parameter $K(t)$, we have, using the relations (28) and (30),
$$ K(t) = -\frac{5}{9t^2} + \frac{8 \pi G a}{3 c^2 t^3}\hskip 5cm$$
$$ = -\frac{5}{9t^2} + \frac{8 \pi}{3c^2 t^3}(-\frac{c^2}{6\pi}t +
3 \frac{EG}{4\pi c^3}) \hskip 1.2cm$$
$$ = -\frac{5}{9t^2} + \frac{8 \pi}{3c^2 t^3}(-\frac{c^2}{6\pi}t +
\frac{3}{4 \pi c^3} \frac{c^5}{9}t) \hskip 0.8 cm$$
$$ = -\frac{5}{9t^2} + \frac{8 \pi}{3c^2 t^3}(-\frac{c^2}{6\pi}t +
 \frac{c^2}{12 \pi}t) \hskip 1.3cm$$
$$ = -\frac{5}{9t^2} -\frac{2}{9t^2} = -\frac{7}{9t^2} \hskip 2.9cm$$
Finally, we have on one hand
$$ \alpha = \frac{K_E e^2}{c \overline{h}} \propto \frac{K_E
e^2}{c K_B^{\frac{4}{3}}}$$and on the other hand
$$ \alpha = \frac{K_E\hskip0.05cm e^2}{c\hskip0.05cm\overline{h}}
= \frac{K_E\hskip0.05cm e^2}{c(K_B\hskip0.05cm
T)^{\frac{4}{3}}\hskip0.05cm t} = \frac{K_E\hskip0.05cm
e^2}{c(K_B\hskip0.05cm T)^{\frac{4}{3}}} \times \frac{c^5}{9GE} =
\frac{c^4 K_E\hskip0.05cm e^2}{9(K_B\hskip0.05cm
T)^{\frac{4}{3}}\hskip0.05cm GE}$$or
$$ \alpha G = \frac{c^4 K_E\hskip0.05cm e^2}{9(K_B\hskip0.05cm
T)^{\frac{4}{3}}\hskip0.05cm E}.$$Let us notice that, since
$\overline{h} \propto t$, we get (using (36)) $K_B T$ is constant
and it is easy to show that the preceding relations imply
$$ \alpha \propto \frac{1}{t} \hskip 0.3cm , \hskip 0.3cm G \propto t \hskip 0.3cm \mbox{and} \hskip 0.3cm K_B
\propto t^{\frac{3}{4}}.$$ 
{\bf Remark}: Knowing that $K_BT\simeq 0.02585$eV at $T=300K$, we presently have $K_B\simeq\frac{0.02585}{300}\simeq 8.62\times 10^{-5}$ and consequently, if $T$ is the cosmic temperature, the constant $K_BT$ is nearly equal to $8.62\times 10^{-5}\times 2.74\simeq 2.361\times 10^{-4}$eV.\\

\noindent At the end, we notice that we have
(within the framework of our model) $R_{t_0} (t) = \frac{t}{t_0}$
and $\frac{dR_{t_0}}{dt} = \frac{1}{t_0}$ and then the Friedmann
equation becomes
$$ \frac{1}{t^2} = \frac{8 \pi \rho G R^2}{3c^2} - K(t)$$which
yields, for $t = t_0$
$$\frac{1}{t_0^2} = \frac{8 \pi \rho_0 G_0}{3c^2} + \frac{7}{9t_0^2}
= \frac{8 \pi}{3c^2} \frac{3E}{4 \pi c^3 t_0^3} \frac{c^5 t_0}{9E}
+ \frac{7}{9t_0^2}$$
$$ = \frac{2}{9t_0^2} + \frac{7}{9t_0^2} \hskip 4.7cm$$
This again shows the validity of our model and the legitimacy of
the time-dependence of the constants $K$ and $G$.

\subsection*{Remarks}
\noindent $1^\circ$) The above relations show that only the
fundamental constants \emph{E}, \emph{A}, \emph{c}, $K_B T$ and
$K_E \hskip0.05cm e^2$ are independent of time, the other ones
\emph{G}, $\overline{h}$, $K$ and $\alpha$ depend on time and are
related
between them and to the first ones.\\
By taking \emph{c} = 1 and taking into account the other
relations, we can state that only \emph{E}, $K_E\hskip0.05cm e^2$,
$K_B T$ and \emph{t} have an intrinsic existence
which shows that the universe reduces to three basic elements:\\

The original energy, electromagnetism and time.\\
This last factor, which is at the same time distance and extent,
determines (with the second factor) the dynamic expansion. So, the
universe is essentially the original energy always expanding.\\\\
$2^\circ$) Let us denote $K_B \hskip0.05cm T$ = $\frac{2}{3}
\langle E_K \rangle$, where $\langle E_K \rangle$ is the
generalized mean kinetic energy, by $E_\ast$ and replace $K_E$ by
\emph{k}, then the previous relations lead, for \emph{c} = 1, to
the following ones:
$$ \alpha \hskip0.05cm \overline{h} = k e^2$$

$$ \;\;\;\overline{h} \propto E^{\frac{4}{3}}_\ast \hskip0.1cm t \hskip 1cm
\mbox{ and } \hskip1cm \alpha \hskip0.05cm E^{\frac{4}{3}}_\ast
\hskip0.1cm t \propto k e^2$$

$$ G = \frac{t}{9E} \hskip 0.8cm \mbox{ and } \hskip 0.8cm \overline{h} = 9 E^{\frac{4}{3}}_\ast \hskip0.1cm EG$$

$$ \alpha = \frac{ke^2}{9 E^{\frac{4}{3}}_\ast \hskip0.1cm
E}\frac{1}{G} \hskip 1cm \mbox{ or } \hskip 1cm \alpha G =
\frac{ke^2}{9E E^{\frac{4}{3}}_\ast}.$$ These relations prove,
particularly, the unity of the
fundamental forces of Nature.\\\\
$3^\circ$) The relation $A = \frac{1}{G}
\frac{\overline{h}^3}{(K_BT)^4} \frac{1}{t^2}$ shows the ultimate
power of the quantum Statistics. The other ones show, on one
hand, that the quantization of every thing is not justified
(\emph{G} and $\overline{h}$, for instance, are continuously
depending on time) and, on the other hand, that the quantization
that is lying on a long and precise experimentation recovers its
global legitimacy by its unification with the theoretical Physics
and Mechanics. In the present case we recover the unification of
the quantum Physics with the reviewed general relativity which are
both unified with the Newton - Lagrange - Hamilton Mechanics. In
particular, it is showed that the universe evolution is thoroughly
described by the (reviewed) Einstein's general
relativity theory.\\\\
$4^\circ$) These same relations joined to the results of section 9
and the other sections institute the bases of the unification of
all Physics' branches: Electromagnetism, general Relativity (i.e.
Cosmology), Thermodynamics, quantum Physics and Mechanics,
Particles' Physics with the (Newton - Lagrange -
Hamilton) Mechanics.\\\\

\section{Commentaries and open issues}
\subsection*{Reflexions about our model and some others}

Our physical and mathematical global model permits to give many
answers, clarifications and precisions concerning open problems of
modern Physics. Moreover, it permits to reformulate some other
problems and to draw some new perspectives for their resolution.\\

The crucial points which led to these possibilities are:\\

\noindent1$^\circ$) The theoretical, mathematical and physical
refutation of some interpretations of the special relativity
second postulate and the replacement of the relativistic
approximate formulas by alternative formulas that reflect more
precisely the Nature's laws although the precision of which could
be improved and more exactly
quantified (see sections $4$ and $7$). These modifications should be generalized to all recesses of modern Physics.\\\\
2$^\circ$) The logical and mathematical clarification of the
experimental and theoretical Quantum theory limits and the
specification of the circumstanciel nature of the Heisenberg
uncertainty principle despite of the practical efficiency of these
tools headed by Schr\"{o}dinger's equations (sections $7$ and
$8$). Nevertheless, the Quantum theory must be used in order to
resolve all problems that does not fall in with an idealized
modeling that permits their resolution by means of classical
Mechanics and Physics.\\\\
3$^\circ$) The geometrization of the three dimensional universe by
means of the physical metric $g_t$ and the setting up of the
Matter - Energy equation ($E^*$) as well as the geometrization of
the four dimensional space - time by means of the evolution metric
$h = dt^2 - g_t$, after the restoration of the natural space and
time notions. Our metrics ensure that the evolution of all quantities with time is taken into account (sections $4,5$ and $6$).
The Einstein's vacuum spacetime metric takes into account only the gravitational field and the Einstein's tensor characterizes
some particular matter fields and electromagnetic fields. Our metric and tensor take into account all forms and effects of matter-energy: Global
gravitational field (included that of black holes), all forms of matter field, global electromagnetic field, global cosmic radiations, pressure and
temperature, all energy evolutions, interactions and singularities.\\\\
4$^\circ$) The logical (and, a posteriori, physical) refutation of
the first Hubble's cosmological principle according to which any
galaxy can be considered as lying at the center of the universe.
This refutation is solidly supported by our model which reveals
its consistence and its extreme compatibility with all definitely
established physical and mechanical laws. When joined to the first
point, this refutation has permitted to modify the Einstein's
general relativity theory, the Friedmann - Einstein's equations
and the Einstein - Hubble - de Sitter - Friedmann's Cosmology in
such a way that (one time adapted to our setting) it is clearly
showed that these theories and equations describe correctly the
universe evolution (i.e. the universe expansion). This adaptation,
joined to the demonstration of the time - dependence of all
fundamental constants, apart from $E,$ $k e^2$, $K_B T$ ,\emph{c}
and \emph{A} (which gives a new dimension to the quantization
process), leads to the
unification of all Physics' branches (sections $10,11$ and $12$).\\\\
5$^\circ$) The matter - energy equation and the use of the Dirac
operator have permitted to yield a new approach to the Physics of
particles by instoring a new classification of the fundamental
particles that has a chance (if being sustained by experimental
results) to lie at the root of this branch of Physics which is
presently deeply related to the Cosmology and has many beneficial
applications for the humanity future (section $9$).\\\\

Along this study, we have given clear answers to many open
questions (such as, for instance, those listed at the end of [ 2]
or those invoked throughout the excellent panoramic book directed
by Paul Davies and wisely entitled: The New Physics) and some
possible answers to other open problems. We have also formulated
other questions and emitted some hypotheses. Nevertheless, we
will, in the following, give some properties and
remarks concerning the above last point.\\

\noindent1$^\circ$) According to our model, the fundamental material
particles are only nine (to which we add the three neutrinos):
$$e \hskip 1cm u
\hskip1cm d \hskip1cm s  \hskip1cm c \hskip 1cm b \hskip 1cm t \hskip 1cm \mu \hskip 1cm \tau \hskip 1cm \nu_e \hskip 1cm \nu_\mu \hskip 1cm \nu_\tau,$$
each of them intrinsically exists with two different spins and each of the six quarks intrinsically exists with three different colors.\\
They constitute, with their antiparticles, the three pure energy
vectors represented by $\Gamma_1,$ $\Gamma_2,$ $\Gamma_3$ and
other $\Gamma$ vectors; each of them having two distinct
polarizations. These latter potentially hold all fundamental
particles to which they give birth. The fundamental particles form
all bound states called hadrons (baryons and mesons) that are more
or less stable.\\The neutrinos are the direct or undirect partners
of a lot of (weak, strong and electromagnetic) interactions,
decays, collisions, nuclear syntheses and annihilations. They are
(originally and continually) produced during these interactions
essentially in order to make them possible and compatible with the
conservation laws. Some significative examples of production and
interactions involving these particles are schematically listed
below (c.f. [2]):
$$W^+ \rightarrow e^+ + \nu_e \hskip 2cmW^- \rightarrow
e^- + \overline{\nu_e}$$
$$\pi^+ \rightarrow \mu^+ + \nu_\mu \hskip 2cm \pi^- \rightarrow
\mu^- + \overline{\nu_\mu}$$
$$K^+ \rightarrow \mu^+ + \nu_\mu \hskip 2cm K^- \rightarrow \mu^-
+ \overline{\nu_\mu}$$
$$\tau^- \rightarrow \nu_\tau + e^- + \overline{\nu_e}$$
$$s \rightarrow u +W^- \rightarrow u + e^- +
\overline{\nu_e}$$
$$c \rightarrow s + W^+ \rightarrow s + e^+ + \nu_e$$
$$c \rightarrow s + W^- \rightarrow s + \mu^+ + \nu_\mu$$
$$b \rightarrow c + W^- \rightarrow c + e^- +
\overline{\nu_e}$$
$$b \rightarrow c +W^- \rightarrow c + \mu^- +
\overline{\nu_\mu}$$
$$b \rightarrow c +W^- \rightarrow c + \tau^- +
\overline{\nu_\tau}$$
$$\nu_\mu + N \rightarrow \mu^- + \mbox{hadrons}$$
$$\overline{\nu_\mu} + N \rightarrow \mu^+ + \mbox{hadrons}$$
$$\nu_\mu + d \rightarrow W^- \rightarrow \mu^- + u$$
$$\overline{\nu_\mu} + u \rightarrow W^+ \rightarrow \mu^+ +
u$$
$$\nu_\mu + q \rightarrow Z^0 \rightarrow \nu_\mu + q$$
$$\overline{\nu_\mu} + q \rightarrow Z^0 \rightarrow \overline{\nu_\mu} +
q$$
$$\nu_\mu + e^- \rightarrow \nu_\mu + e^-$$
$$\mu^- \rightarrow e^- + \overline{\nu_e} + \nu_\mu$$
$$\mu^+ \rightarrow e^+ + \nu_e + \overline{\nu_\mu}$$
$$\tau^- \rightarrow e^- + \overline{\nu_e} + \nu_\mu$$
$$\tau^- \rightarrow  \mu^- + \overline{\nu_\mu} + \nu_\tau$$
$$\tau^- \rightarrow  \nu_\tau + \mbox{hadrons}$$
$$\tau^+ \rightarrow e^+ + \nu_e + \overline{\nu_\tau}$$
$$\tau^+ \rightarrow  \overline{\nu_\tau} + \mbox{hadrons}$$
These interactions and many others show that the two leptonic
families $\mu$ and $\tau$ are produced by the interactions of
neutrinos with several types of fundamental (and other) particles
as well as by other interactions and by several decays. They are
produced with a very short lifetimes. We consider that these
families are potentially composite particles, although they are
fundamental particles (i.e. involved into solutions to the wave
equation associated with the Dirac operator). They have electrical
charges and their very fast decays give birth to stable neutrinos
and unstable hadrons beside of stable electrons or positrons
(which
interact and annihilate very quickly). This is roughly the case of the five more massive quarks which are fundamental particles although they are not absolutely stable and they end up by giving birth to the $u$ quark.\\\\
2$^\circ$) It is well known that neutrinos are left handed
particles (i.e. have a negative helicity) and that antineutrinos
are right handed particles. These properties have to be related to
the two different polarizations of the electromagnetic waves and
to the existence of two electron's types with opposite spins
$e_{\frac{1}{2}}$ and $e_{-\frac{1}{2}}$. They can be considered
as intrinsic characteristics of these particles. The $\beta$ -
decay of the $W^-$ particle produces a negatively polarized
electron and an antineutrino that can not be other than positively
polarized. The collision - annihilation $p - \overline{p}$ or
rather $q - \overline{q}$ of a left handed quark with a right
handed antiquark produces (by the intermediate of a $W$ particle)
one right handed positron and a neutrino that can not be other
than left handed.\\Can we then talk about parity
violation?\\\\
3$^\circ$) Our model is based on the notion of field theory which
does not require the existence of strong charges (in the same
manner as gravity does not require gravitational charge) nor the
existence of intermediate charge conveyors. Gluons are in fact
similar particles to photons and like them do not carry charges;
they only are mediators for quarks and nucleons' strong
interactions (in the same manner as photons are mediators for
electromagnetic interactions between charged particles).\\Unlike
photons which, on one hand, form a permanently evolving sea inside
atoms and, on the other hand, propagate throughout the universe,
gluons only form a sea of particles (among other particles) inside
hadrons (and nuclei) that evolve and transform permanently bearing
and causing all sorts of interactions. Otherwise, gluons are
fundamentally different from $W$ and $Z^0$ massive particles which
are at the origin of weak interactions from their formation until
their decays and their effects. Besides, we notice that neither
gravitons nor balls of glu exist within our model. Gravitational
field does exist; it curves the space and need neither
gravitational charges nor intermediary particles for conveying
"gravitational force" and it has an unlimited range inside the
universe. The strong interactions exist only inside hadrons and
nuclei and are conveyed
by gluons.\\\\
4$^\circ$) The only non hadronic stable particles inside our model
are electrons, photons, neutrinos and the $u$ quark. The only
presumely stable hadron is the proton. The other hadrons (except
the neutron) have ephemeral existence. This fact sustains our
hypothesis about the non existence of particular strong forces.
The hadronic bound states are achieved by the ephemeral and
episodic attractions caused by the gravitation and electromagnetic
attractions. This fact is supported by comparing, for instance,
the following three bound states: the proton state $u \hskip
0.05cm u \hskip 0.05cm d$, the neutron state $u \hskip 0.05cm d
\hskip 0.05cm d$ and the $\Delta^{++}$ state $u \hskip 0.05cm u \hskip
0.05cm u$. The first two states have a spin of $\frac{1}{2}$ and
the third one has a spin of $\frac{3}{2}$. This shows that only
two constituent quarks have aligned spins for the former states
whereas all three spins are aligned for the latter one. Moreover,
the first bound state is formed with two positively charged quarks
(+$\frac{2}{3}$) and one negatively charged one (-$\frac{1}{3}$),
the second state is formed with two negatively charged quarks
(-$\frac{1}{3}$) and one positively charged one (+$\frac{2}{3}$)
whereas the third state is formed with three positively charged
quarks (+$\frac{2}{3}$). The above two factors contribute to make
the $\Delta^{++}$ particle extremely unstable, compared to the
other two states, with its infinitely short lifetime. The
extremely larger lifetime of the proton compared to that of the
neutron could be explained by the fact that the quark which is
antialigned with the two others inside the proton is the quark
\emph{u} which has a larger electrical charge than the quark
\emph{d} which is antialigned with the other two quarks inside the
neutron and so the electromagnetic attraction inside the proton is
larger than that inside the neutron (whereas it does not exist
inside the $\Delta^{++}$ particle). The mass difference between the
quarks \emph{d} and \emph{u} and consequently between the neutron
and the proton has a determinant role concerning the stability
difference; the neutron has the possibility of decaying in order
to give birth to a proton, the decaying of which is apparently
forbidden. The transformation of the quark \emph{d} into the quark
\emph{u} inside the neutron has no major constraints. The quark
\emph{u}
can not naturally give birth to another quark. We can conclude that the energy equilibrium between the (gravitational and electromagnetic) potential energy, the (vibrational and rotational) kinetic energy and the mass energy which is established (at very short distances) between the three quarks $u,u$ and $d$ inside the proton is extremely more stable than that established between the quarks $u,d$ and $d$ inside the neutron which itself is extremely more stable than between quarks and antiquarks inside all other hadrons.\\\\
5$^\circ$) The non existence of three different strong charges and
three or eight differently charged gluons (even though the
existence of each quark with three different colors has been
proved and the existence of several differently colored gluons is
not excluded) does not prevent the possibility that strong
interactions could be associated with the symmetry $SU$(3).
Likewise, the weak interactions could also be associated with the
symmetry $SU(2).$ Moreover the weak interactions and the
electromagnetic interactions could be unified by means of a gauge
theory associated with the group $SU$(2) $\times$ \emph{U}(1)
inside an electroweak theory. It is equally possible to construct
a grand unification theory based on the group $SU$(5) and to ask
for the specific conditions concerning the different symmetry
breakings. All this could allow us to construct the table of all
particles which could ever exist starting from our $24$
fundamental particles and to characterize all particles that are
qualified as gauge particles, including
particles of Higgs' type.\\\\
On other respects, we could give in to the mathematical charm of
the supersymmetry theory and associate to it the supergravity
theory. But the existence of garvitons, Higgs particles and their
supersymmetric partners (gravitinos and Higgsinos) is excluded
inside our model. Likewise, we have to admit that the Kaluza -
Klein compactification of the five dimensional space (four of them
constitute the spacetime related to general relativity and the
fifth invisible compactified one corresponds to Electromagnetism)
and that the theory of everything (TOE) attempted by Cremmer and
Julia and its association with the supersymmetry $N=8$ theory
(with its pyrgons) do not have anything to do
with our model.\\
There is also the ten dimensional supersymmetric string theory
which is associated with a symmetry group that owns a gauge group
\emph{G} of rank 16 (which could be $SO$(32)$/Z_2$ or \emph{E8}
$\times$ \emph{E8}) that reproduces 496 Yang - Mills gauge
particles, 480 of which are solitons. The association of the
beautiful topological results of Witten and other mathematical
ingredients with this theory does not constitute for us other than
a fascinating intellectual gymnastic exceptionally esthetic which,
after long theoretical and experimental confrontations and
appropriate modifications could contribute to enlighten the last
part of the knowledge path
leading to the ultimate laws of Nature.\\\\
We finally notice that inflation and vacuum energy and its
polarization have no place in our model. The vacuum that is
crowded with all sorts of  real (material or immaterial) particles
which are created and are interacting, annihilating and decaying
more or less fastly is not a real vacuum. These particles can
really exist and participate to many energy transformations and
can also be specifically polarised but could we then speak about
polarised vacuum? The vacuum negative energy may only be explained
by the radiational pressure (resulting from a radiational sea when
it exists) that acts in the expansion direction. This certainly is
not
a matter of a negative gravity associated with a false vacuum as they are usually qualified by inflation theories.\\\\
6$^\circ$)  When we admit the existence of the original energy that was concentrated at some point, our model explains the universe creation and formation process by the following splittings:\\
\begin{itemize}
\item The energy splitting into two polarizations at the propagation time.
\item The matter-antimatter splitting together with the electrical charge splitting.
\item The spin splitting of the electron and the other two-spins particles.\\
\end{itemize}
These three phenomena are tightly related to the temperature: Temperature = Energy (via frequencies) = (radiational) pressure. The propagation-expansion, the interactions and the subsequent evolution can be well explained and obey precise laws.\\

We can briefly state that the universe formation process essentially reduces to\\

\noindent$1.\;$ The propagation (related to the polarization splitting) of electromagnetic waves starting from the original energy.\\

\noindent$2.\;$ The matter-antimatter formation and particularly the electron-positron and the u quark-antiquark creation. The other quarks-antiquarks and leptons-antileptons (with the neutrinos-antineutrinos) creation has led to the state qualified as quarks and leptons soup that preceded the hadrons (and the others) formation.\\
This scheme indicates that there exist only three privileged ultra-fundamental particles (with their antiparticles): The photon (with its two polarizations) that is essentially an energy particle into movement, the electron (with its two opposite spins) and the u quark (with its two spins and three colors).\\
The other fundamental particles end up all by giving birth to these three particles and to neutrinos.\\
The real universe is essentially made up with the most stable five
particles: Photons, Electrons, Neutrinos, Protons and Neutrons.
The last two particles are made up by the u and d quarks. The
$uud$ bound state being more stable than the udd state because the
d quark can naturally transform into the u quark and because of the larger attraction between the two u quarks inside proton than the two d quarks inside neutron in view of their larger charge and their opposite movement, and their opposite spins. The other
leptons and hadrons are extremely less stable and have an
ephemeral existence. The bound state uu (or dd) does not exist and the
state uuu can exist only during an insignificant infinitesimal
time. Moreover, the disappearance of the antimatter could be
explained by the extreme unstability of all particles that are
partially made up with antimaterial fundamental particles on one
hand and, on the other hand, by the absorption of antineutrinos by
a large number of "absolutely" stable protons in order to give
birth to the relatively
unstable neutrons which are yet less numerous than protons. All antiparticles have quickly annihilated giving birth to photons and gluons.\\

\subsection*{Beyond the Big Bang}
From our mathematical and physical modeling of the expanding universe (published at the arxiv and entitled `The expanding universe and energy problems") results that there exists originally the virtual theoretical three dimensional space that extends indefinitely in all directions such that the choice of an arbitrary point $O$ as being the origin of space identifies it with $\mathbb{R}^3$. There also exists another dimension representing the continual flow of time such that the choice of an origin corresponding to the time $t=0$ identifies it with $\mathbb{R}=]-\infty,+\infty[$. As every notion referring to the infinity notion is a matter for the metaphysics because the infinity, although mathematically is well defined, can not be physically realizable and remains something that is fundamentally abstract. Starting from this, our model schematizes the universe, at the time referred to as $t=0$, as being a well determined eternal quantity of energy concentrated in $O$ the calculation of which, by using our conventional units, gives $E_0=9.57\times 10^{70}$ J. The matter-energy distribution through the physical universe at time $t>0$, denoted by $E_t(X)=E(t,X)$, is then reduced, at the time referred to as $t=0$, to $E_0(X)=\delta E_0$ where $\delta$ is the Dirac measure at the spatial origin $O$. It also results from our model that, at time $t>0$, the universe is modeled by
$$U(t)=\left(B(O,R(t)),g_t\right),$$
where $B(O,R(t))$ is the ball of Euclidean radius $R(t)$ and $g_t$ is a Riemannian metric defined on $B(O,R(t))$ that reflects, at every time $t>0$, the influence of the matter-energy distribution filling the physical space $B(O,R(t))$. The regularized metric $g_t$ will be called the real physical metric.\\

By considering the canonical volume form $dv_{g_t}$ associated with $g_t$ and defining the regularized measure $\nu_t=E(t,X)\,dX$, we obtain
$$dv_{g_t}=dX-\nu_t=(1-E(t,X))\,dX.$$
Thus, $dv_{g_t}$ measures the real physical volume into the physical space $B(O,R(t))$ and $\nu_t$ measures the failure of volume, caused by the existence of the physical consistency of a domain $D$ included in $B(O,R(t))$ (i.e. the existence of the matter-energy distribution $E(t,X)$ in that domain as well as the influence of $E(t,X)$ throughout all of the physical universe on it), in order that the physical volume of $D$ be the same as the Euclidean volume of $D$ that is conventionally measured by the Lebesgue measure $dX$ and used in the case where this domain is absolutely empty of matter, under all its forms, and of its influences.

Just after the time that is labeled by $t=0$, the physical universe is formed by the propagation, through the theoretical space $\mathbb{R}^3$, of the electromagnetic waves which is governed by the wave equation
$$\frac{1}{v^2(t)}\frac{\partial^2 }{\partial t^2}E(t,X)-\Delta E(t,X)=0$$ 
where $v(t)$ is the propagation speed, which is extremely small at the beginning in view of the immensity of the central gravity caused by the gigantic density of matter-energy for $t$ near $0$. Actually, what we have considered as corresponding to the initial time $t=0$ extends to a time interval that could be very large and even infinite. This interval can be qualified as being ``beyond the Planck era". This latter corresponds to the situation where the length of $R(t)$ becomes significative but still extremely small. Along the above interval, we can consider that $R(t)\cong0$ in view of the extreme smallness of the speed $v(t)$. After that, the intensity of gravity continually decreases when $t>0$ increases and $v(t)$ increases progressively in order to become, later on, very close to $c=1$ and tends firmly to $1$ when $t$ would tend to $+\infty$. During this interval, which can also be qualified as the Big Bang era, the speed $v(t)$ and the radius $R(t)$ were extremely small, the wave frequency was extremely large, the wave length was extremely small and the universe temperature was extremely high. Nevertheless, when we consider that, during all of this time interval, $E(t,X)=\delta E_0$, we actually have assumed that $v(t)\equiv0$, $R(t)\equiv0$, $\lambda(t)\equiv0$, $\omega=+\infty$, $T=+\infty$ and  that the density of matter-energy and the curvature of the physical space (reduced to a single point) were $+\infty$. This can be justified by the fact that $R(t)$ was then infinitely small when compared to the present radius $R(t_0)$ where $t_0\simeq 2.365\times 10^{19}$s is the time calculated starting from the time when $v(t)$ becomes close to $1$. The above situation, where everything could have been reduced to $0$ or $\pm\infty$ (except for the energy $E_0$), could really exist well before the time interval labeled as $t=0$. This situation could exist at some instant $t_1<0$, which is impossible to determine, or during some finite interval $[t_1,0[$ or an infinite interval which could be denoted, when using our mathematical conceptions, by $]-\infty,0[$. We will qualify this point or this interval by ``beyond the Big Bang era". During the two intervals successively qualified by ``beyond the Planck era" (or `the Big Bang era") which corresponds to $t=0$ and by ``beyond the Big Bang" which would correspond to $\{t_1\}$, $[t_1,0[$ or $]-\infty,0[$, the physical universe (that is dynamic for the first interval and static for the other one) is supposed to reduce to one dimension that corresponds to that of the time; the three dimensions were inexistent for the negative interval and could be considered as being inexistent for the zero interval. This result could be equally deduced from the string theory.\\

 An other eventuality could be seriously considered in the case where $t_1$ is finite, which could then be taken as origin of time, $t_1=0$, and in the case where the total energy $E_0$ of the universe was concentrated only at the time $t_1=0$, at  a point O considered as being the origin of the virtual three dimensional space $\mathbb{R}^3$. We could then consider the evolution of the universe in the two directions of time: the positive and negative times ($t>0$ and $t<0$).\\
So, it results from the discussion of the section ``Universal time and proper time'' that, if $\tau$ is the expansion proper time, we could consider the ball $B(O,r_0)$ that represents the original quasi-balck hole. $B(O,r_0)$ intersects then all the directions of the space $\mathbb{R}^3$ (i.e. the axes through O) along intervals $]-r_0,r_0[$.\\
According to the characterization of the black holes metrics we have $\tau=|t|$ inside $B\setminus O$ because $\|X'(t)\|_{g_0}=0$ and for $t=0$ we have $\tau=t=0$ because $\tau=1-v_{g_0}^2=1-\|X'(0)\|_{g_0}=1-1=0$ considering that the total energy of the universe as the unit of energy.\\
Therefore, we obtain
$$\tau'=\left\{\begin{array}{ll}
\;\;1&\;\;\mbox{for}\;t>0,\\ 
{}\\ 
-1&\;\;\mbox{for}\;t<0,\\
\end{array}
\right.
$$  
and consequently
$$\tau'=H(t)-\check{H}(t)\qquad\mbox{where}\qquad \check{H}(t)=H(-t).$$
In view of the relation 
$<\tau'',\varphi>=-<\tau',\varphi'>=-\left(\int_0^\infty\varphi'(t)\,dt-\int_{-\infty}^0\check{\varphi}'(t)\,dt\right)=-\left(\int_0^\infty\varphi'(t)\,dt+\int_{+\infty}^0\varphi'(t)\,dt\right)=-(-\varphi'(0)+\varphi'(0))=0$, for every $\varphi\in C^\infty_c(\mathbb{R}^3)$ along all directions of $\mathbb{R}^3$, we obtain
$$\tau''=\delta_{\mathbb{R}^+}+\delta_{\mathbb{R}^-}=\Gamma^{+}+\Gamma^{-}=F^++F^-=0$$
and then $F^+=-F^-$ and $\tau'$ is the Euclidean speed $v_e$ meanwhile $\tau$ is the Euclidean distance inside the ball $B(O,r_0)$.\\
If we consider, starting from $t_1=0$ the negative time continuously decreasing to $-\infty$ and the positive time continuously increasing to $+\infty$, we obtain two opposit semi-cone; the first one corresponds to the positive times and the second one to the negative times.\\
The first, studied along this paper, corresponds to the matter, characterized by the positive matter an energy densities and by negative radiational density and pressure, that constitutes the physical and real universe. The other one would correspond to the antimatter characterized by negative antimatter and energy densities and by positive radiational density and pressure that constitutes the virtual universe about which we now practically nothing.\\
I we admit the relations $\displaystyle E_0=\lim_{t\rightarrow0}h(t)f(t)$ and $\displaystyle v_0=\lim_{t\rightarrow0}\lambda(t)f(t)$, it follows, from $\displaystyle \lim_{t\rightarrow0}h(t)=0$ and $E_0$ and $v_0$ are both finite, that $\displaystyle \lim_{t\rightarrow0}f(t)=+\infty$ and $\displaystyle \lim_{t\rightarrow0}\lambda(t)=0$ which conform with the results previously obtained.\\
In this precise eventuality, we have $B(O,t_0)=B(O,r_0)$ and the interval $]-t_0,t_0[$ is well the interval that represents the interval of time referred to as being $t=0$. The evolution of the real and the virtual universes starts from $t=t_0$ and is induced by the appearance of the material and antimaterial particles that have two opposite electrical charges creating two opposite force fields.\\
The space and time cone can be sketched, in $3$ dimensions, as being of the form of figure $16$. Finally, we think that it is possible to determine the volume of $B(O,r_0)$, being calculated $E_0$, by comparing it with ordinary black holes, and consequently to determine $t_0$. It is also possible to determine the law that governs, starting from $t_0$, the evolution of the speeds (Euclidean and with respect to $g_t$) of the electromagnetic waves (i.e. the expansion speeds) still by studying the ordinary black holes.\\
We denote by $\tau_t$ and $\tau_e$ the proper times of the expansion with respect to the two metrics $g_t$ and $g_e$ as well as $\gamma_t:=(1-v_t^2)^{-1/2}$ and $\gamma_e:=(1-v_e^2)^{-1/2}$ where $v_t:=\|X'(t)\|_{g_t}$ and $v_e:=\|X'(t)\|_{g_e}$. We have
$$\frac{\tau_t^2}{t^2}=\frac{1}{\gamma_t^2}=1-v_t^2\qquad\mbox{and}\qquad \frac{\tau_e^2}{t^2}=\frac{1}{\gamma_e^2}=1-v_e^2.$$
So, we obtain
$$\tau_t^2\gamma_t^2=t^2=\tau_e^2\gamma_e^2\qquad\mbox{and}\qquad \gamma_t^2(1-v_t^2)=1=\gamma_e^2(1-v_e^2).$$
Therefore, we have
$$\frac{\tau_t^2}{\tau_e^2}=\frac{\gamma_e^2}{\gamma_t^2}=\frac{1-v_t^2}{1-v_e^2}$$
and when $g_t$ decreases from $g_e$ to $g_0=0$, then $v_t$ decreases from $v_e$ to $0$ and consequently $\displaystyle \frac{\tau_t^2}{\tau_e^2}=\frac{\gamma_e^2}{\gamma_t^2}$ increases from $1$ to $\displaystyle \frac{1}{1-v_e^2}$.\\
Inside $B\setminus O$, we have $\tau_t=t$ as $v_t=0$, $\tau_e=0$ as $v_e=1$, $\gamma_t=1$ and $\gamma_e=+\infty$.\\ 

Thus, our model leads to the conclusion that the original real universe reduces to a unique gigantic black hole (as it could result from the ring theory) with the difference that the physical space did not exist originally whereas, for an ordinary black hole at a time $t>0$, we have $E(t,X)=\delta e$ on a ball $B(I,r)\subset B(O,R(t))$ which does exist physically and inside which the gravity is extremely intense and the distance between two points is practically null. As for the existence of the original energy $E_0$ (for $t=-\infty$ or $t=t_1$) that was concentrated at the point $O$ and of the original instant at which started the expansion process as well as the reason for which $v_{g_t}(t)$ became non null (if only it has been null at some finite instant), all of this is a matter for metaphysics or religion because it involves the null and the infinity notions which are beyond our conception of the physical reality that is at the origin of our consciousness. Nevertheless, all physical laws of the universe $U(t)$ from the time when it is modeled by $(B(O,R(t)),g_t)$ (i.e. for $t\geq t_0$) are perfectly reachable by our comprehension or technical means (which evolves continually) and for our intelligence. Actually, our model rigorously establishes some coherent mathematical and physical laws that govern the evolution of the universe and lead to the unification of all branches of Physics: General Relativity, Quantum Theory, Electromagnetism, Thermodynamics and Newton-Lagrange-Hamilton's Mechanics. In particular, our model confirms that the (slightly modified) general relativity instituted by Einstein thoroughly governs the evolution of our universe and the cosmology.\\

According to the preceding, we conclude that the discovery and the understanding of the Nature's laws and the subsequent evolution of the energy-matter as well as the universe evolution constitutes the Sciences' domain, the existence of the original energy and the original splitting and movement's reason constitutes the Metaphysics' domain, whereas the reflexion of the consequences of these laws and this evolution on the humanity constitutes the Philosophy and the human reason's domain.\\\\

\textbf{Bibliography}\\\\
This paper is written along the period: January 2007-March 2009.
Several parts of it are published at the Arxiv of Mathematics.\\\\
$[$1$]$ Nazih Moukaddem\\\\
\hskip 2cm $\bullet$ Mod\'{e}lisation physico-math\'{e}matique de
l'expansion de l'Univers, Arxiv, Mai 2007, corrig\'{e} et compl\'{e}t\'{e} \`{a} plusieurs reprises.\\\\\
\hskip 2cm $\bullet$ Universe expansion and Energy problems,
Arxiv, October
2007, continuously updated.\\\\
\hskip 2cm $\bullet$ L'Univers en expansion et probl\`{e}mes
d'\'{e}nergie, Arxiv, Mars 2008.\\\\
\hskip 2cm $\bullet$ This study is published in 2009 in a book that has the same title and the ISBN number: 978 - 2 - 7466 - 1247 - 1.\\\\
$[$2$]$ James William Rohlf, Modern Physics, John Willy and Sons,
Inc.\\\\
$[$3$]$ Thomas Taylor, Mechanics: Classical and Quantum, Pergamon press.\\\\
$[$4$]$ Robert Wald, General relativity, University of Chicago
press, 1984.\\\\
$[$5$]$ Michael Taylor, Partial differential equations, Applied
Mathematical Sciences, vol.117.\\\\
$[$6$]$ Paul Davies, La nouvelle Physique, Sciences,
Flammarion.\\\\
$[$7$]$ J.P P\'{e}rez, Relativit\'{e}, Fondements et applications, Dunod.\\\\
$[$8$]$ Gallot-Hulin-Lafontaine, Riemannian Geometry, Springer-Verlag.\\\\
\textbf{e.mail}:  hmoukadem@hotmail.com

\end{document}